\documentclass[useAMS,usenatbib]{mn2e}
\usepackage{times}
\usepackage{natbib}
\usepackage{graphics}
\usepackage{graphicx}
\usepackage{color}
\usepackage{verbatim}
\usepackage{amsmath}
\usepackage{mathtools}
\usepackage{subfigure}
\usepackage{float}
\usepackage{tabularx}
\usepackage{supertabular}
\usepackage{rotating}
\usepackage{longtable}
\usepackage{dpfloat, booktabs}
\usepackage{color}
\usepackage{multicol}
\usepackage{graphicx}
\usepackage{multirow}               
\usepackage{amsfonts}

\usepackage{pifont}

\usepackage{wrapfig}
\usepackage{subfig}   

\definecolor{Mygrey}{gray}{0.75}

\title[Molecular Line Diagnostics in NGC~4710 \& NGC~5866]
{Molecular Gas Kinematics and Line Diagnostics in Early-type
  Galaxies: NGC~4710 \& NGC~5866}
\author[S.\ Topal et al.]
{Sel\c{c}uk Topal,$^{1}$\thanks{E-mail: selcuk.topal@astro.ox.ac.uk.}
  Martin Bureau,$^{1}$ Timothy A.\ Davis,$^{2}$ Melanie
  Krips,$^{3}$ \newauthor Lisa M.\ Young,$^{4, 5}$ and Alison
  F. Crocker$^{6}$\\ 
  $^{1}$Sub-department of Astrophysics, University of Oxford, Denys
  Wilkinson Building, Keble Road, Oxford OX1~3RH, U.K.\\
  $^{2}$School of Physics \&\ Astronomy, Cardiff University, Queens
  Buildings, The Parade, Cardiff, CF24 3AA, UK,\\
  $^{3}$Institut de Radio Astronomie Millimetrique (IRAM), 300 rue de
  la Piscine, Domaine Universitaire, F-38406 Saint Martin d'H\'eres,
  France\\
  $^{4}$Physics Department, New Mexico Institute of Mining and
  Technology, Socorro, NM 87801, U.S.A.\\
  $^{5}$Academia Sinica Institute of Astronomy \& Astrophysics, PO Box
  23-141, Taipei 10617, Taiwan, R.O.C.\\
  $^{6}$Reed College, 3203 SE Woodstock Blvd., Portland, OR 97202, U.S.A.}
\begin{document}
\date{Accepted . Received ; in original form }
\pagerange{\pageref{firstpage}-\pageref{lastpage}} \pubyear{2016}
\maketitle
\label{firstpage}
%
%
\begin{abstract}
  We present interferometric observations of CO lines ($^{12}$CO(1-0,
  2-1) and $^{13}$CO(1-0, 2-1)) and dense gas tracers (HCN(1-0),
  HCO$^+$(1-0), HNC(1-0) and HNCO(4-3)) in two nearby edge-on barred
  lenticular galaxies, NGC~4710 and NGC~5866, with most of the gas
  concentrated in a nuclear disc and an inner ring in each galaxy. We
  probe the physical conditions of a two-component molecular
  interstellar medium in each galaxy and each kinematic component by
  using molecular line ratio diagnostics in three complementary
  ways. First, we measure the ratios of the position-velocity diagrams
  of different lines, second we measure the ratios of each kinematic
  component's integrated line intensities as a function of projected
  position, and third we model these line ratios using a non-local
  thermodynamic equilibrium radiative transfer code. Overall, the
  nuclear discs appear to have a tenuous molecular gas component that
  is hotter, optically thinner and with a larger dense gas fraction
  than that in the inner rings, suggesting more dense clumps immersed
  in a hotter more diffuse molecular medium. This is consistent with
  evidence that the physical conditions in the nuclear discs are
  similar to those in photo-dissociation regions. A similar picture
  emerges when comparing the observed molecular line ratios with those
  of other galaxy types. The physical conditions of the molecular gas
  in the nuclear discs of NGC~4710 and NGC~5866 thus appear
  intermediate between those of spiral galaxies and starbursts, while
  the star formation in their inner rings is even milder.
\end{abstract}
\begin{keywords}
  galaxies: lenticulars ~- galaxies: individual: NGC~4710 and NGC~5866
  ~- galaxies: ISM ~- ISM: molecules
\end{keywords}
%
%
\section{Introduction}
\label{sec:intro}
Molecular clouds are the stellar nurseries of galaxies, and probing
their physical properties in different galaxy types has the potential
to answer many important open questions regarding star formation
processes and galaxy evolution. Actively star-forming spiral galaxies,
including our own Milky Way, are rich in cold gas and their molecular
gas reservoirs have been studied for many years. However, since
so-called 'red and dead' early-type galaxies (ETGs; lenticulars and
ellipticals) are generally thought to be very poor in molecular gas,
star formation within them is thought to have largely
stopped. Nevertheless, roughly $10$ years after the first detection of
molecular gas in external spiral galaxies \citep{r75,s75}, different
phases of the interstellar medium (ISM) of ETGs were also studied,
through observations of X-rays \citep[e.g.][]{fct85}, optical emission
lines \citep[e.g.][]{col84}, H\,{\small I} \citep[e.g.][]{ktc85} and
CO \citep[e.g.][]{w86,ws03,sa07}.

\cite{y11} carried out the most extensive survey of molecular gas
($^{12}$CO(1-0)) in ETGs so far, in the $260$ galaxies of the
volume-limited ATLAS$^{\rm 3D}$
sample\footnote{http://www-astro.physics.ox.ac.uk/atlas3d/}
\citep{cap11}. They obtained a $22\%$ detection rate, with H$_2$
masses ranging from $10^7$ to $10^9$~$M_\odot$. The CO-rich ETGs in
the ATLAS$^{\rm 3D}$ sample were further studied to probe the
molecular gas properties in more details. Interferometric observations
of $^{12}$CO(1-0) in $40$ objects were presented in \citet{al13},
revealing a variety of CO morphologies (discs, rings, bars, and spiral
arms), with sizes smaller than in spirals in absolute terms but
similar when compared to their optical extent \citep{d13}. The
molecular gas kinematics is generally regular \citep{d13}, allowing
one to easily probe the Tully-Fisher (luminosity-rotational velocity;
\citealt{t77}) relation of ETGs \citep{d11}. However, gas-star
kinematic misalignments indicate that the molecular gas has an
external origin in at least one third of the systems, with significant
field--cluster environmental differences \citep{dav11}.

The molecular gas can also be used to study the physical conditions
(temperature, density, column density, opacity, excitation mechanism,
etc) within the dense cold gas of ETGs, where star formation takes
place. For example, different transitions of a given molecule are good
proxies for the gas temperature (e.g.\
$^{12}$CO(2-1)/$\,^{12}$CO(1-0)), isotopologues probe the gas optical
depth and column density (e.g.\ $^{13}$CO/$\,^{12}$CO), and complex
molecules (e.g.\ HCN, HCO$^+$, HNC, and HNCO) require much larger
critical densities (up to $n_{\rm crit}\approx10^{6}$~cm$^{-3}$) to be
excited compared to simpler ones (typically
$n_{\rm crit}\approx10^{3}$~cm$^{-3}$). More subtle effects also
exist. HCN and its isotopomer HNC trace respectively the warm-dense
and cool-slightly less dense parts of a cloud, while HCO$^+$ traces
even more tenuous regions \citep{hu95}. HCO$^+$ can also be enhanced
in shocks associated with young supernova remnants (SNRs), due to
cosmic rays (CRs) in the shocked material \citep{dic80,wo81,el83}, and
is therefore also an important tracer of CR-dominated regions of the
ISM.  HNCO, on the other hand, is a good tracer of shocked gas
\citep{mei05, mei12, rod10, ot14}, and it correlates well with SiO, a
well-known shock tracer \citep{zin00}.

Focusing on the physical conditions of the ISM through single-dish
observations of several $^{12}$CO transitions, the $^{13}$CO
isotopologue, and other molecules, \citet{ka10} and later \citet{c12}
(see also \citealt{tim13}) found that the molecular line ratios of
ETGs are generally similar to those of spirals and Seyferts, but
different from those of starbursts and (ultra-) luminous infrared
galaxies ((U)LIRGs). Interestingly, the line ratios are statistically
correlated with several other ISM and stellar properties, e.g.\ the
molecular-to-atomic gas ratio, dust temperature, dust morphology,
K-band absolute magnitudes and stellar population age \citep{c12}. The
$^{13}$CO(1-0)/$\,^{12}$CO(1-0) ratio of ETGs also seems to depend on
environment \citep{ala15}.

Here, we conduct interferometric observations of two typical edge-on
lenticular galaxies (with large bulges and central dust lanes),
NGC~4710 and NGC~5866 (morphological type $-0.9$ and $-1.2$,
respectively; HyperLEDA\footnote{http://leda.univ-lyon1.fr/}). Both
galaxies are fast rotating and their ionised gas is
kinematically-aligned with the stellar kinematics, indicating that the
gas is likely supplied by internal processes (e.g.\ stellar mass loss;
\citealt{dav11}) or is left over from the galaxy formation event
itself. NGC~4710 is a member the Virgo Cluster \citep{kra82} at a
distance of $16.8$~Mpc \citep{t98}, while NGC~5866 is in a small group
including two spirals at a distance of $15.3$~Mpc
\citep{t98}. However, since NGC~4710 is located in the outskirts of
the cluster and the distance between NGC~5866 and its nearest
companion is rather large, both galaxies are unlikely to have had
recent significant interactions with other galaxies or to have been
affected by environmental effects. The general properties of the
galaxies are listed in Table~\ref{tab:gprop}.

NGC~4710 and NGC~5866 also happen to be CO-bright, unusual for
supposedly 'red and dead' systems, allowing the first
spatially-resolved study of multiple molecular tracers in an
early-type galaxy. We map the entire discs of NGC~4710 and NGC~5866 in
common low-$J$ CO lines such as $^{12}$CO(1-0), $^{12}$CO(2-1),
$^{13}$CO(1-0) and $^{13}$CO(2-1), thus probing relatively tenuous
molecular gas, as well lines of more complex molecules such as
HCN(1-0), HCO$^+$(1-0), HNC(1-0) and HNCO(4-3) (the latter two lines
being detected for the first time in those galaxies), thus probing
denser gas.

%
%
\begin{table}
  \begin{center}
    \caption{General properties of NGC~4710 and NGC~5866.}	
    \begin{tabular}{@{}l l r r r r r@{}} \hline
      Galaxy & Property & Value & Reference \\ \hline
      NGC~4710 & Type & S0$_3$(9) & a\\
      & RA~(J2000) & $12^{\rm h}49^{\rm m}38.8^{\rm s}$ & b\\
      & Dec~(J2000) & $15^{\rm d}09^{\rm m}56^{\rm s}$ & b\\
      & Distance (Mpc) & $16.8$ & c\\
      & $\log(M_{{\rm H}_2}/M_\odot)$ & $8.72\,\pm\,0.01$ & d\\
      & SFR$_{\rm 22\micron}$~($M_\odot$~yr$^{-1}$) & $0.11\,\pm\,0.02$ & e\\
      & V$_{\rm sys}$~(km~s$^{-1}$) & $1102$ & f\\
      & Major diameter & $4\farcm9$ & c\\
      & Minor diameter & $1\farcm2$ & c\\
      & Position angle & $207\degr$ & g\\
      & Inclination & $86\degr$ & h\\		
      \hline
      NGC~5866 & Type & S0$_3$(8) & a\\
      & RA~(J2000) & $15^{\rm h}06^{\rm m}29.5^{\rm s}$ & b\\
      & Dec~(J2000) & $55^{\rm d}45^{\rm m}48^{\rm s}$ & b\\
      & Distance (Mpc)& $15.3$ & c\\
      & $\log(M_{{\rm H}_2}/M_\odot)$ & $8.47\,\pm\,0.01$ & f\\
      & SFR$_{\rm 22\micron}$~($M_\odot$~yr$^{-1}$) & $0.21\,\pm\,0.04$ & e\\
      & V$_{\rm sys}$~(km~s$^{-1}$) & $755$ & f\\
      & Major diameter & $4\farcm7$ & c\\
      & Minor diameter & $1\farcm9$ & c\\
      & Position angle & $127\degr$ & g\\
      & Inclination & $89\degr$ & h\\
      \hline
    \end{tabular}
    \label{tab:gprop}
  \end{center}
  References: 
  $^{\rm a}$ \citet{san94};
  $^{\rm b}$ Nasa/Ipac Extragalactic Database (NED);
  $^{\rm c}$ \citet{t98};
  $^{\rm d}$ \citet{y11};
  $^{\rm e}$ \citet{tim14};
  $^{\rm f}$ \citet{cap11};
  $^{\rm g}$ \citet{dav11};
  $^{\rm h}$ \citet{d11}.
\end{table}

The most abundant tracers reveal an X-shaped position-velocity diagram
(PVD) in both galaxies, indicating the presence of an edge-on barred
disc \citep[see][]{ba99,ab99,ma99}, with gas concentrated in a nuclear
disc within the inner Lindblad resonance and in an inner ring around
the end of the bar (corotation) and possibly farther out. Infrared
observations also support a nuclear disc surrounded by a ring-like
structure in the disc of NGC~5866. Indeed, \citet{xi04} found that the
$6.75$ and $15$~$\micron$ emission in NGC~5866 peaks in the centre and
at $\approx4$~kpc on either side of it (see their Fig.~10). The barred
nature of NGC~4710 is consistent with its box/peanut-shaped bulge, but
that of NGC~5866 is more surprising given its classical bulge. Should
the ring-like structure in NGC~5866 have a different origin, however,
the barred nature of NGC~5866 would have to be
revisited. Nevertheless, we discuss our empirical and model results in
light of these facts throughout.

The first goal of our study is thus to exploit the variations of the
molecular line ratios along the galaxy discs (as a function of
projected radius and velocity), to study the physical properties of
the molecular gas in each of those two dynamical components
independently (i.e.\ nuclear disc and inner ring).
We also do this quantitatively by modeling the molecular line ratios
with a non-local thermodynamic equilibrium (non-LTE) code (RADEX;
\citealt{van07}). A secondary goal is to compare the star formation
activity in ETGs to that in other types of galaxies. For this, the
observed line ratios are compared with those at the centre of spirals,
starbursts, Seyferts and other lenticulars, as well as with those of
some giant molecular clouds (GMCs) in the spiral arms and inter-arm
regions of nearby galaxies \citep{so02,br05,baa08,ka10,c12}.

The paper is divided as follows. Section~\ref{sec:obsredu} describes
the observations and data reduction, while \S~\ref{sec:iman} presents
the results and a basic analysis of the data. We discuss the line
ratio diagnostics in \S~\ref{sec:rat}, both empirically and through
modelling, the latter detailed in Appendix~\ref{sec:LVG}. A detailed
discussion is presented in \S~\ref{sec:result} and we conclude briefly
in \S~\ref{sec:conc}.
%
%
\section{Observations \& Data Reduction}
\label{sec:obsredu}
%
%
\subsection{Observations}
\label{sec:obs}
\subsubsection{NGC~4710}
NGC~4710 was primarily observed using the Combined Array for Research
in Milimeter-wave Astronomy (CARMA), which includes $15$ antennae
($6\times10.4$~m, $9\times6.1$~m) and thus $105$ baselines. The
$^{12}$CO(1-0) and $^{13}$CO(1-0) observations were carried out in
April 2009, with $3$ spectral windows per line ($186$~MHz or
$\approx450$~km~s$^{-1}$ bandwidth, with $1$~MHz or
$\approx2.5$~km~s$^{-1}$ channels), in the D configuration (maximum
baseline of $150$~m, yielding a synthesized beam and thus angular
resolution of $\approx3\farcs8$ at $115$~GHz). The CO(2-1)
observations were carried out in January 2011, with $3$ spectral
windows per line ($1500$~MHz or $\approx2000$~km~s$^{-1}$ bandwidth,
with $5$~MHz or $\approx7$~km~s$^{-1}$ channels), using the E
configuration (maximum baseline of $66$~m, yielding a resolution of
$\approx4\farcs2$ at $230$~GHz). The $^{12}$CO(2-1) and $^{13}$CO(2-1)
observations were obtained with a mosaic of $7$ pointings, a central
pointing surrounded by $6$ pointings in an hexagonal pattern (see
Fig.~\ref{fig:n4710mom}). Simultaneous observations of HCN(1-0),
HCO$^+$(1-0), HNC(1-0) and HNCO(4-3) were obtained in October
2011. For each of HCN(1-0) and HCO$^+$(1-0), $3$ spectral windows per
line were used ($375$~MHz or $\approx1200$~km~s$^{-1}$ bandwidth, with
$0.4$~MHz or $\approx1.3$~km~s$^{-1}$ channels), using the D
configuration (yielding an average resolution of $\approx5\farcs5$ for
these lines). Observations of HNC(1-0) and HNCO(4-3) used a single
spectral window for each line ($500$~MHz or $\approx1600$~km~s$^{-1}$
bandwidth, with $5$~MHz or $\approx16$~km~s$^{-1}$ channels). All the
lines except HNCO(4-3) were detected with a signal-to-noise ratio
$S/N\ge3$.

All four dense gas tracers, namely HCN(1-0), HCO$^+$(1-0), HNC(1-0)
and HNCO(4-3), were also observed using the Institut de Radio
Astronomie Millimetrique (IRAM) Plateau de Bure Interferometer (PdBI),
with six $15$~m antennae and thus $15$ baselines in the 6ant-Special
configuration (yielding an average resolution of $\approx6\farcs0$ for
these lines). A $3.6$~GHz or $\approx12\,\,500$~km~s$^{-1}$ bandwidth
with a spectral resolution of $2$~MHz or $\approx7$~km~s$^{-1}$ was
used.  The observations of the dense gas tracers were obtained with a
mosaic of $2$ pointings, centred at offsets ($+5\farcs3$,
$+11\farcs3$) and ($-5\farcs3$, $-11\farcs3$) with respect to the
galaxy centre (and thus along the galaxy major axis; see
Fig.~\ref{fig:n4710mom}).

At the distance of NGC~4710, $1\arcsec$ corresponds to
$\approx81$~pc. The resolution of the CARMA observations thus
corresponds to a linear scale of $\approx300$ and $\approx450$~pc for
the tenuous (CO) and dense gas tracers, respectively, while that of
the PdBI observations corresponds to $\approx480$~pc. The main
observational parameters for NGC~4710 are listed in
Table~\ref{tab:obs}.
\begin{table*}
  \begin{center}
    \caption{Main observational parameters for NGC~4710 and NGC~5866.}
    \begin{tabular}{llrlrcccc} \hline
      Galaxy & Transition & Rest Freq. & Obs.\ Date & Total Obs.\ Time
      & Interferometer & Beam & Conversion factor & Noise\\ 
      & & (GHz) & & (hours) & & (arcsec) & (K~Jy$^{-1}$~beam) & (mJy~beam$^{-1}$)\\
      \hline
      NGC4710&$^{12}$CO(1-0)& $115.271$ & \multirow{2}{*}{$02$~APR~$2009$} & \multirow{2}{*}{6.42} &\multirow{2}{*}{CARMA}&$3.9\times3.2$&$7.3$ & $9$\\
      &$^{13}$CO(1-0)& $110.201$ &&&&$4.4\times3.9$&$5.8$ & $5$\\ \\
      &$^{12}$CO(2-1) & $230.538$ &\multirow{2}{*}{$30/31$~JAN~$2011$} & \multirow{2}{*}{$5.03$} & \multirow{2}{*}{CARMA} &$4.1\times3.5$&$1.6$ & $28$\\
      &$^{13}$CO(2-1) & $220.398$ &&&&$4.6\times3.9$&$1.4$ & $23$\\ \\
      & HCN(1-0) & $88.633$ & \multirow{3}{*}{$15/18/23/24/25$~OCT~$2011$} & \multirow{3}{*}{$16.62$} & \multirow{3}{*}{CARMA} &$5.1\times4.6$&$6.7$ &\\
      & HCO$^+$(1-0) & $89.188$ &&&&$5.5\times4.8$&$5.8$ & $2$\\
      & HNC(1-0) & $90.663$ &&&&$5.1\times4.6$&$6.3$ &\\ \\
      & HCN(1-0) & $88.633$ & \multirow{4}{*}{$30$~NOV~2011, $19$~JAN~$2012$} & \multirow{4}{*}{$7.90$} & \multirow{4}{*}{PdBI} &$6.1\times5.3$&$4.9$ &\\
      & HCO$^+$(1-0) & $89.188$ &&&&$6.1\times5.2$&$4.9$ & $1$\\
      & HNC(1-0) & $90.663$ &&&&$5.7\times5.0$&$5.2$ &\\
      & HNCO(4-3) & $87.925$ &&&&$6.3\times5.4$&$4.7$ &\\
      \hline	
      NGC~5866 &$^{12}$CO(1-0) & $115.271$ & $12/13/16$~AUG~$2010$ & $6.67$ & CARMA & $3.6\times2.9$&$8.7$ & $11$\\\\
      &$^{13}$CO(1-0) &$110.201$ & $28/30$~APR \& $03$~MAY~$2011$ & $11.55$ & PdBI &$4.9\times3.6$&$5.8$ & $1$\\ \\
      &HCN(1-0) & $88.633$ & \multirow{4}{*}{$03/04$~MAY~$2011$} & \multirow{4}{*}{$11.33$} & \multirow{4}{*}{PdBI} &$6.3\times5.2$&$4.8$ &\\
      &HCO$^+$(1-0) & $89.188$ &&&&$6.3\times5.2$&$4.7$ & $0.5$\\
      &HNC(1-0)& $ 90.663$ &&&&$6.2\times5.1$&$4.8$ &\\
      &HNCO(4-3)& $ 87.925$ &&&&$6.3\times5.2$&$4.8$ &\\
      \hline
    \end{tabular}
    \label{tab:obs}
  \end{center}
\end{table*}
\subsubsection{NGC~5866}
NGC~5866 was primarily observed with PdBI. Observations of
$^{13}$CO(1-0) and the dense gas tracers HCN(1-0), HCO$^+$(1-0),
HNC(1-0) and HNCO(4-3), were carried out using the 6Dq configuration
during April--May 2011, yielding a resolution of $\approx4\farcs9$ at
$110$~GHz and $\approx6\farcs5$ for the dense gas tracers. The total
bandwidth was $3.6$~GHz or $\approx12\,\,500$~km~s$^{-1}$, with a spectral
resolution of $2$~MHz or $\approx7$~km~s$^{-1}$.  $^{12}$CO(1-0)
observations were obtained at CARMA in the D configuration, yielding a
resolution of $\approx3\farcs8$ at $115$~GHz. Three spectral windows
were used ($375$~MHz or $\approx1000$~km~s$^{-1}$ bandwidth, with a
spectral resolution of $0.4$~MHz or $\approx1$~km~s$^{-1}$)

At the distance of NGC~5866, $1\arcsec$ corresponds to
$\approx74$~pc. The resolution of the PdBI observations thus
corresponds to a linear scale of $\approx360$ and $\approx480$~pc for
$^{13}$CO(1-0) and the dense gas tracers, respectively, while that of
the CARMA observations corresponds to $\approx280$~pc. The main
observational parameters for NGC~5866 are listed in
Table~\ref{tab:obs}.
\subsection{Data reduction}
\label{sec:redu}
%
%
\subsubsection{CARMA data reduction}
\label{sec:carmadata}
The CARMA data were reduced using the Multichannel Image
Reconstruction, Image Analysis and Display (MIRIAD) package
\citep{st95}. First, for each track, initial data corrections were
applied (i.e.\ line-length calibration, baseline and rest frequency
corrections). The temporal behaviour of the phase calibrator was then
checked to flag de-correlations when necessary. Second, the bandpass
and phase calibrations were performed using a bright calibrator,
usually a quasar within $20^\circ$ of the source (3C273, 1224+213 or
1419+513). The gain solutions were then derived and applied to the
source. Flux calibration was carried out using the latest calibrator
flux catalog maintained at CARMA (typically using a planet). After
successfully calibrating the source data for each track and each
observed line, all data for a given line were combined into one
visibility file and imaged using the MIRIAD task \emph{invert}. As
CARMA is a heterogeneous array, \emph{invert} was run with the
mosaicking option, to take into account the different primary
beams. All data cubes were created with a pixel size of
$1\arcsec\times1\arcsec$ and $1\farcs5\times1\farcs5$ for the CO lines
and dense gas tracers, respectively, typically yielding $\approx4$
pixels across the synthesized beam major axis. The dirty cubes were
cleaned to a threshold equal to the rms noise of the dirty channels in
regions devoid of emission. The cleaned components were then added
back and re-convolved using a Gaussian beam of full-width at
half-maximum (FWHM) equal to that of the dirty beam. A fully
calibrated and reduced data cube was thus obtained for each molecular
line.

We note that to simplify the discussion, the fully-calibrated and
cleaned data cubes were rotated using the MIRIAD task $\emph{regrid}$
with the keyword $\emph{rotate}$ and the molecular gas position angle
(see Table~\ref{tab:gprop}), so that the emission (galaxy major-axis)
is horizontal. The data presented in
  Figures~\ref{fig:n4710mom} and \ref{fig:n5866mom} therefore do not
represent the true orientation of the galaxies on the sky.
\subsubsection{PdBI data reduction}
\label{sec:pdbidata}
The PdBI data were reduced using the Grenoble Image and Line Analysis
System (GILDAS) packages Continuum and Line Interferometer Calibration
(CLIC) and MAPPING\footnote{http://www.iram.fr/IRAMFR/GILDAS}.  CLIC
was used for the initial data reduction, including the bandpass, phase
and flux calibrations, and the creation of the $uv$ tables. MAPPING
was then used to create fully calibrated and cleaned cubes, with the
same procedure and pixel sizes as above. Those cubes were then
converted into Flexible Image Transport System (FITS) files for
further analysis in MIRIAD and Interactive Data Language (IDL)
environments.
%
%
\section{Imaging \& Analysis}
\label{sec:iman}
%
%
\subsection{Emission regions and moment maps}
\label{sec:cont}
\begin{figure*}
\centering
  \includegraphics[width=5.8cm,clip=]{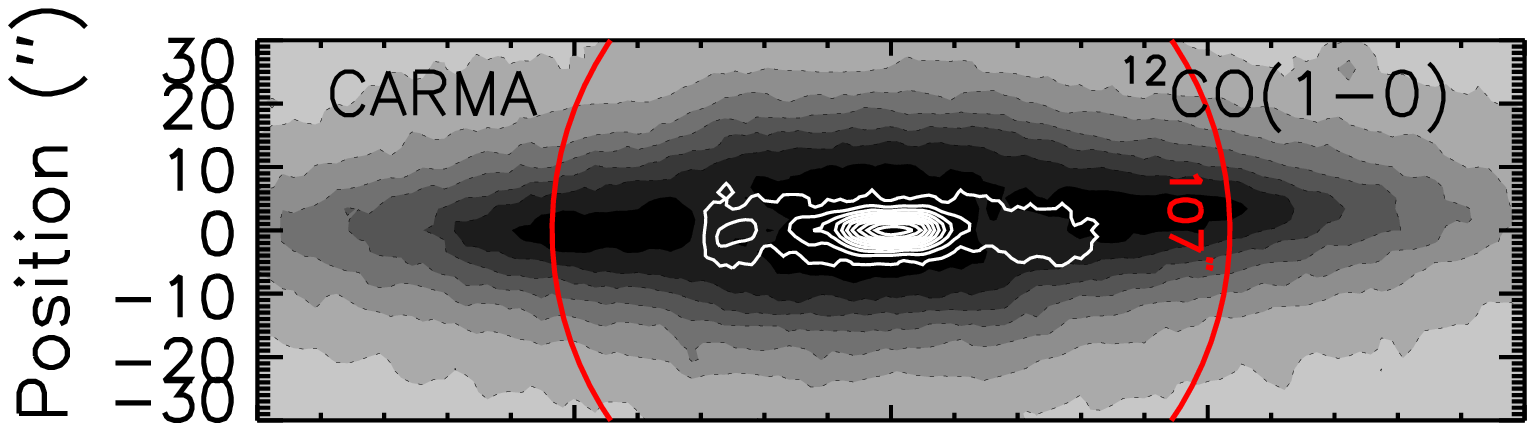}
  \includegraphics[width=5.8cm,clip=]{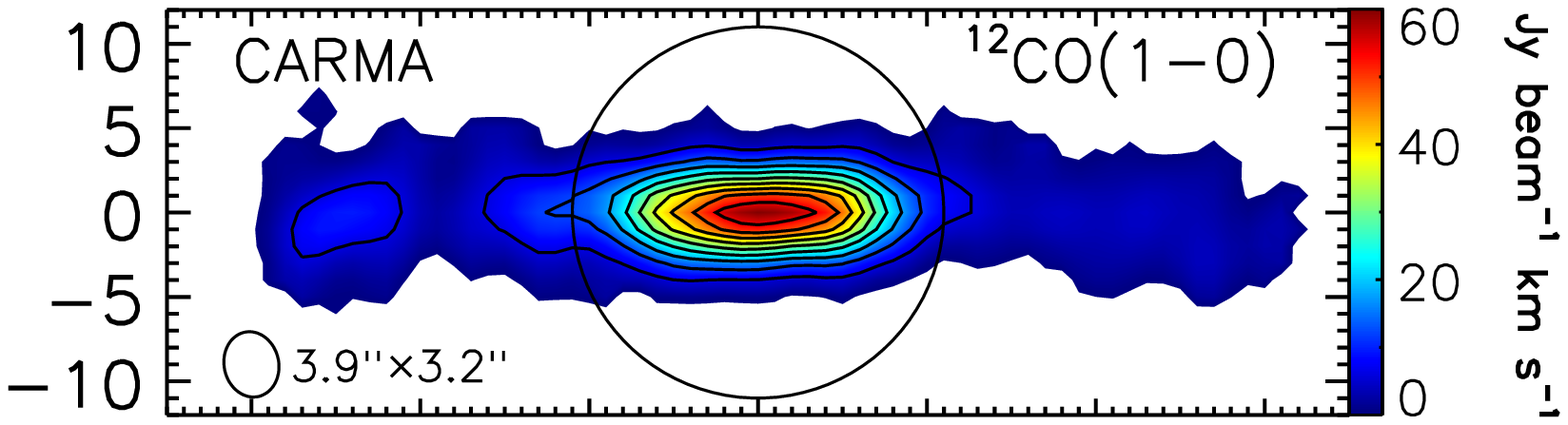}
  \includegraphics[width=5.8cm,clip=]{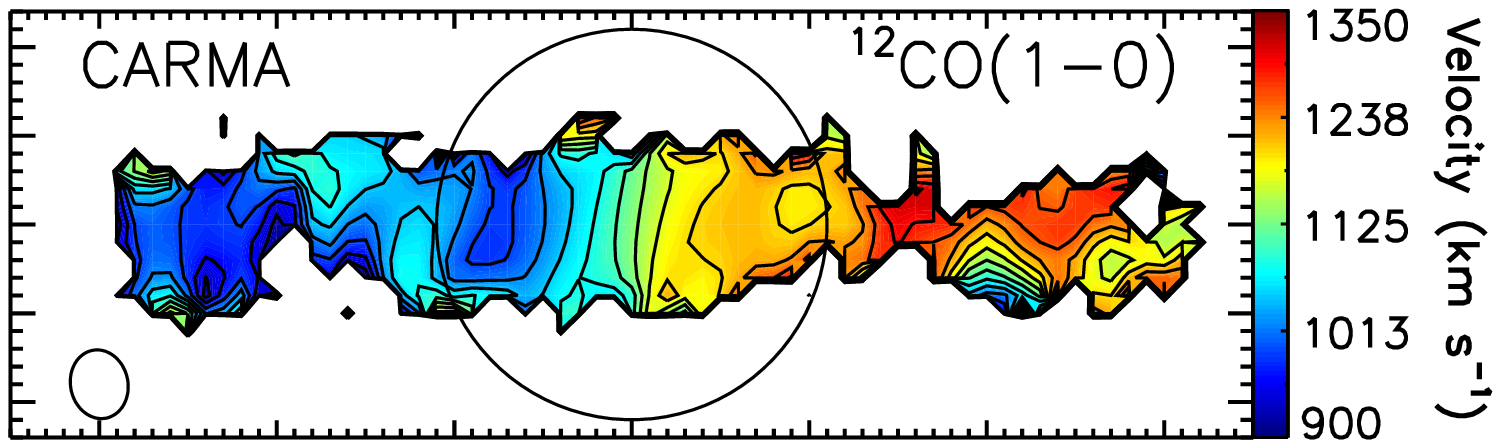}\\
  \vspace{-15pt}
  \includegraphics[width=5.8cm,clip=]{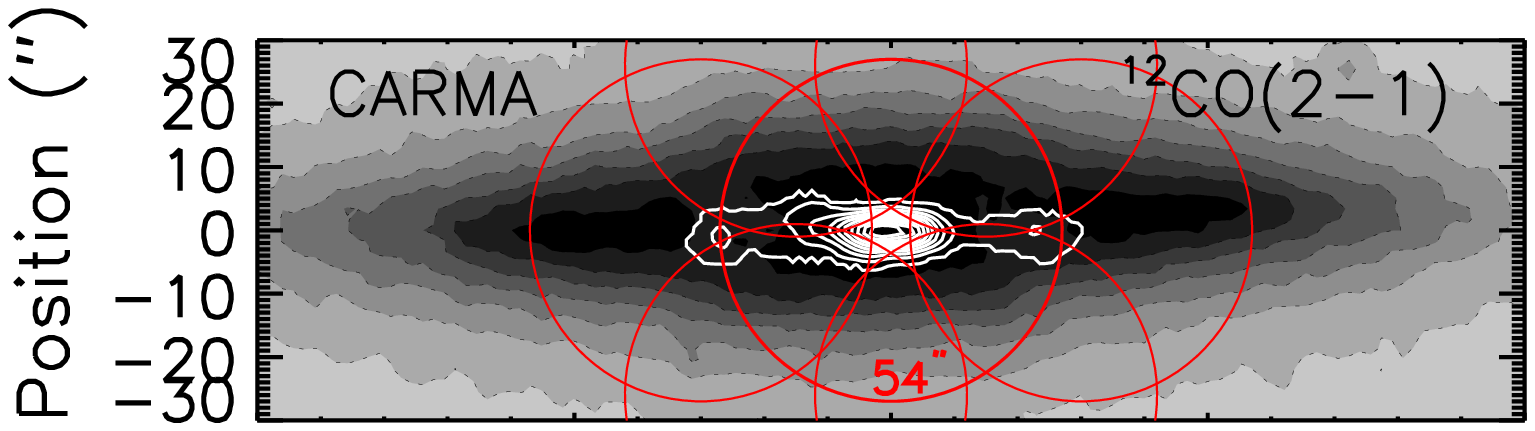}
  \includegraphics[width=5.8cm,clip=]{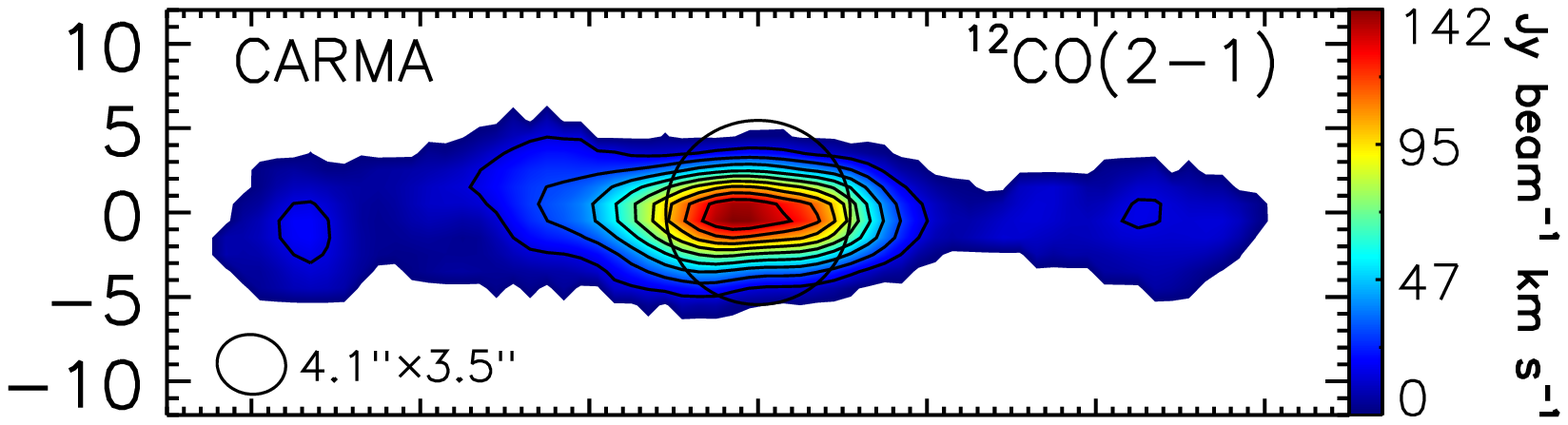}
  \includegraphics[width=5.8cm,clip=]{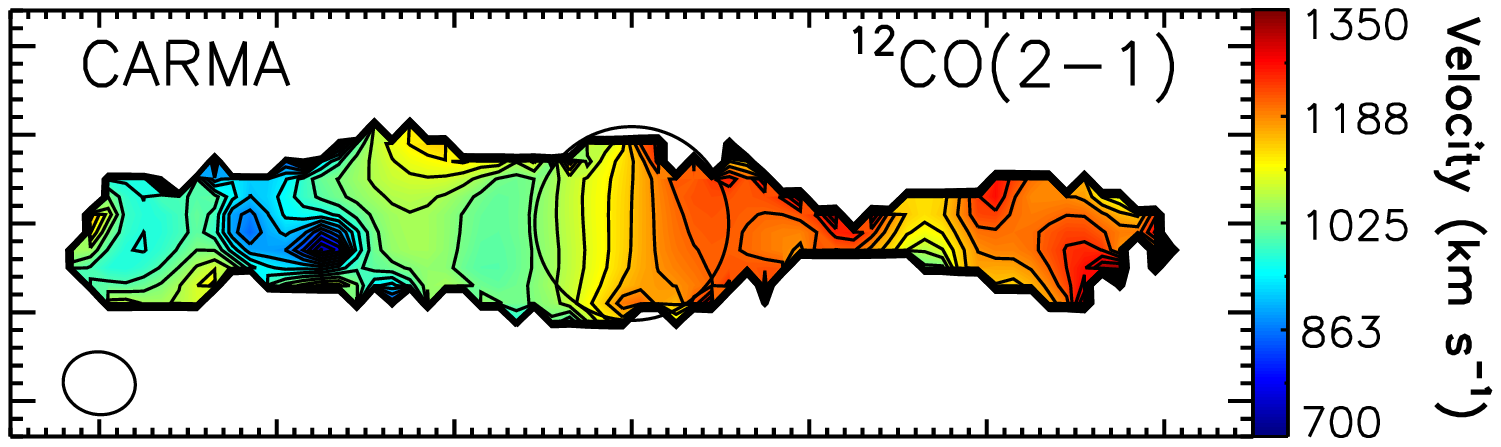}\\
  \vspace{-15pt}
  \includegraphics[width=5.8cm,clip=]{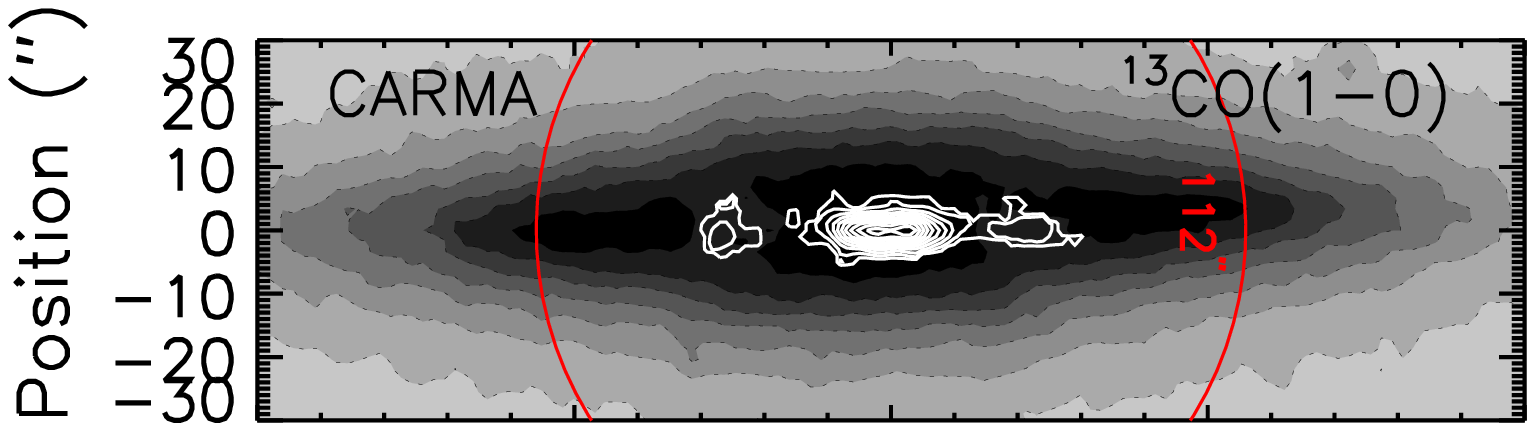}
  \includegraphics[width=5.8cm,clip=]{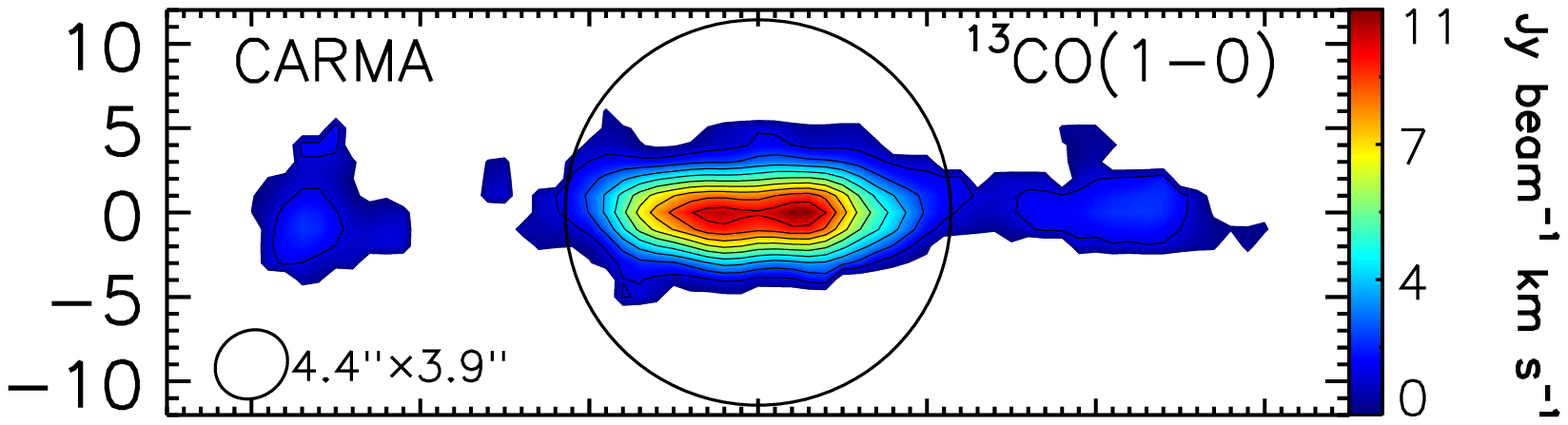}
  \includegraphics[width=5.8cm,clip=]{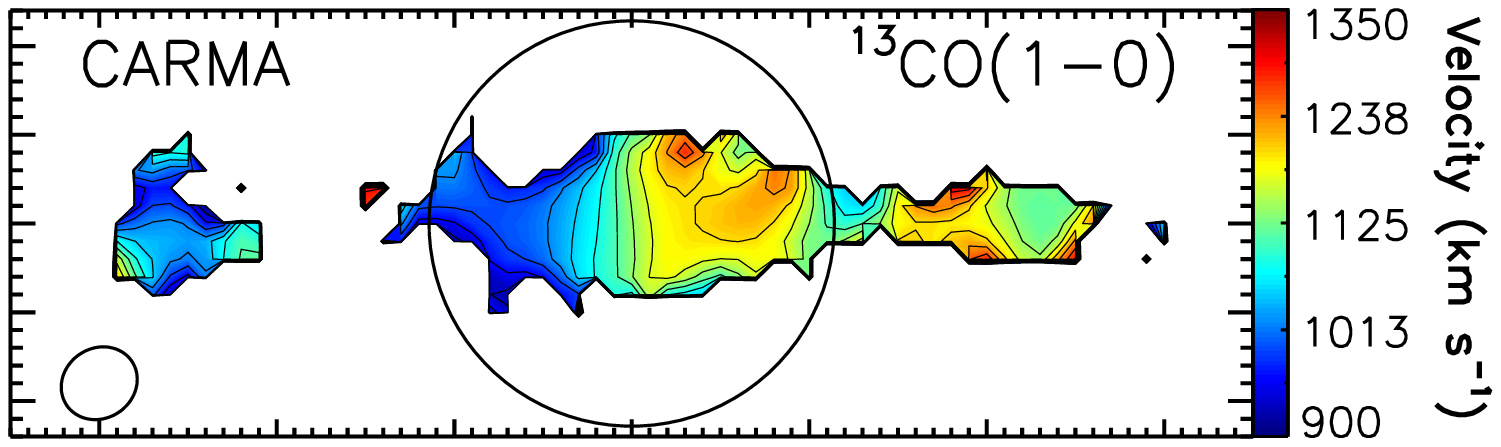}\\
  \vspace{-15pt}
  \includegraphics[width=5.8cm,clip=]{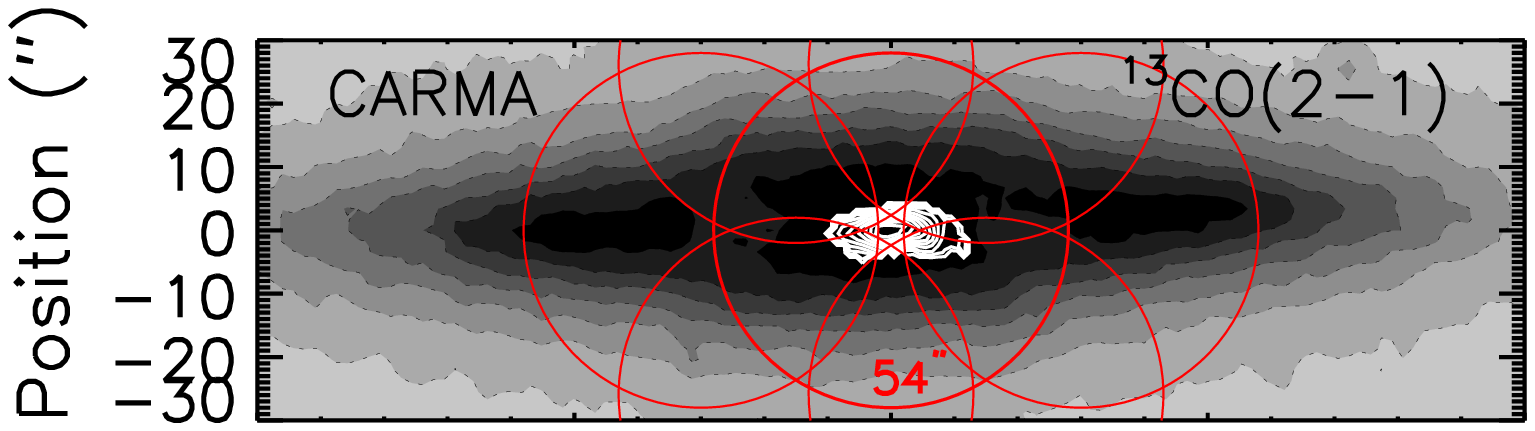}
  \includegraphics[width=5.8cm,clip=]{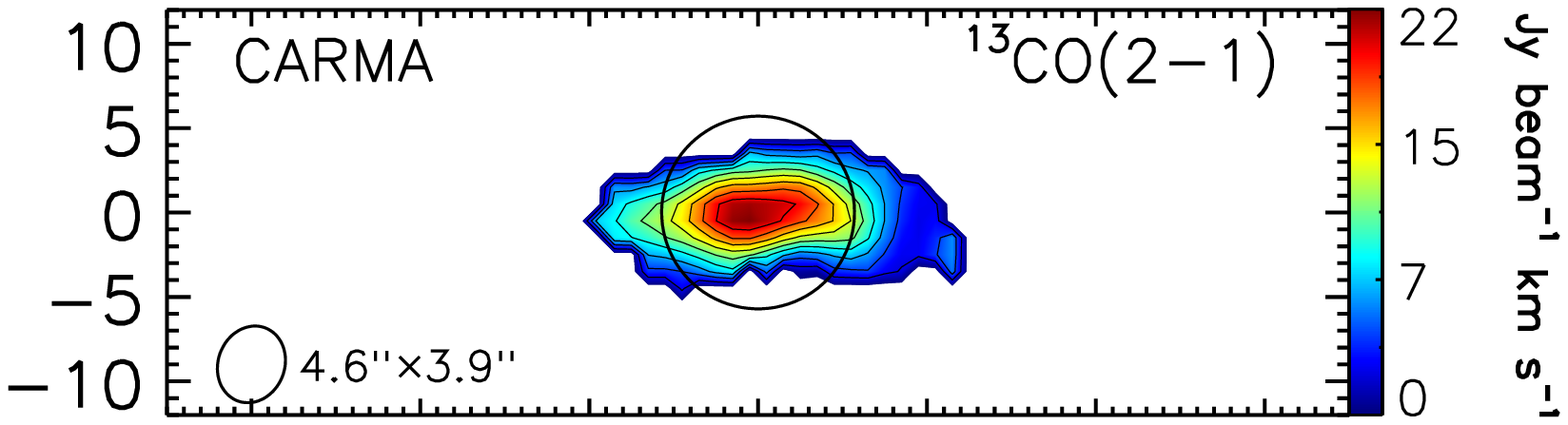}
  \includegraphics[width=5.8cm,clip=]{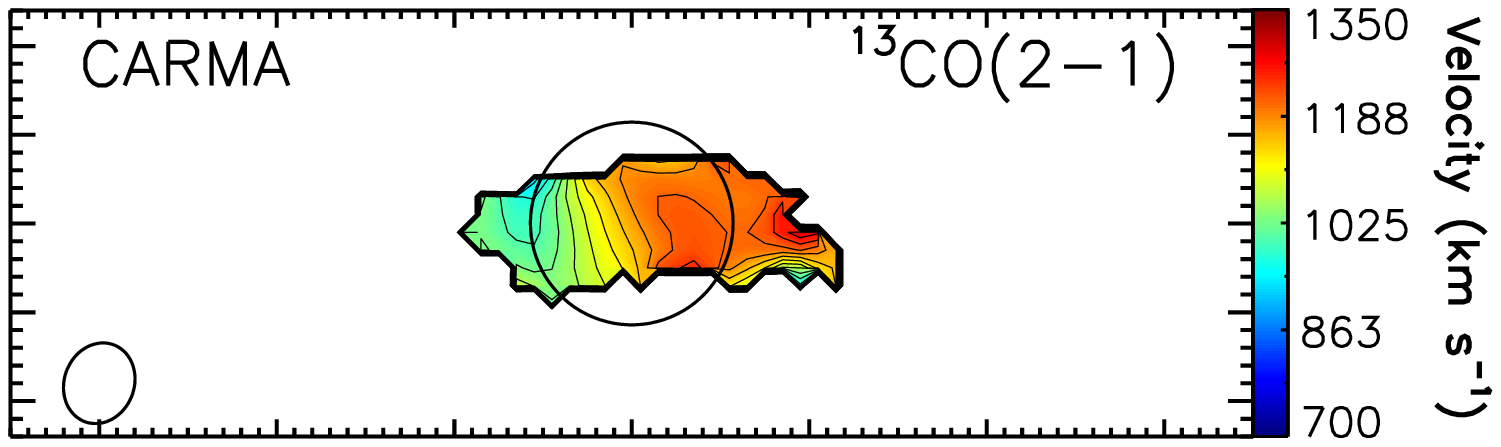}\\
  \vspace{-15pt}
  \includegraphics[width=5.8cm,clip=]{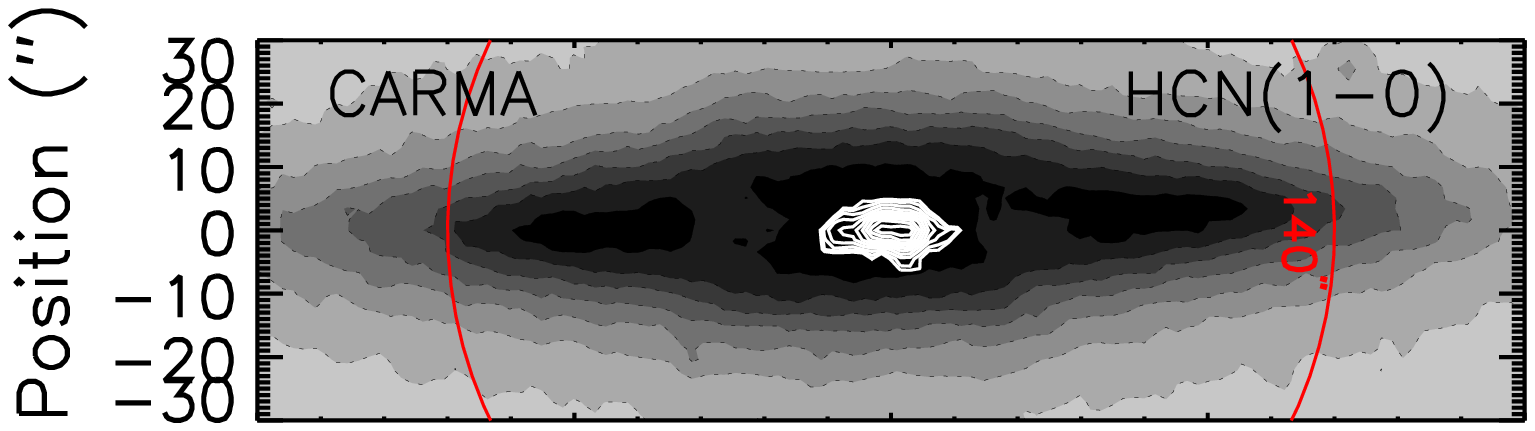}
  \includegraphics[width=5.8cm,clip=]{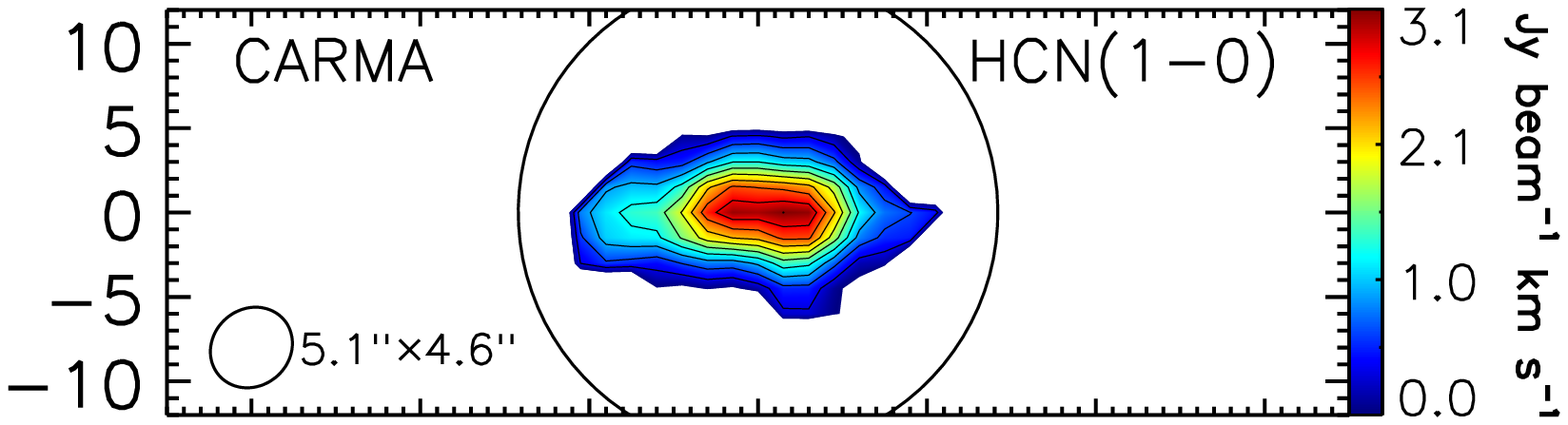}
  \includegraphics[width=5.8cm,clip=]{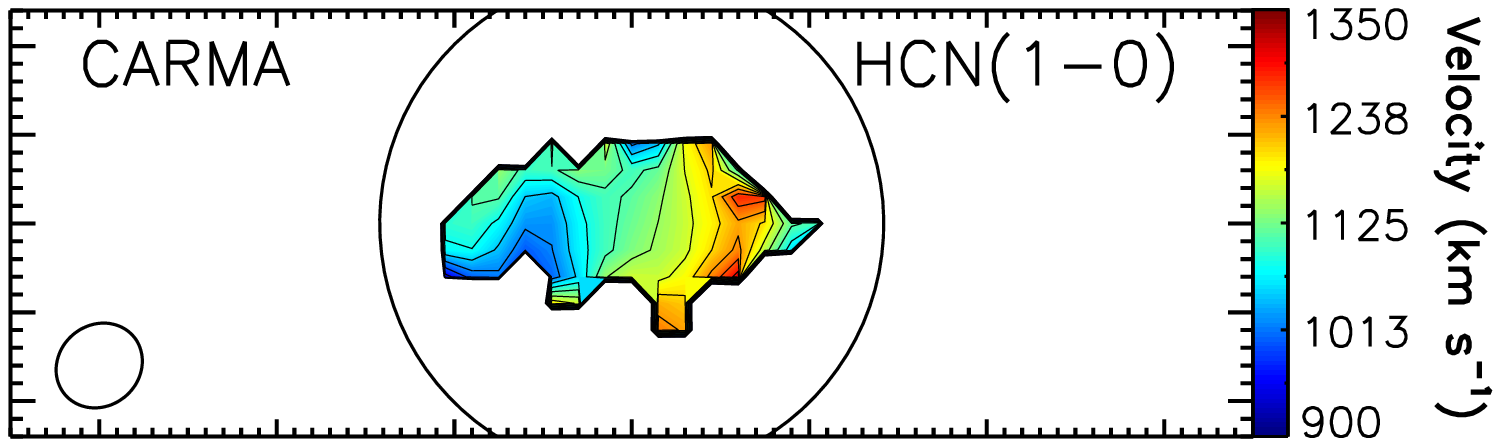}\\
  \vspace{-15pt}
  \includegraphics[width=5.8cm,clip=]{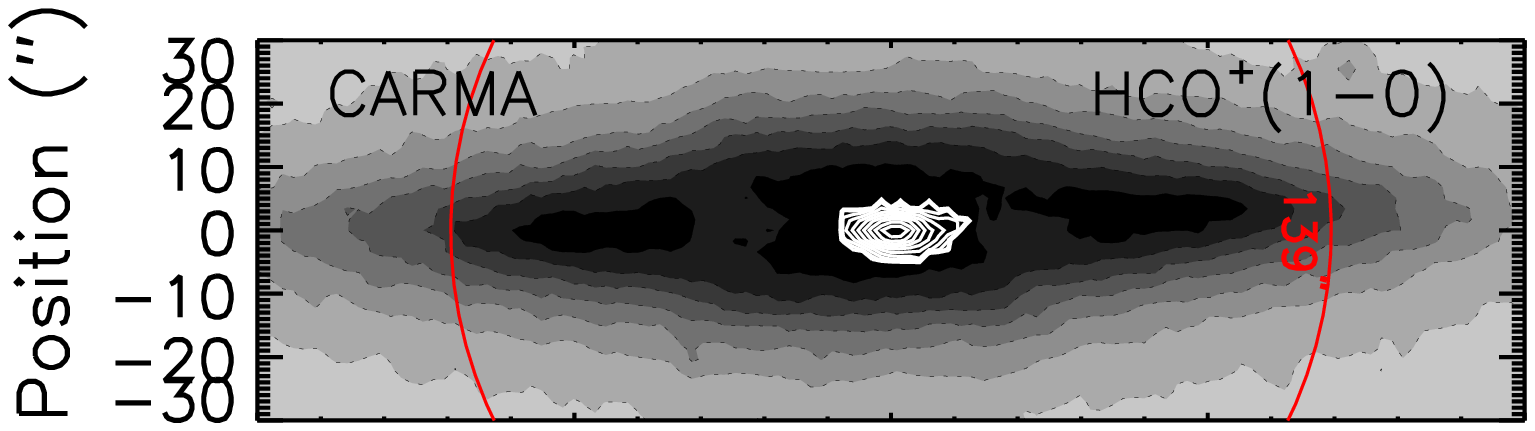}
  \includegraphics[width=5.8cm,clip=]{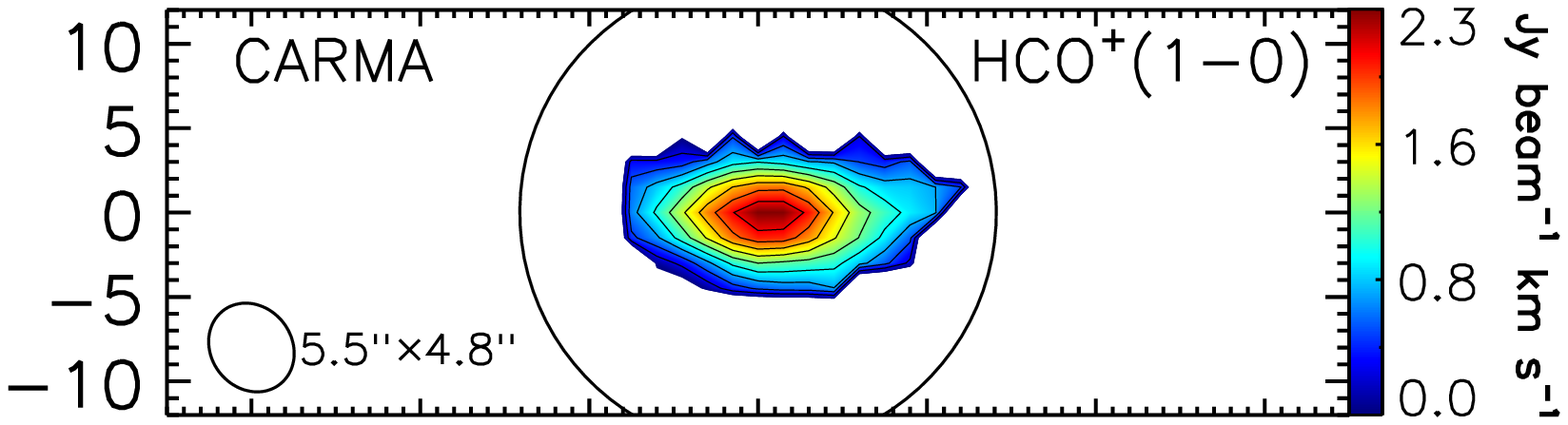}
  \includegraphics[width=5.8cm,clip=]{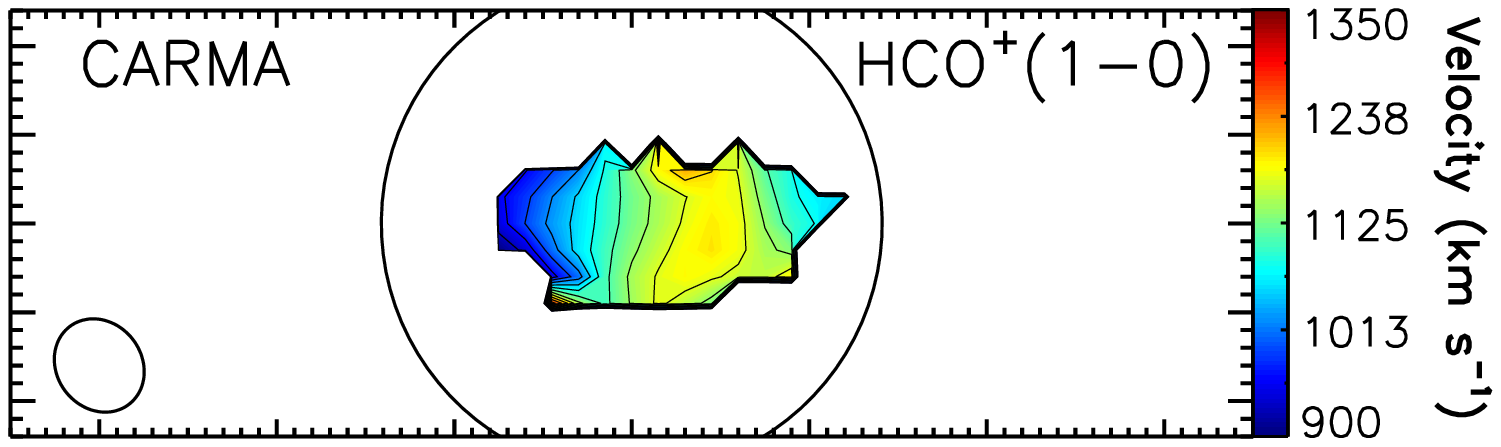}\\
  \vspace{-15pt}
  \includegraphics[width=5.8cm,clip=]{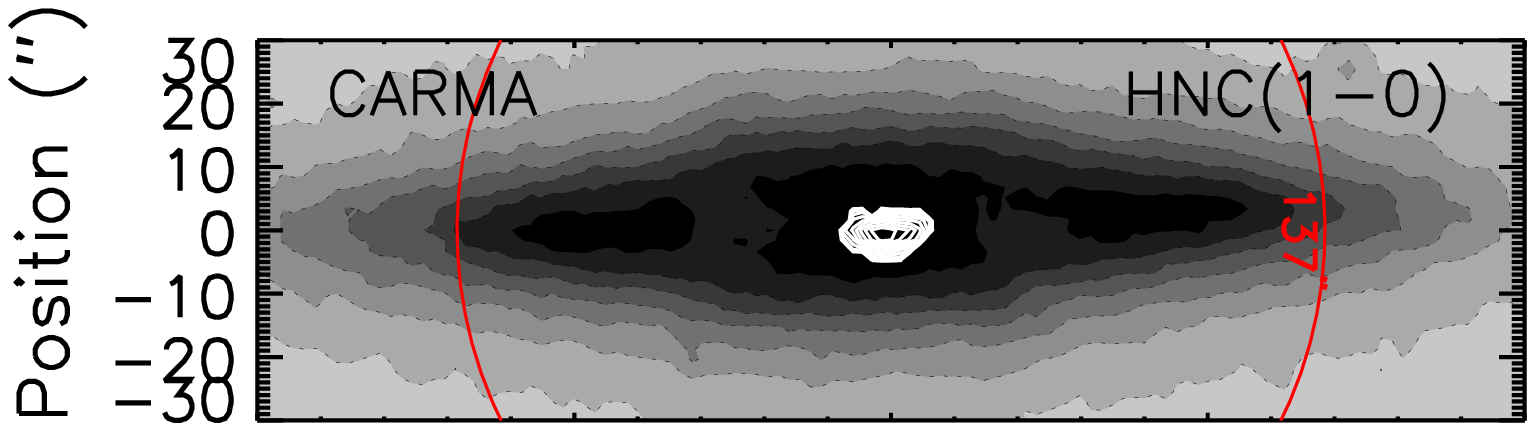}
  \includegraphics[width=5.8cm,clip=]{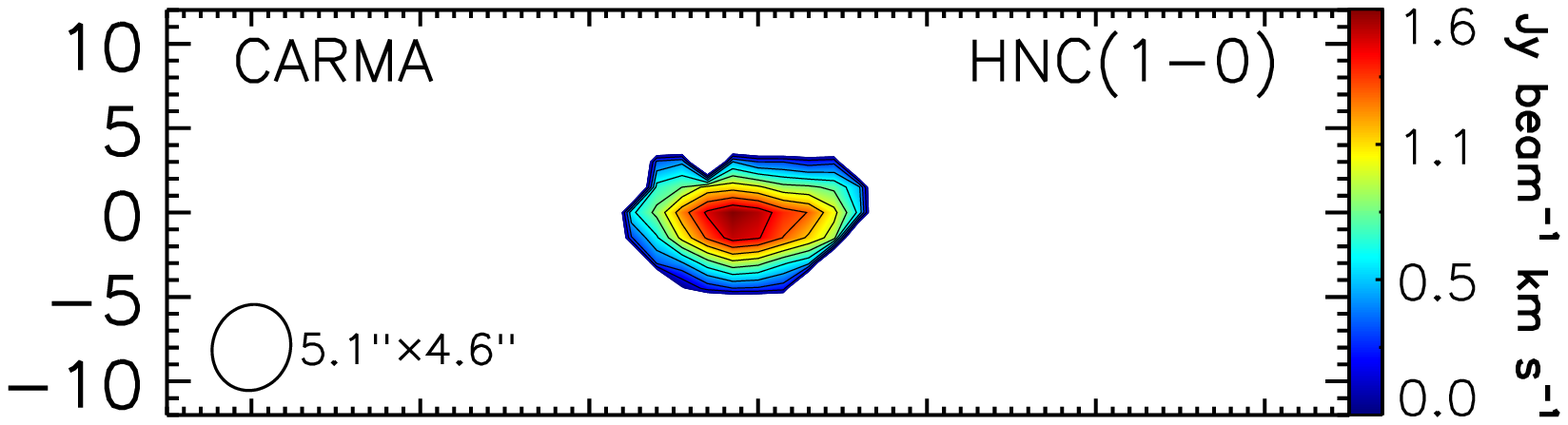}
  \includegraphics[width=5.8cm,clip=]{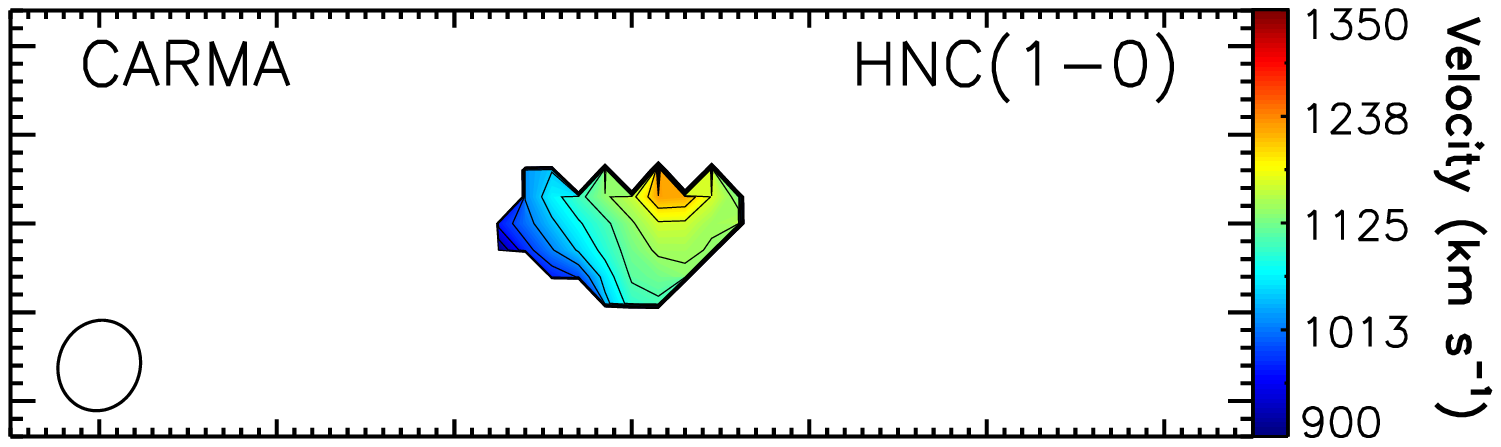}\\
  \vspace{-15pt}
  \includegraphics[width=5.8cm,clip=]{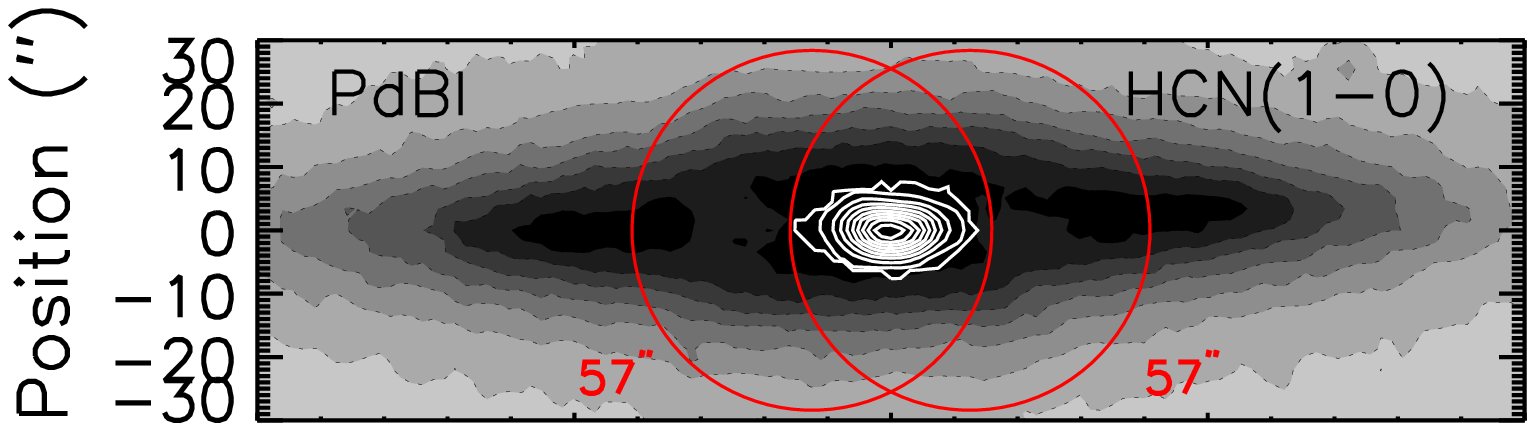}
  \includegraphics[width=5.8cm,clip=]{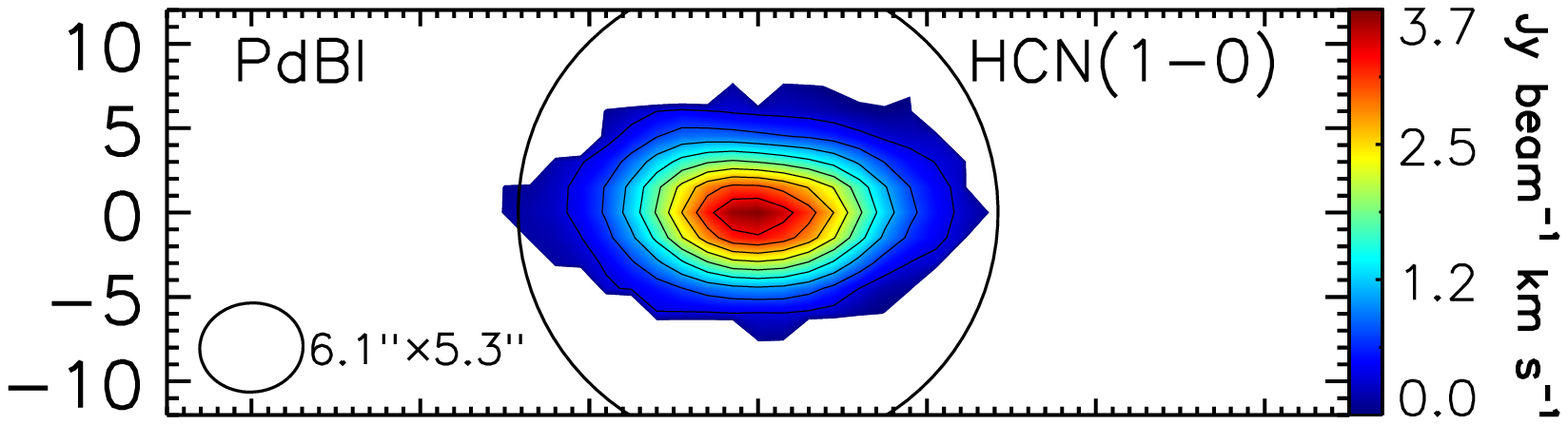}
  \includegraphics[width=5.8cm,clip=]{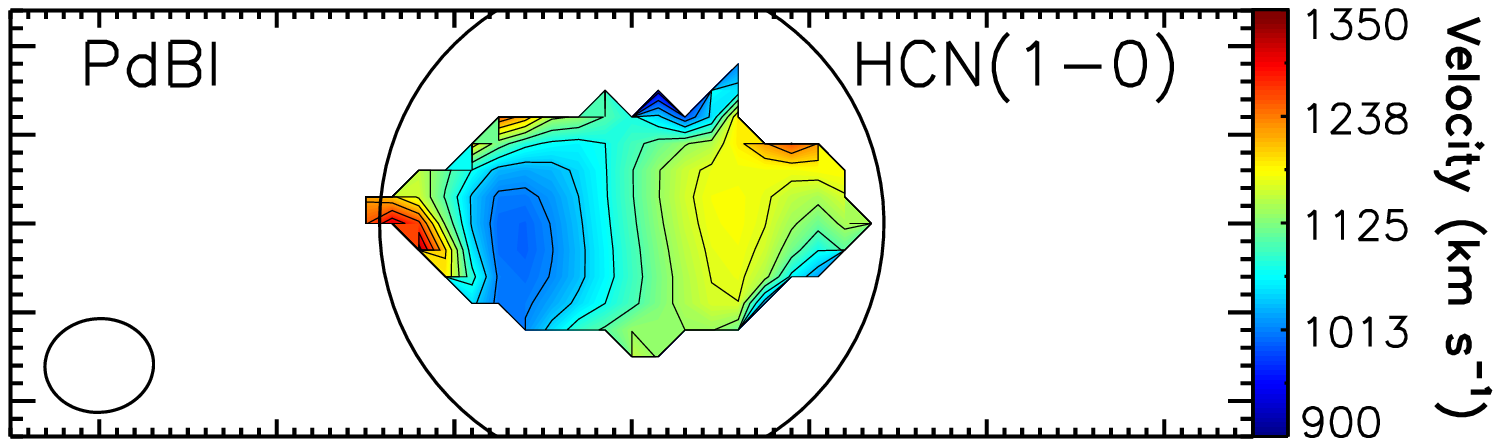}\\
  \vspace{-15pt}
  \includegraphics[width=5.8cm,clip=]{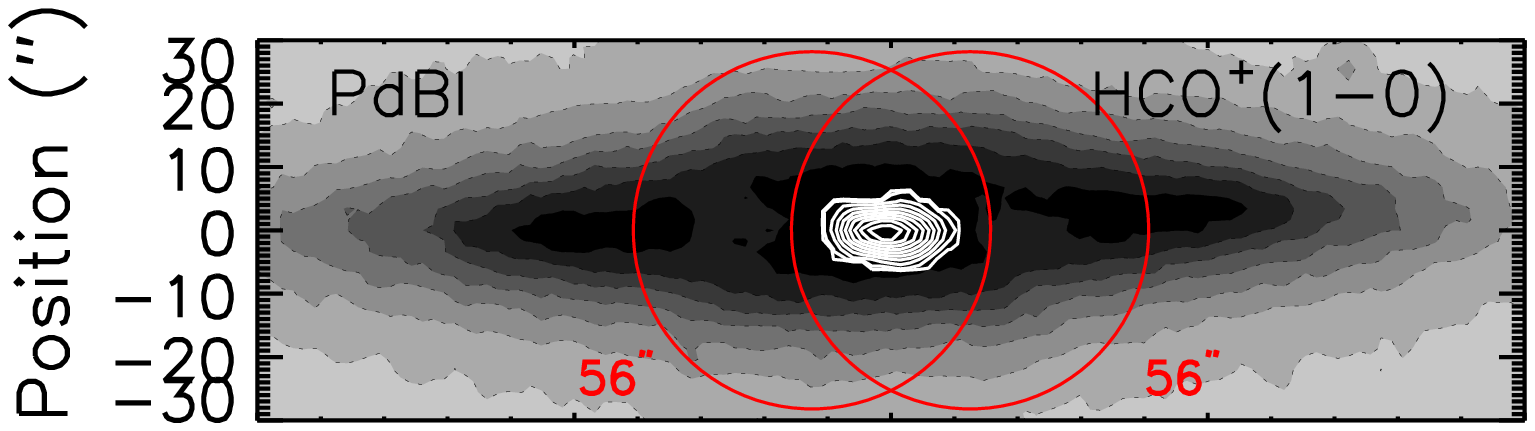}
  \includegraphics[width=5.8cm,clip=]{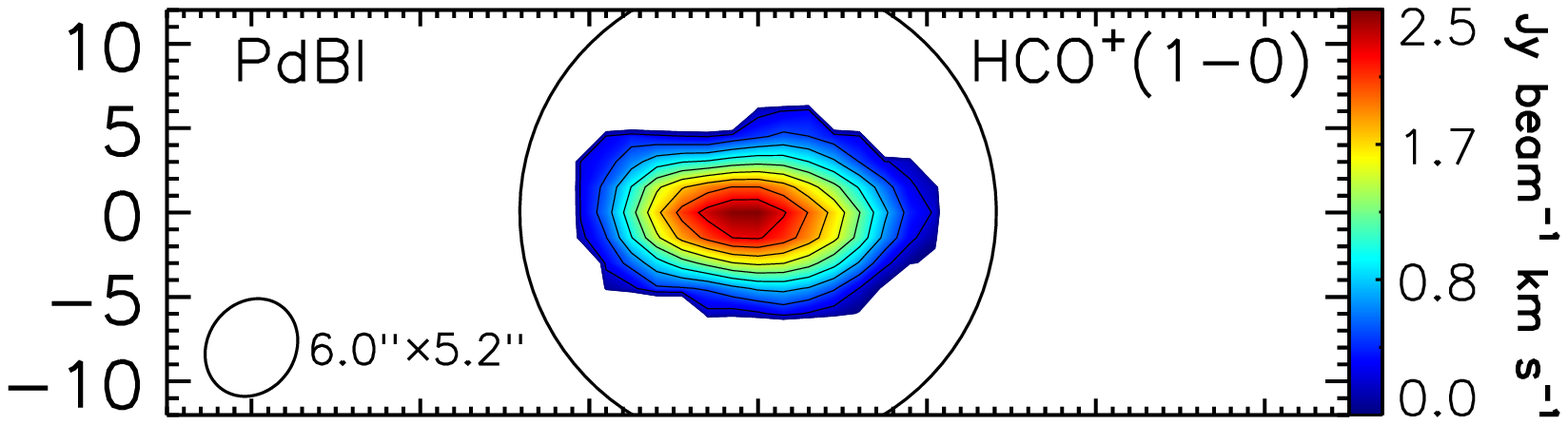}
  \includegraphics[width=5.8cm,clip=]{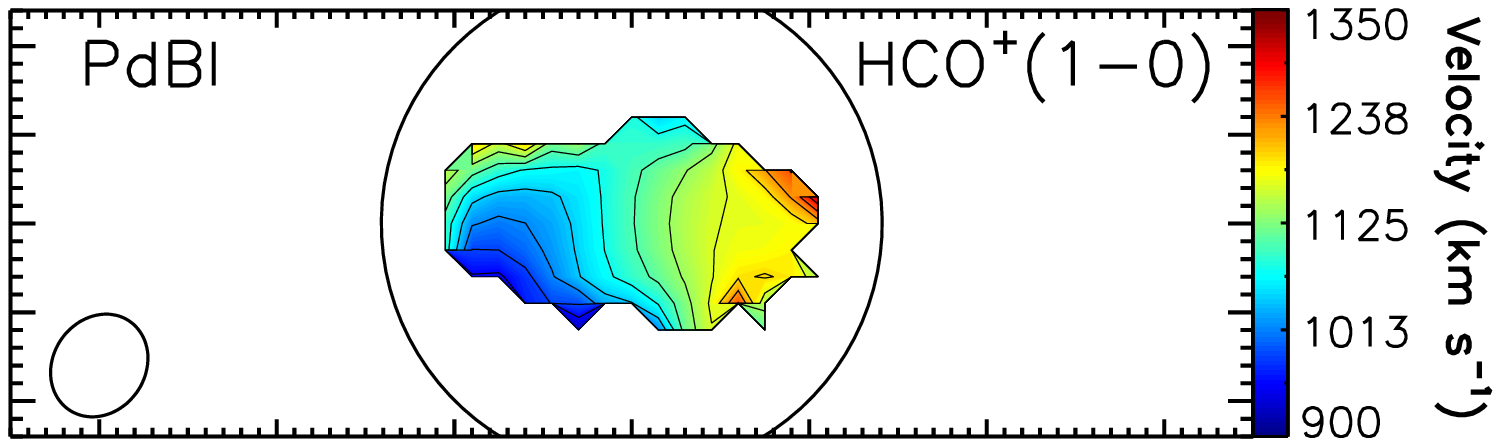}\\
  \vspace{-15pt}
  \includegraphics[width=5.8cm,clip=]{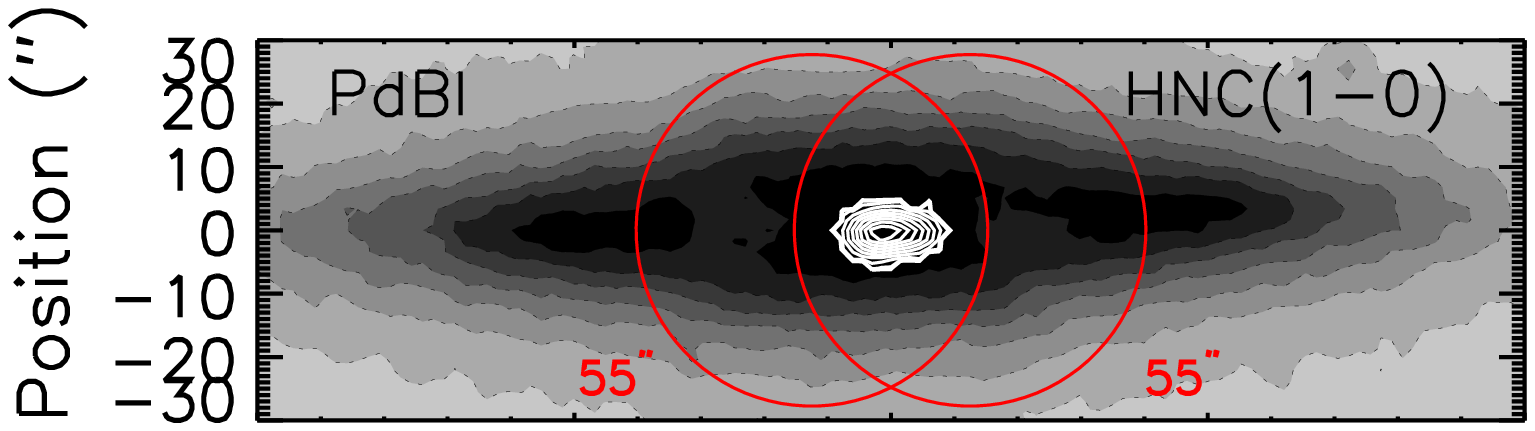}
  \includegraphics[width=5.8cm,clip=]{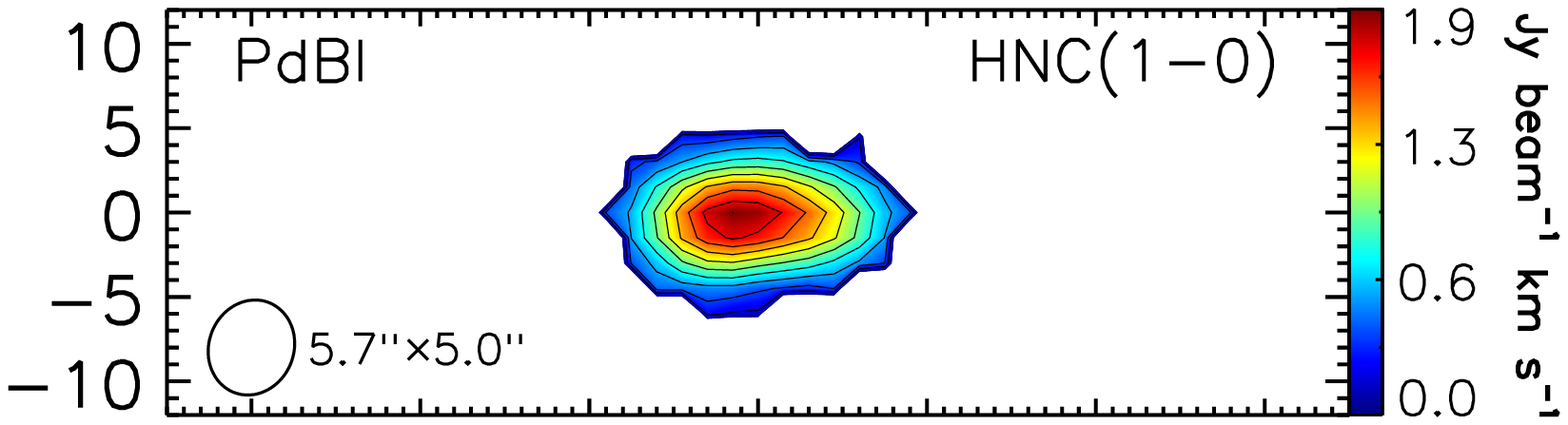}
  \includegraphics[width=5.8cm,clip=]{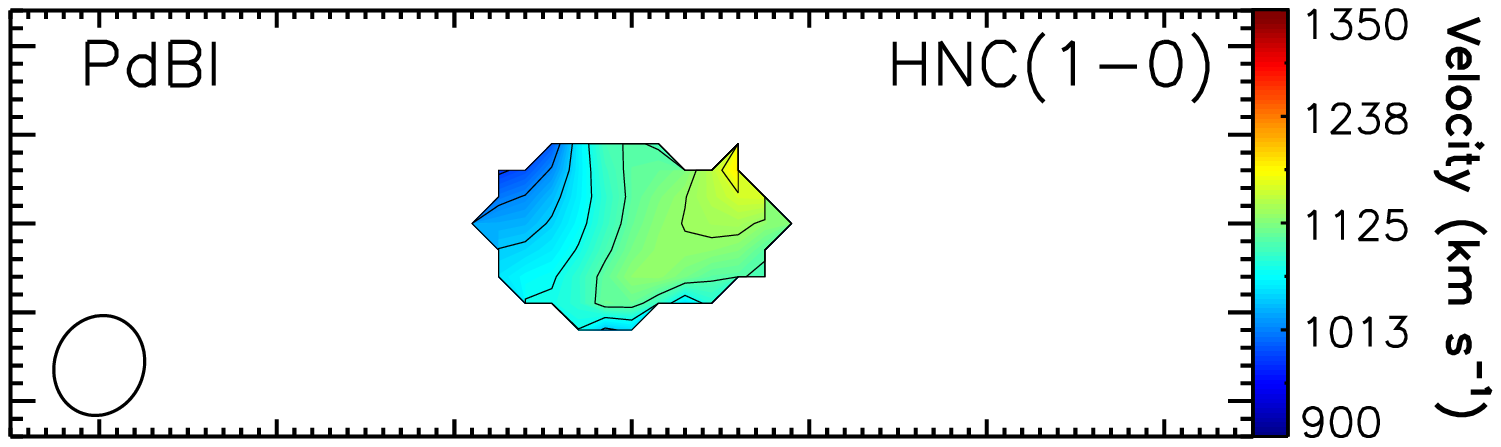}\\
  \vspace{-15pt}
  \includegraphics[width=5.8cm,clip=]{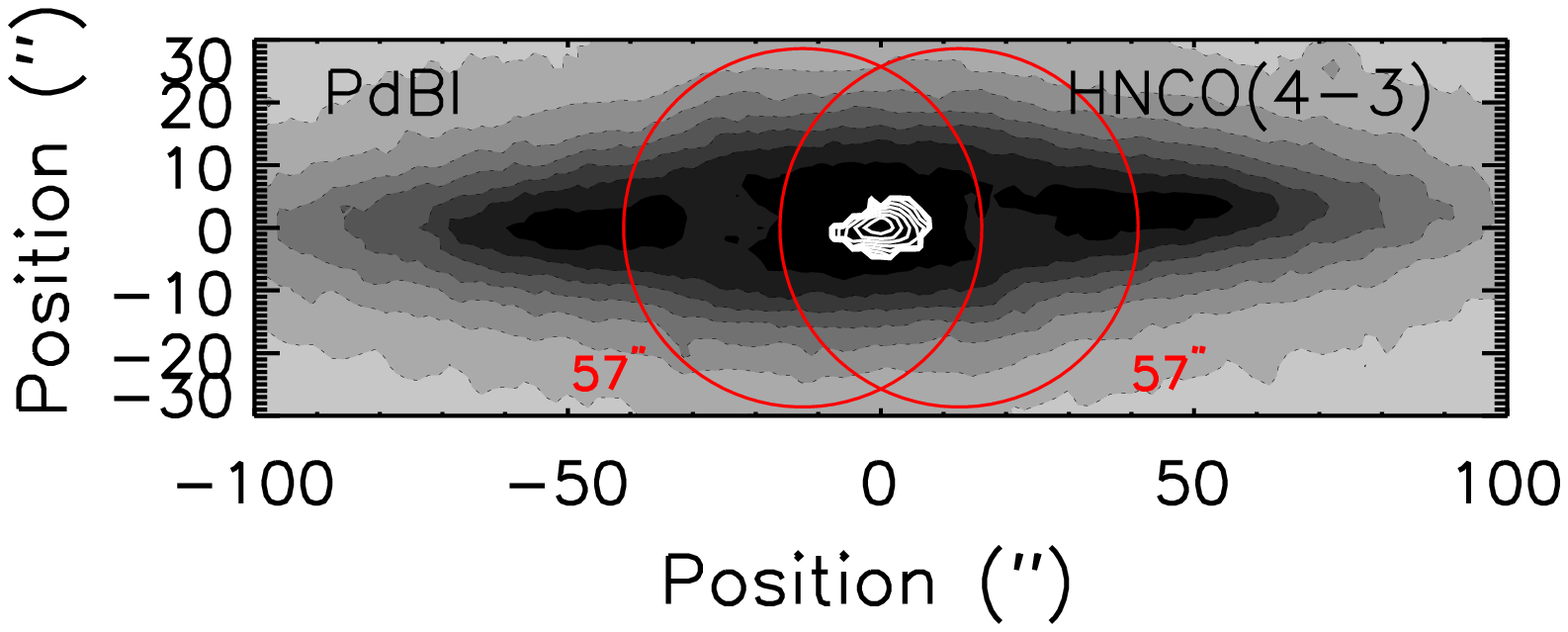}
  \includegraphics[width=5.8cm,clip=]{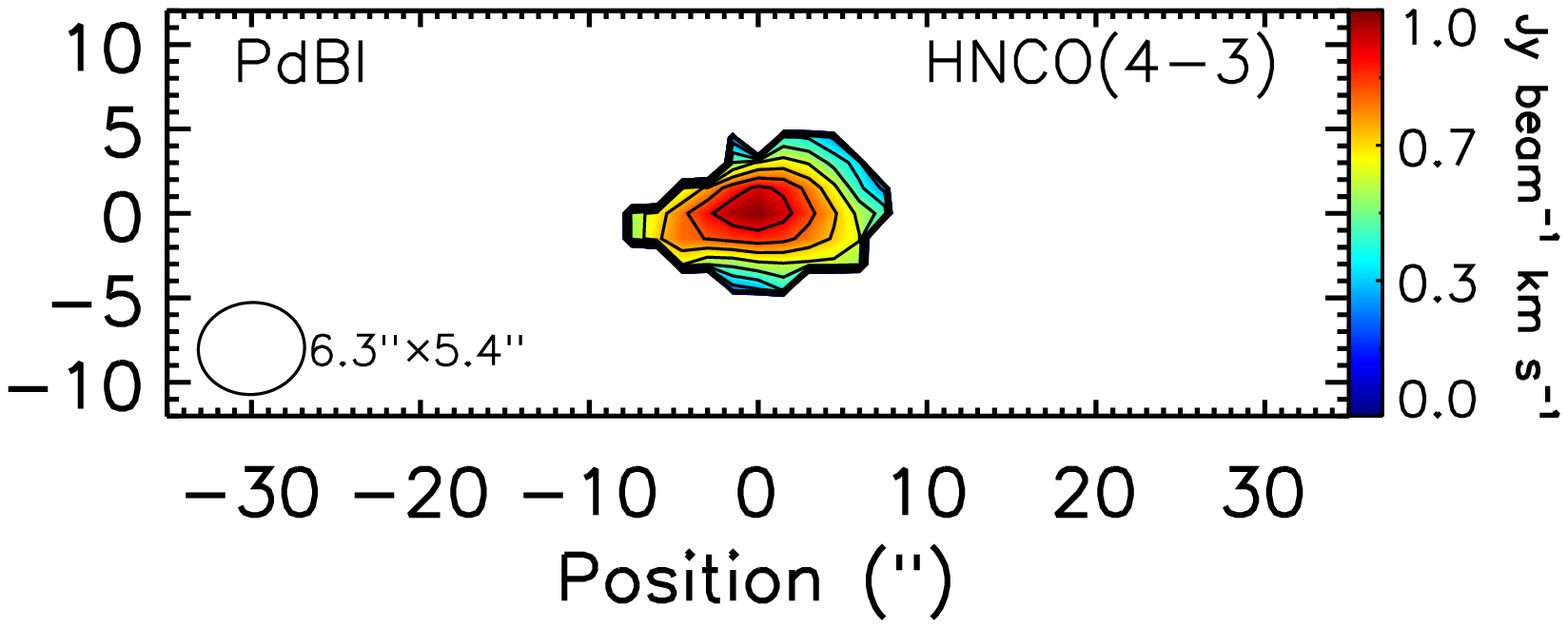}
  \includegraphics[width=5.8cm,clip=]{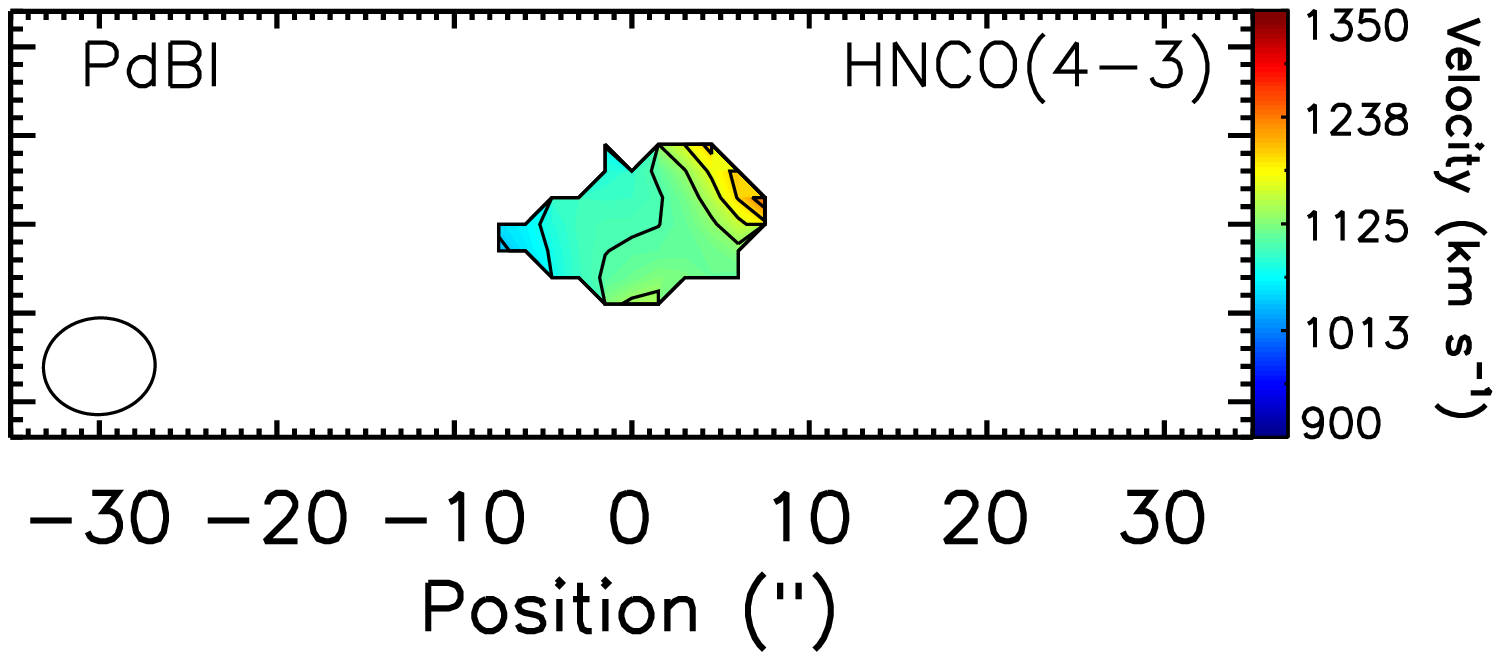}
  \caption{NGC~4710 moment maps. {\bf Left}: Moment~0 contour maps
    (white) of the detected lines overlaid on an optical image of the
    galaxy (greyscale) from SDSS. Red circles show the primary beam of
    CARMA and/or PdBI at the respective frequency of each line. The
    observations obtained with multiple pointings (mosaicking) are
    shown with multiple primary beams. The beam sizes are also
    indicated. {\bf Centre}: Moment~0 maps with overlaid isophotal
    contours. {\bf Right}: Moment~1 maps with overlaid isovelocity
    contours. The large black circles on the moment maps show the IRAM
    30~m telescope beam for comparison (HNC(1-0) and HNCO(4-3) were
    not observed; see \citealt{c12}). Contour levels on the moment~0
    maps are from $10$ to $100\%$ of the peak integrated line
    intensity in steps of $10\%$. The moment~0 peaks are (from top to
    bottom) $59.8$, $145.8$, $10.8$, $22.3$, $3.0$, $2.3$, $1.7$,
    $3.7$, $2.5$, $1.9$ and $1.0$ Jy~beam$^{-1}$~km~s$^{-1}$. Contour
    levels on the moment~1 maps are spaced by $30$~km~s$^{-1}$. The
    array used, molecular line displayed, and synthesized beam are
    also indicated in each panel.}
  \label{fig:n4710mom}
\end{figure*}

\begin{figure*}
  \centering
  \hspace{-15pt}
  \includegraphics[width=6.0cm,clip=]{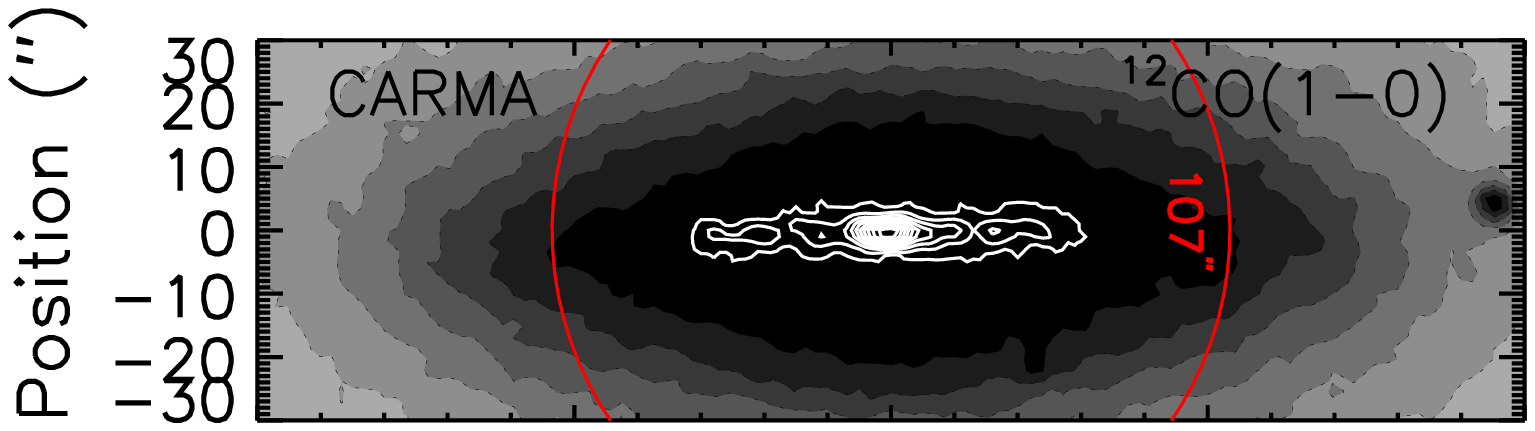}
  \includegraphics[width=6.0cm,clip=]{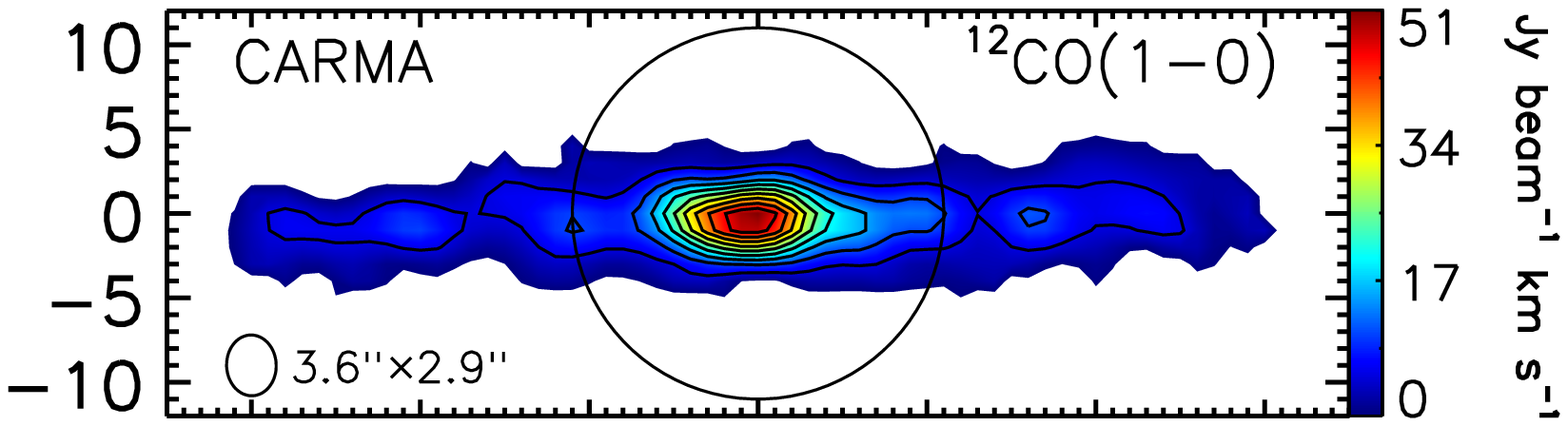}
  \includegraphics[width=6.0cm,clip=]{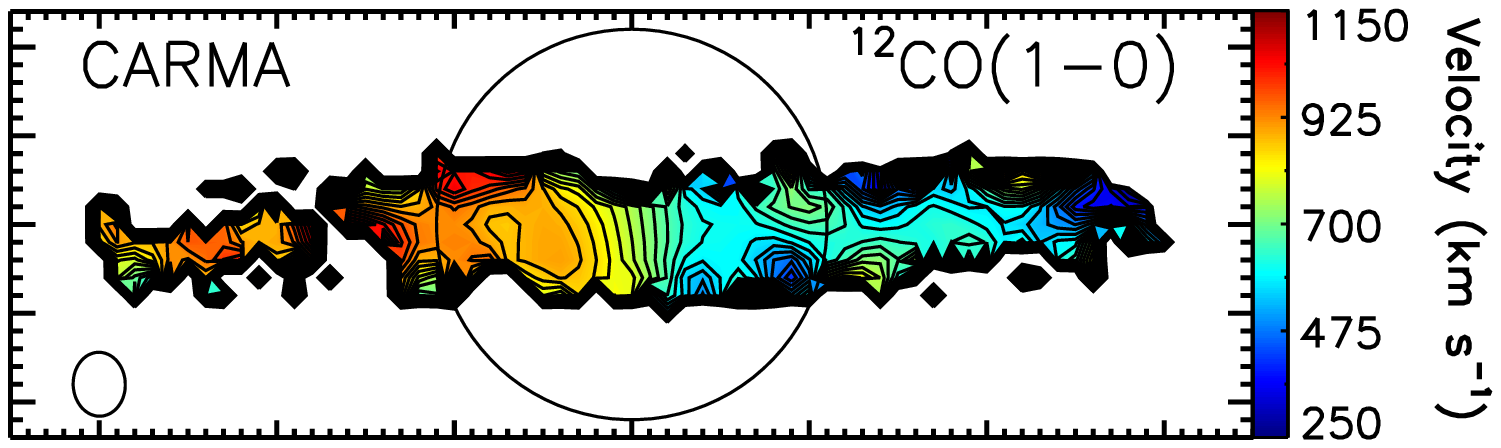}\\
  \vspace{-10pt}
  \hspace{-15pt}
  \includegraphics[width=6.0cm,clip=]{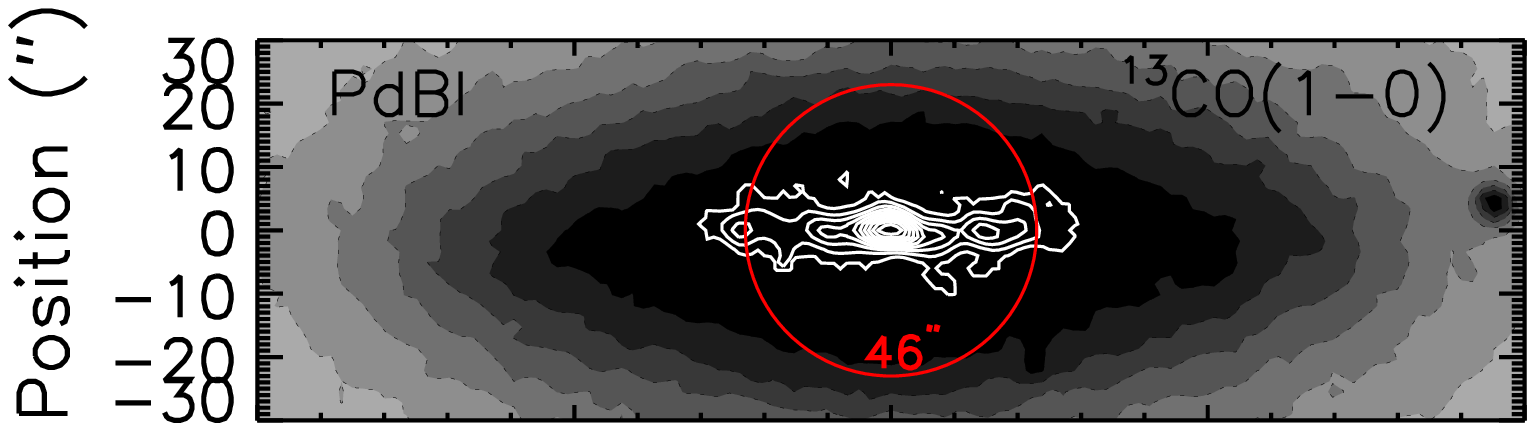}
  \includegraphics[width=6.0cm,clip=]{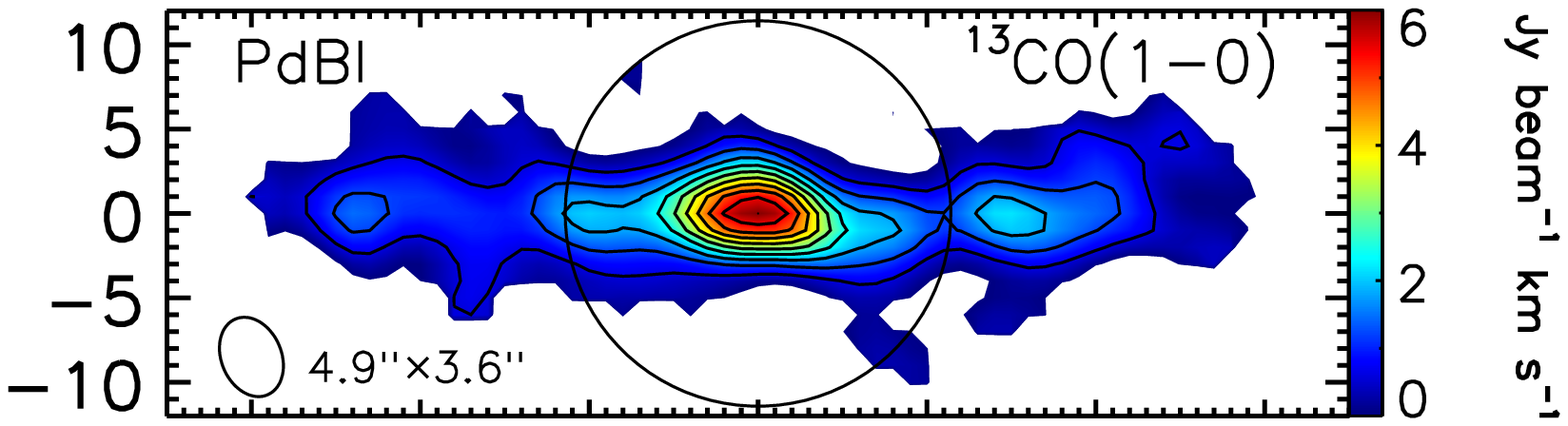}
  \includegraphics[width=6.0cm,clip=]{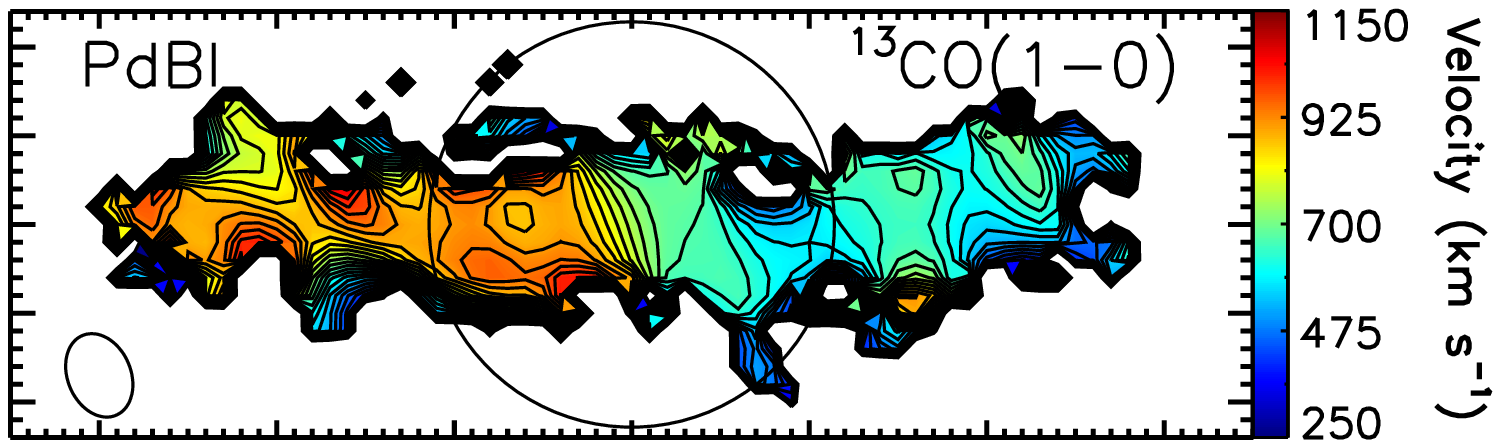}\\
  \vspace{-10pt}
  \hspace{-15pt}
  \includegraphics[width=6.0cm,clip=]{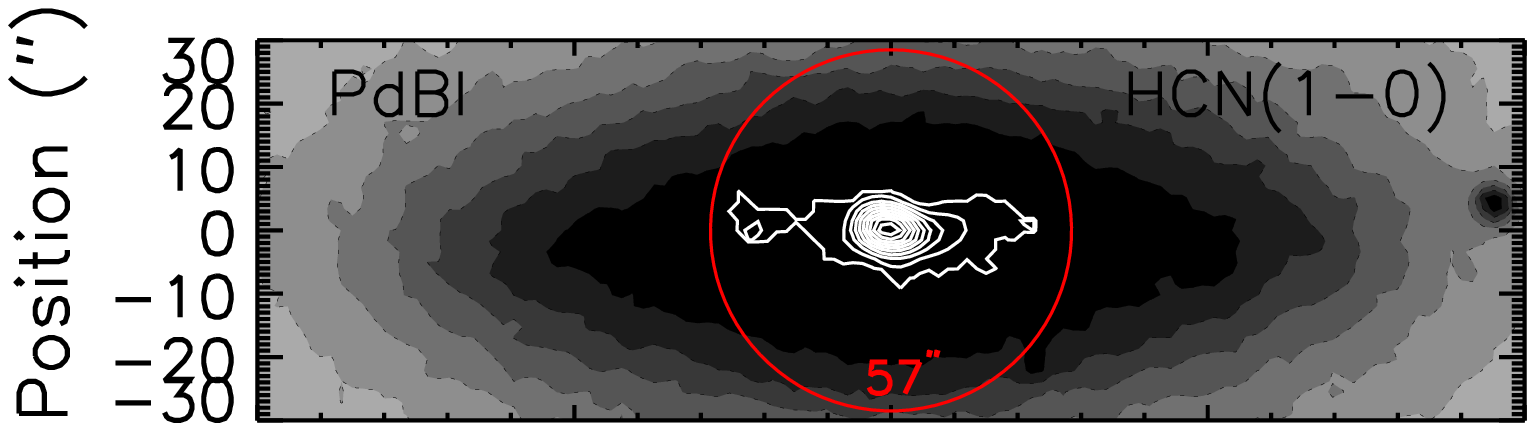}
  \includegraphics[width=6.0cm,clip=]{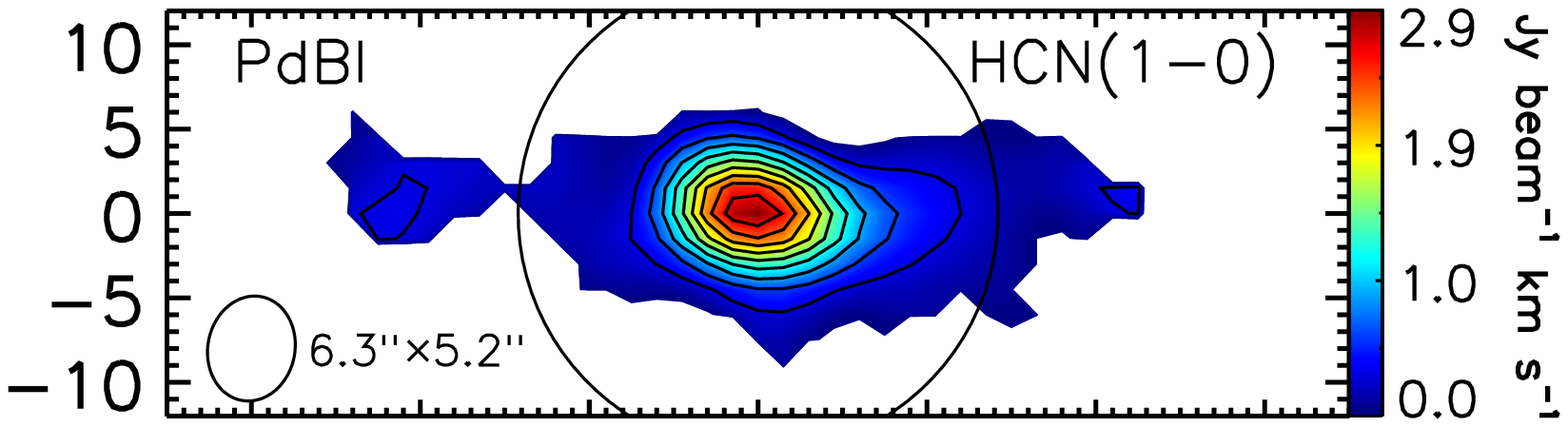}
  \includegraphics[width=6.0cm,clip=]{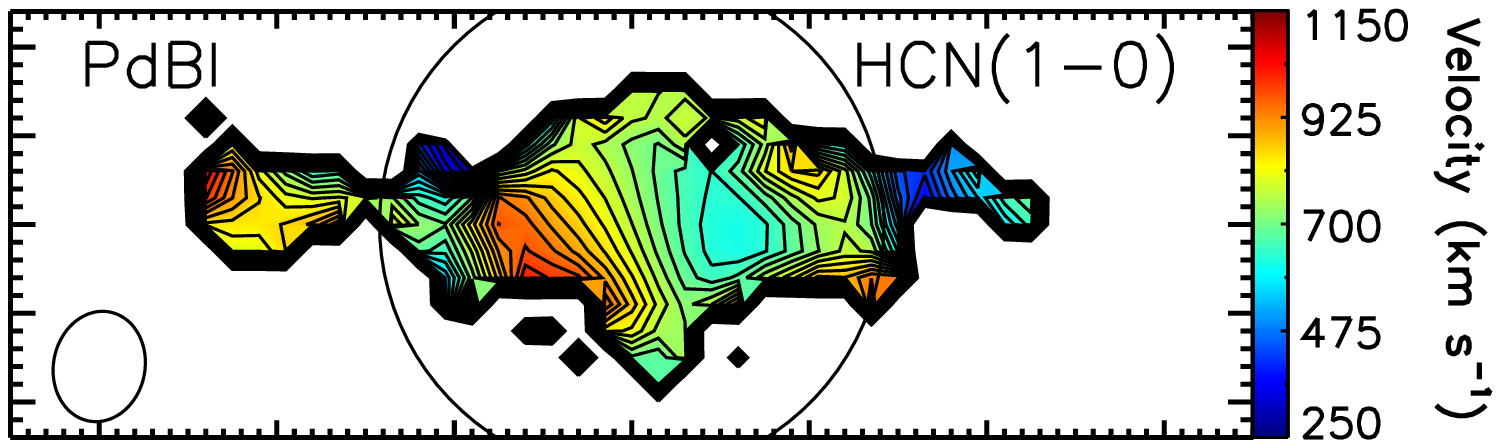}\\
  \vspace{-10pt}
  \hspace{-15pt}
  \includegraphics[width=6.0cm,clip=]{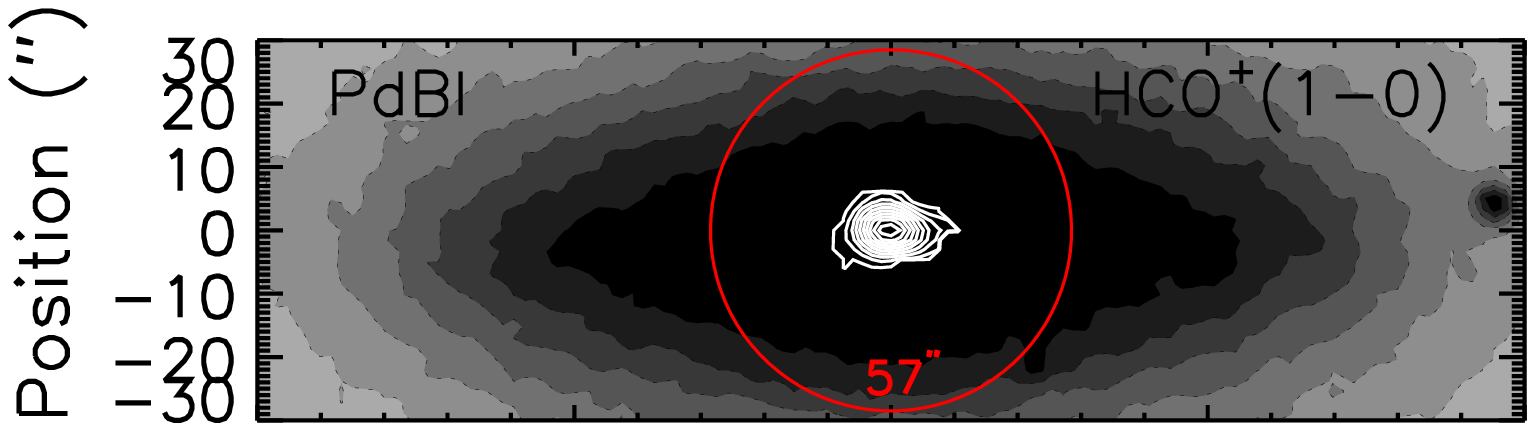}
  \includegraphics[width=6.0cm,clip=]{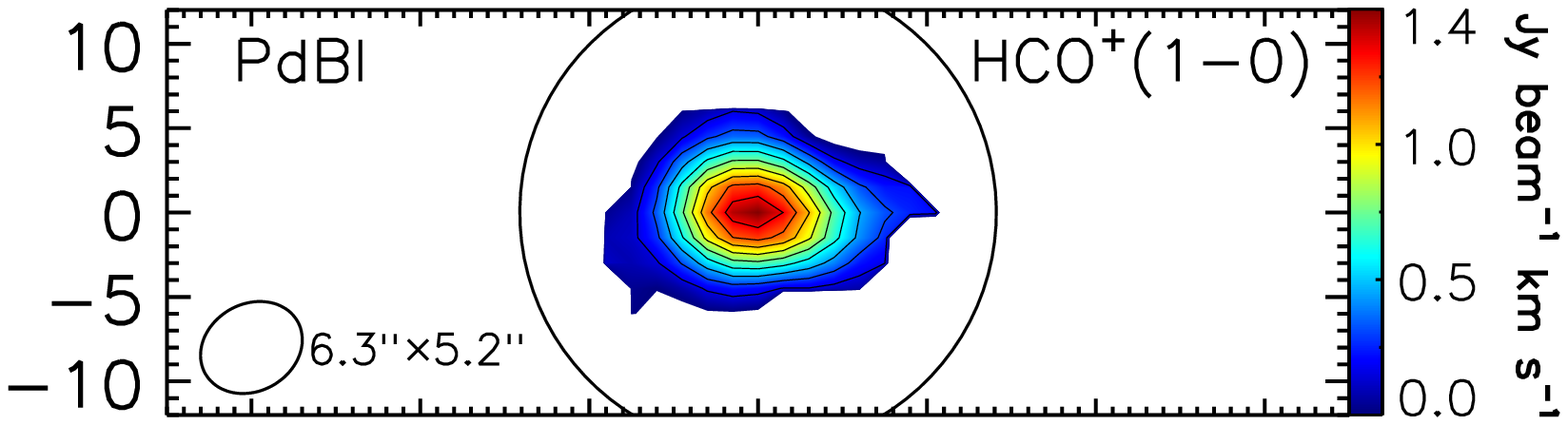}
  \includegraphics[width=6.0cm,clip=]{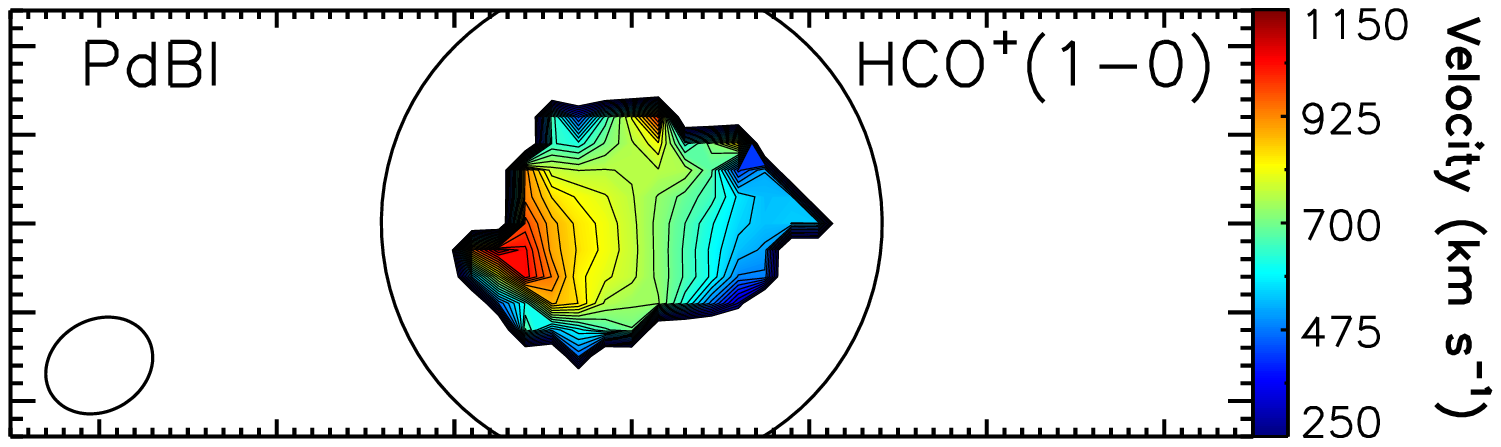}\\
  \vspace{-10pt}
  \hspace{-15pt}
  \includegraphics[width=6.0cm,clip=]{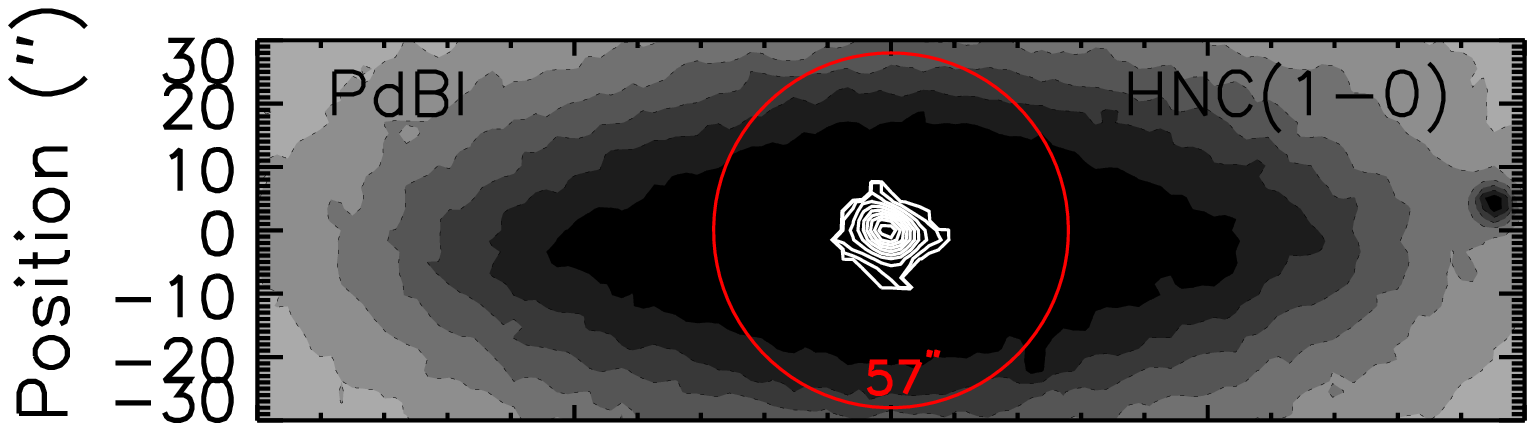}
  \includegraphics[width=6.0cm,clip=]{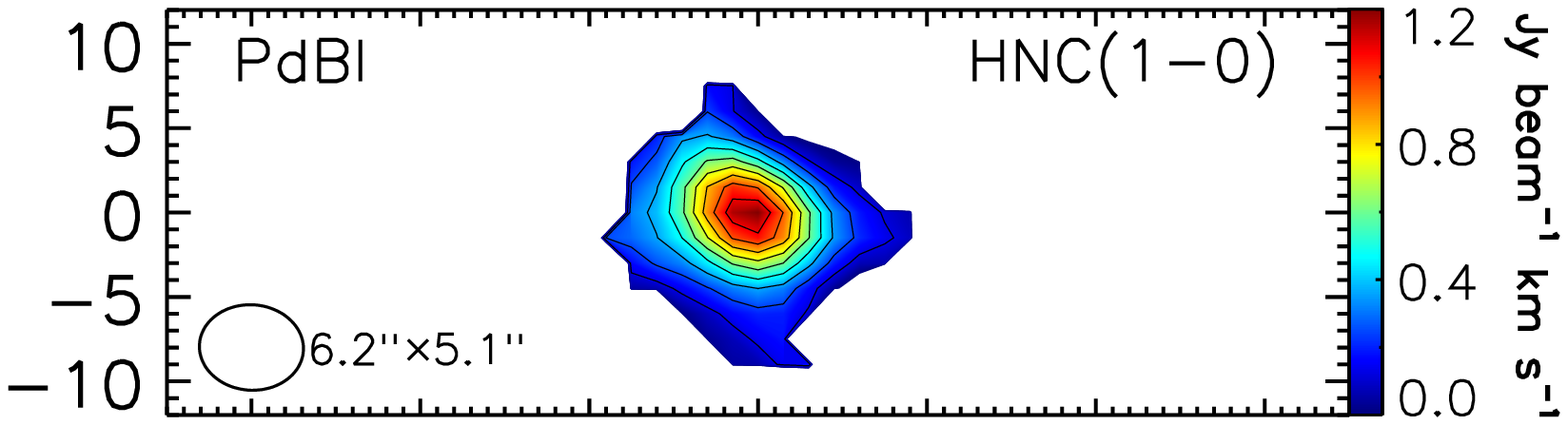}
  \includegraphics[width=6.0cm,clip=]{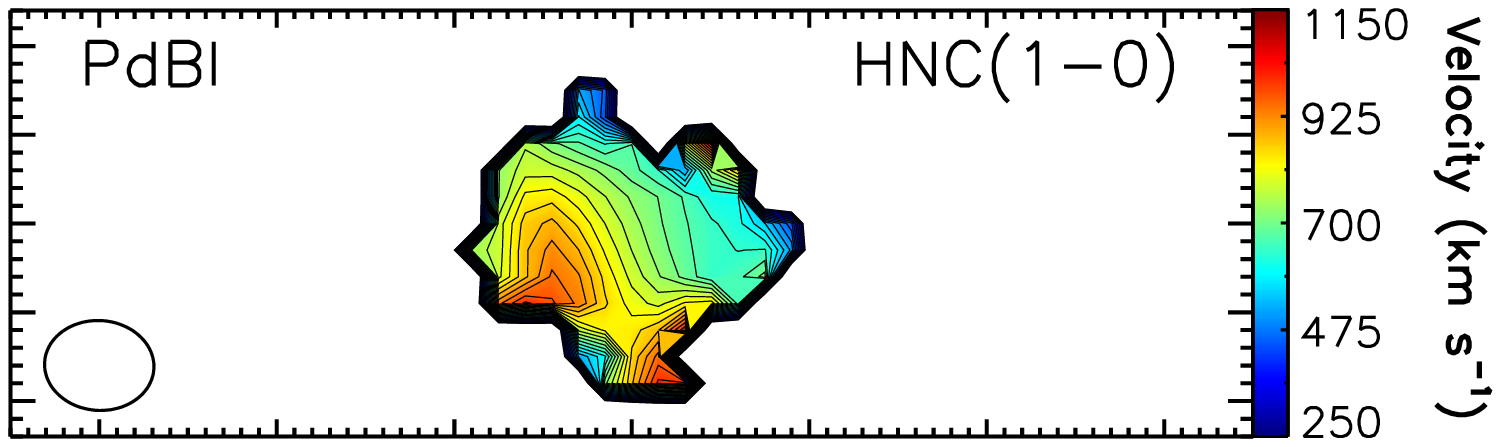}\\
  \vspace{-10pt}
  \hspace{-15pt}
  \includegraphics[width=6.0cm,clip=]{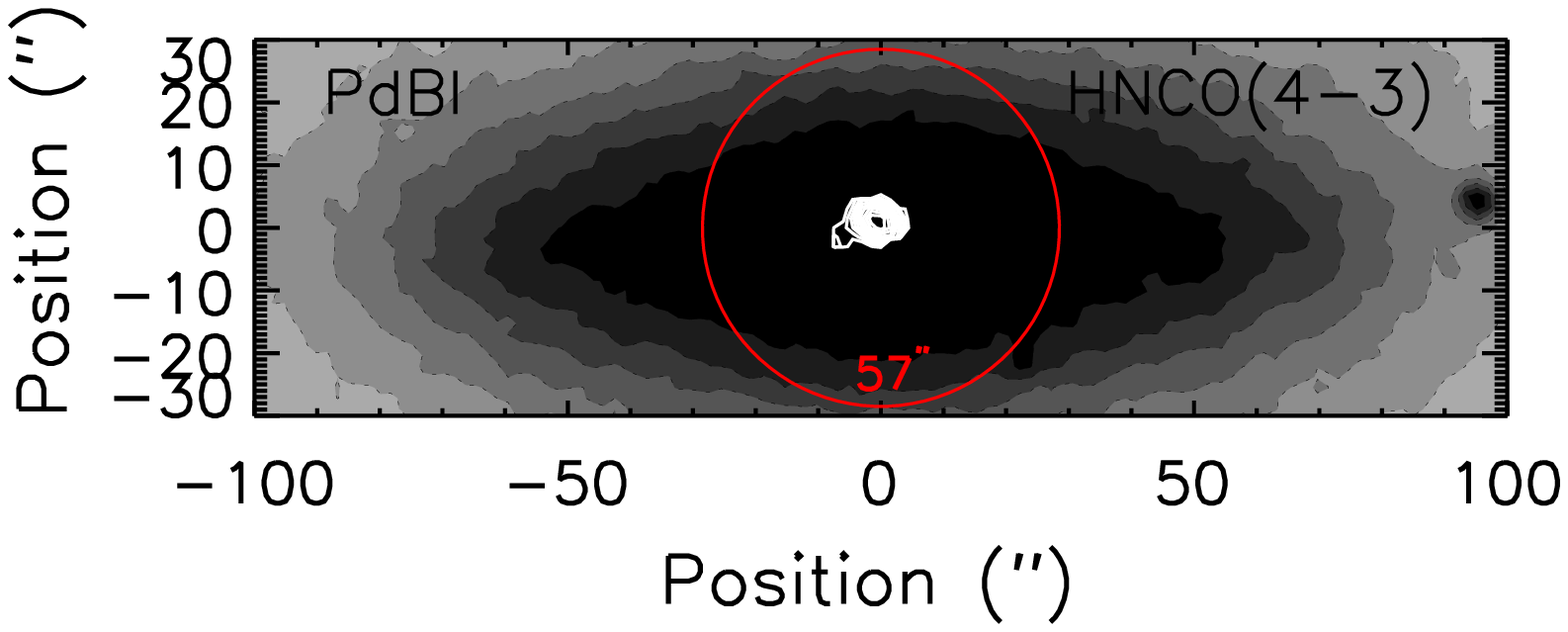}
  \includegraphics[width=6.0cm,clip=]{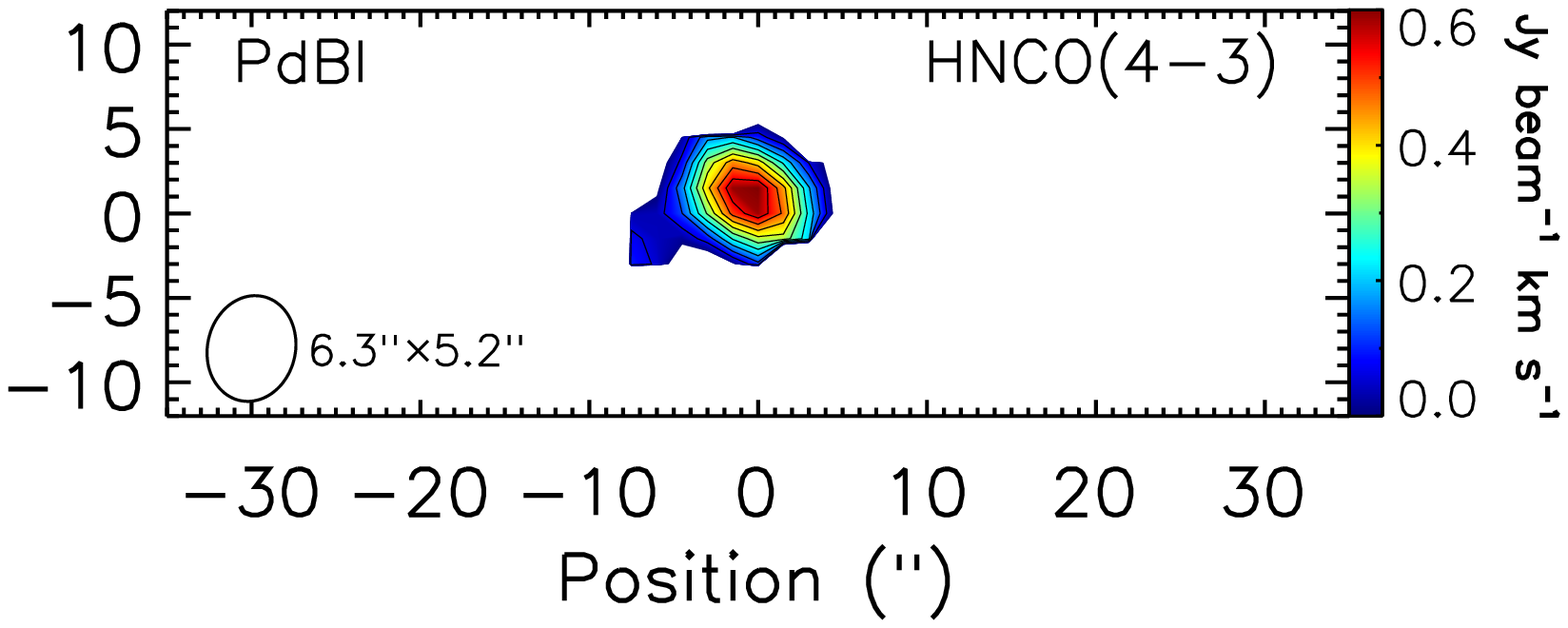}
  \includegraphics[width=6.0cm,clip=]{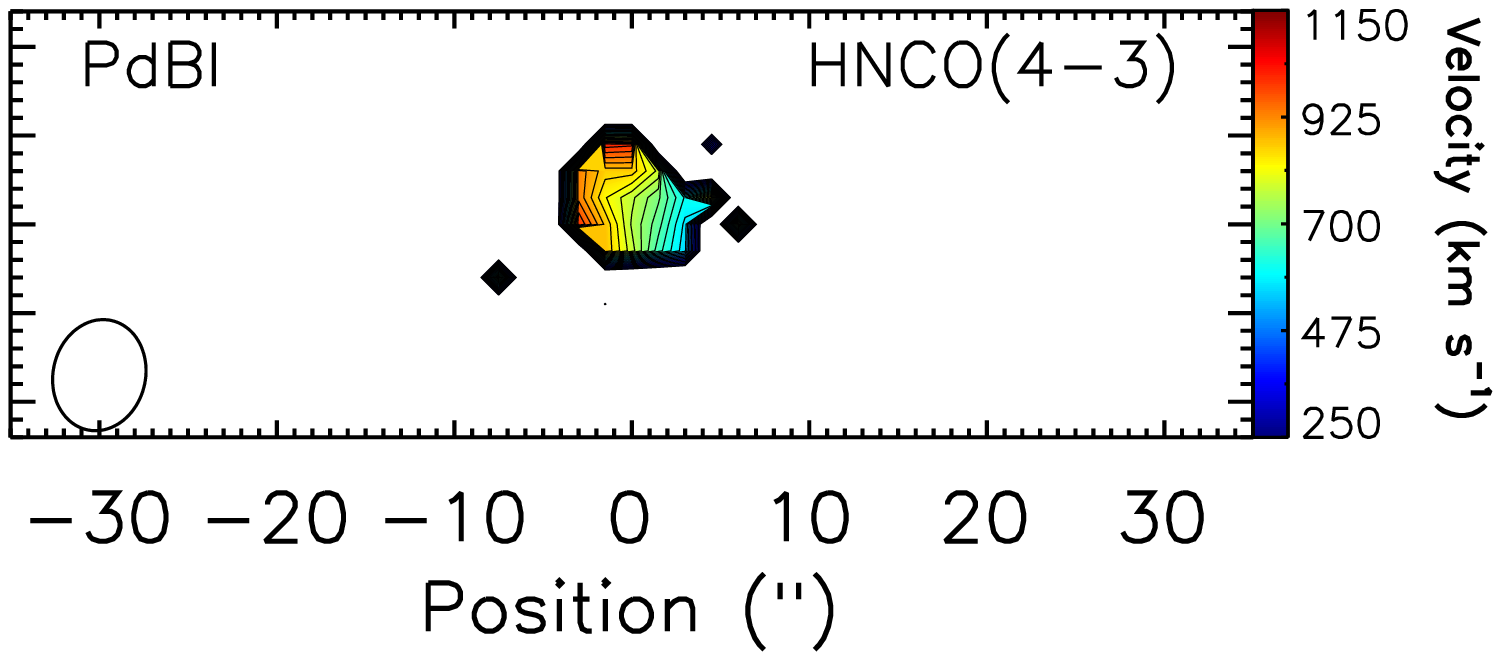}
  \caption{Same as Figure~\ref{fig:n4710mom} but for NGC5866. The
    moment~0 peaks are (from top to bottom) $51.5$, $5.5$, $2.9$,
    $1.4$, $1.2$, and $0.6$ Jy~beam$^{-1}$~km~s$^{-1}$.}
  \label{fig:n5866mom}
\end{figure*}

To derive the moment maps of NGC~4710 and NGC~5866, the spatial extent
of the emission must first be defined in the fully-calibrated and
cleaned cube of each line. The data cubes were therefore first
Hanning-smoothed spectrally and Gaussian-smoothed spatially with a
FWHM equal to that of the beam. The smoothed cubes were then clipped
(in three dimensions) at a $3\sigma$ threshold (where $\sigma$ is the
rms noise of the smoothed cube) and (smoothed) moment maps
created. Regions of contiguous emission for each line were then
defined using the IDL region-growing algorithm \emph{label\_region}
and the (smoothed) moment~0 maps. The largest central contiguous
emission region for each line was then adopted as a two-dimensional
mask, and used to derive the moments of the original (unsmoothed and
unclipped) cube in that line. The moment maps are
  shown in Figures~\ref{fig:n4710mom} and \ref{fig:n5866mom} for all
  lines.
%
%
%
%

As the $^{12}$CO(1-0) is the most spatially extended line in both
galaxies, the contiguous emission region for the other lines cannot be
more extended than this. Furthermore, as all $^{12}$CO(1-0) emission
lies along the major axis of the galaxies, we assume that this must be
the case for the other lines as well. These two criteria were thus
also used to define the real extent of the line emission when this
emission is discontinuous, typically in the centre (nuclear disc) and
nearly symmetric regions farther along the disc on either side (inner
ring edges; see, e.g., the $^{13}$CO(1-0) emission in
Fig.~\ref{fig:n4710mom}). The rms noise levels listed in
Table~\ref{tab:obs} were calculated using all pixels outside of the
identified emission regions in the original cubes.
%
%
\subsection{Position-velocity diagrams}
\label{sec:posvel}
\begin{figure*}
  \includegraphics[width=6.3cm,clip=]{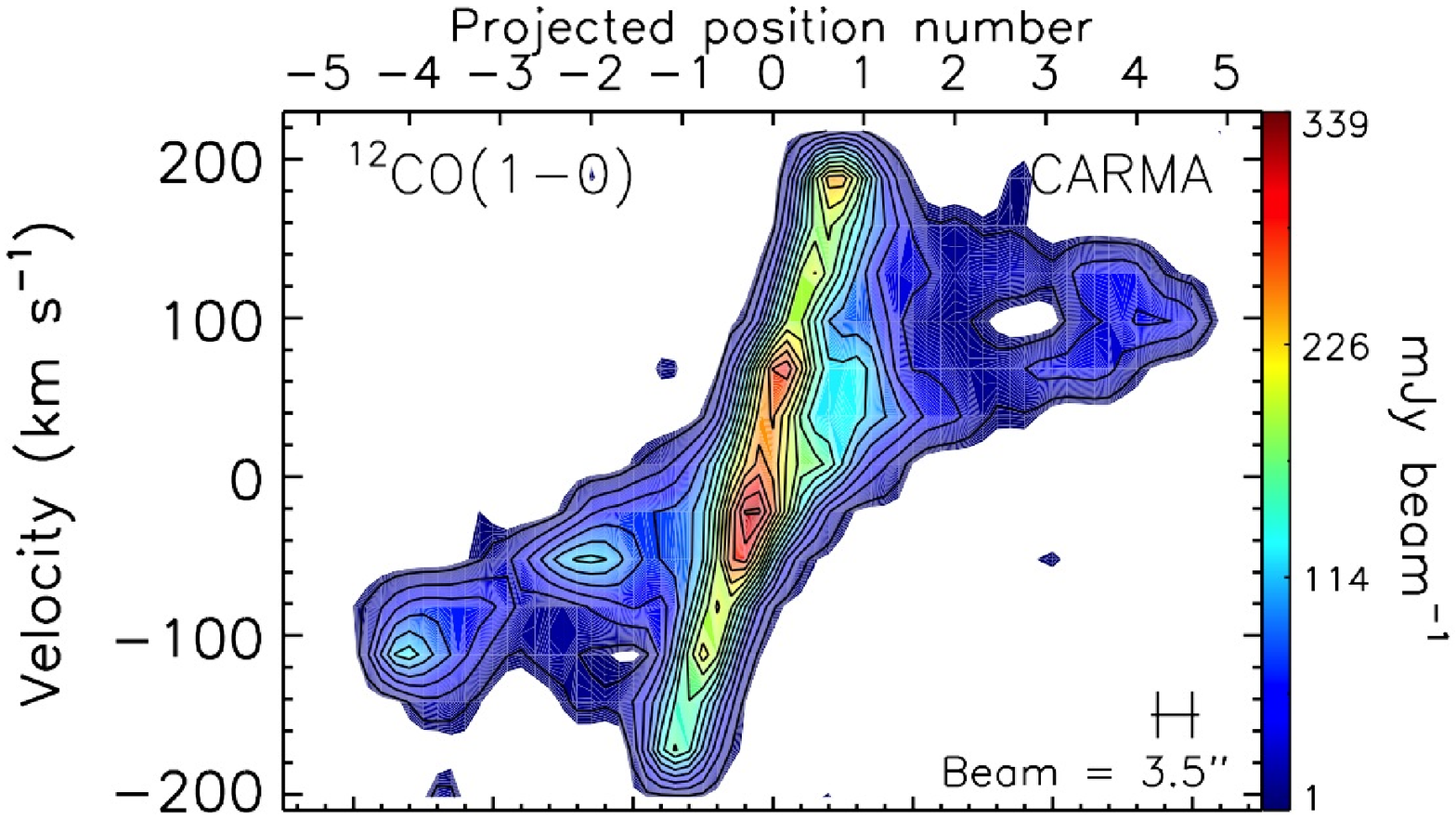}
  \includegraphics[width=5.6cm,clip=]{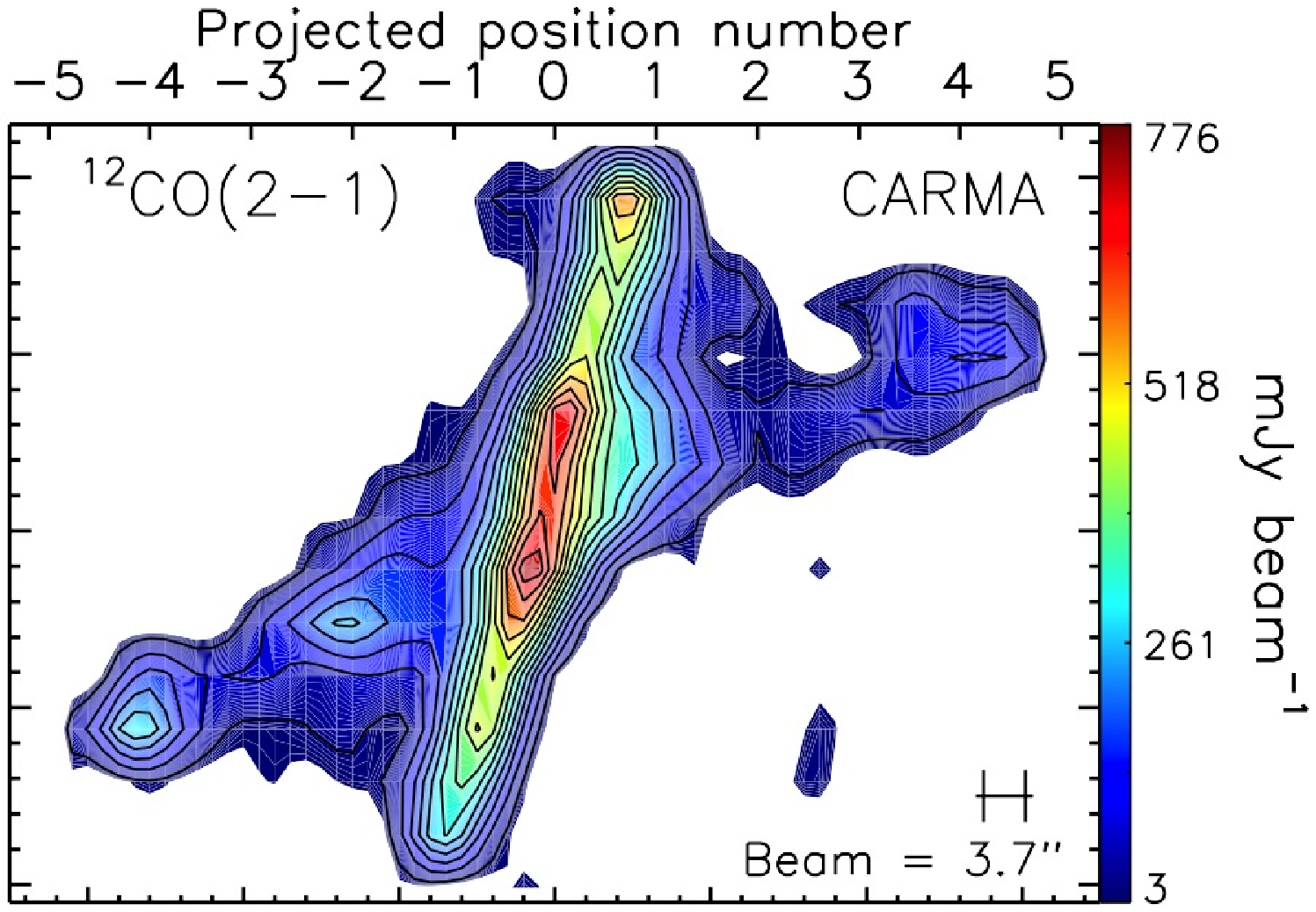}
  \includegraphics[width=5.6cm,clip=]{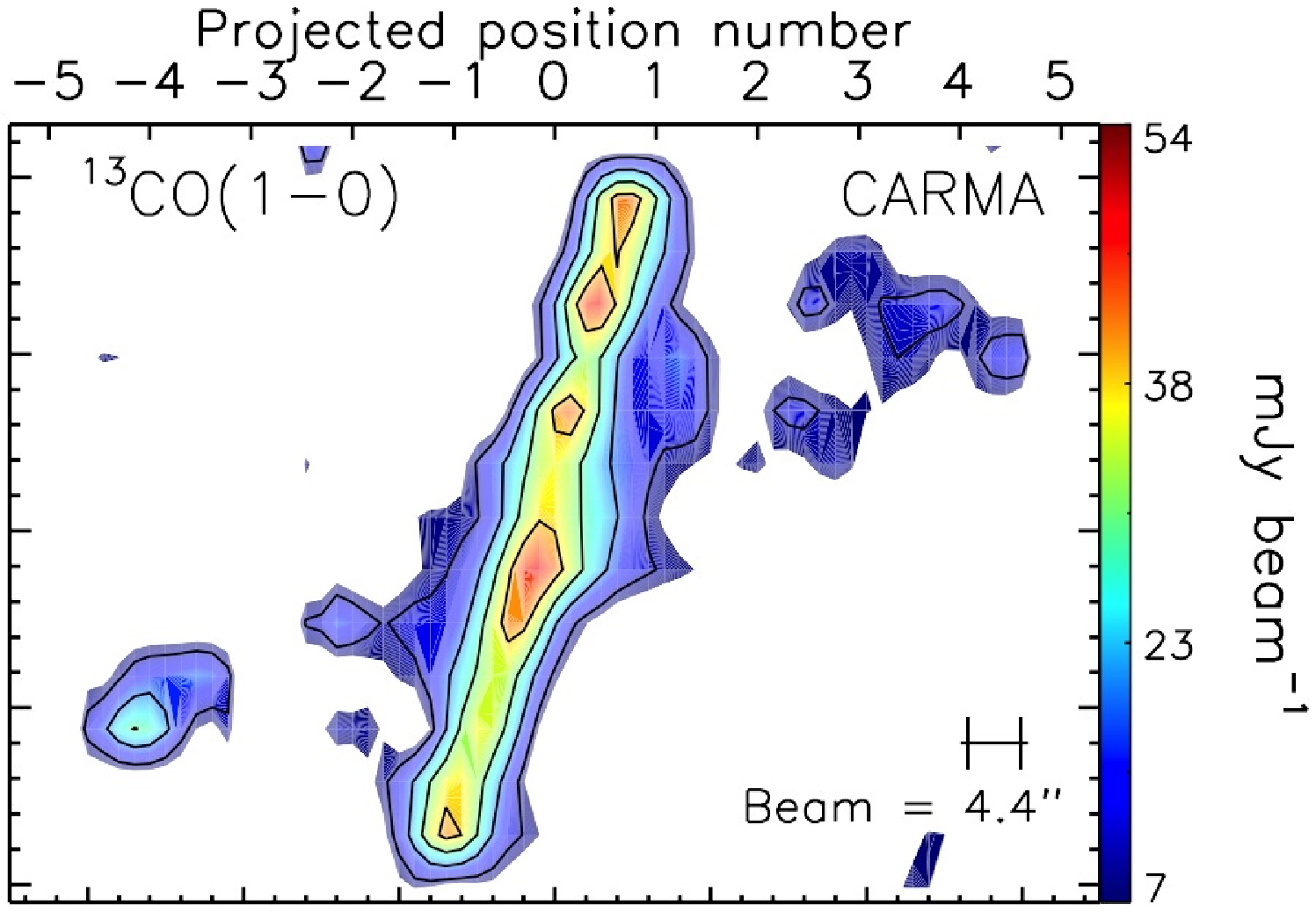}\\
  \vspace{-5pt}
  \includegraphics[width=6.3cm,clip=]{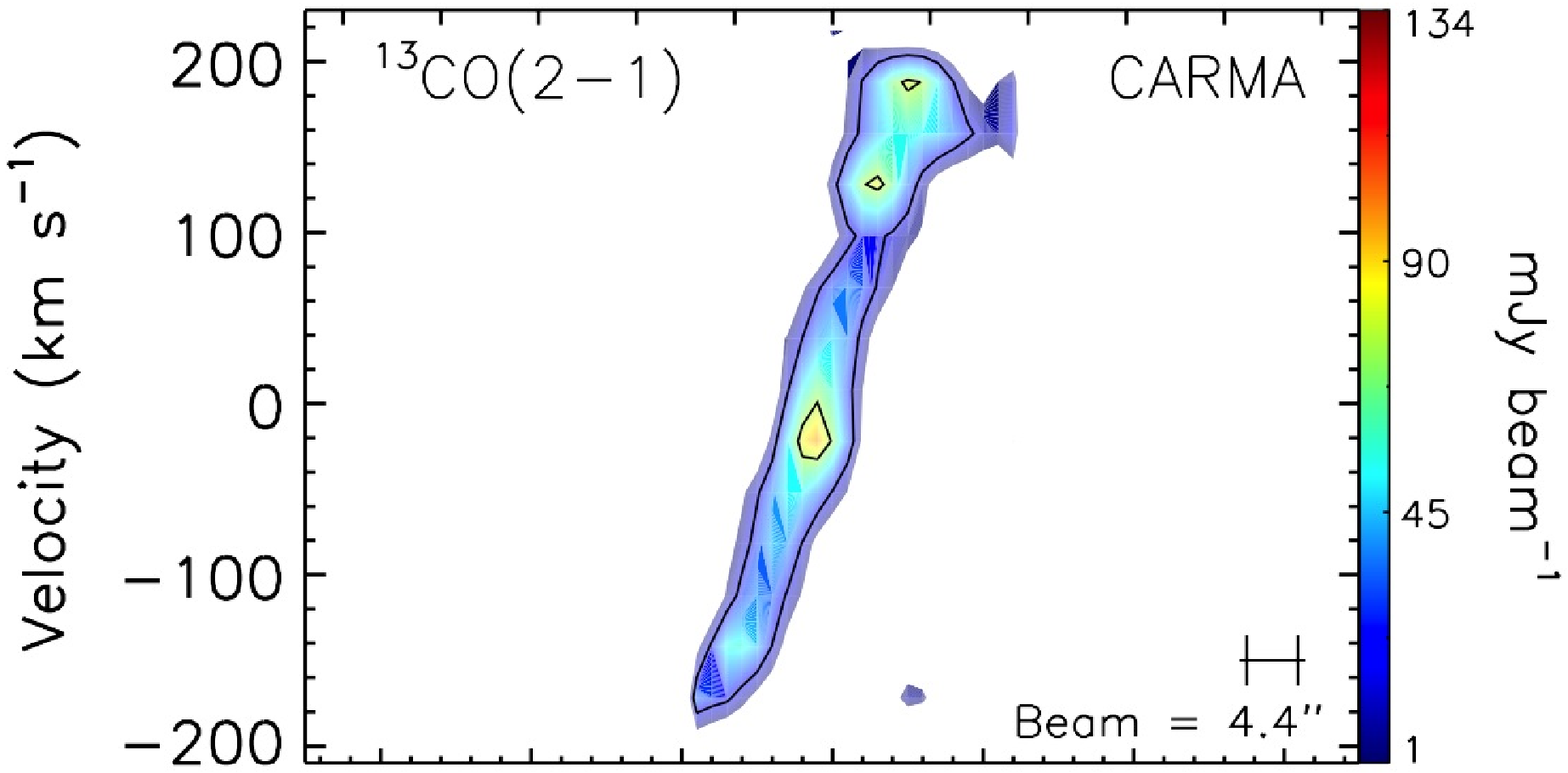}
  \includegraphics[width=5.6cm,clip=]{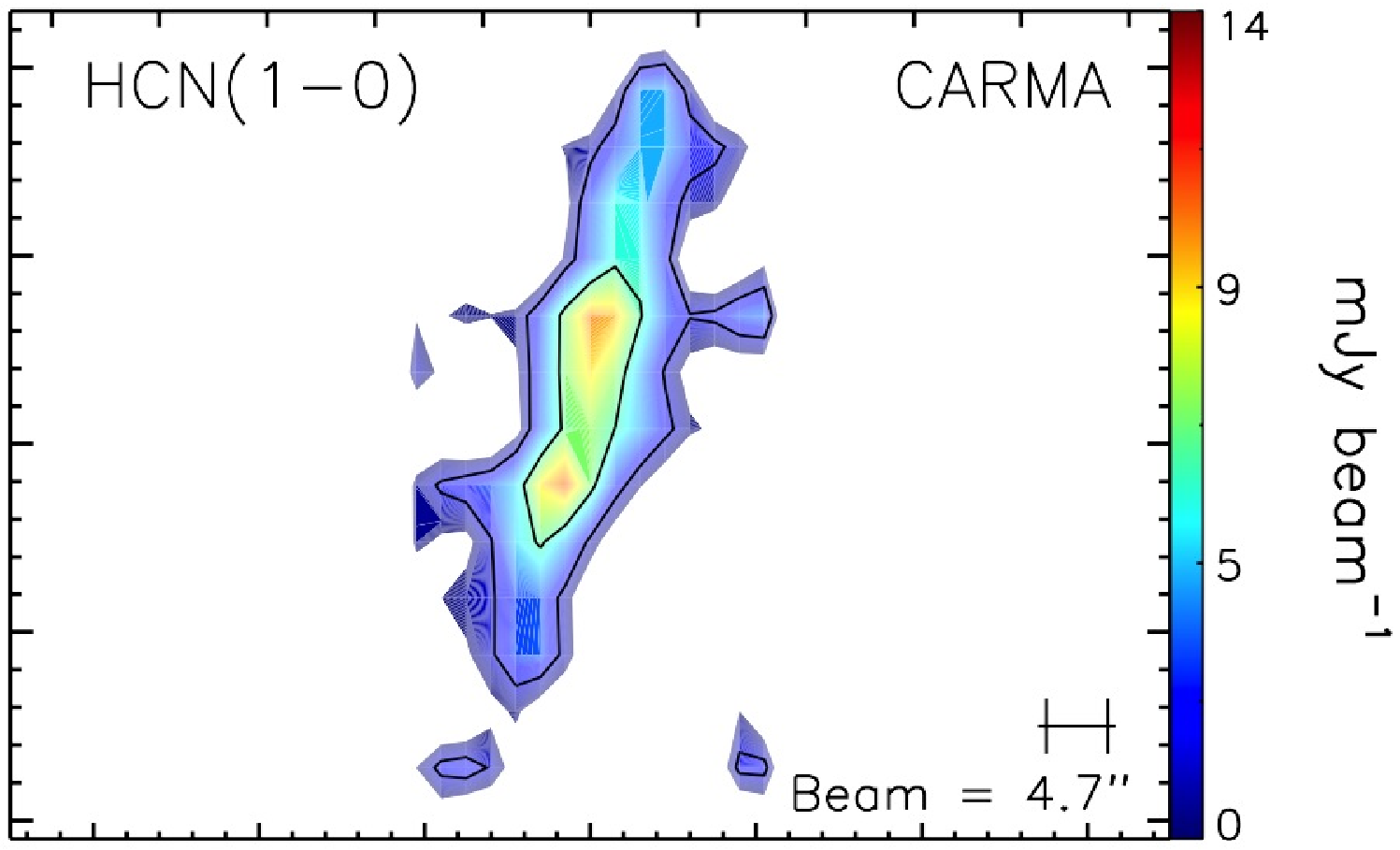}
  \includegraphics[width=5.6cm,clip=]{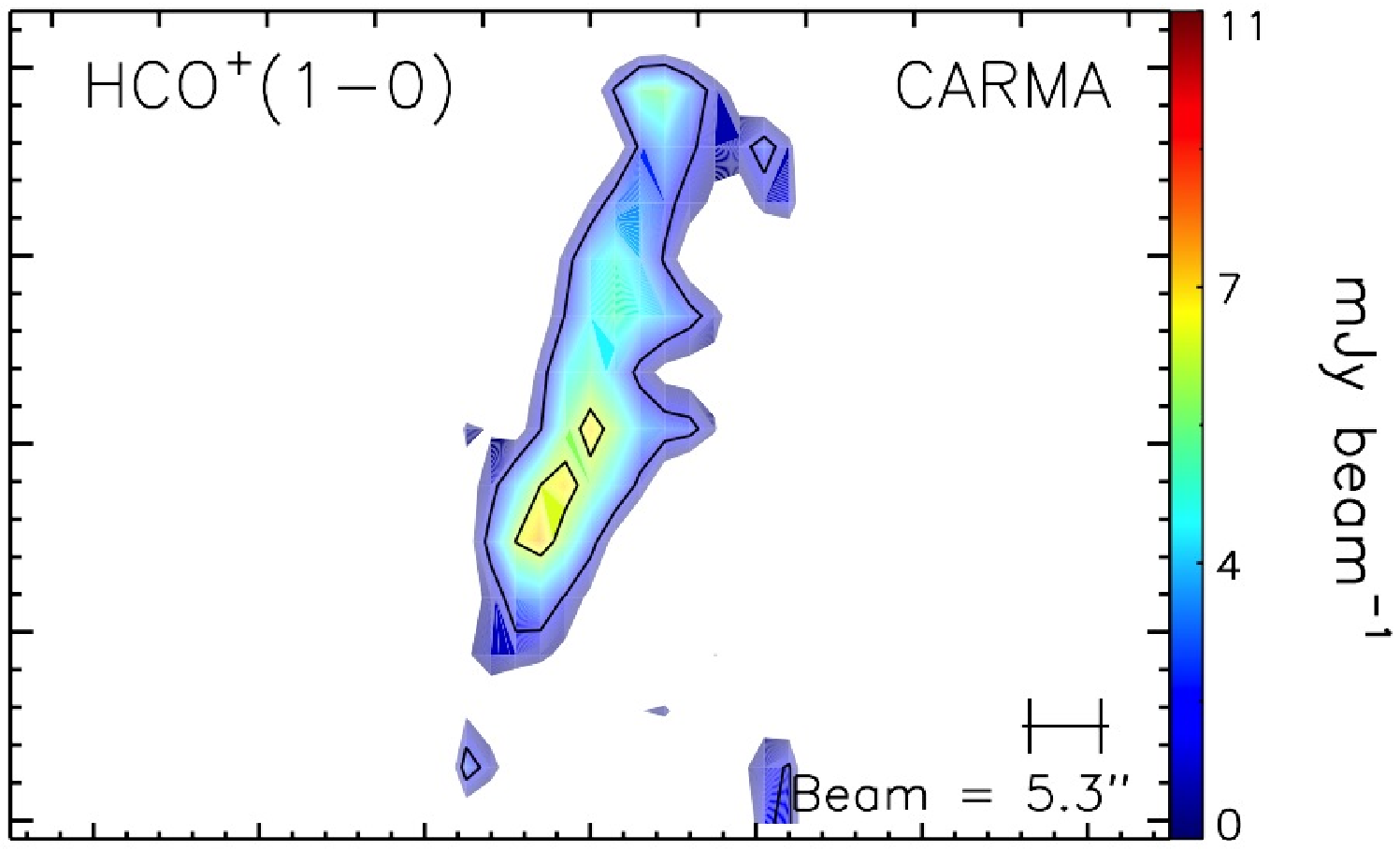}\\
  \vspace{-20pt}
  \includegraphics[width=6.3cm,clip=]{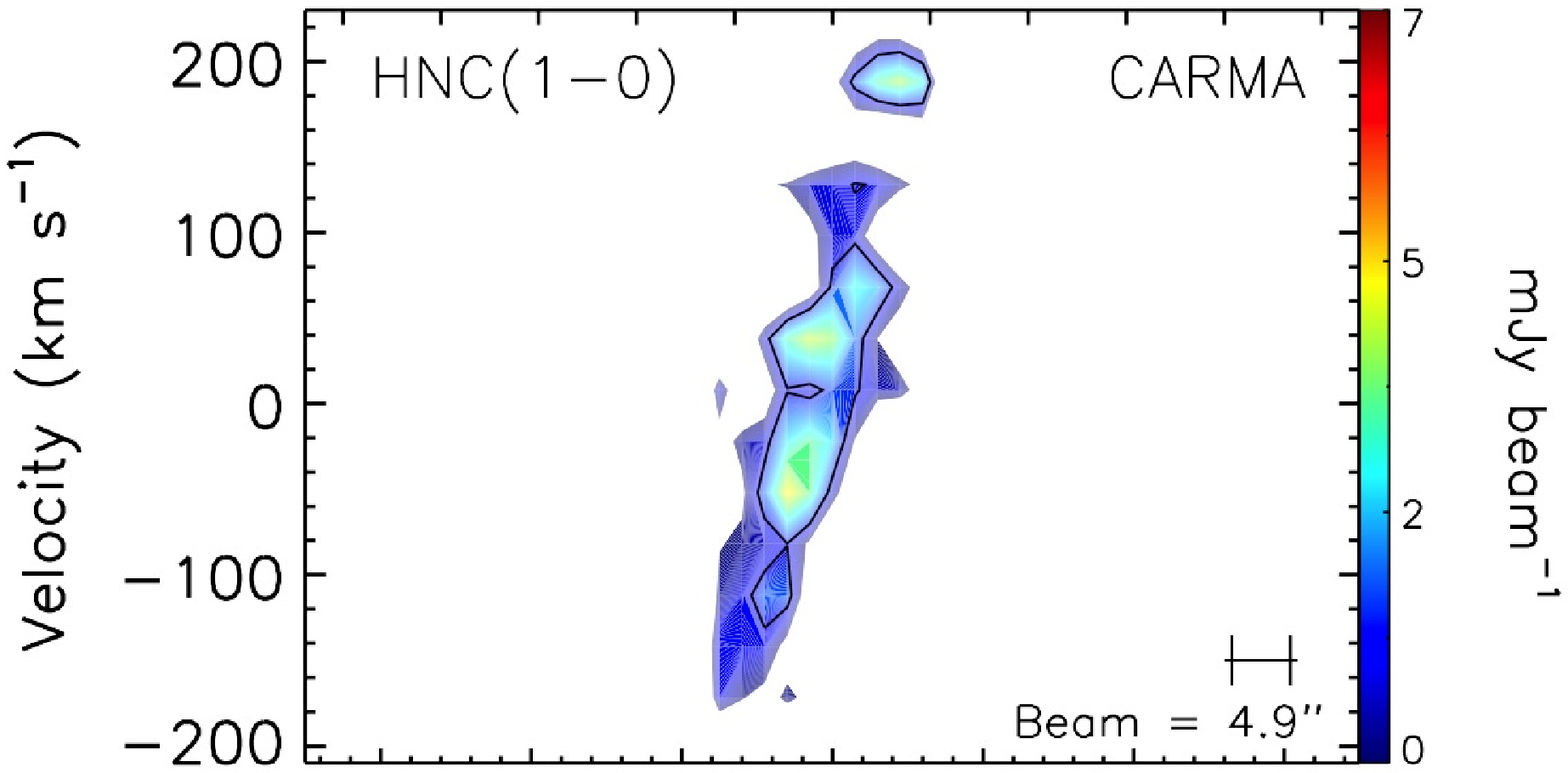}
  \includegraphics[width=5.6cm,clip=]{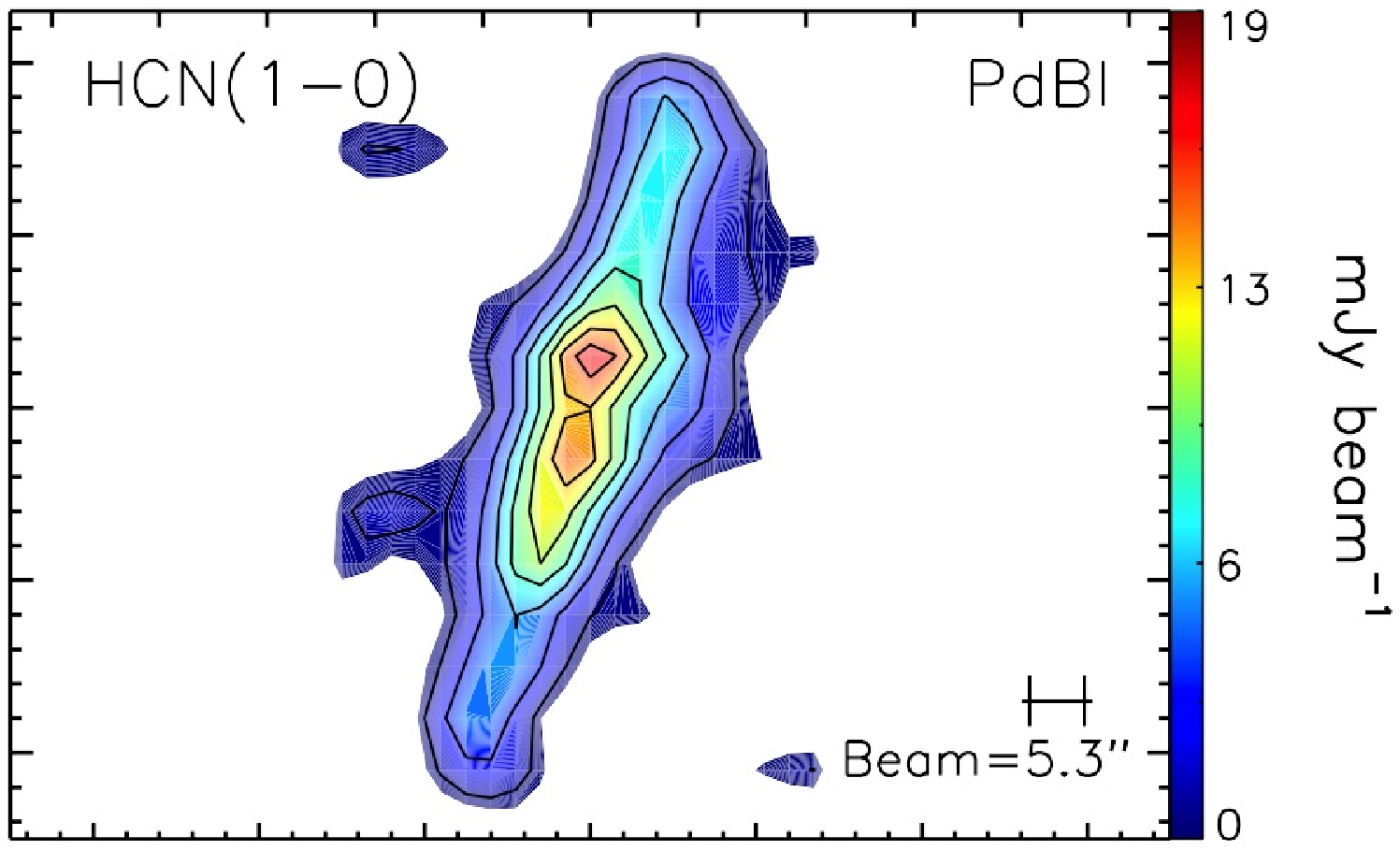}
  \includegraphics[width=5.6cm,clip=]{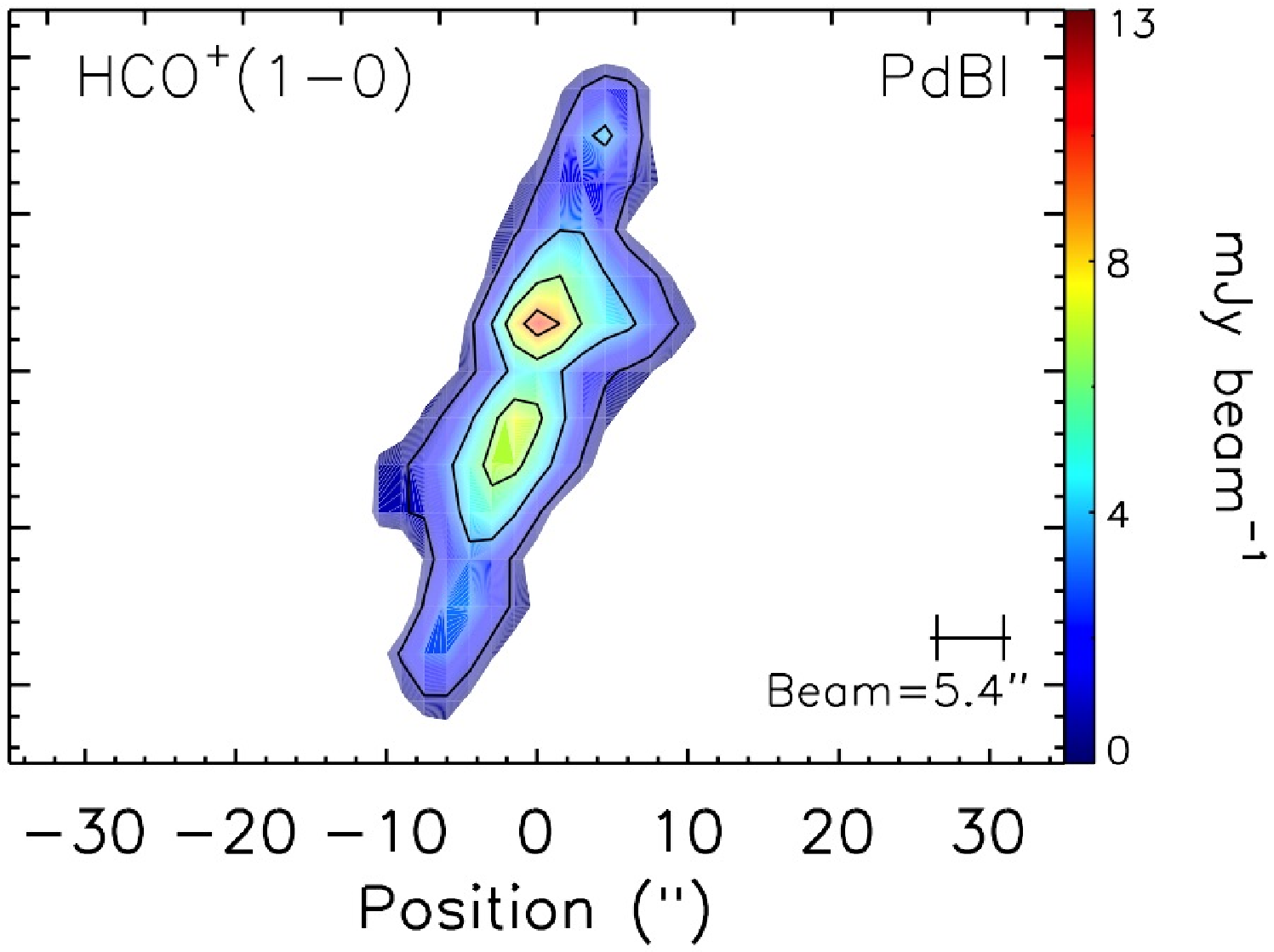}\\
  \vspace{-20pt}
  \hspace{-163pt}
  \includegraphics[width=6.3cm,clip=]{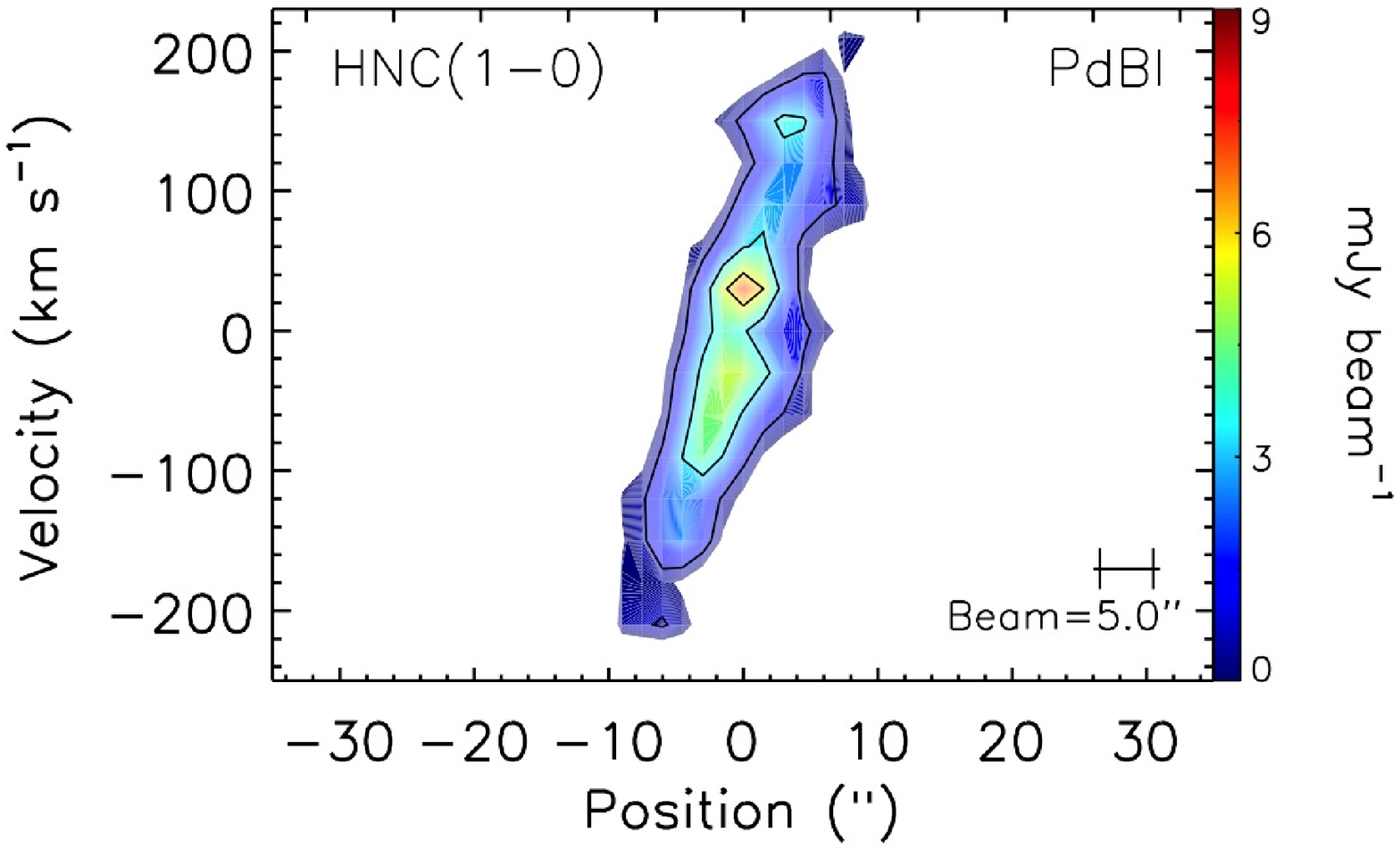}
  \includegraphics[width=5.6cm,clip=]{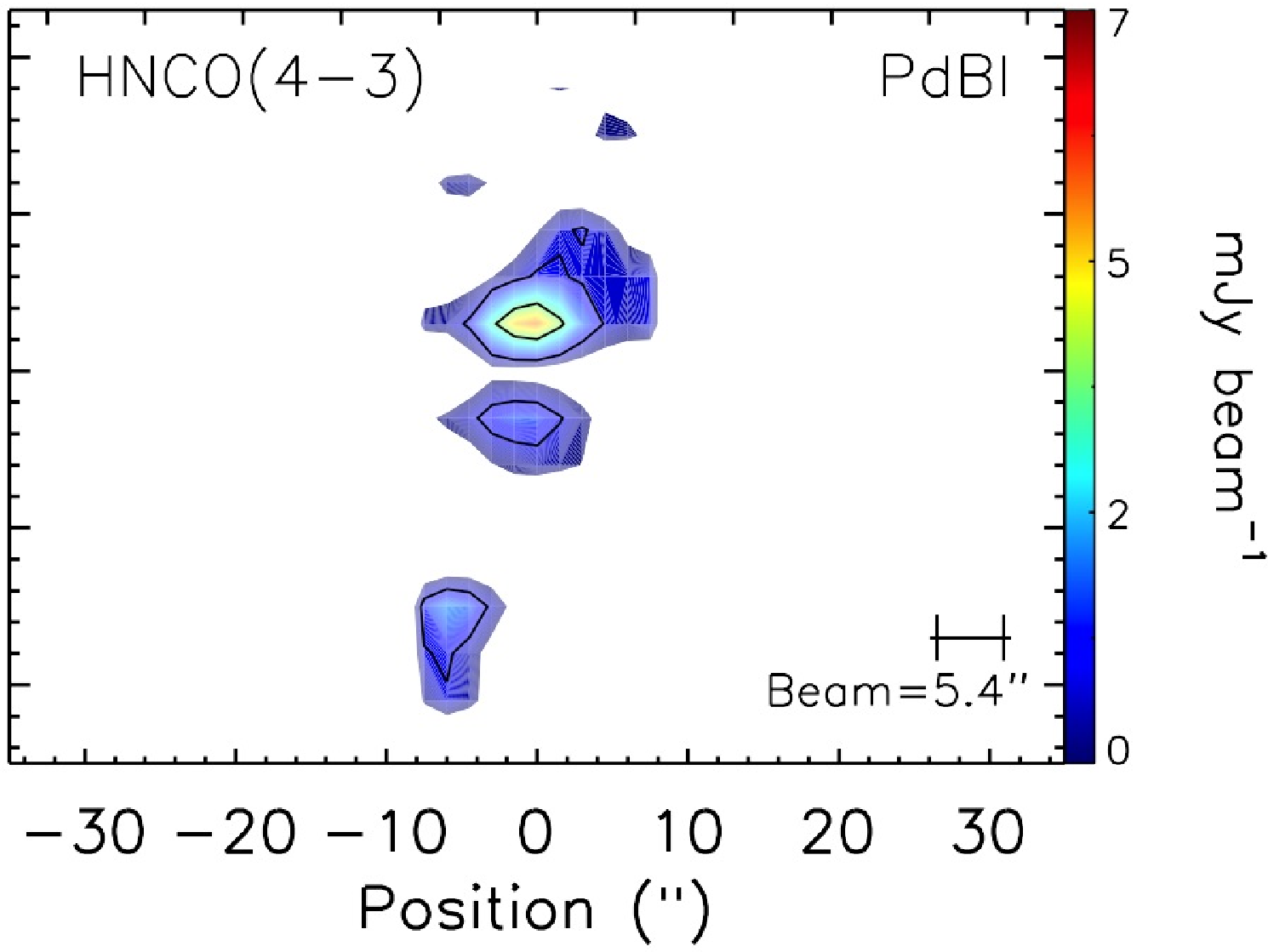}
  \caption{PVDs of all the lines detected in NGC~4710. The PVDs are
    overlaid with contours (black lines) spaced by $3\sigma$ and
    starting at $3\sigma$, while the colour scales start at
    $1\sigma$. The rms noise for the CARMA observations of
    $^{12}$CO(1-0), $^{13}$CO(1-0), $^{12}$CO(2-1), $^{13}$CO(2-1),
    HCN(1-0), HCO$^+$(1-0) and HNC(1-0) is $7.5$, $3.7$, $19.0$,
    $19.0$, $1.5$, $1.6$ and $1.2$~mJy~beam$^{-1}$, respectively,
    while that for the PdBI observations of HCN(1-0), HCO$^+$(1-0),
    HNC(1-0) and HNCO(4-3) is $0.8$, $0.9$, $0.9$ and
    $0.9$~mJy~beam$^{-1}$, respectively. The projected position
    numbers, as discussed in \S~\ref{sec:extract} and illustrated in
    Figure~\ref{fig:pos}, are indicated on the top axes. The array
    used, molecular line displayed, and angular resolution along the
    major axis are also indicated in each panel.}
  \label{fig:n4710pvd}
\end{figure*}

\begin{figure*}
  \includegraphics[width=6.3cm,clip=]{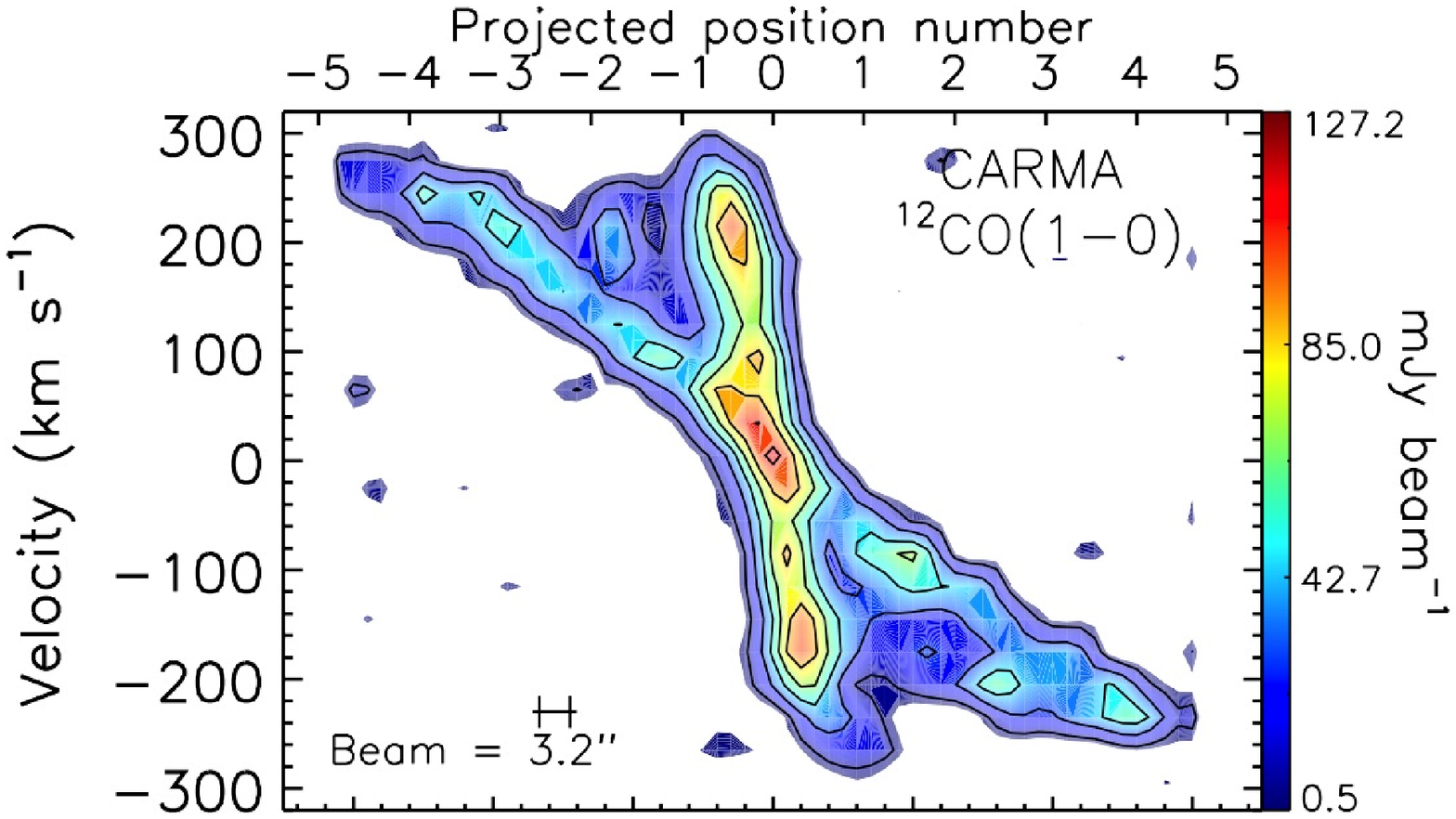}
  \includegraphics[width=5.6cm,clip=]{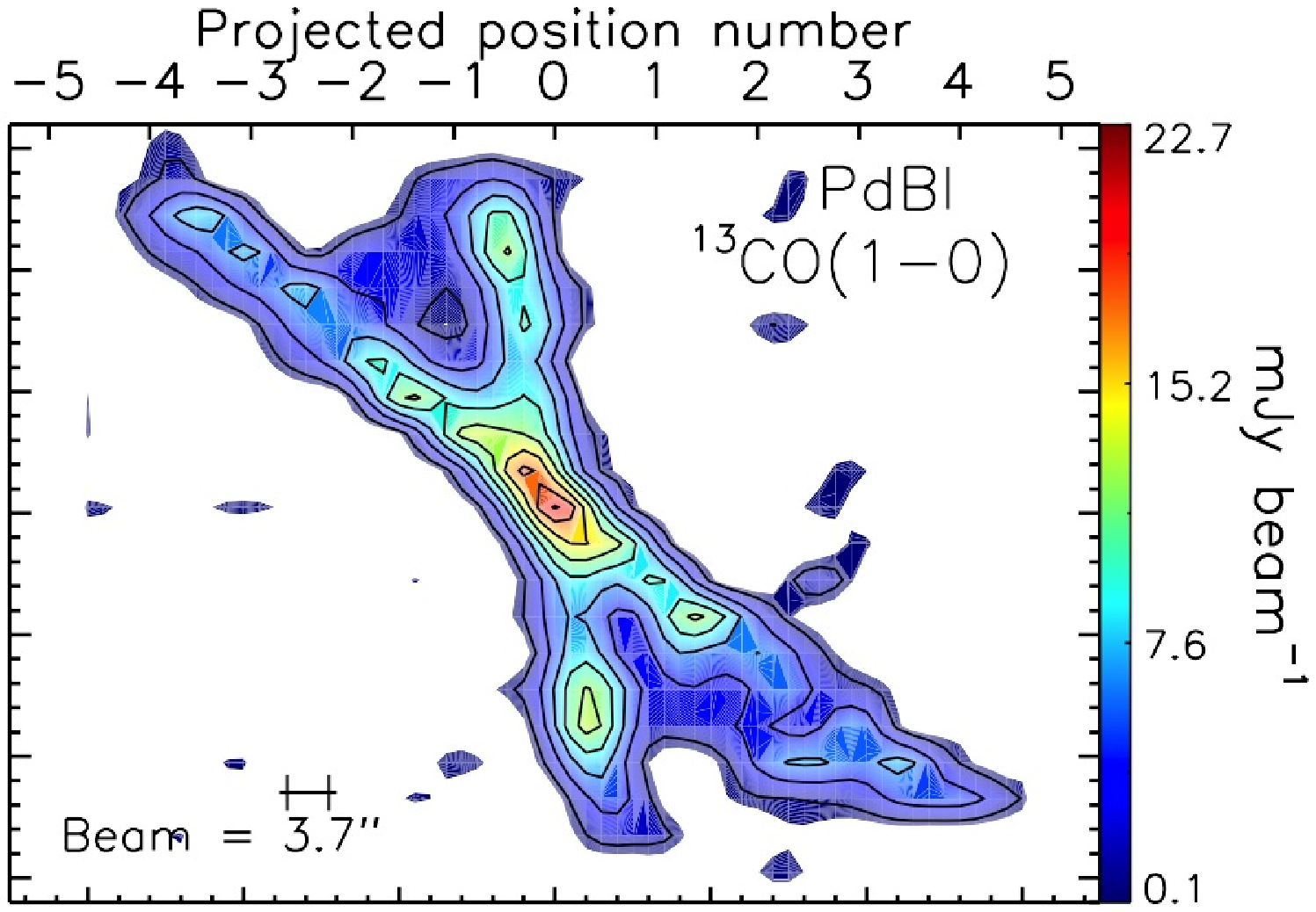}
  \includegraphics[width=5.6cm,clip=]{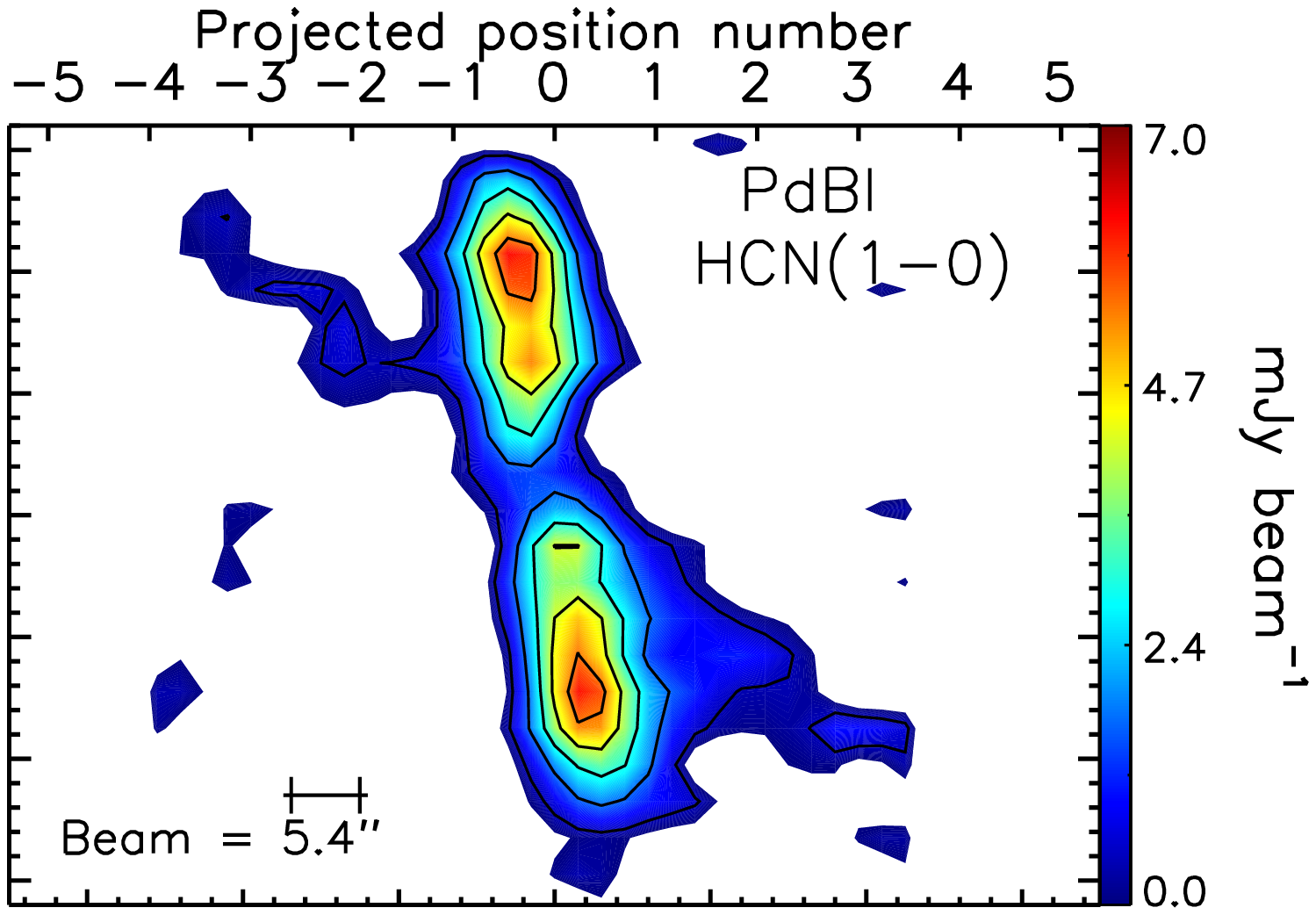}\\
  \vspace{-5pt}
  \includegraphics[width=6.3cm,clip=]{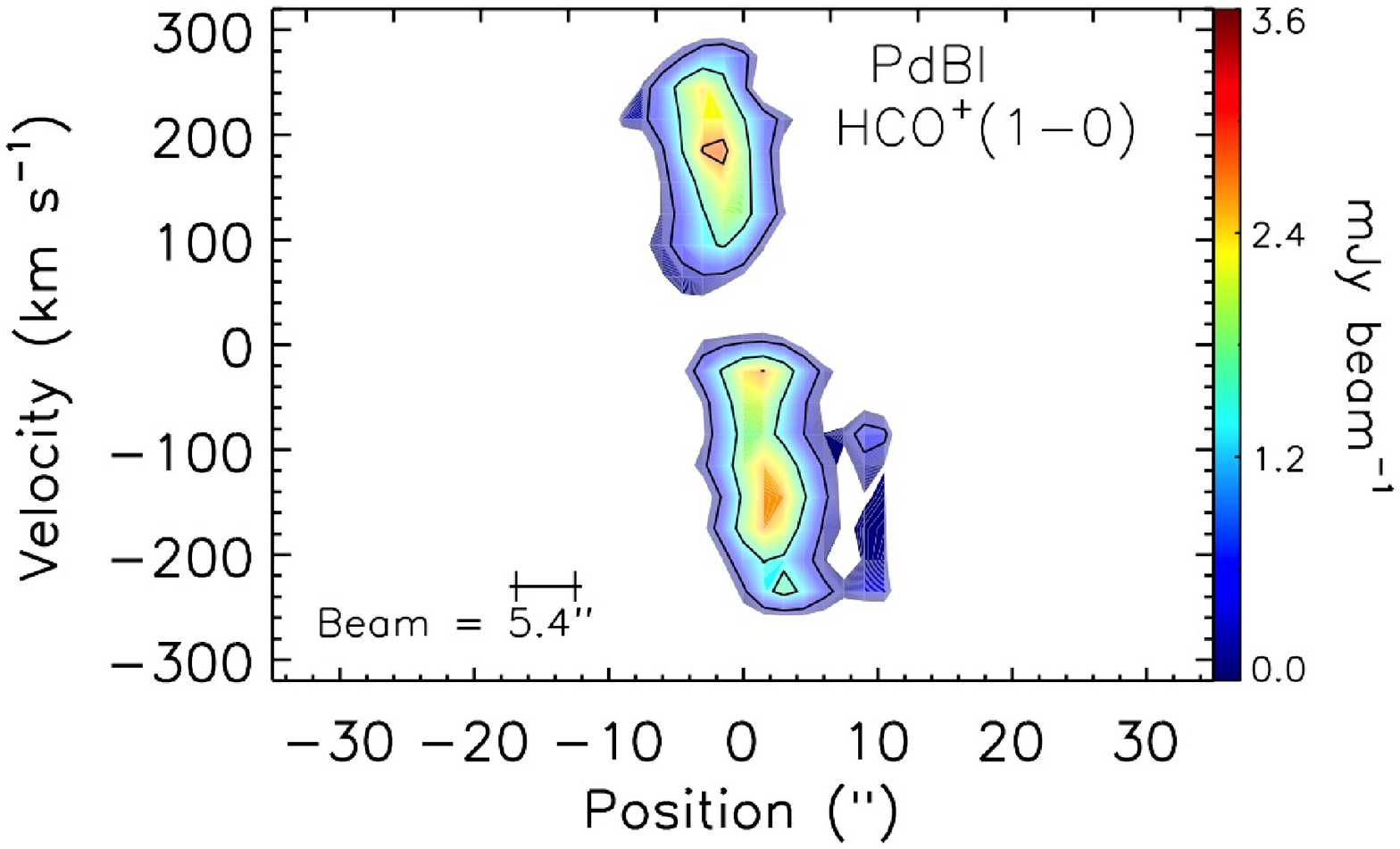}
  \includegraphics[width=5.6cm,clip=]{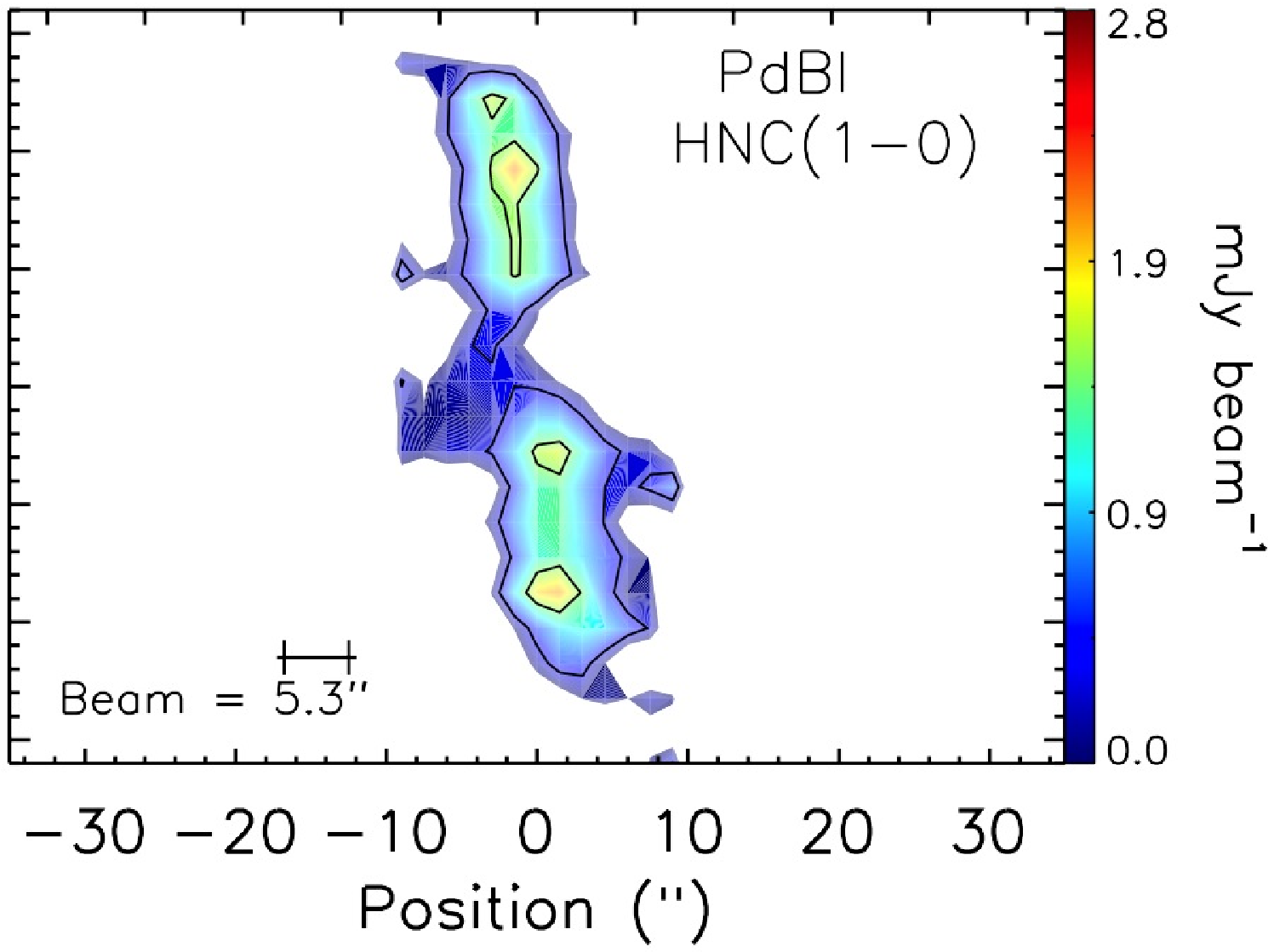}
  \includegraphics[width=5.6cm,clip=]{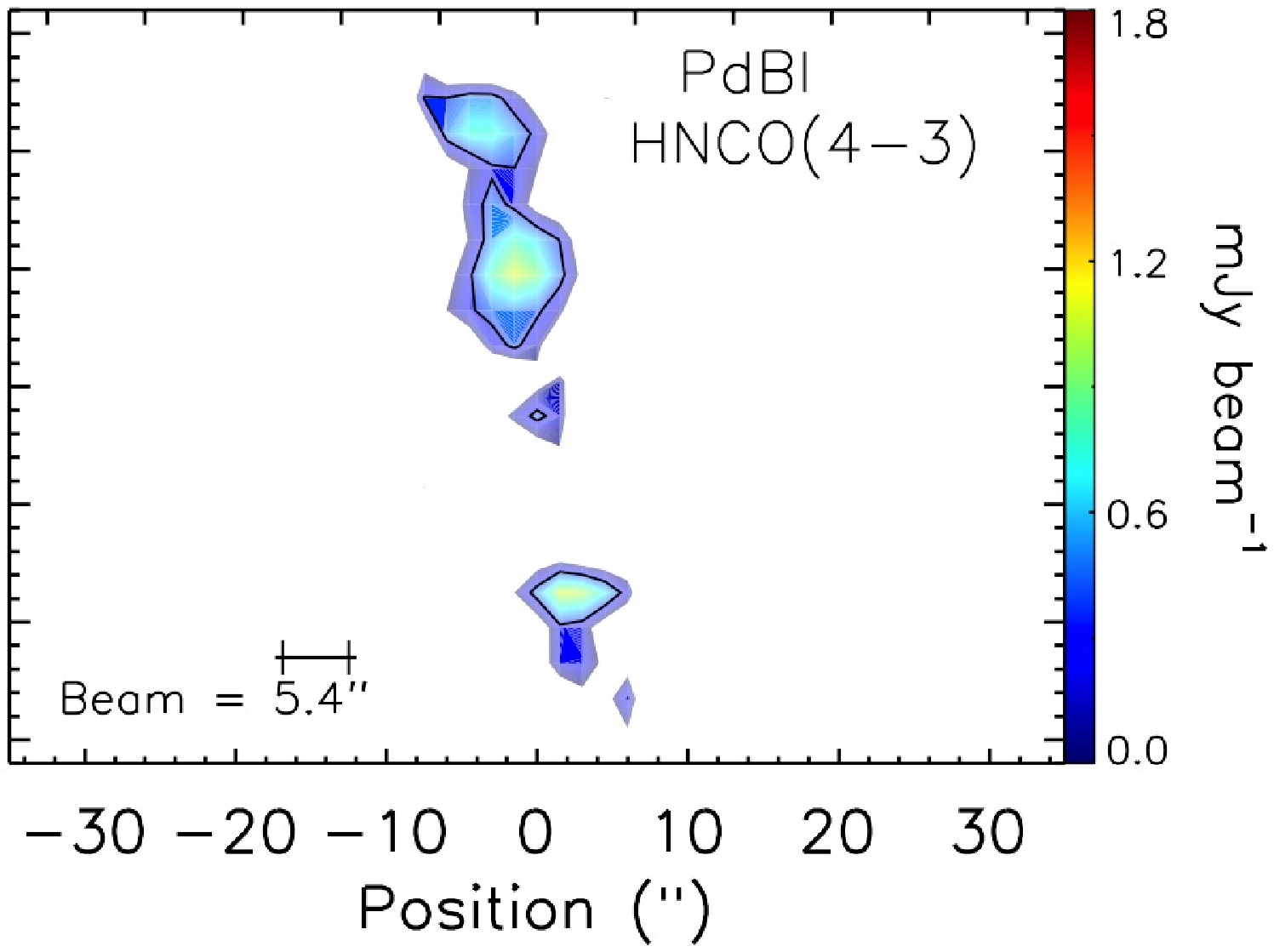}\\
  \caption{Same as Figure~\ref{fig:n4710pvd} but for NGC~5866. The rms
    noise for the observations of $^{12}$CO(1-0), $^{13}$CO(1-0),
    HCN(1-0), HCO$^+$(1-0), HNC(1-0) and HNCO(4-3) are $6.7$, $0.9$,
    $0.4$, $0.4$, $0.4$ and $0.4$~mJy~beam$^{-1}$, respectively.}
  \label{fig:n5866pvd}
\end{figure*}

To create a PVD for each molecular line detected, we simply took a
slice along the major axis of the galaxies in the fully-calibrated and
cleaned data cubes, averaging $5$ pixels in the perpendicular
direction (i.e.\ along the galaxy minor axis; slightly larger than the
synthesised beam, maximising $S/N$ as the emission is generally not
resolved perpendicular to the disc). The PVDs of all the lines
detected in NGC~4710 and NGC~5866 are shown in
Figures~\ref{fig:n4710pvd} and \ref{fig:n5866pvd}, respectively.
The spatial resolution of each PVD, i.e.\ the size of
  the beam along the major axis of the galaxy, was calculated by
  taking into account the position angle, major axis and minor axis of
  synthesised beam (see \citealt{d13}).

%
%
%
%

As seen in Figures~\ref{fig:n4710pvd}\,--\,\ref{fig:n5866pvd}, the
PVDs of the CO lines in both galaxies (except $^{13}$CO(2-1) in
NGC~4710) and the PVD of HCN(1-0) in NGC~5866 reveal X-shape patterns,
with two distinct velocity components: a central rapidly-rising
velocity component (hereafter nuclear disc) and an outer slowly-rising
velocity component (hereafter inner ring). Both can easily be
understood in the context of barred galaxy dynamics (see
\citealt{sell93} for a general review; \citealt{ba99}, \citealt{ab99}
and \citealt{ma99} for the observed kinematics of edge-on
systems). Bar-driven inflows cause gas to accumulate on $x_2$ orbits
(elongated perpendicular to the bar) within the inner Lindblad
resonance (ILR; when present) at the centre of the galaxy, creating a
feature known as a nuclear disc (or ring), and giving rise to the
central rapidly-rising velocity component seen in the PVDs. Gas also
accumulates just beyond the end of the bar near corotation (and
possibly beyond), creating a feature known as an inner ring, and
giving rise to the outer slowly-rising velocity component of the
PVDs. The intermediate region occupied by $x_1$ orbits (elongated
parallel to the bar) is swept free of gas, creating a gap between the
nuclear disc and inner ring (both morphologically and in the
PVDs). However, there is some material there, at least in NGC~4710,
where the X shape of the $^{12}$CO(1-0) PVD resembles a figure of
eight at positive projected radii (see position $3$ in
Fig.~\ref{fig:pos}; less so for other tenuous gas transitions). The
emission coming from the intermediate region is less clear at position
$2$ and $4$ since higher S/N emission coming respectively from the
central disc and inner ring mostly contributes to the flux at these
positions. These three morphological and kinematic features
(nuclear disc, inner ring and empty intermediate
  region) generally constitute reliable bar signatures.

The boxy/peanut-shaped bulge of NGC4710 confirms that it is barred,
while the dominant classical bulge of NGC5866 makes the characteristic
kinematic bar signatures observed somewhat surprising (see the optical
images in Figures~\ref{fig:n4710mom} and
\ref{fig:n5866mom}). The X-shaped PVD (and thus kinematic bar
signature) was first observed in NGC~4710 by \citet{wr92} and in
NGC~5866 by \citet{al13}, both from $^{12}$CO(1-0) observations.

The nuclear disc has a radial extent of $\approx12\arcsec$
($\approx1$~kpc) in NGC~4710 and $\approx8\arcsec$ ($\approx0.6$~kpc)
in NGC~5866, while the inner ring has a radius of $\approx32\arcsec$
($\approx2.6$~kpc) in NGC~4710 and $\approx32\arcsec$
($\approx2.4$~kpc) in NGC~5866 (see Figs.~\ref{fig:n4710pvd} and
\ref{fig:n5866pvd}).
%
%
\subsection{Comparisons with IRAM 30m data}
\label{sec:comp}
Some lines detected with CARMA and PdBI in this work were also
previously detected with the IRAM 30m single-dish telescope. For
NGC~4710, the relevant lines are $^{12}$CO(1-0), $^{12}$CO(2-1)
\citep{y11}, $^{13}$CO(1-0), $^{13}$CO(2-1), HCN(1-0) and HCO$^+$(1-0)
\citep{c12}, while those for NGC~5866 are $^{12}$CO(1-0) \citep{ws03},
$^{13}$CO(1-0), HCN(1-0) and HCO$^+$(1-0) \citep{c12}. As the IRAM~30m
beam is smaller than the primary beams of CARMA and PdBI at any
frequency (see Figs.~\ref{fig:n4710mom} and \ref{fig:n5866mom}), CARMA
and PdBI are better able to recover the true integrated molecular gas
content of the galaxies (as long as the interferometer does not filter
out diffuse, extended emission).

To check the consistency of the datasets from the IRAM 30m single-dish
telescope and the interferometers, we simulated IRAM 30m integrated
spectra using our CARMA and PdBI interferometric data cubes. To do
this, for each line we summed the flux in our cubes spatially using a
Gaussian weighting function of FWHM equal to that of the single-dish
beam at the given frequency (and centred on the galaxies, thus
assuming no 30m pointing error). The IRAM~30m beam size adopted for
the $^{12}$CO(1-0), $^{13}$CO(1-0), $^{12}$CO(2-1) and $^{13}$CO(2-1)
line was $22\arcsec$, $23\arcsec$, $11\arcsec$ and $11\farcs5$,
respectively, while that for the HCN(1-0), HCO$^+$(1-0), HNC(1-0) and
HNCO(4-3) lines was $28\arcsec$. When the IRAM 30m fluxes were listed
in Kelvin (either antenna temperature $T_{\rm a}^*$ or main beam
temperature $T_{\rm mb}$), we converted the fluxes using the
conversion factors given in the associated papers or listed on the
IRAM
website\footnote{http://www.iram.es/IRAMES/mainWiki/Iram30mEfficiencies}. For
values at frequencies not specified there, we linearly interpolated
between the two nearest values.

\begin{figure*}
  \centering
  \includegraphics[width=6.0cm,clip=]{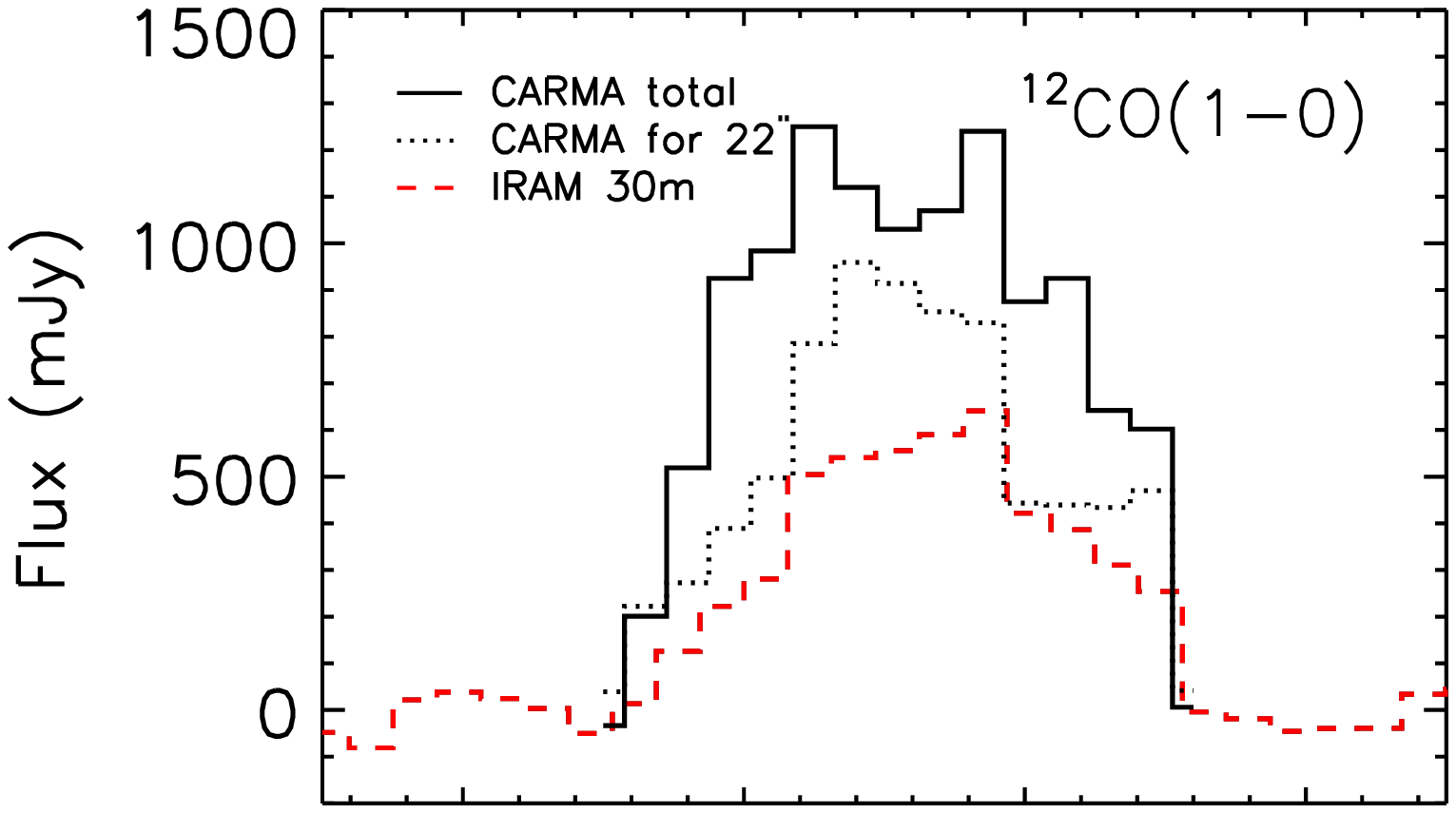}
  \hspace{-20pt}
  \includegraphics[width=6.0cm,clip=]{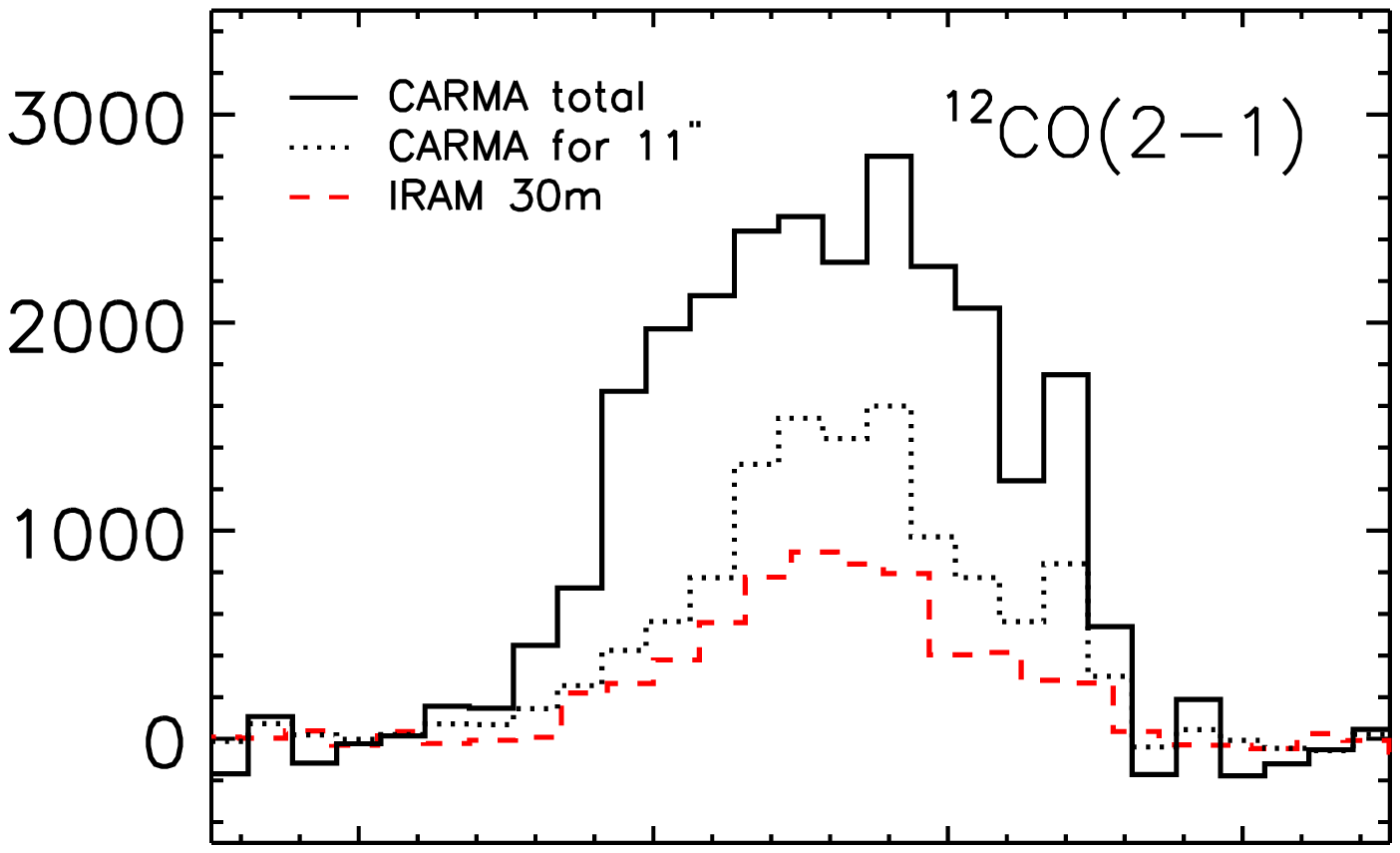}
  \hspace{-20pt}
  \includegraphics[width=6.0cm,clip=]{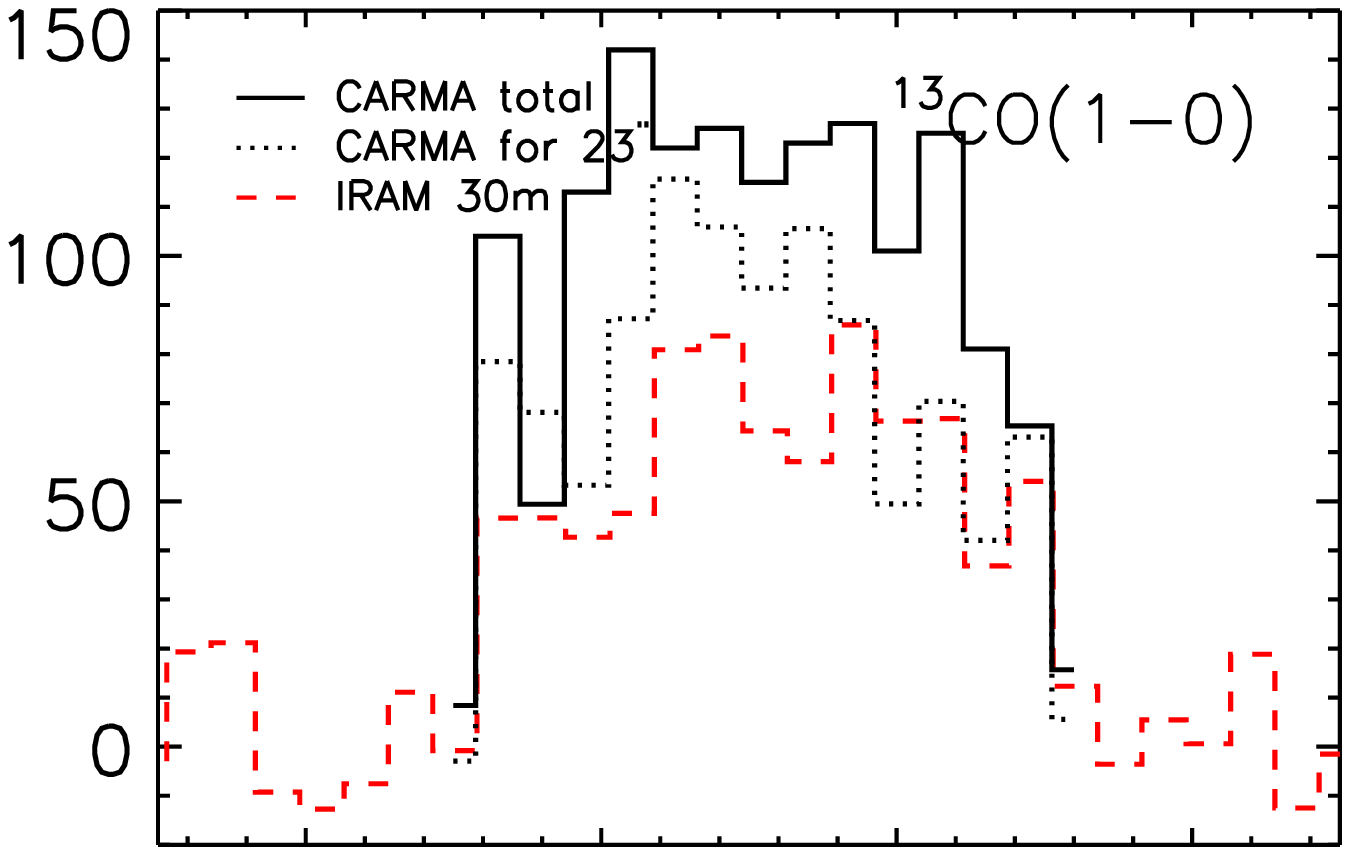} \\
  \vspace{-35pt}
  \includegraphics[width=6.0cm,clip=]{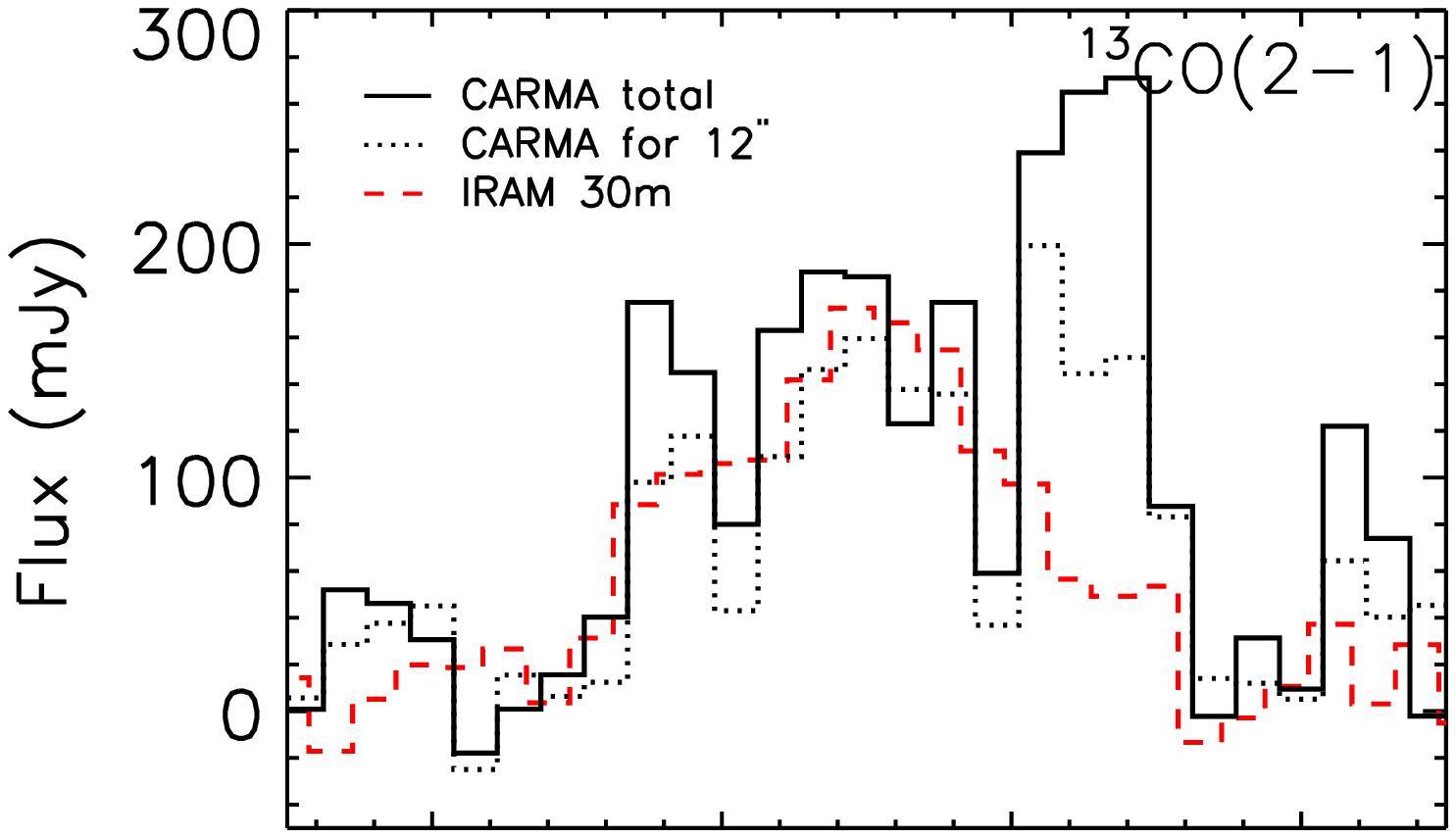}
  \hspace{-20pt}
  \includegraphics[width=6.0cm,clip=]{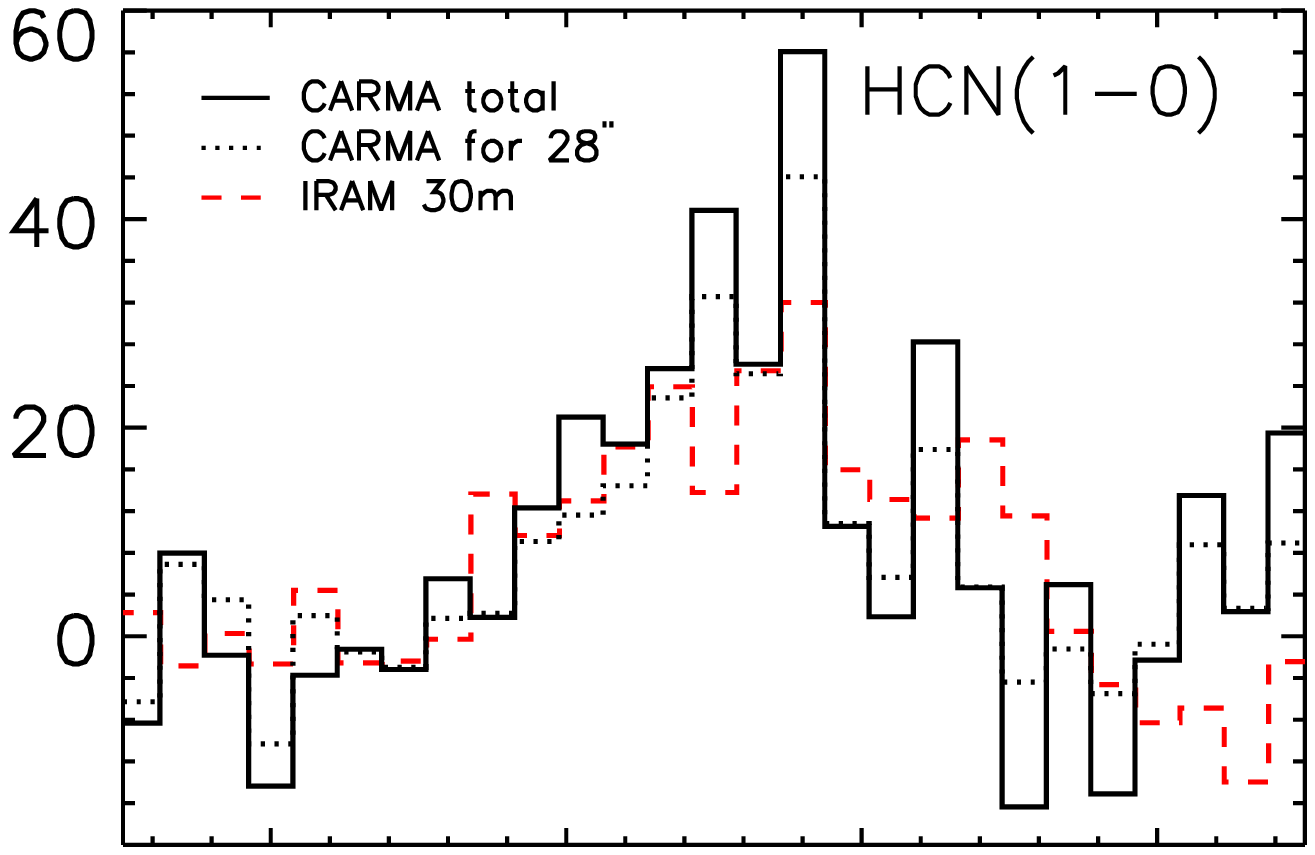}
  \hspace{-20pt}
  \includegraphics[width=6.0cm,clip=]{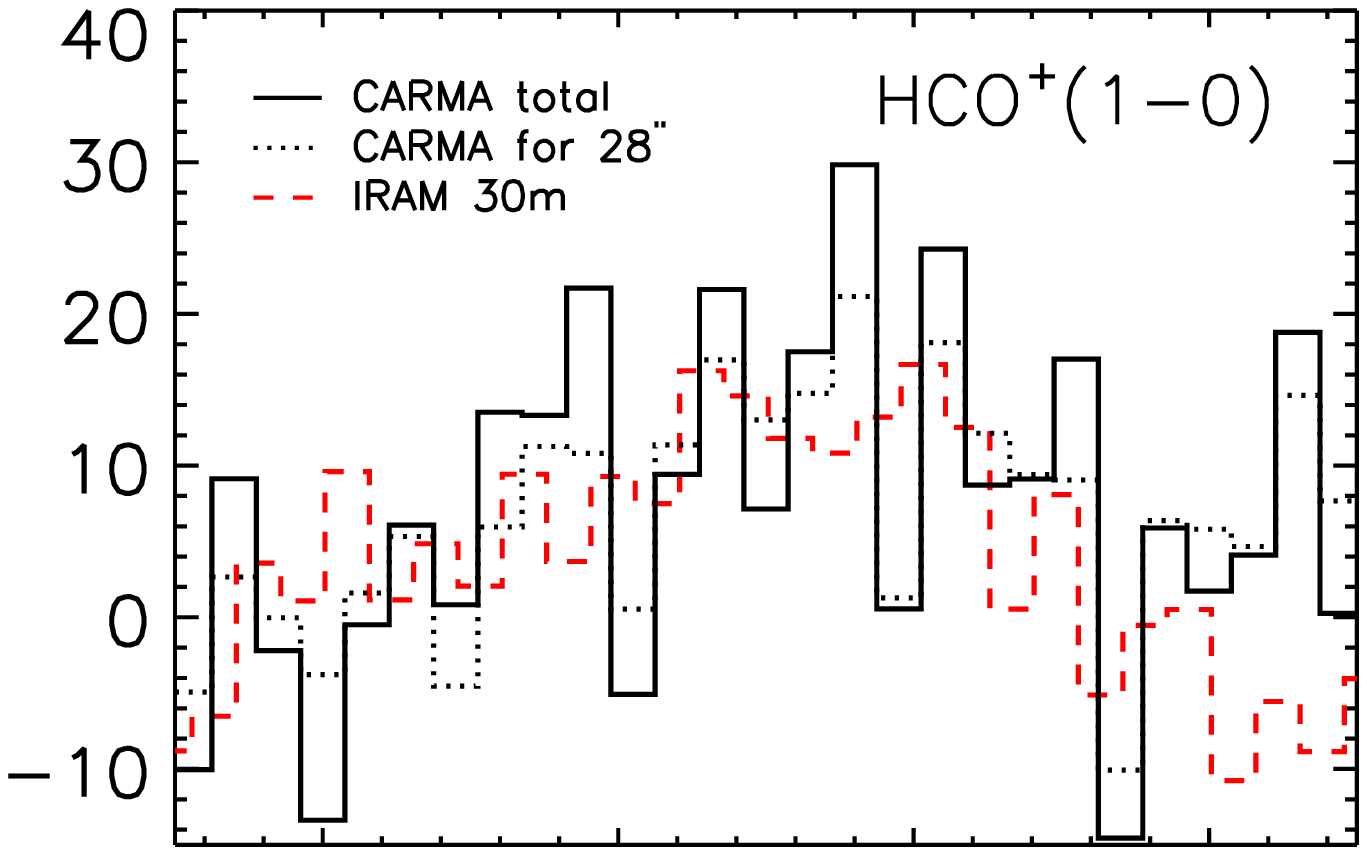} \\
  \vspace{-35pt}
  \includegraphics[width=6.0cm,clip=]{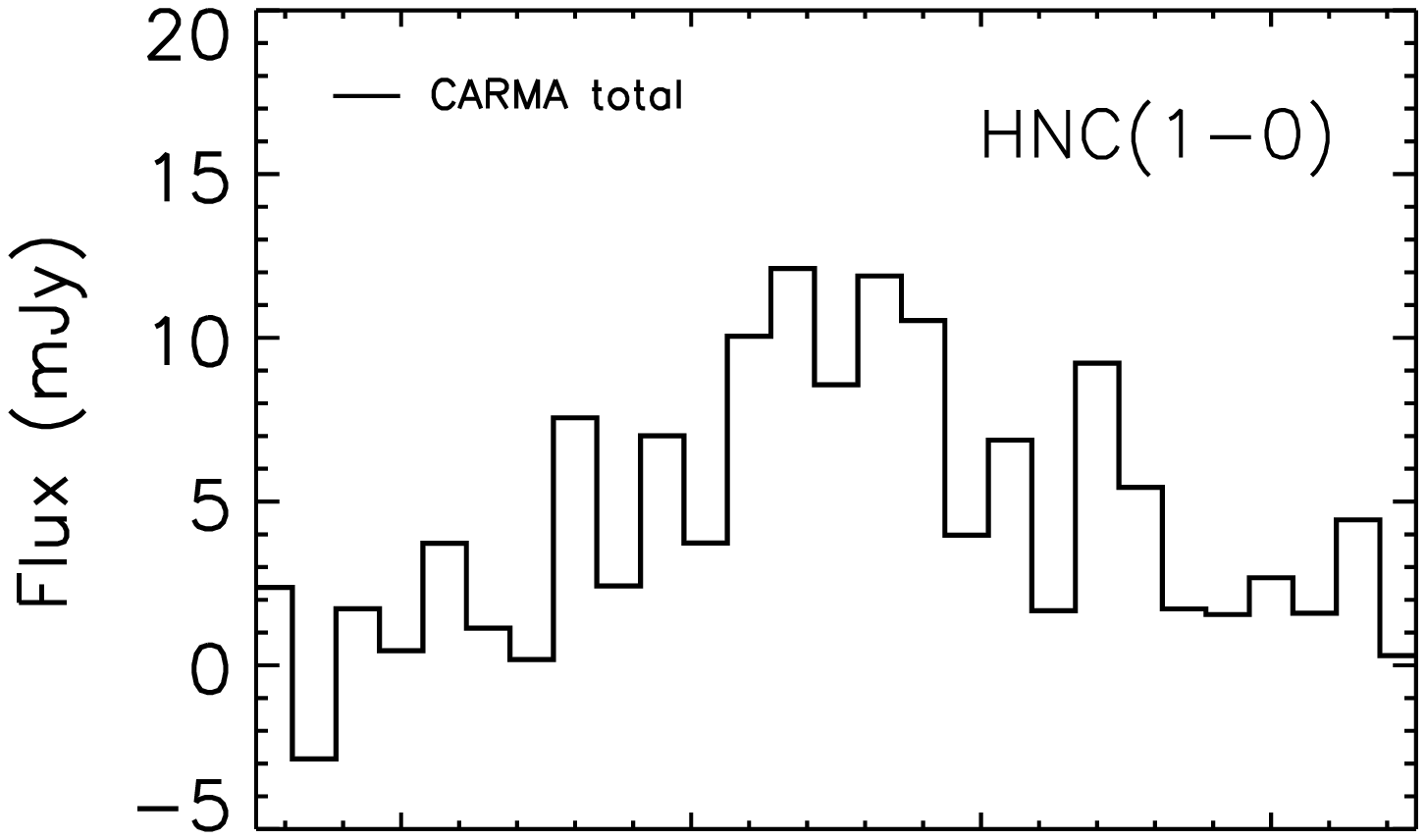}
  \hspace{-20pt}
  \includegraphics[width=6.0cm,clip=]{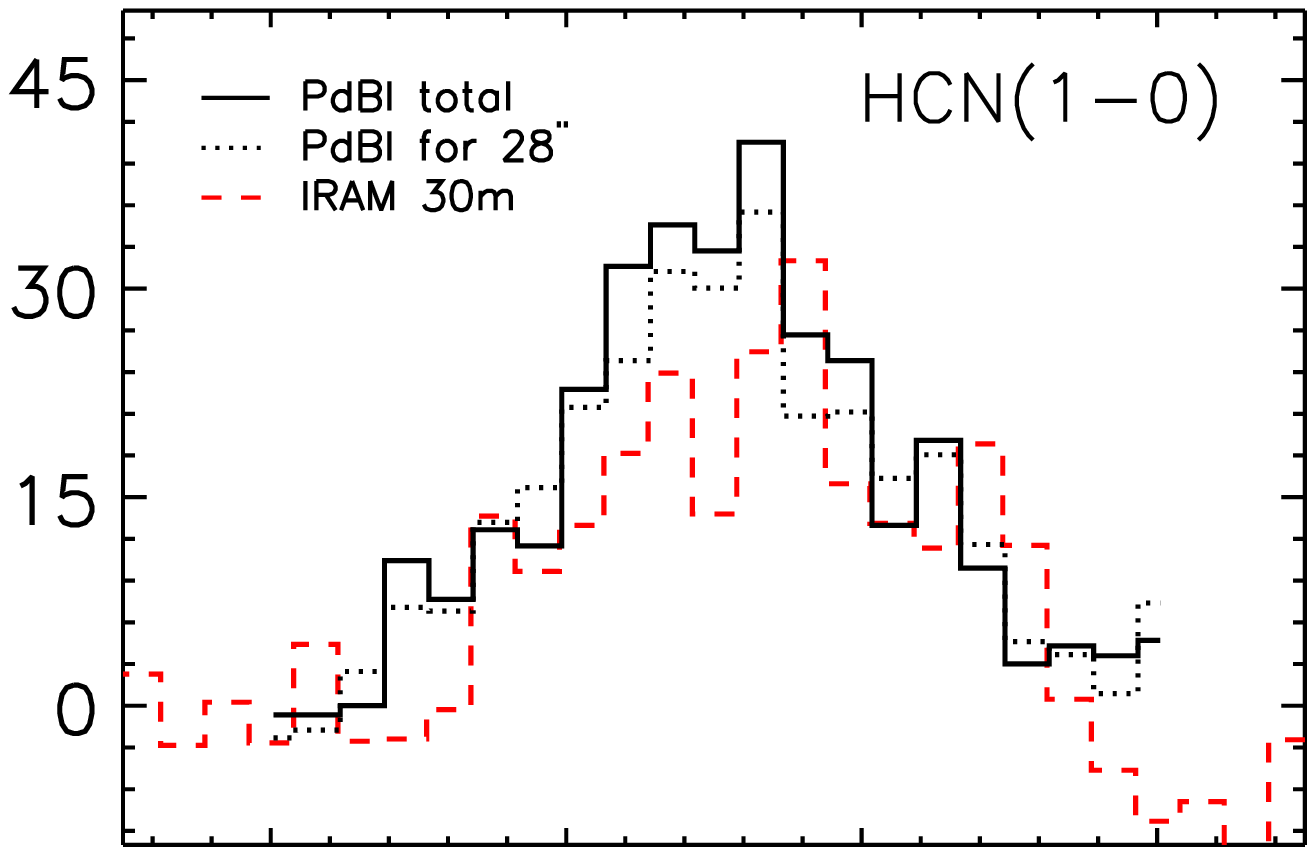}
  \hspace{-20pt}
  \includegraphics[width=6.0cm,clip=]{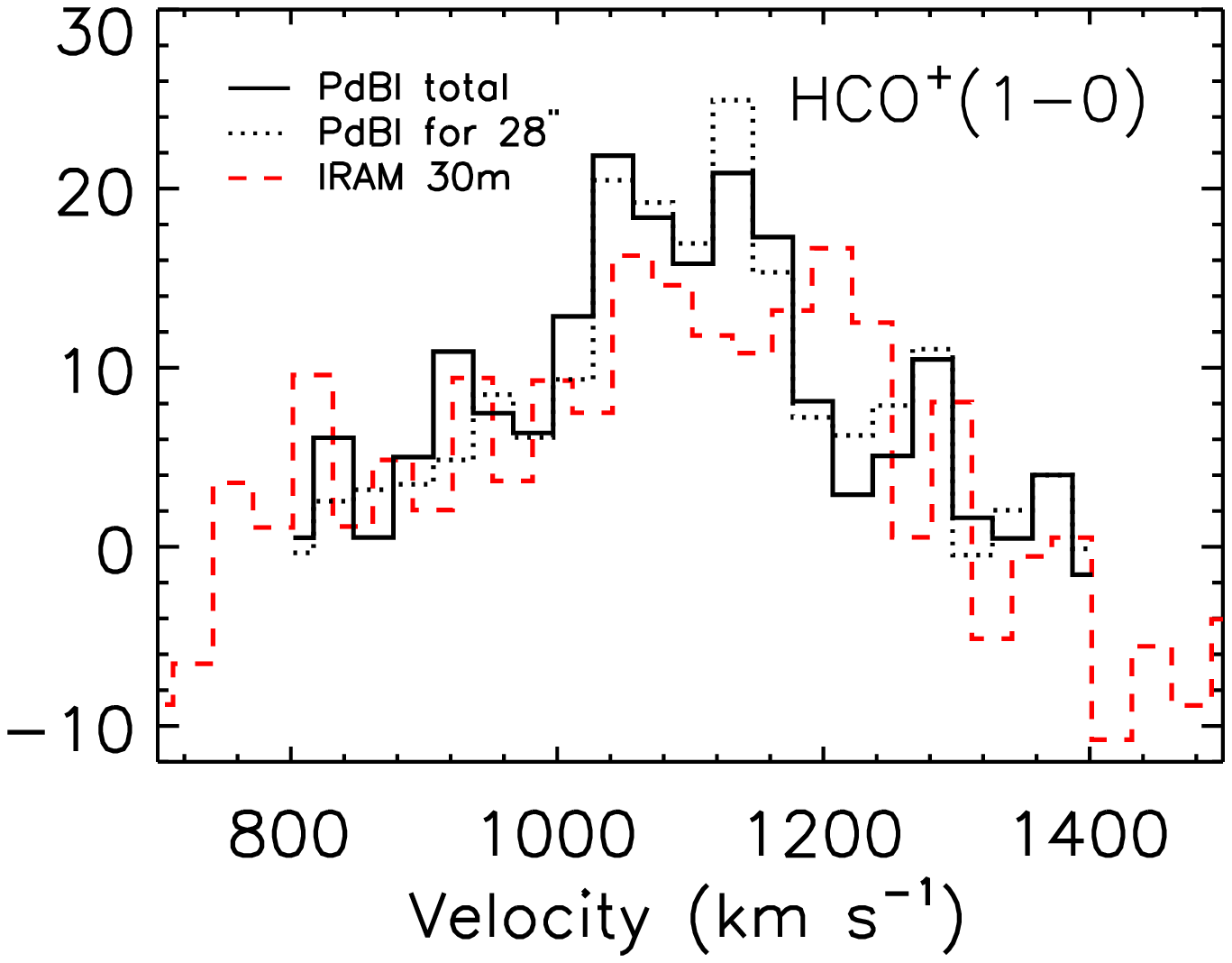} \\
  \vspace{-35pt}
  \hspace{-157pt}
  \includegraphics[width=6.0cm,clip=]{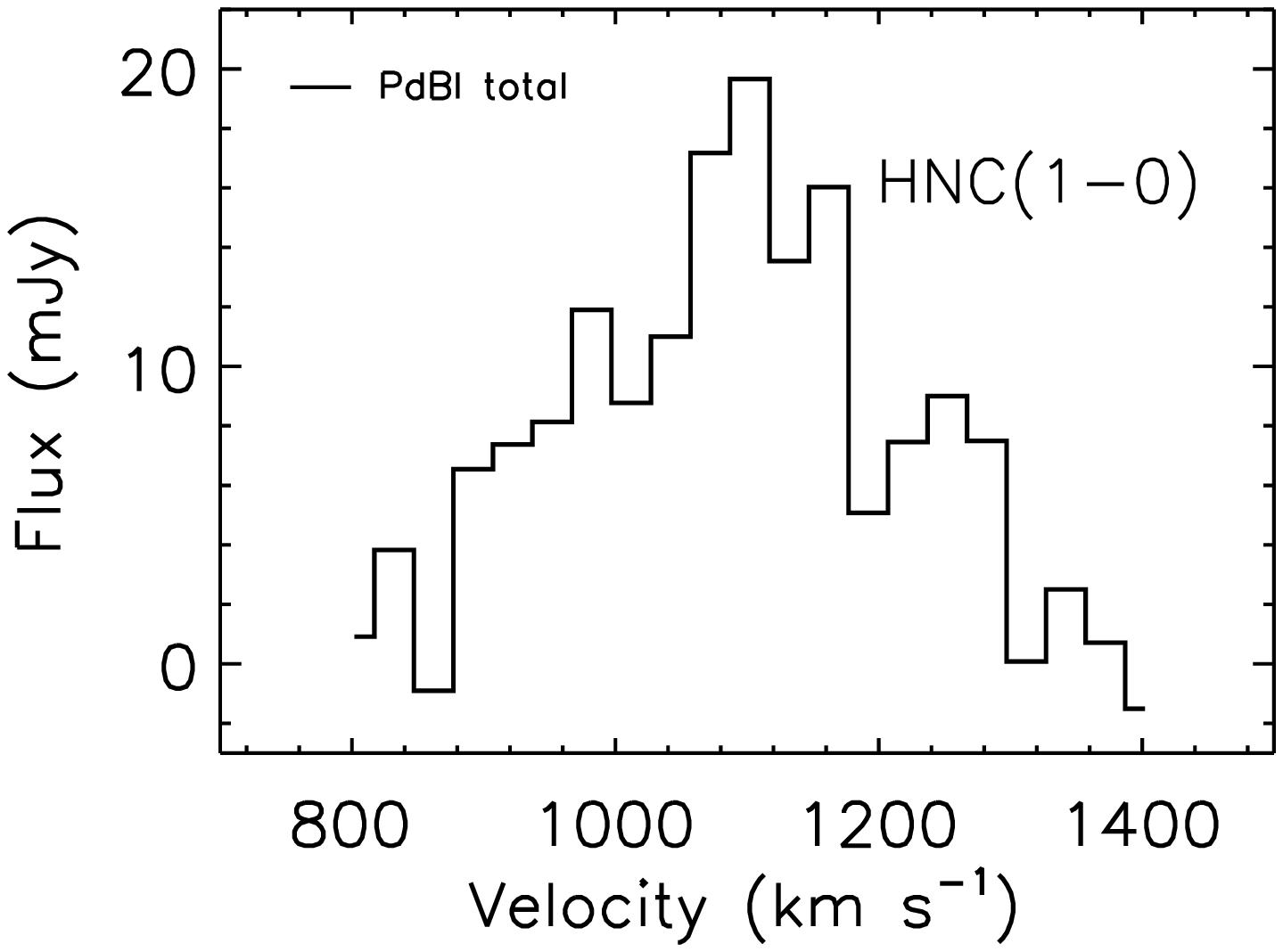}
  \hspace{-20pt}
  \includegraphics[width=6.0cm,clip=]{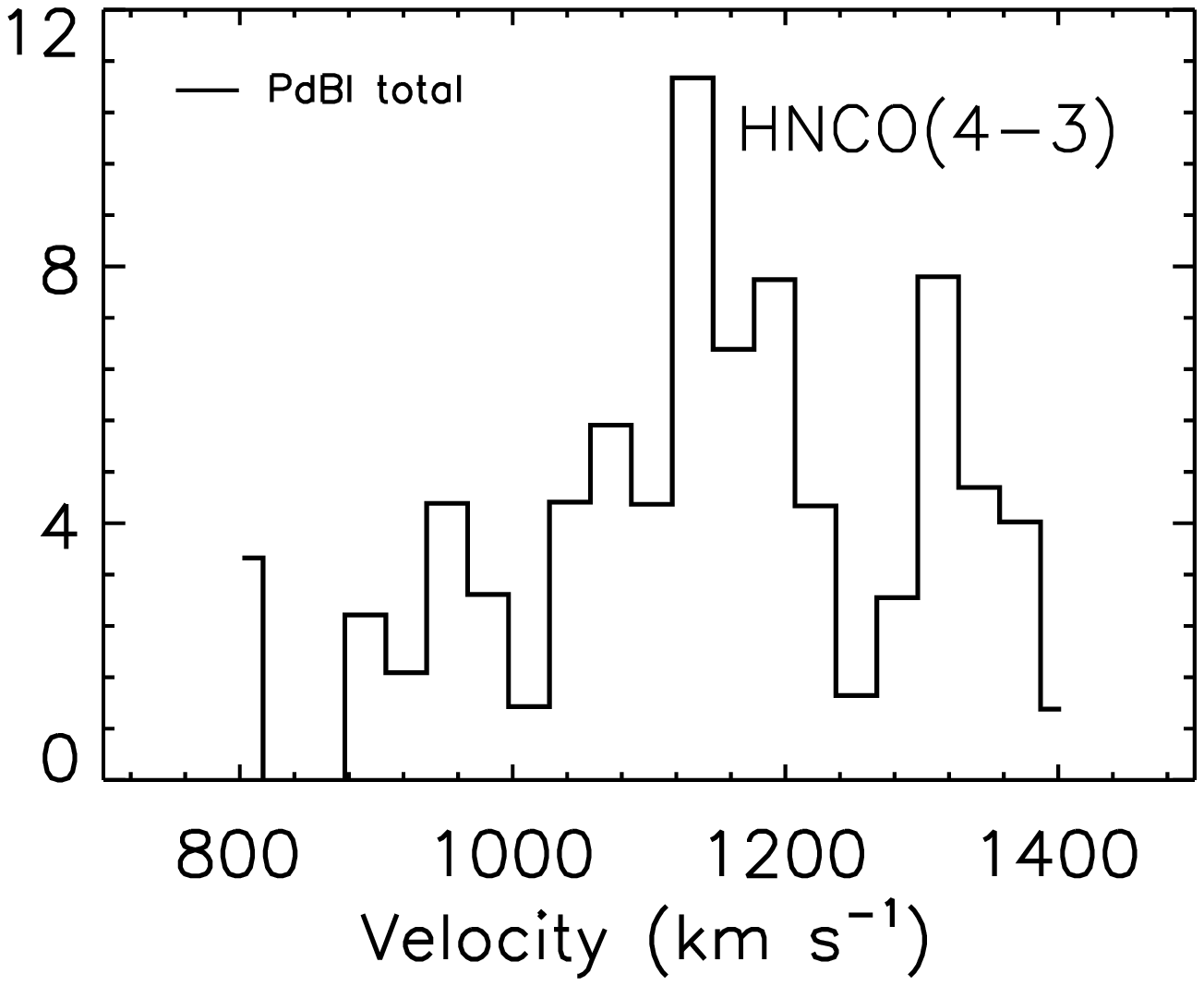}             
  \caption{Comparison of integrated molecular line spectra for
    NGC~4710. Black solid lines show the integrated spectra of our
    CARMA and PdBI observations, with no spatial weighting nor
    limit. Black dotted lines show the IRAM 30m integrated spectra
    simulated from our CARMA and PdBI observations (with Gaussian
    spatial weighting; see \S~\ref{sec:comp}). Red dashed lines show
    original IRAM 30m integrated spectra from the literature. The
    HNC(1-0) and HNCO(4-3) lines were never observed previously, so
    their integrated spectra are shown here for the first time.}
  \label{fig:spec1}
\end{figure*}

\begin{figure*}
  \includegraphics[width=6.0cm,clip=]{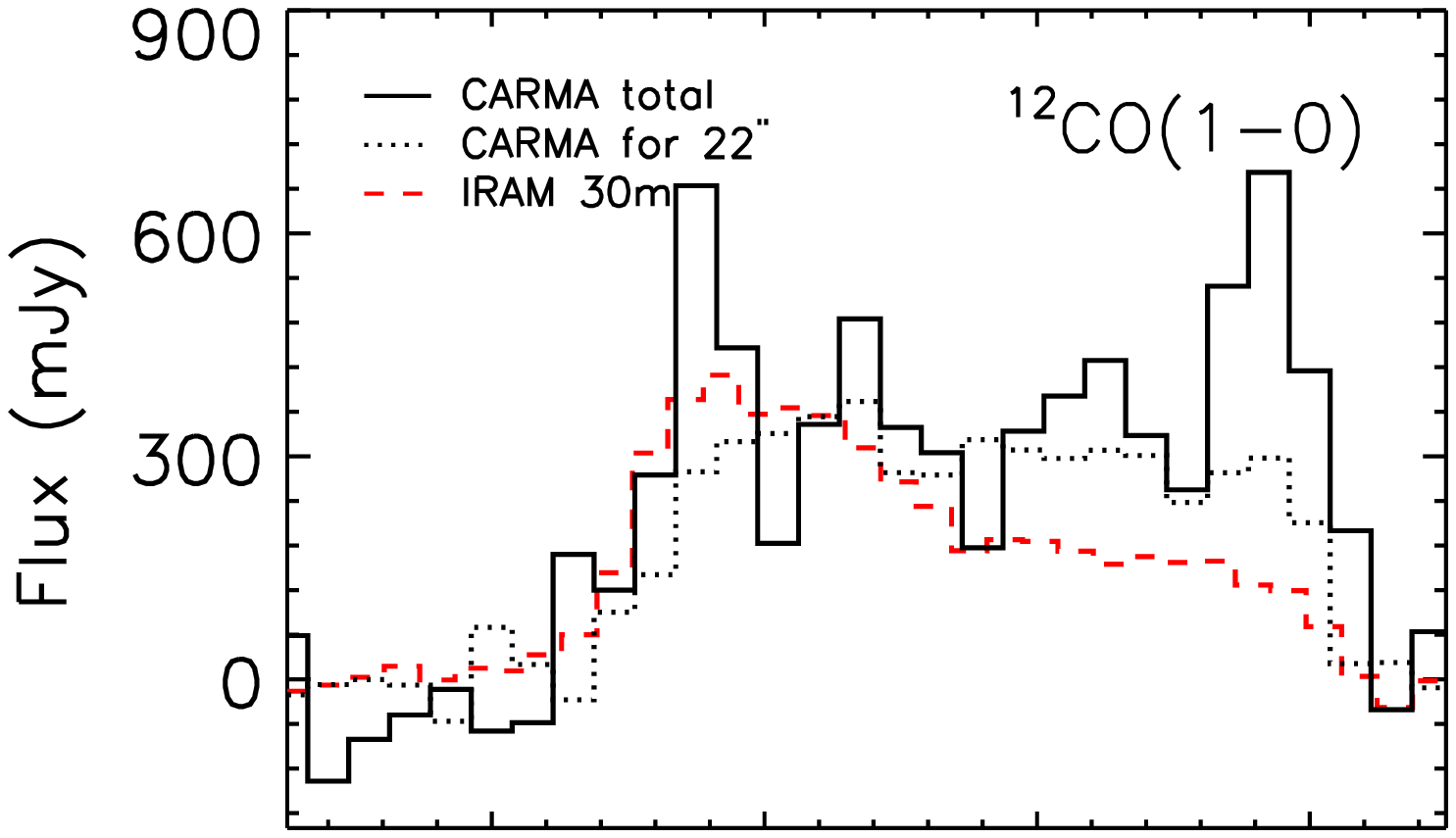}
  \hspace{-20pt}
  \includegraphics[width=6.0cm,clip=]{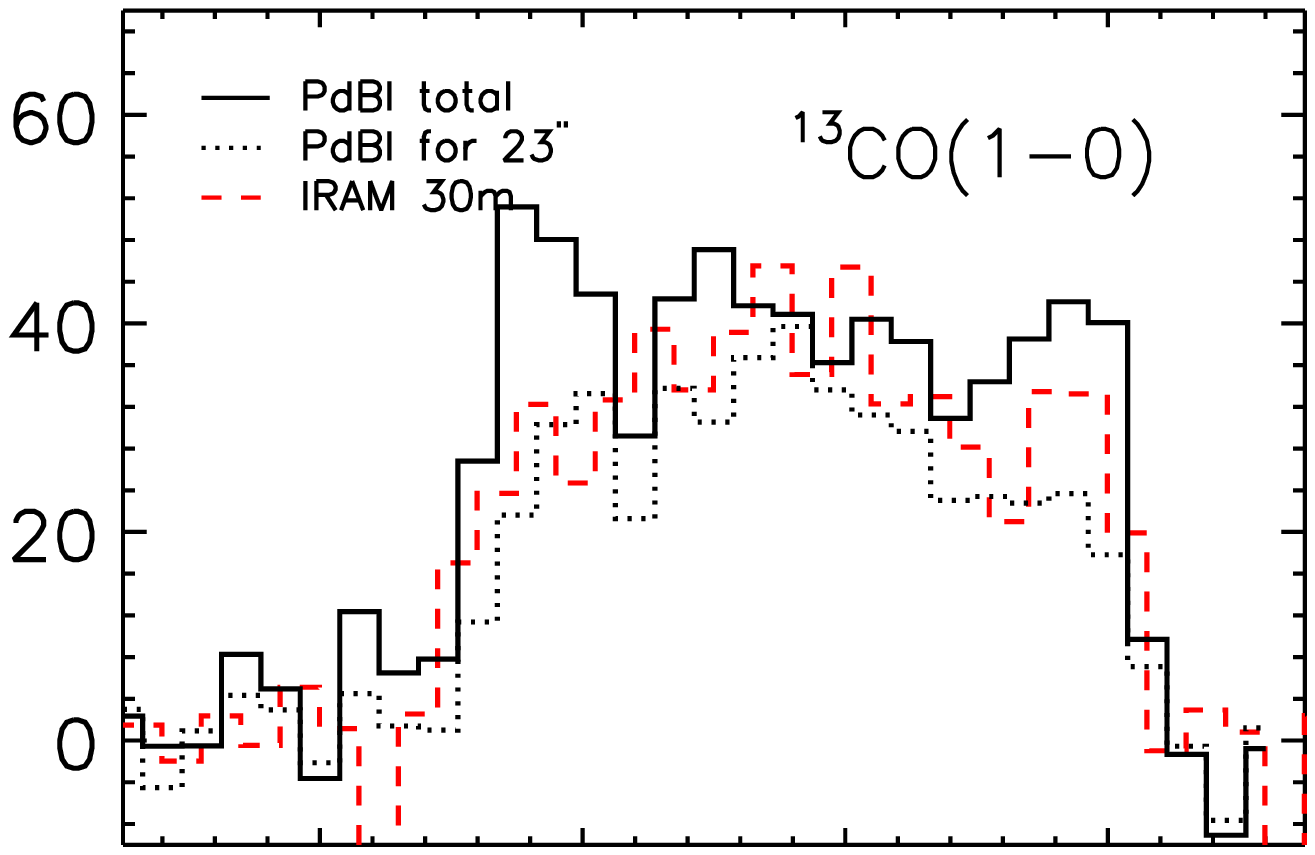}
  \hspace{-20pt}
  \includegraphics[width=6.0cm,clip=]{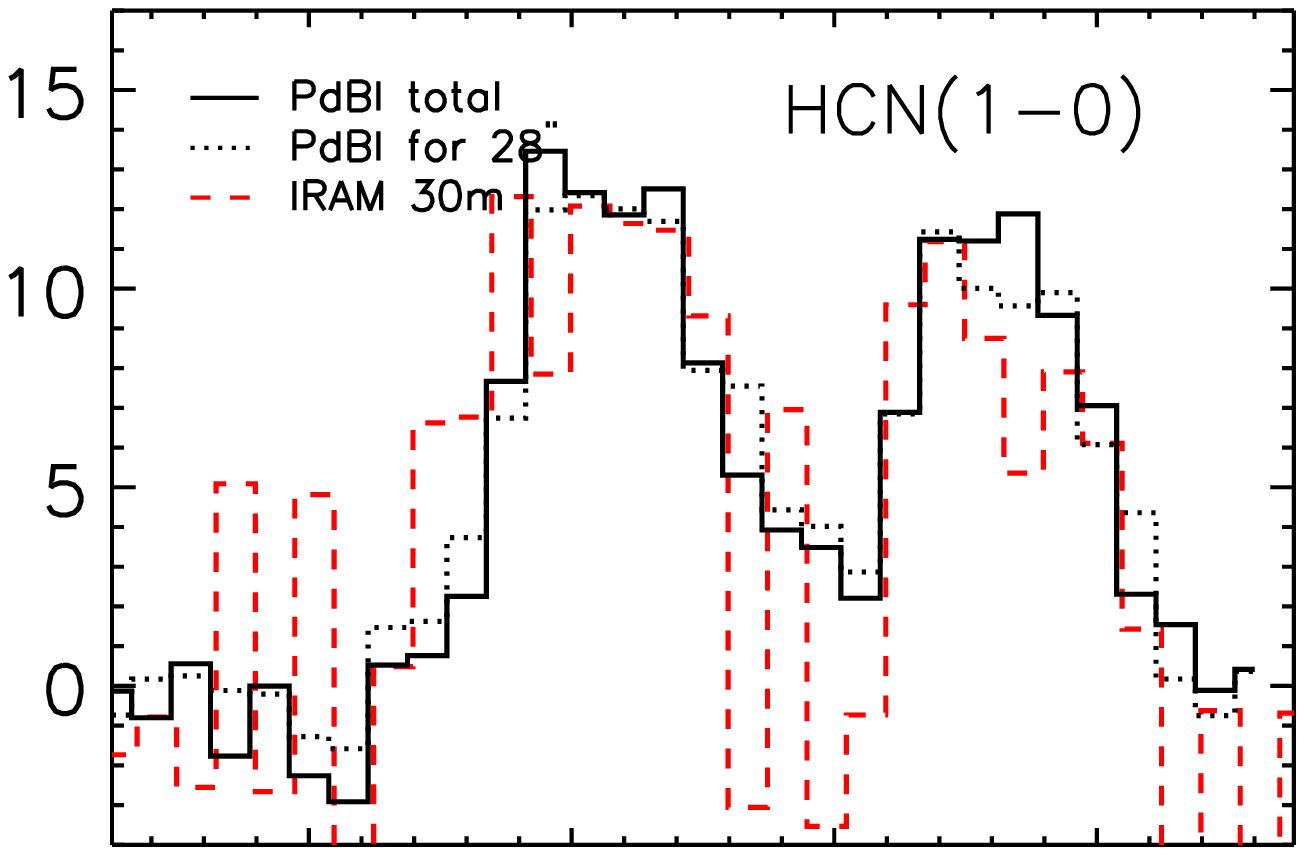}\\
  \vspace{-35pt}
  \includegraphics[width=6.0cm,clip=]{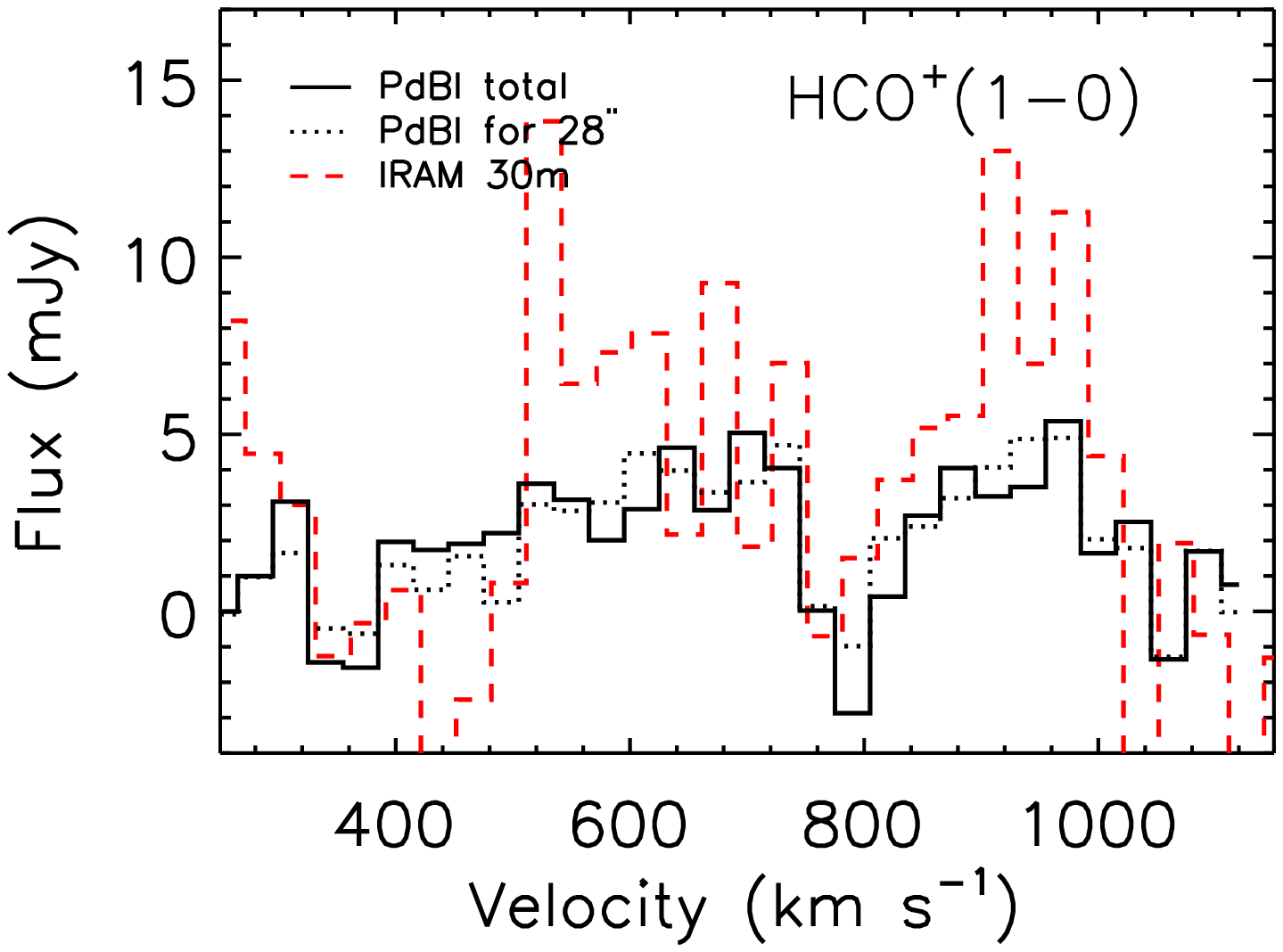}
  \hspace{-20pt}
  \includegraphics[width=6.0cm,clip=]{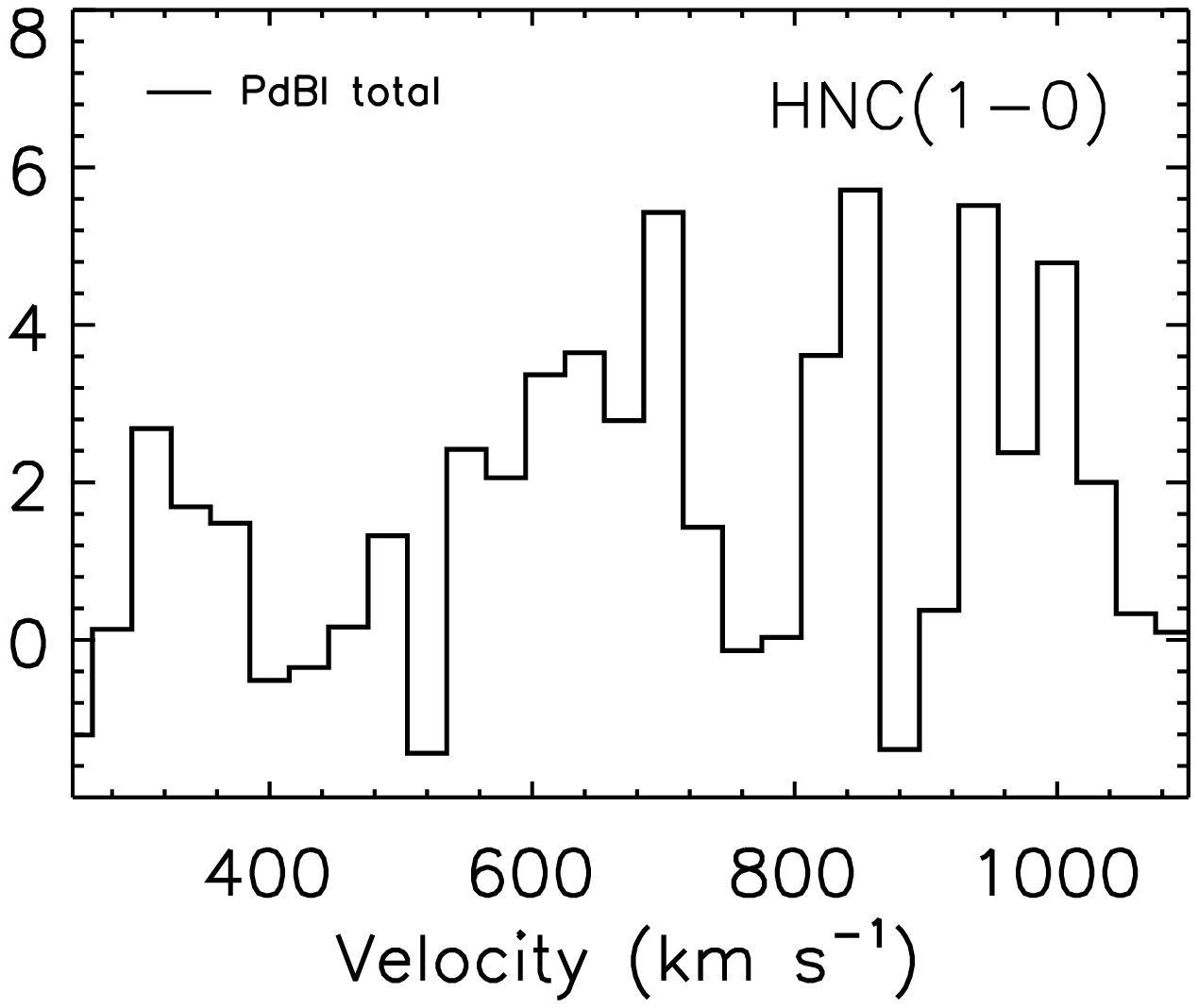}
  \hspace{-20pt}
  \includegraphics[width=6.0cm,clip=]{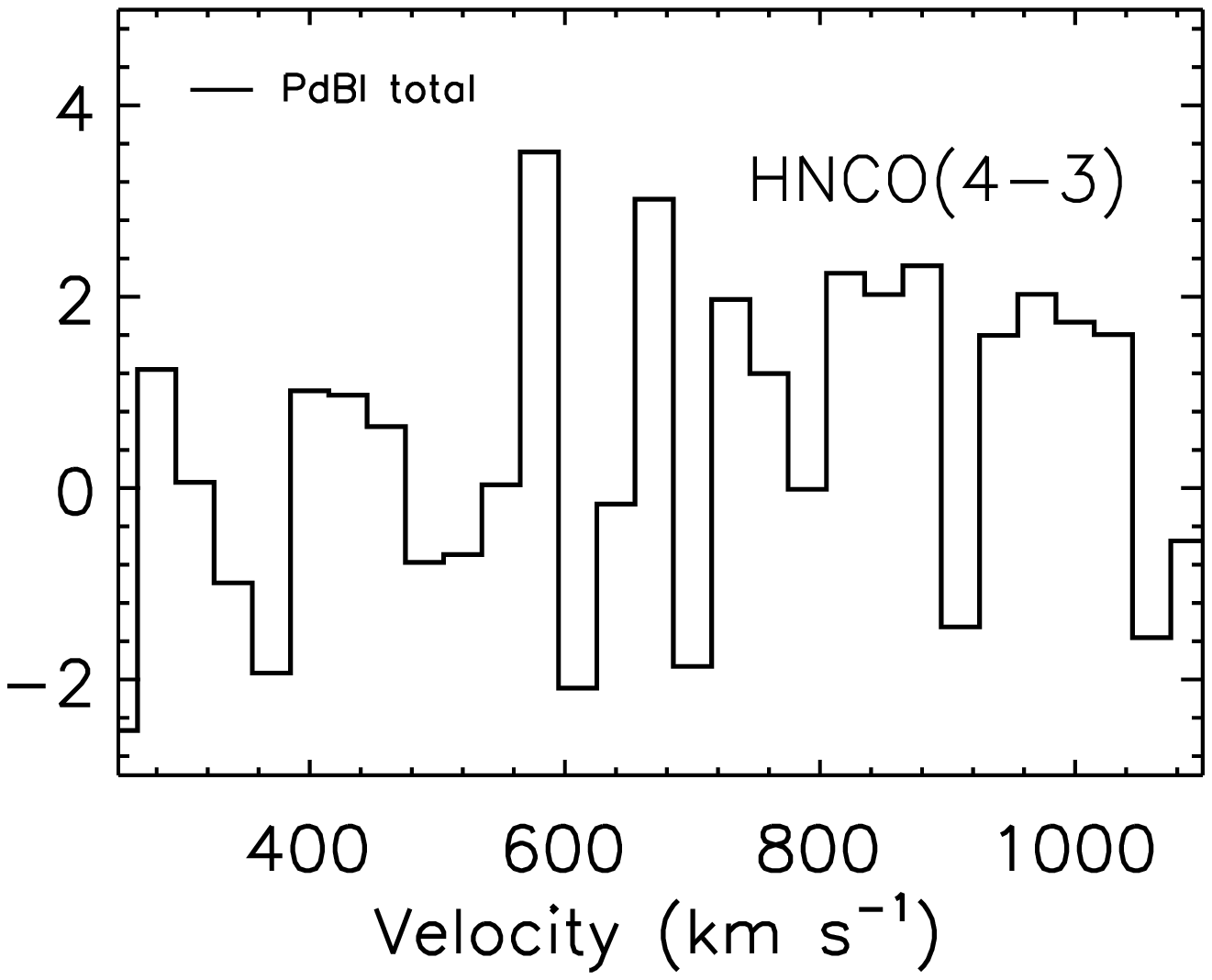}             
  \caption{Same as Figure~\ref{fig:spec1} but for NGC~5866.}
  \label{fig:spec2}
\end{figure*}

The true IRAM 30m spectra, simulated IRAM 30m spectra and
spatially-integrated CARMA and PdBI spectra (the latter without
beam weighting applied, thus recovering more flux if the emission
extends beyond the 30m beam) are shown in Figure~\ref{fig:spec1} and
\ref{fig:spec2} for NGC~4710 and NGC~5866, respectively (see also
\S~\ref{sec:compiram}). 
%
%
%
%

%
%
\subsection{Continuum emission}
\label{sec:contem}
NGC~4710 and NGC~5866 do not reveal spatially extended continuum
emission, but they do harbour a continuum point source at their
respective centre. We thus measured the continuum flux of each galaxy
by fitting a point source model in the $uv$ plane, using only channels
free of line emission. To do this, for PdBI data we used the MAPPING
task \emph{uv\_fit}, and for CARMA data we used the MIRIAD task
\emph{uvfit}.

In NGC~4710, we found a continuum flux of $3.23\,\pm\,0.88$~mJy at
$230$~GHz and $3\sigma$ upper limits of $2.46$ and $3.80$~mJy at $220$
and $90$~GHz, respectively. We were not able to estimate a continuum
flux at $115$ and $110$~GHz, as there is no line free channel, but
\citet{al13} list a $3\sigma$ upper limit of $5.20$~mJy at
$115$~GHz. In NGC~5866, we found a continuum flux of
$5.99\,\pm\,1.28$~mJy at $115$~GHz, $3.71\,\pm\,0.07$~mJy at $110$~GHz
and $3.55\,\pm\,0.04$~mJy at $90$~GHz.
%
%
\section{Line ratio diagnostics}
\label{sec:rat}
Different molecular lines require different physical conditions to be
excited, so a multitude of line ratios are required to probe complex
media. In this work, we perform three types of line ratio analyses,
providing complementary information on the physical conditions of the
gas along the disc (major-axis) of NGC~4710 and NGC~5866. First, we
analyse direct ratios of the major-axis PVDs (see \S~\ref{sec:rpvd}),
providing a qualitative view of the physical conditions in both
kinematic components of the galaxies (nuclear disc and inner ring).
Second, we calculate the ratios of integrated line intensities as a
function of projected radius along the galaxy discs, by extracting
integrated spectra as a function of projected position for each
kinematic component separately (see \S~\ref{sec:extract}). Third, we
model these line ratios using a non-LTE radiative transfer code, thus
estimating a number of physical parameters describing a two-component
molecular ISM (see \S~\ref{sec:model} and Appendix~\ref{sec:LVG}).

When the density exceeds a given molecular transition critical density
$n_{\rm crit}$, collisions become the dominant excitation and
de-excitation mechanism. In dense clouds, gas excitation is thus
dominated by collisions with H$_{2}$, by far the most abundant
species. Low-$J$ CO lines, such as $^{12}$CO(1-0) and its isotopologue
$^{13}$CO(1-0), have $n_{\rm crit}\approx10^{3}$~cm$^{-3}$, whereas
high density tracers, such as HCN(1-0), HCO$^+$(1-0), HNC(1-0) and
HNCO(4-3), have critical densities up to
$n_{\rm crit}\approx10^{6}$~cm$^{-3}$. We note however that when the
emitting gas gets optically thick ($\tau\gg1$), radiative-trapping
occurs (as some spontaneously-emitted photon are absorbed within the
cloud), resulting in a reduction of the critical density by a factor
$\sim1/\tau$ and a higher excitation temperature than that expected
due to collisions with H$_{2}$ only \citep{s13}.

In this paper, we separate our line ratios in three different groups,
to better probe the physical conditions of different phases of the
molecular ISM. The three groups are: i) ratios of low-$J$ CO lines
only, tracing the temperature, opacity and column density of the
relatively tenuous molecular gas; ii) ratios of dense gas tracers
only, tracing the density, chemical state and dominant
excitation/ionisation mechanisms of the dense molecular gas (e.g.\ UV
and X-ray radiation, stellar winds and supernova explosions); and iii)
ratios of CO to dense gas tracers, tracing the dense gas fraction. As
our PdBI observations of HCN(1-0), HCO$^+$(1-0), HNC(1-0) and
HNCO(4-3) in NGC~4710 are roughly twice as deep as our CARMA
observations of the same lines, we will exclusively use the PdBI data
of those lines for the line ratio analyses of this galaxy.
%
%
\subsection{PVD ratios}
\label{sec:rpvd}
We start all line ratio analyses by first creating identical data
cubes for all molecular lines. These identical cubes have the same
number of channels, channel width, start and end velocities, and pixel
size. They are also convolved to a common circular beam size of
$6\farcs5$, the largest synthesised beam in our dataset (i.e.\ the
beam of the high density tracers observed at PdBI; see
Table~\ref{tab:obs}). The pixel size of these data cubes was chosen to
be $1\farcs3$, yielding $5$ pixels across the beam. We then converted
our fluxes from Jy~beam$^{-1}$ to Kelvin (K) using the conversion
factors calculated by the MIRIAD task \emph{imstat} (see
Table~\ref{tab:obs}), so that by convention and throughout this paper
we take the ratio of line intensities expressed in K.

\begin{figure*}
  \centering
  \includegraphics[width=5.8cm,clip=]{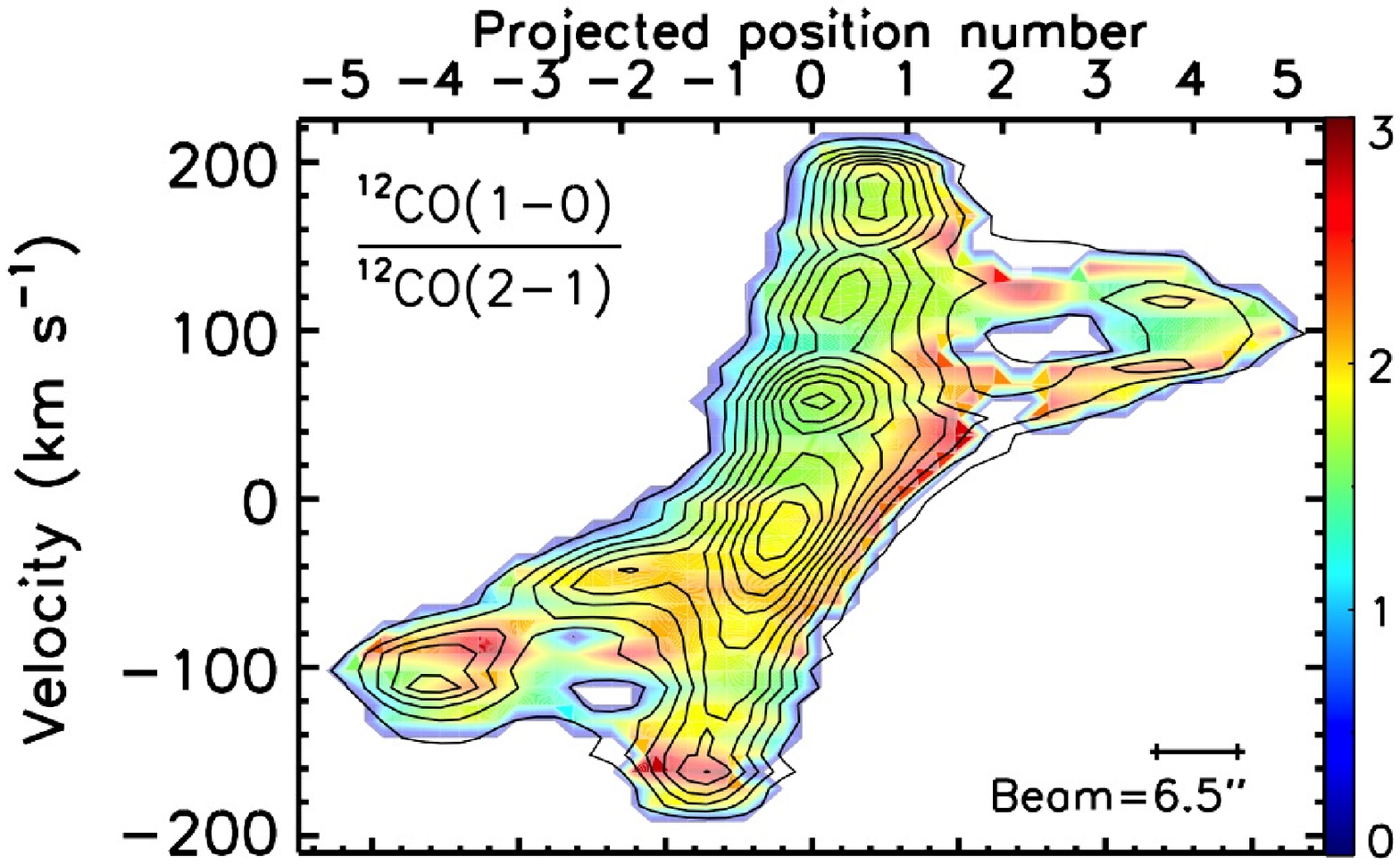}
   \hspace{-10pt}
  \includegraphics[width=5.8cm,clip=]{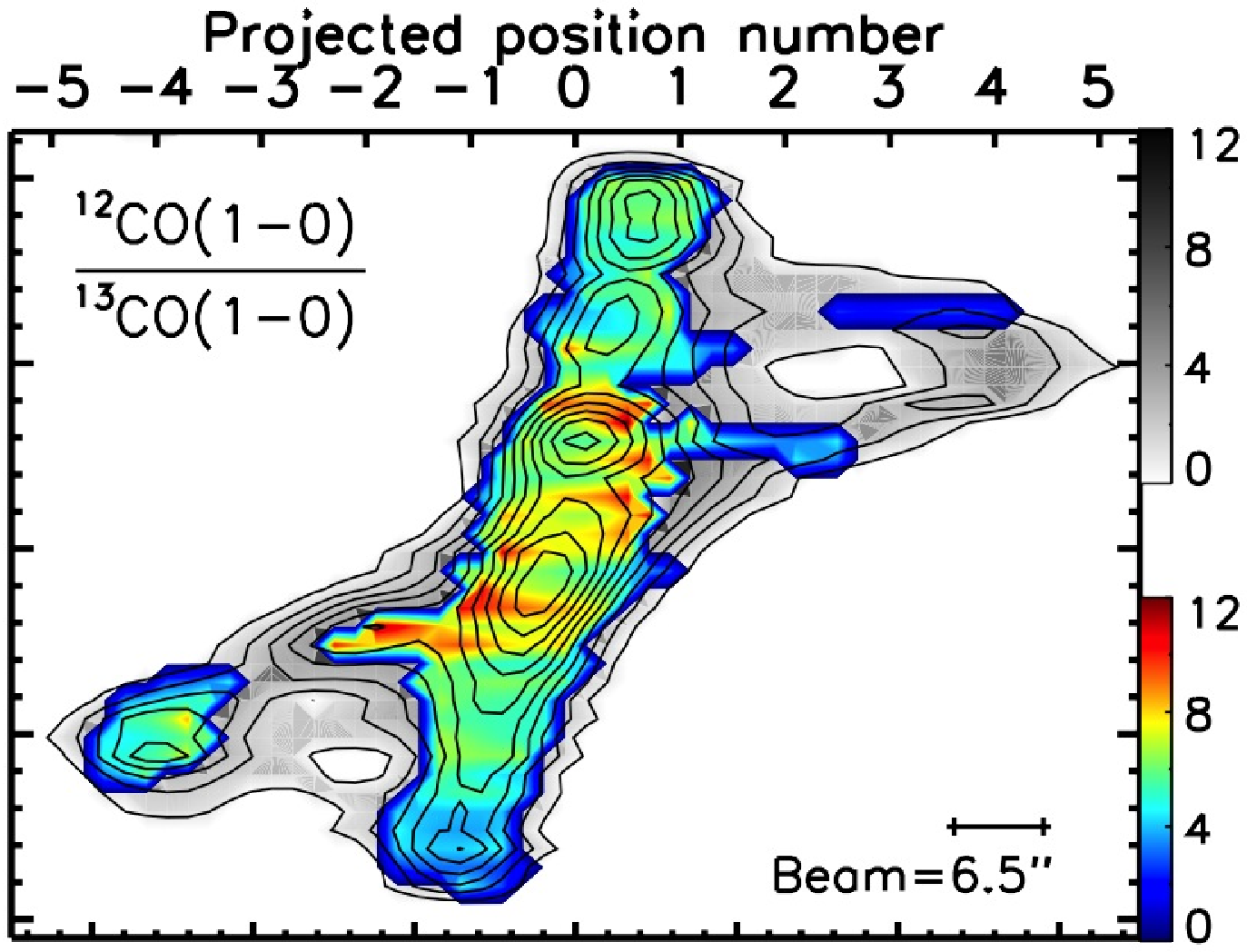}
   \hspace{-10pt}
  \includegraphics[width=5.8cm,clip=]{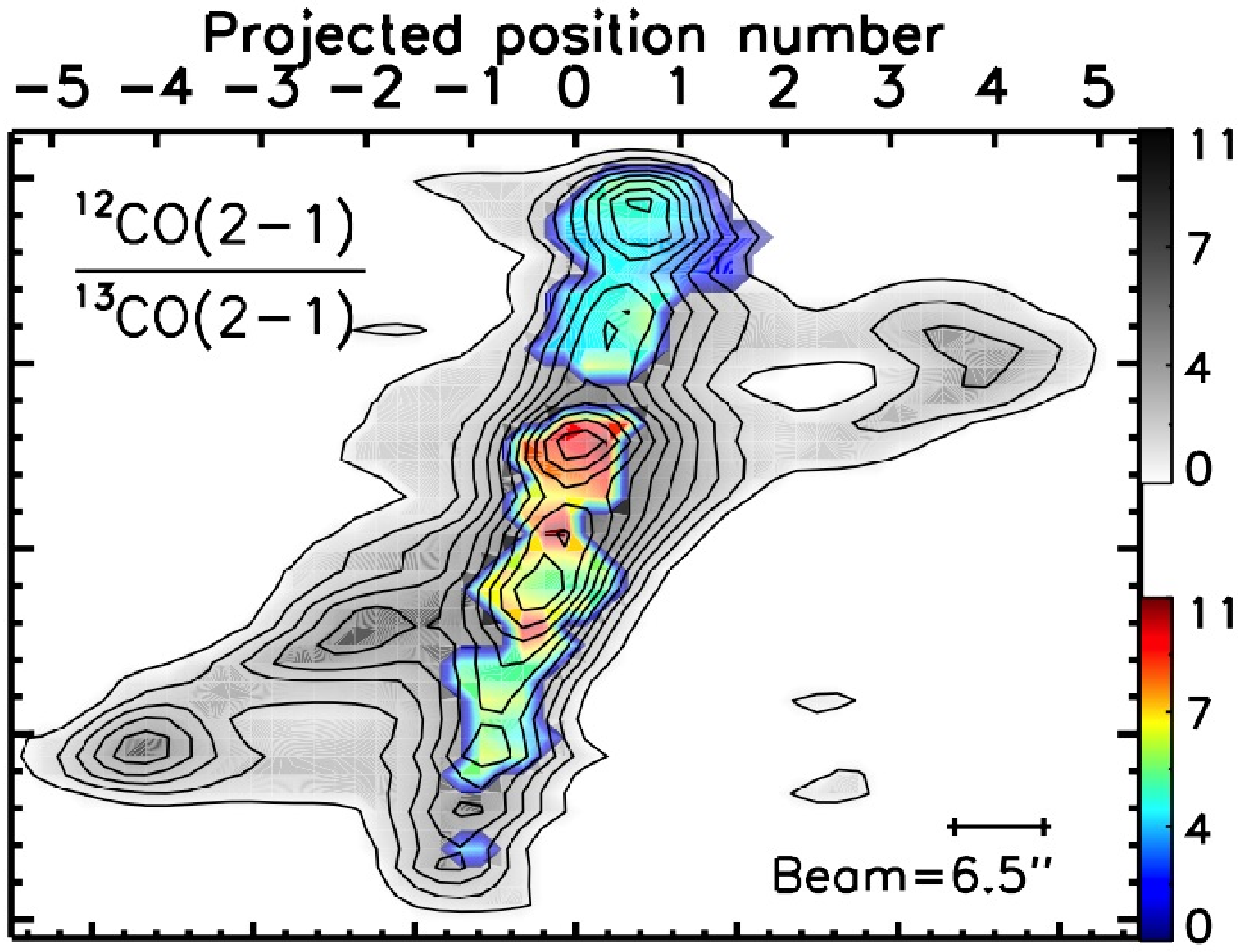}\\
  \vspace{-10pt}
  \includegraphics[width=6.1cm,clip=]{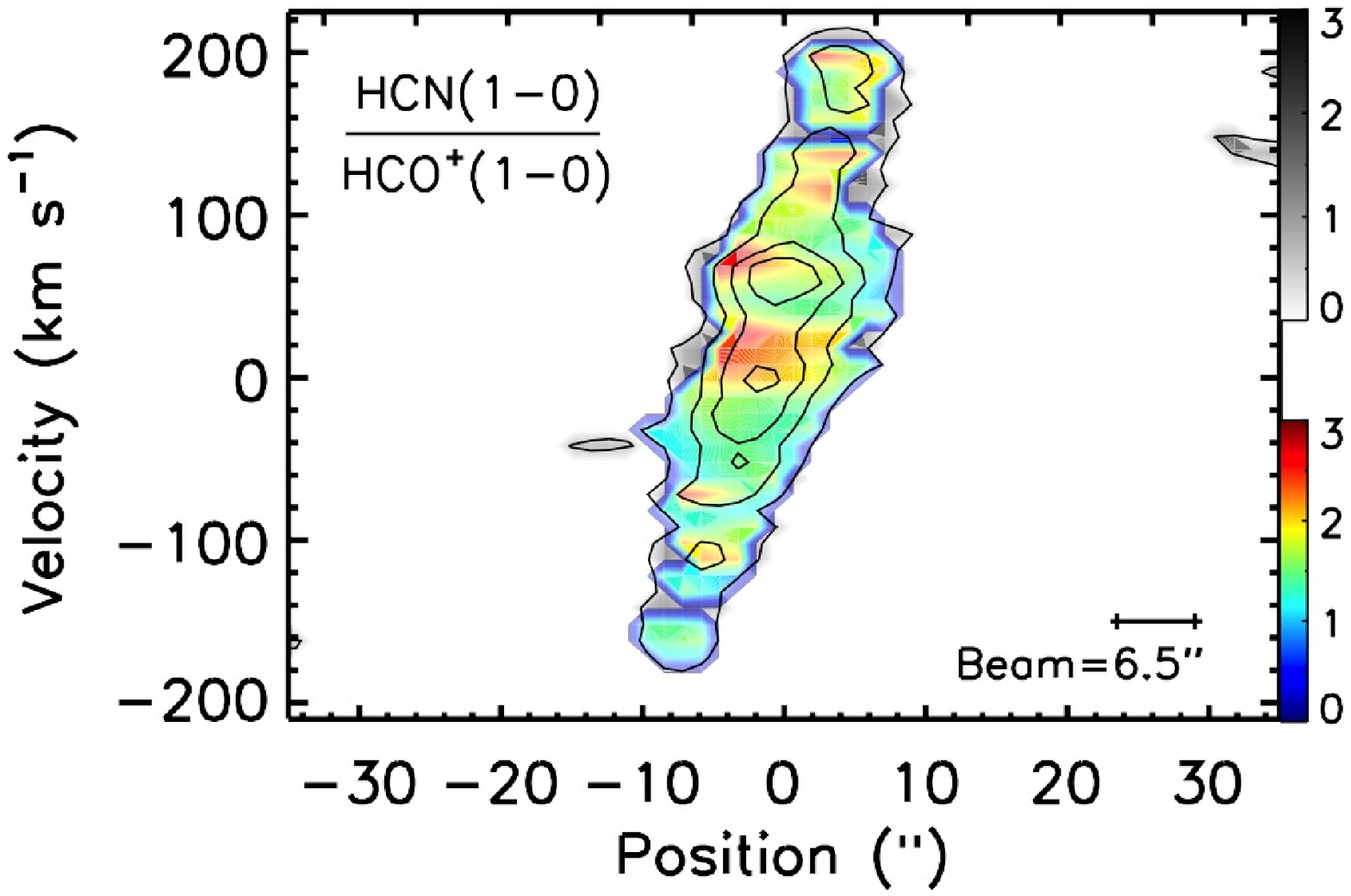}
  \hspace{-10pt}
  \includegraphics[width=5.5cm,clip=]{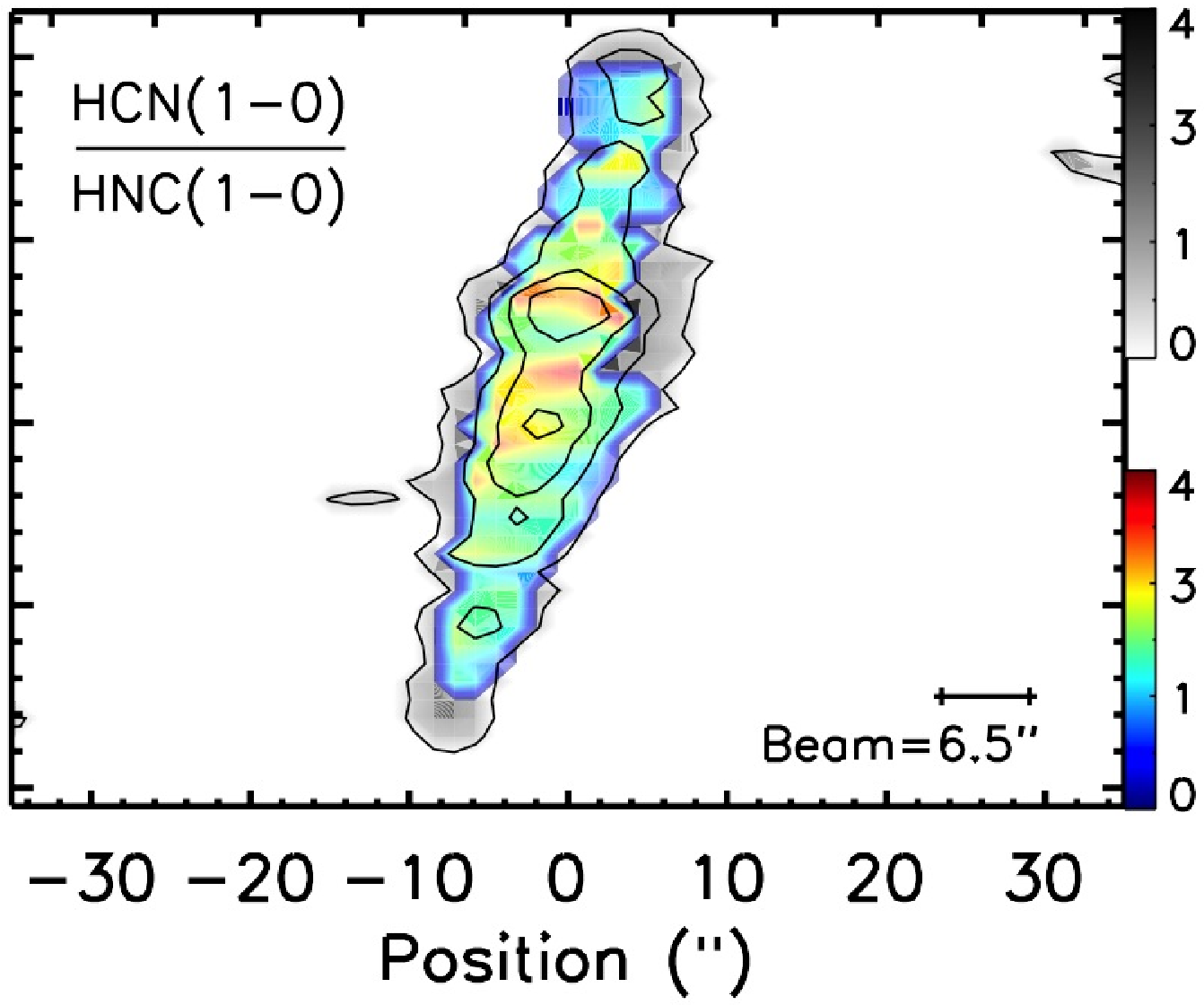}
  \includegraphics[width=5.5cm,clip=]{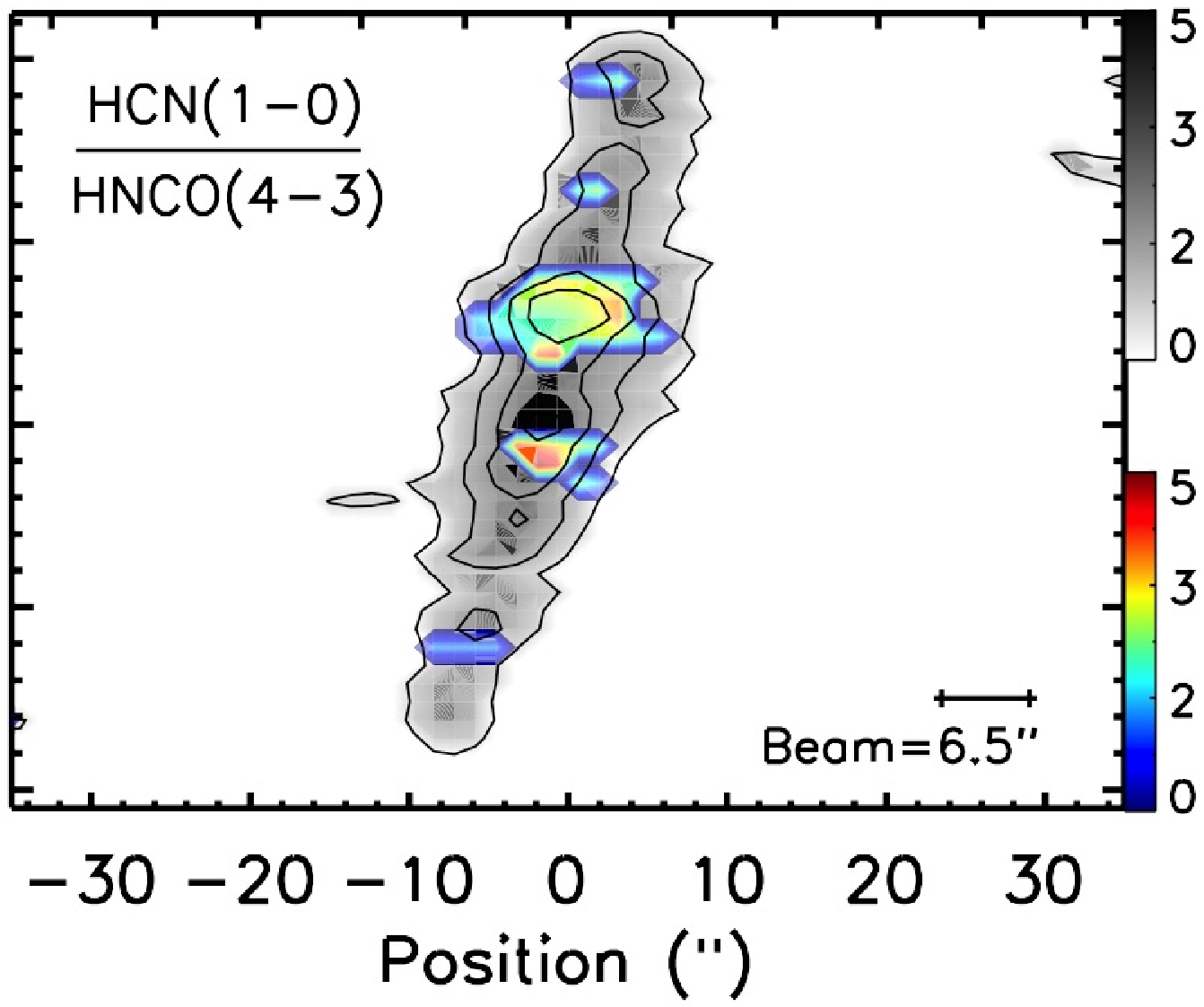}\\
  \caption{PVD ratios of CO lines only and dense gas tracer lines only
    in NGC~4710. {\bf Top row:} PVD ratios of CO lines only, with the
    relevant $^{12}$CO(1-0) or $^{12}$CO(2-1) PVD contours overlaid
    (black lines). {\bf Bottom row:} PVD ratios of dense gas tracer
    lines only, with the HCN(1-0) PVD contours overlaid (black
    lines). Contour levels are spaced by $3\sigma$ starting at
    $3\sigma$. Greyscales indicate lower limits to the line ratios
    (see \S~\ref{sec:rpvd}). The projected position numbers, as
    discussed in \S~\ref{sec:extract} and illustrated in
    Figure~\ref{fig:pos}, are indicated on the top axes.}
  \label{fig:n4710ratioPVD1}
\end{figure*}

For our first approach to line ratio analysis, PVDs were created from
these identical cubes as before, and ratios of these PVDs were
calculated for relevant line pairs in each of the three aforementioned
line ratio groups (CO lines only, dense gas tracers only, and CO
versus dense gas tracers). For the high density tracer and $^{13}$CO
lines that remain undetected in some regions of the discs, a $3\sigma$
flux upper limit was adopted, yielding a lower limit on the line
ratios considered. The PVD ratios are shown in
Figures~\ref{fig:n4710ratioPVD1}\,--\,\ref{fig:n5866ratioPVD}, where the
greyscales indicate lower limits.

%
%
%
%
\begin{figure*}
  \centering
  \hspace{5pt}
  \includegraphics[width=5.1cm,clip=]{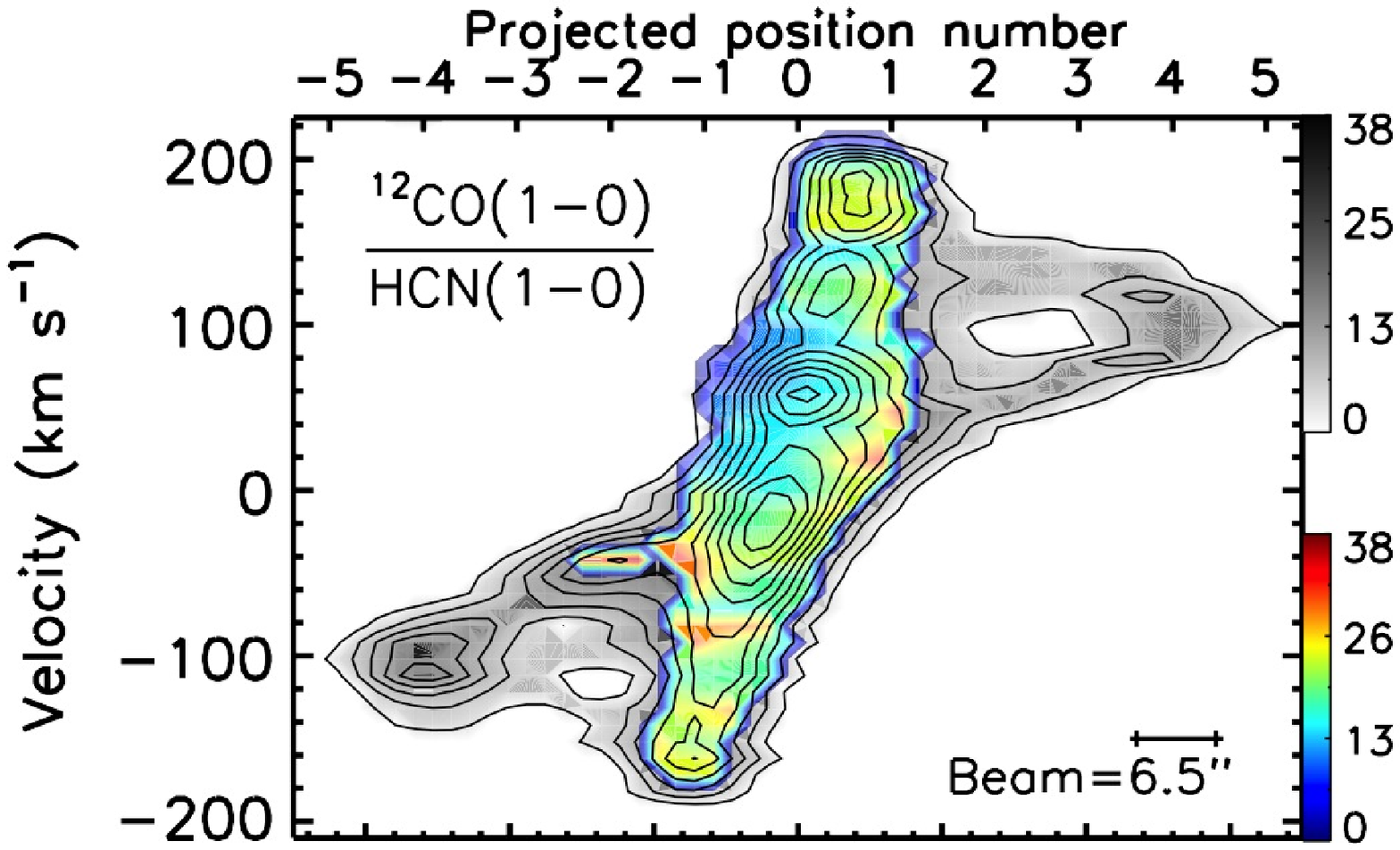}
  \hspace{-14pt}
  \includegraphics[width=4.2cm,clip=]{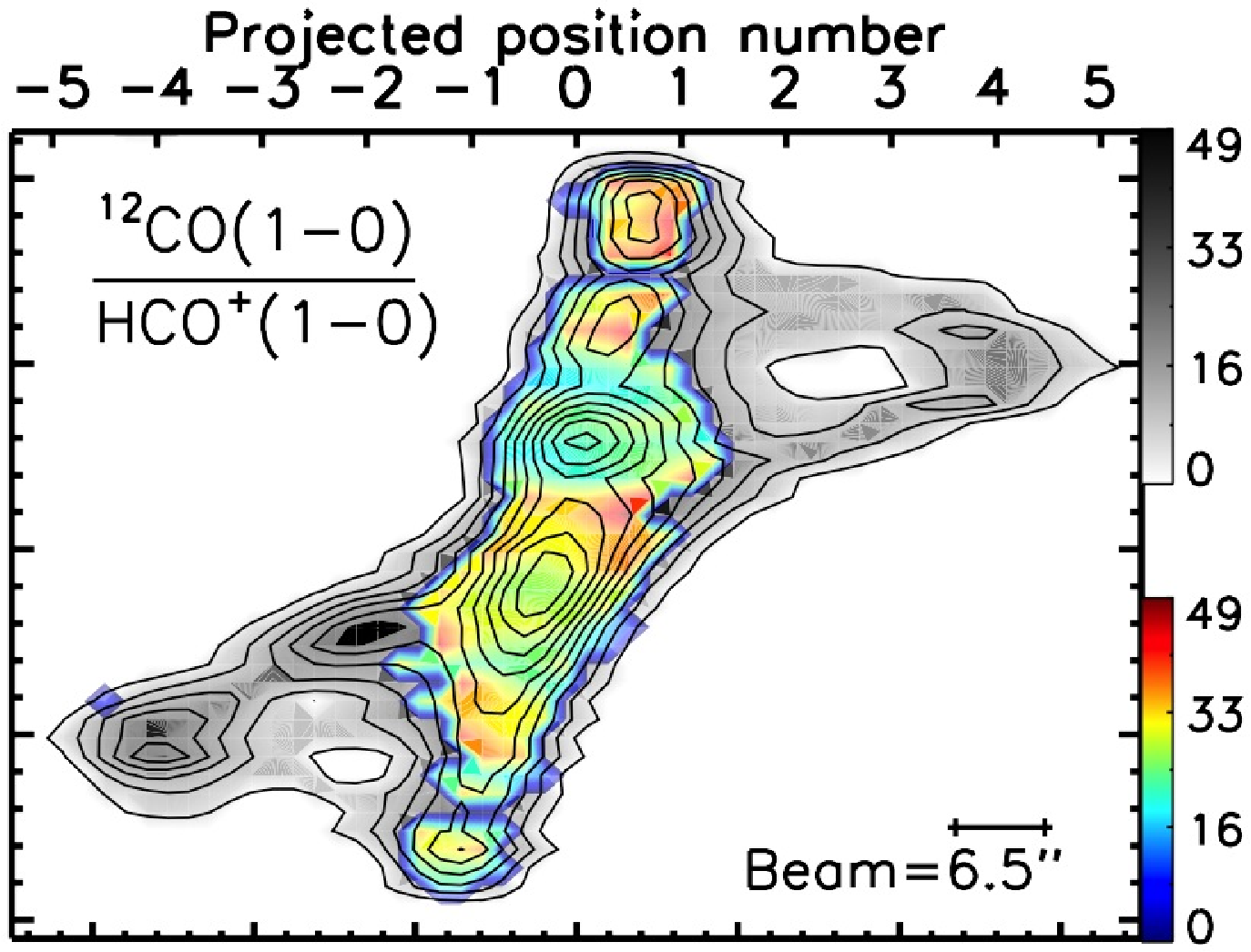}
  \hspace{-14pt}
  \includegraphics[width=4.2cm,clip=]{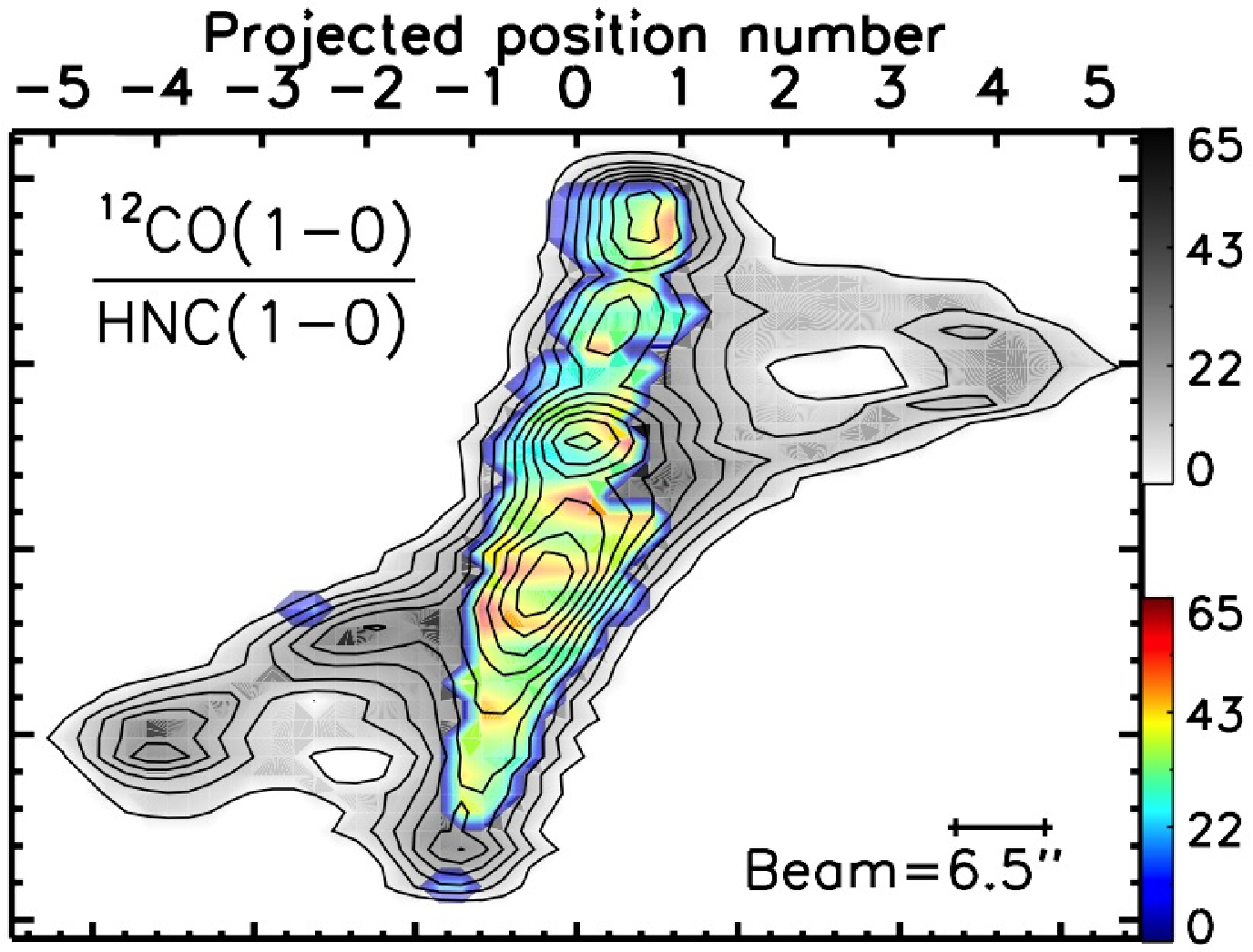}
  \hspace{-14pt}
  \includegraphics[width=4.2cm,clip=]{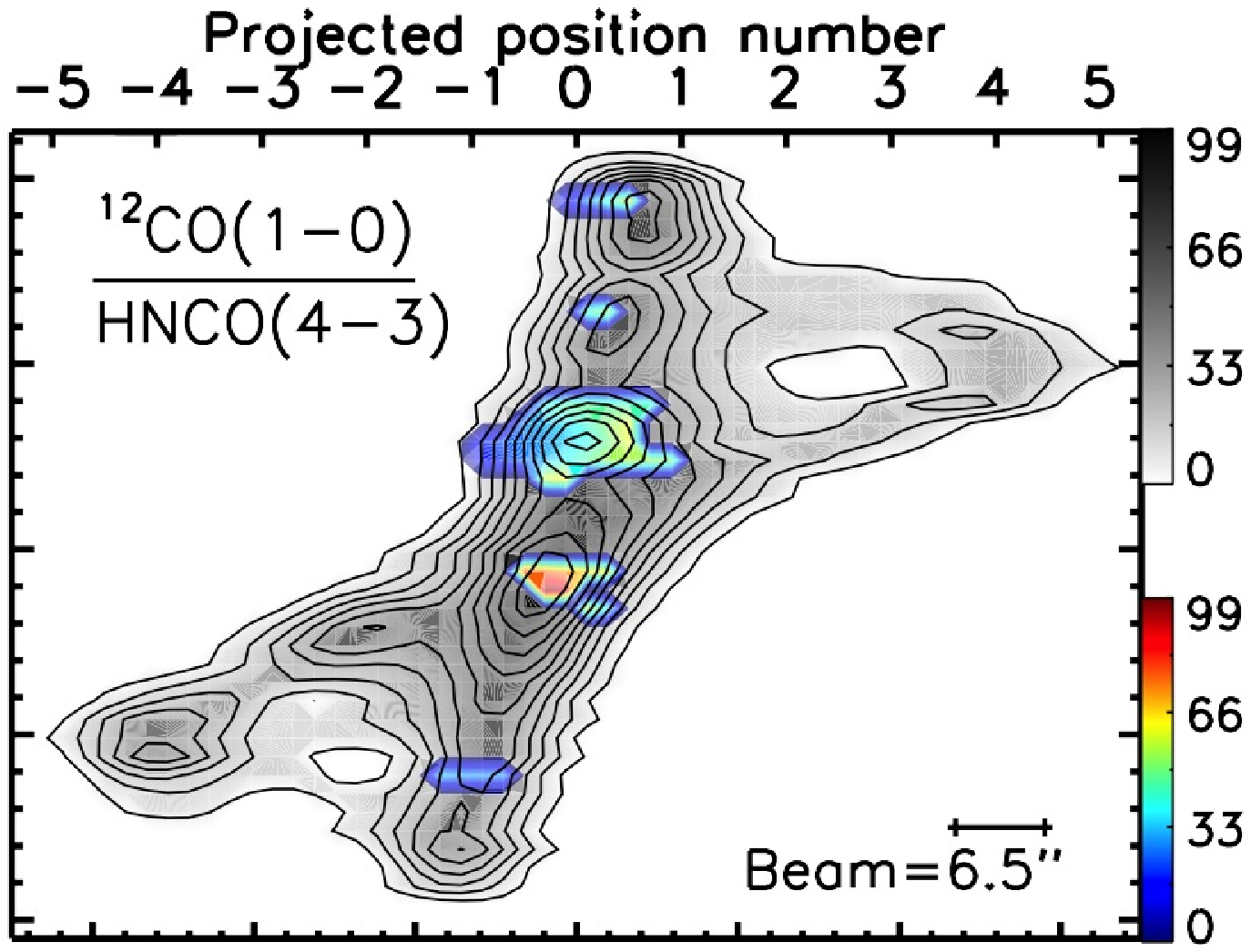}\\
  \vspace{-10pt}
  \includegraphics[width=5.1cm,clip=]{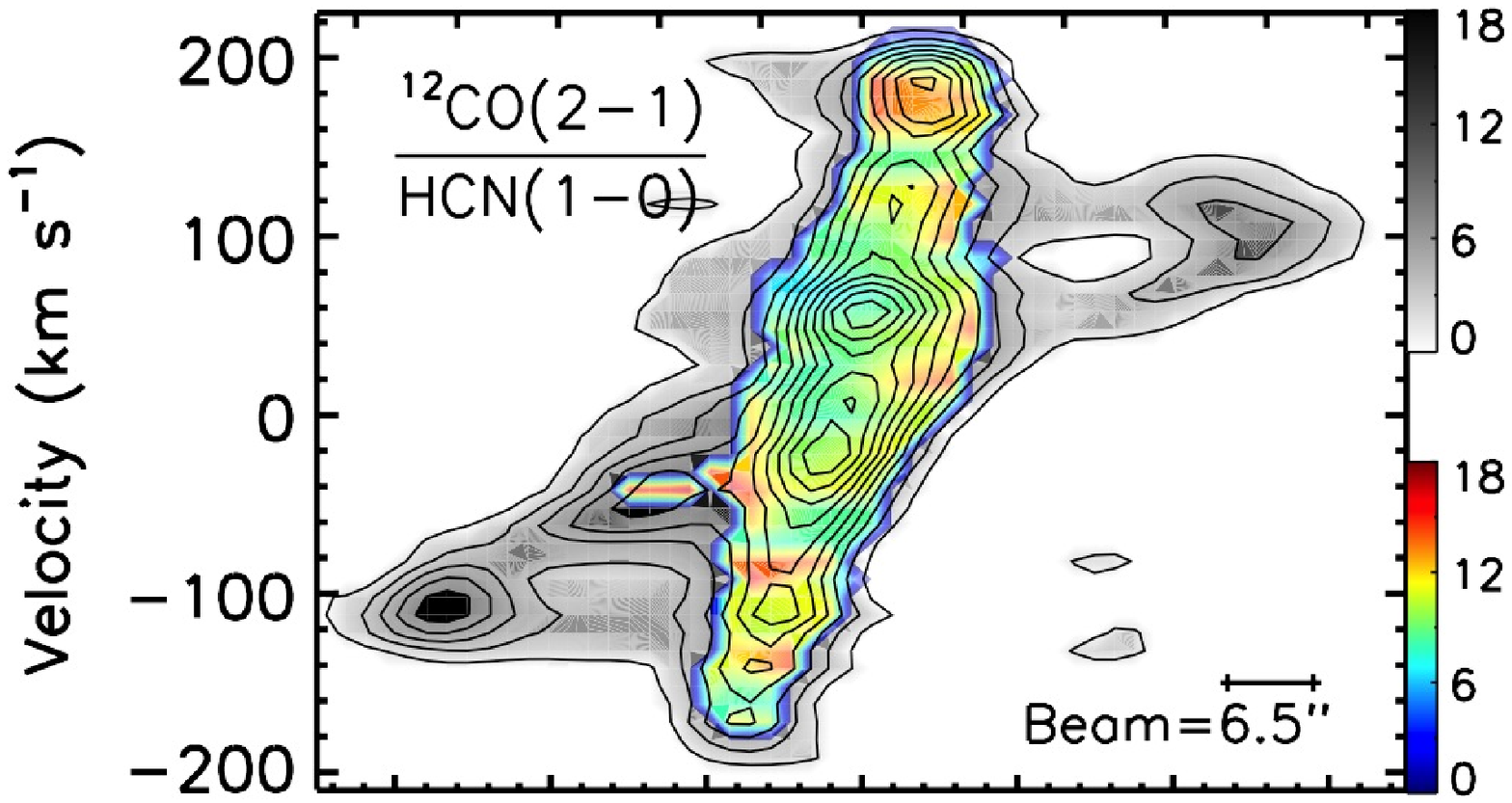}
  \hspace{-13pt}
  \includegraphics[width=4.0cm,clip=]{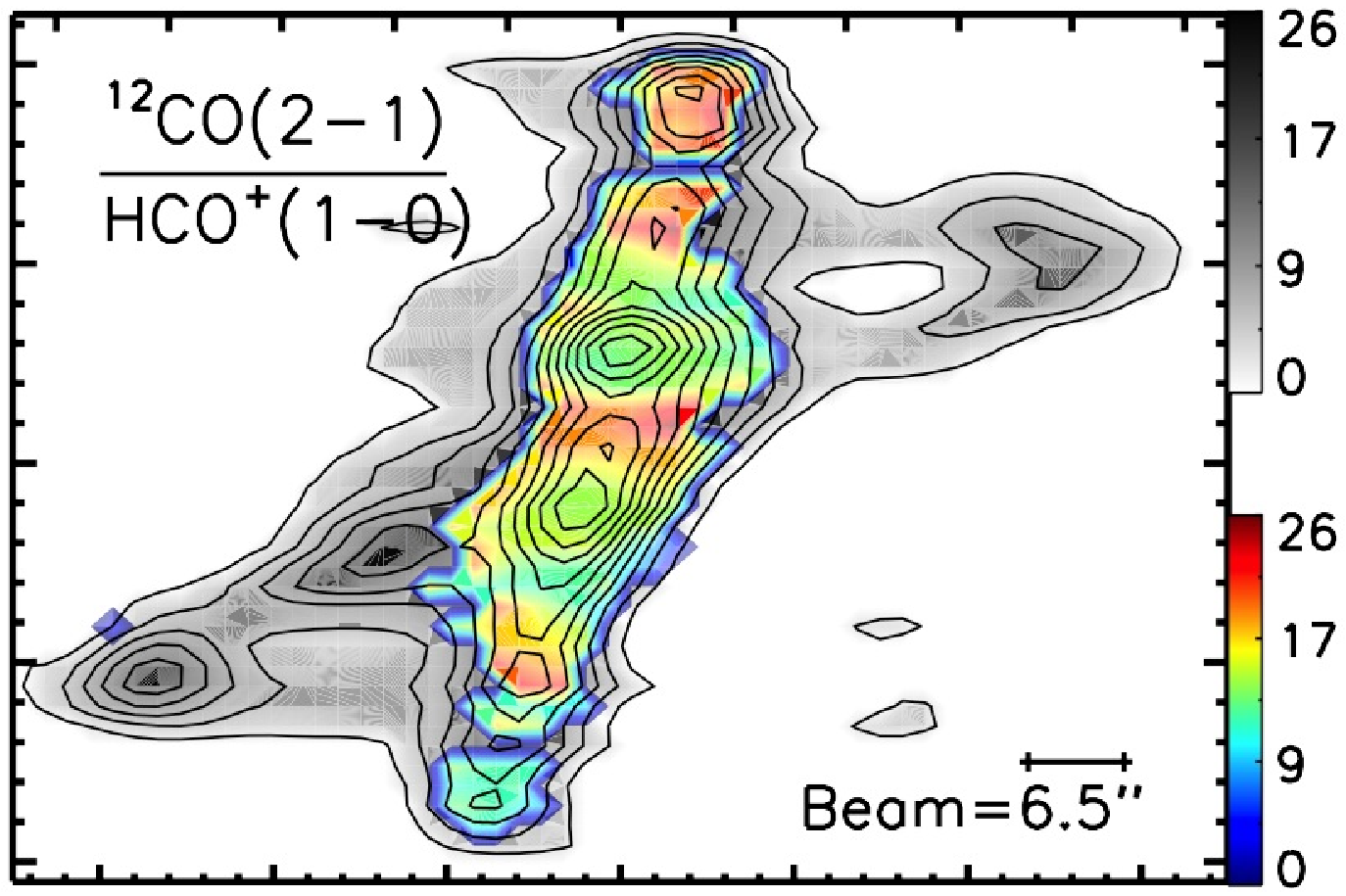}
  \hspace{-8pt}
  \includegraphics[width=4.0cm,clip=]{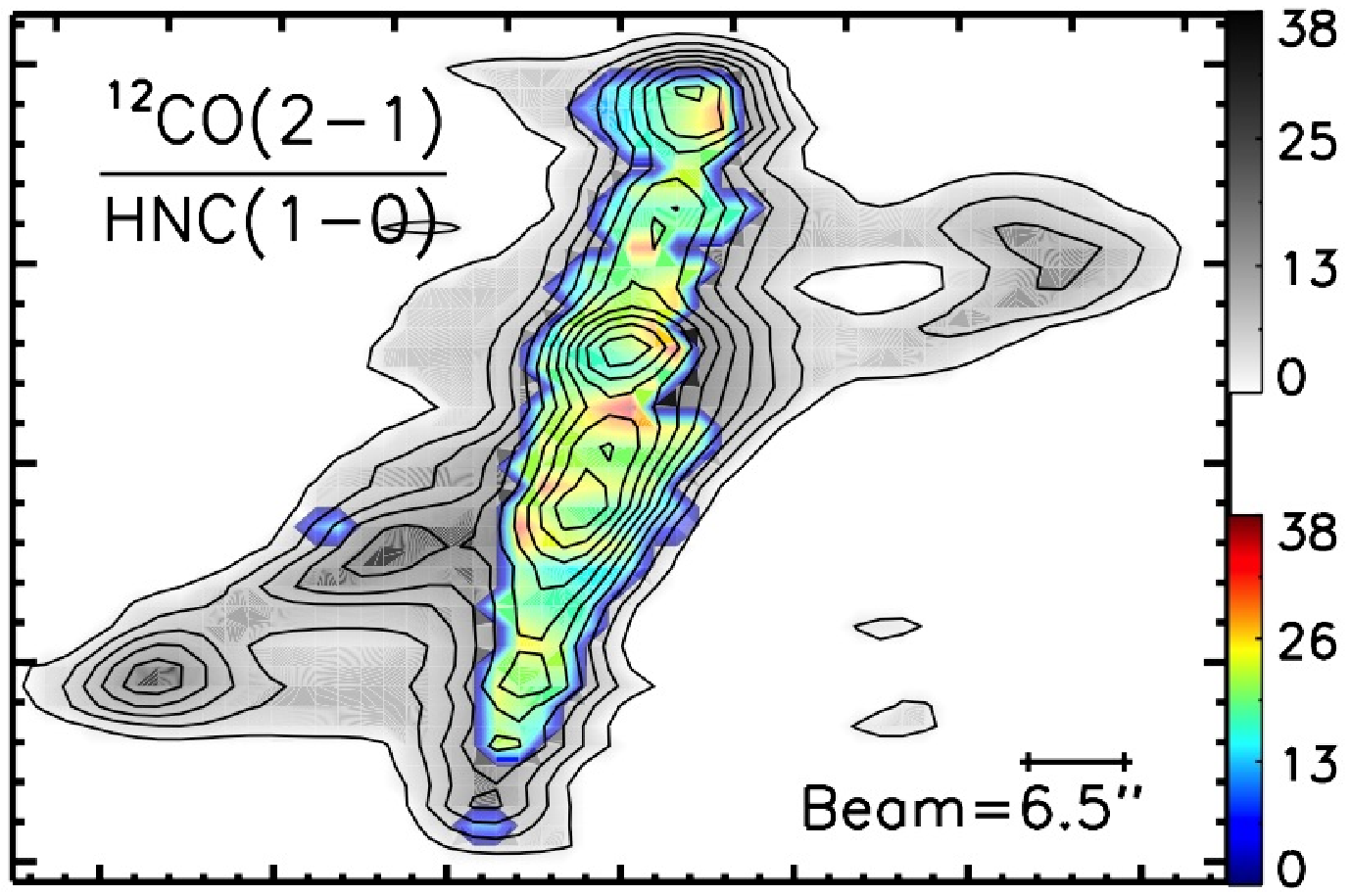}
  \hspace{-8pt}
  \includegraphics[width=4.0cm,clip=]{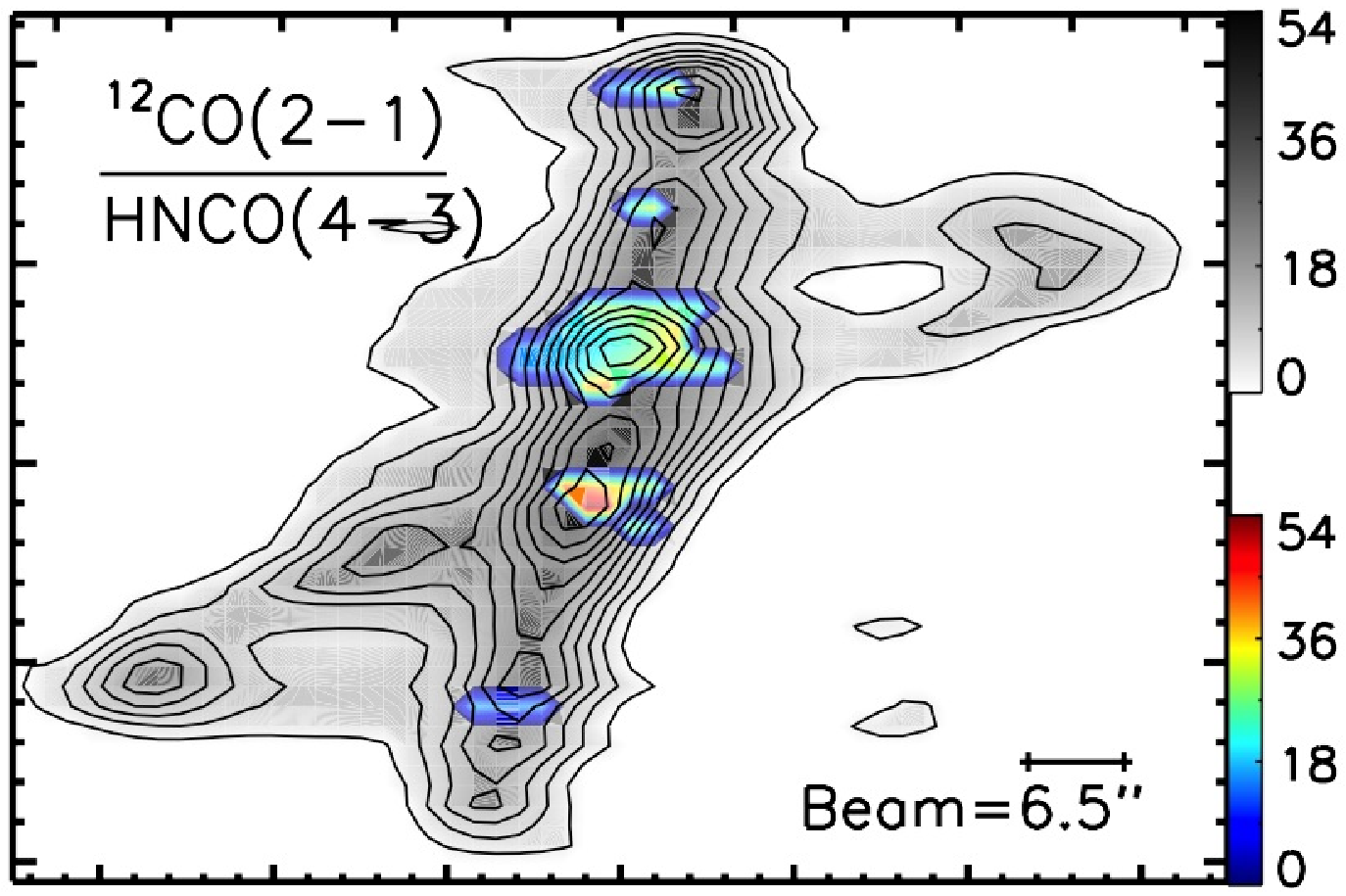}\\
  \vspace{-10pt}
  \includegraphics[width=5.1cm,clip=]{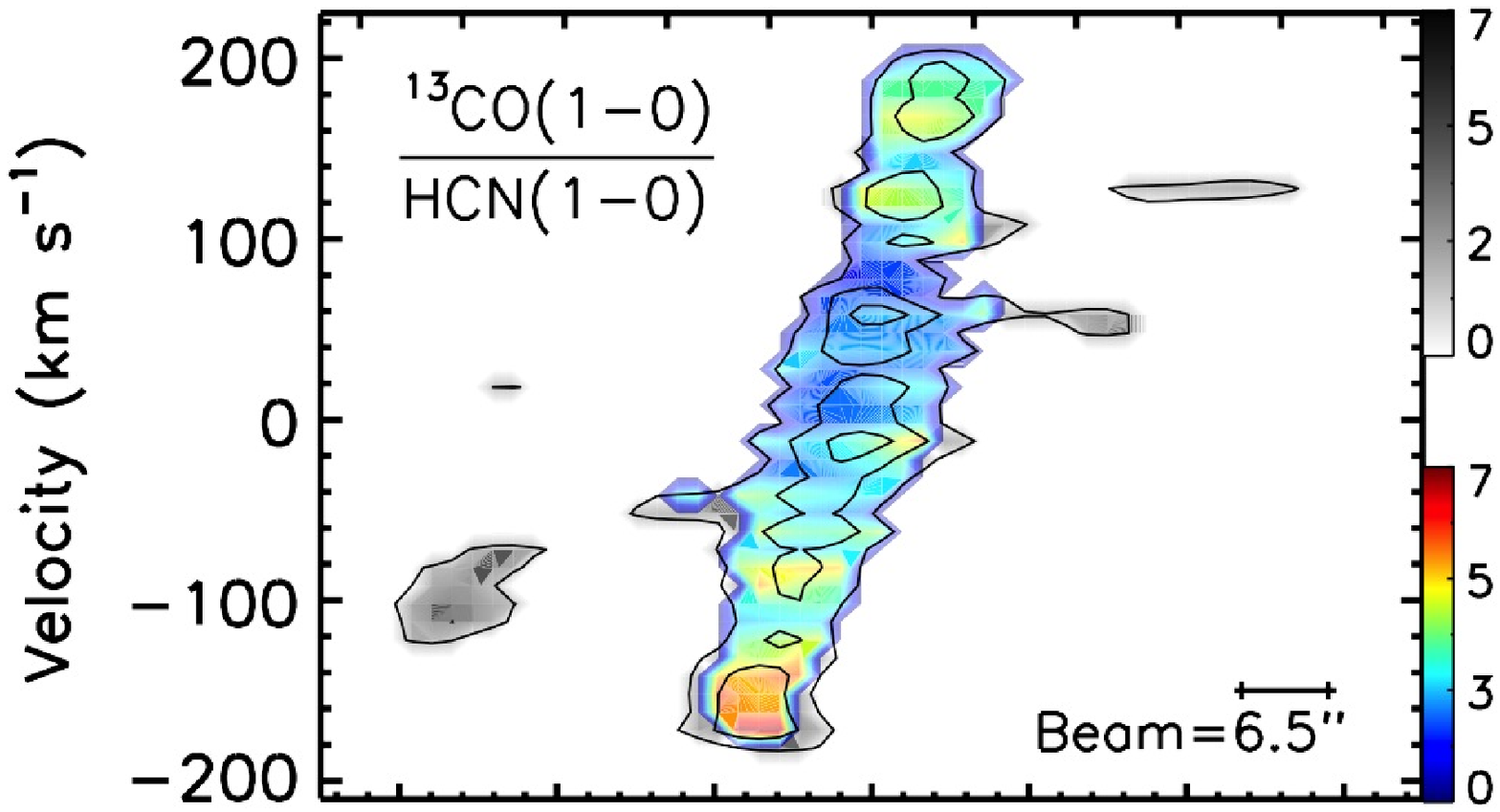}
  \hspace{-13pt}
  \includegraphics[width=4.0cm,clip=]{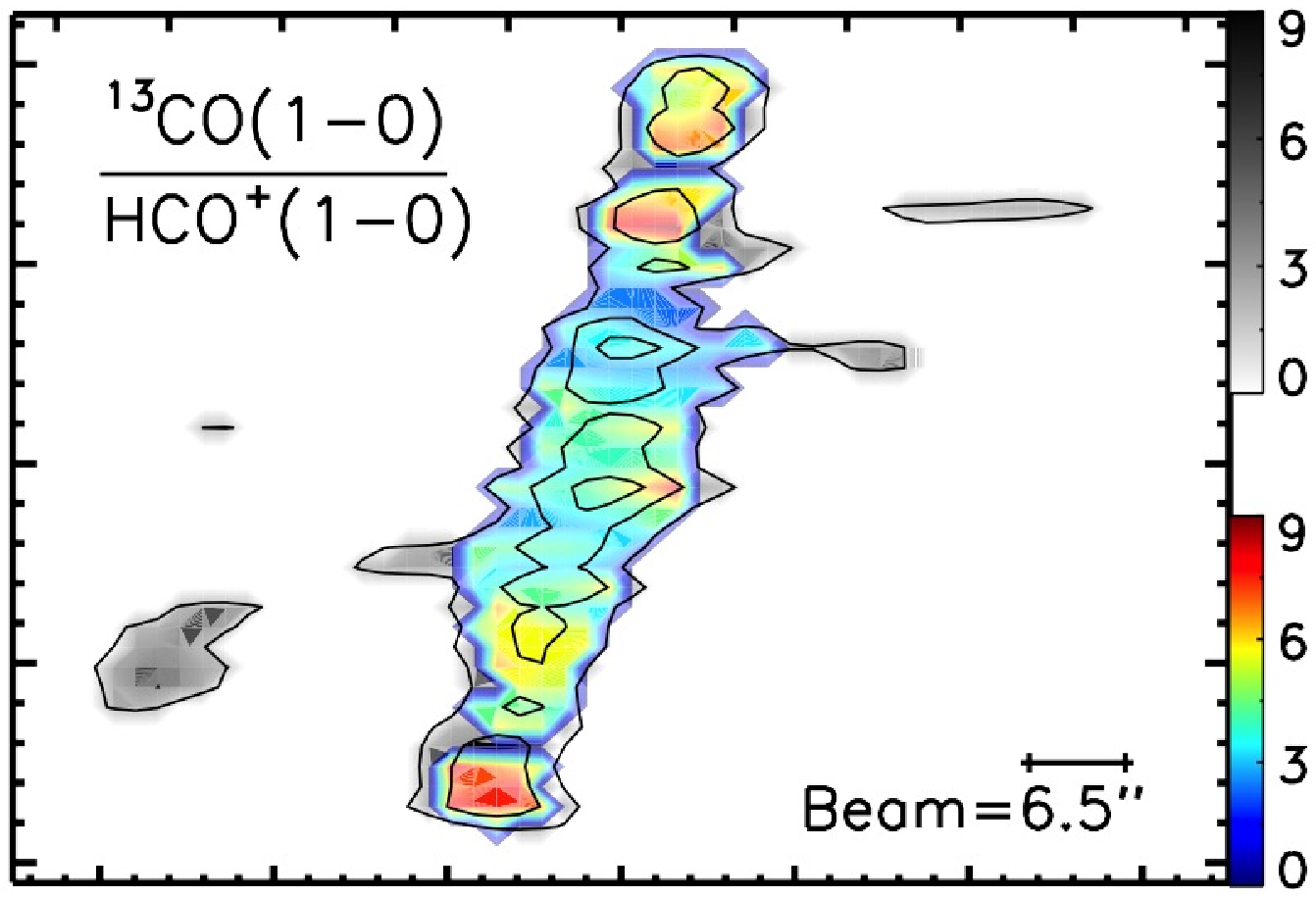}
  \hspace{-8pt}
  \includegraphics[width=4.0cm,clip=]{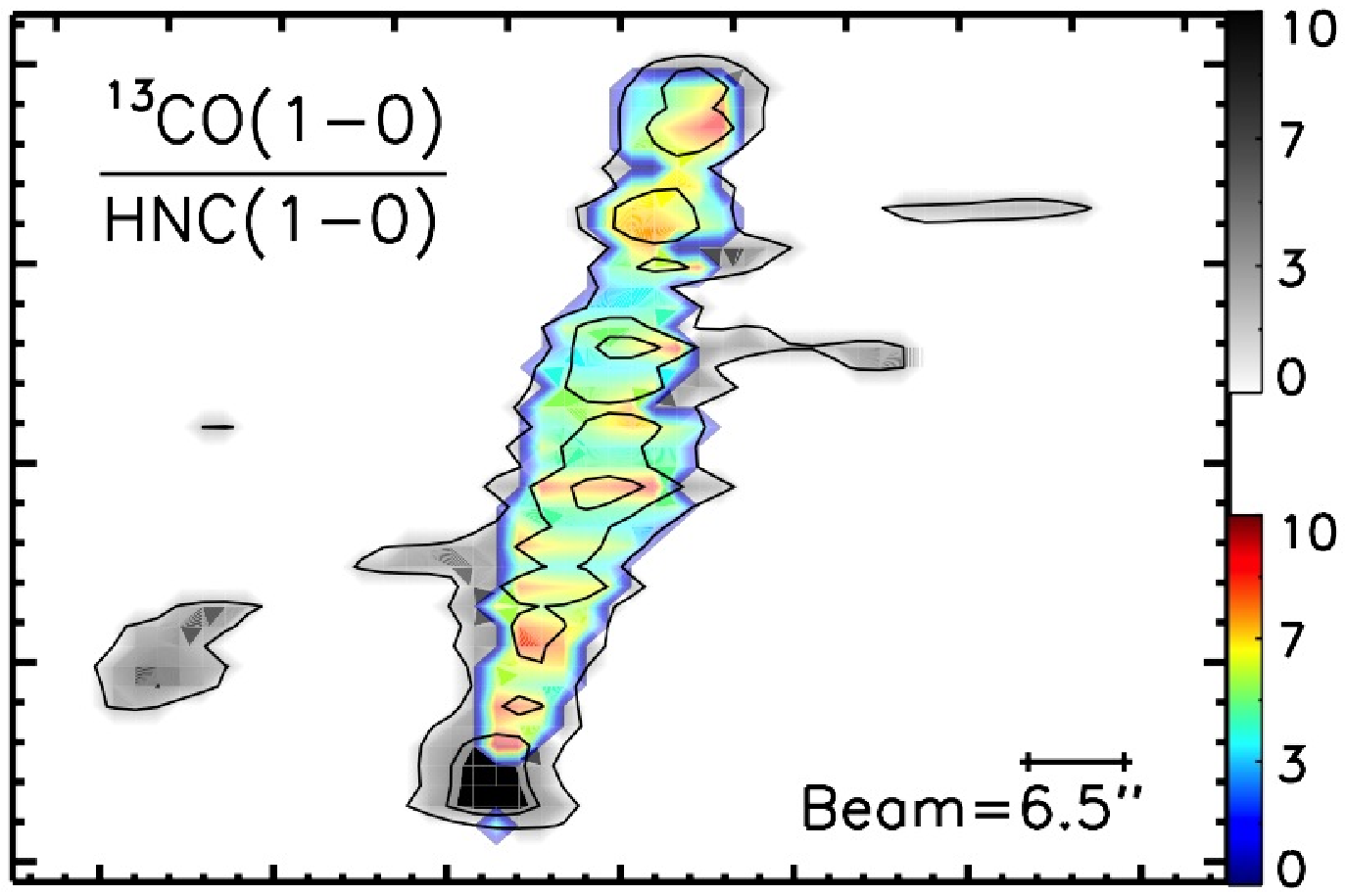}
  \hspace{-8pt}
  \includegraphics[width=4.0cm,clip=]{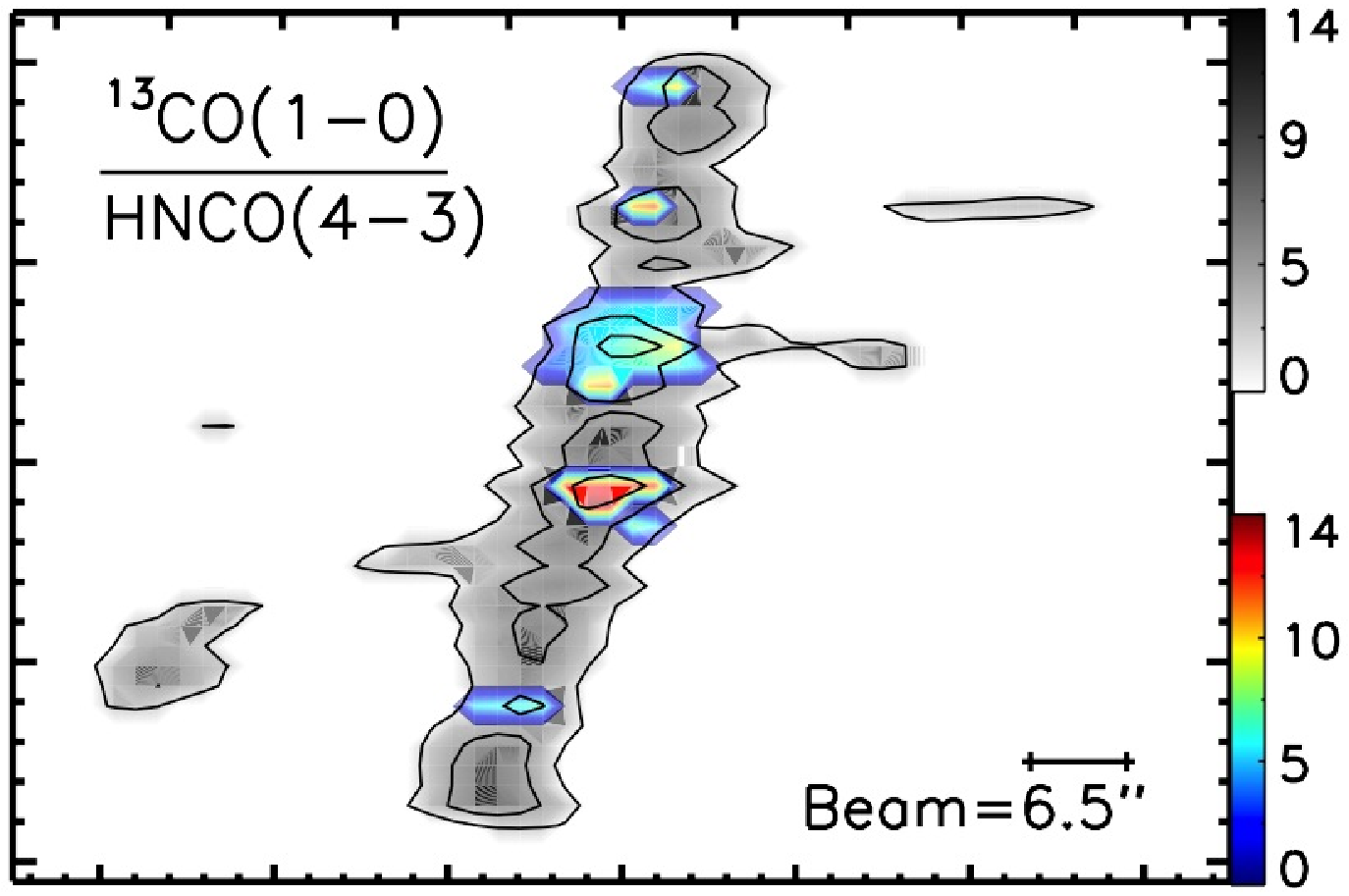}\\
  \vspace{-10pt}
  \includegraphics[width=5.1cm,clip=]{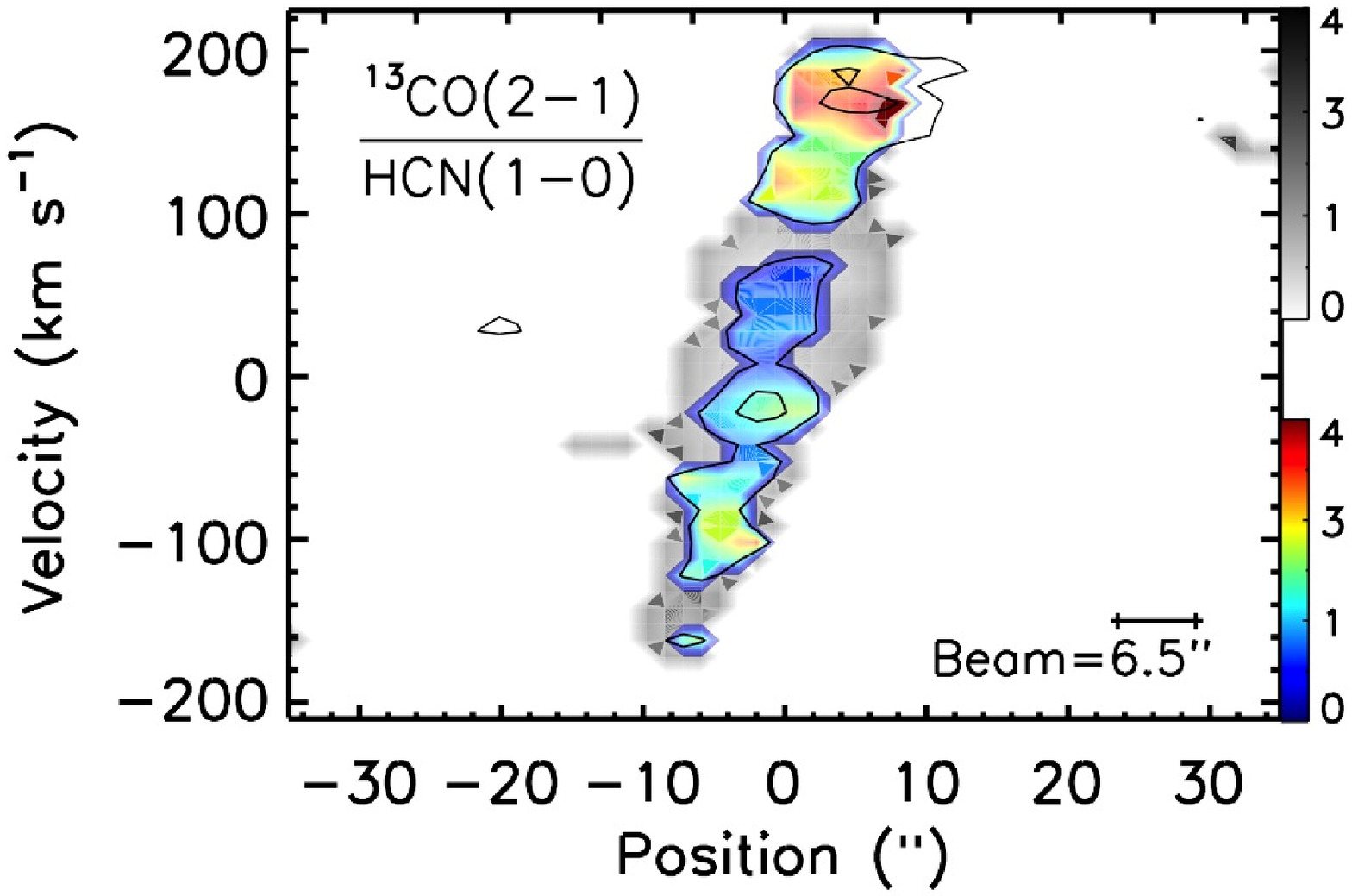}
  \hspace{-13pt}
  \includegraphics[width=4.0cm,clip=]{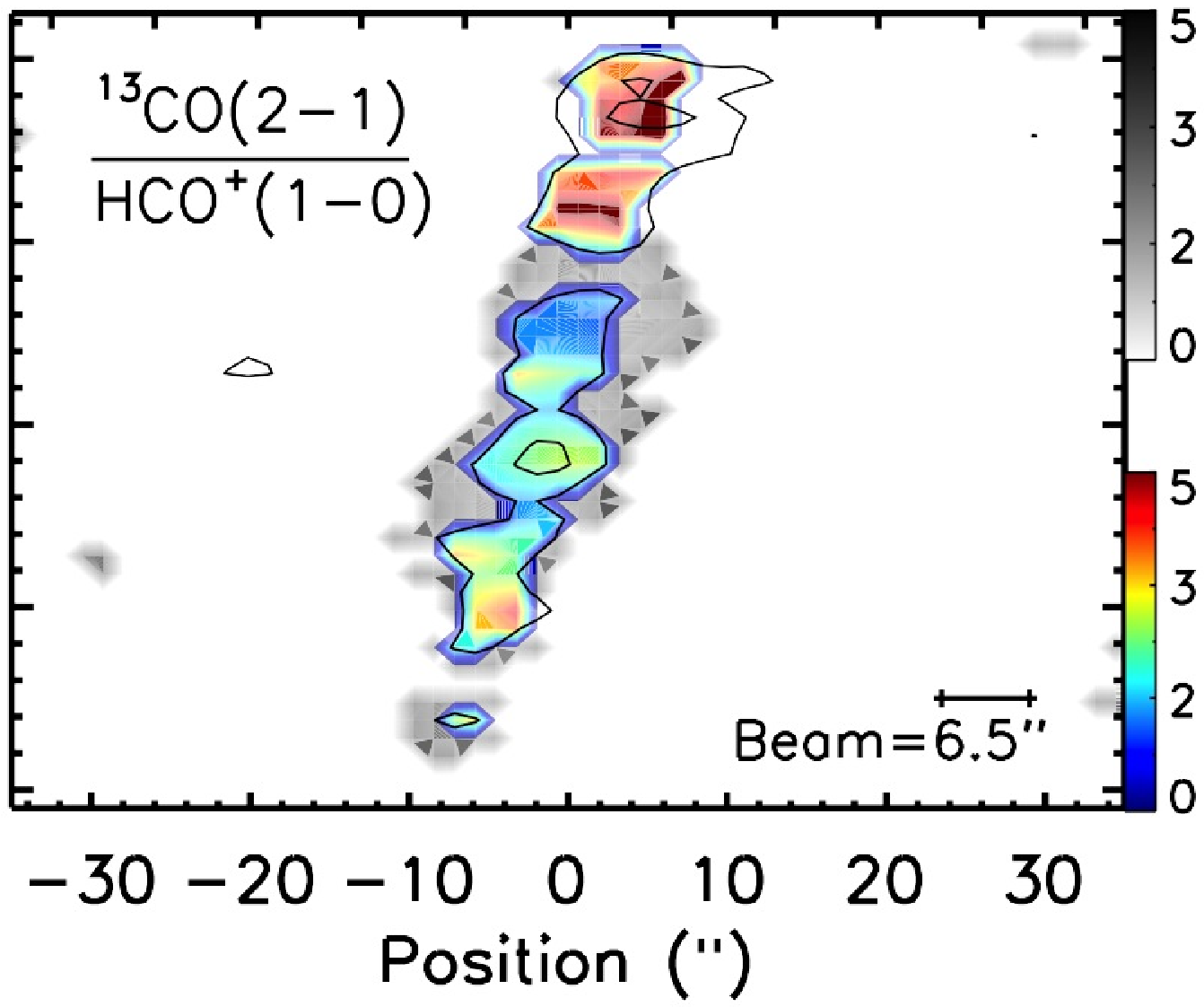}
  \hspace{-8pt}
  \includegraphics[width=4.0cm,clip=]{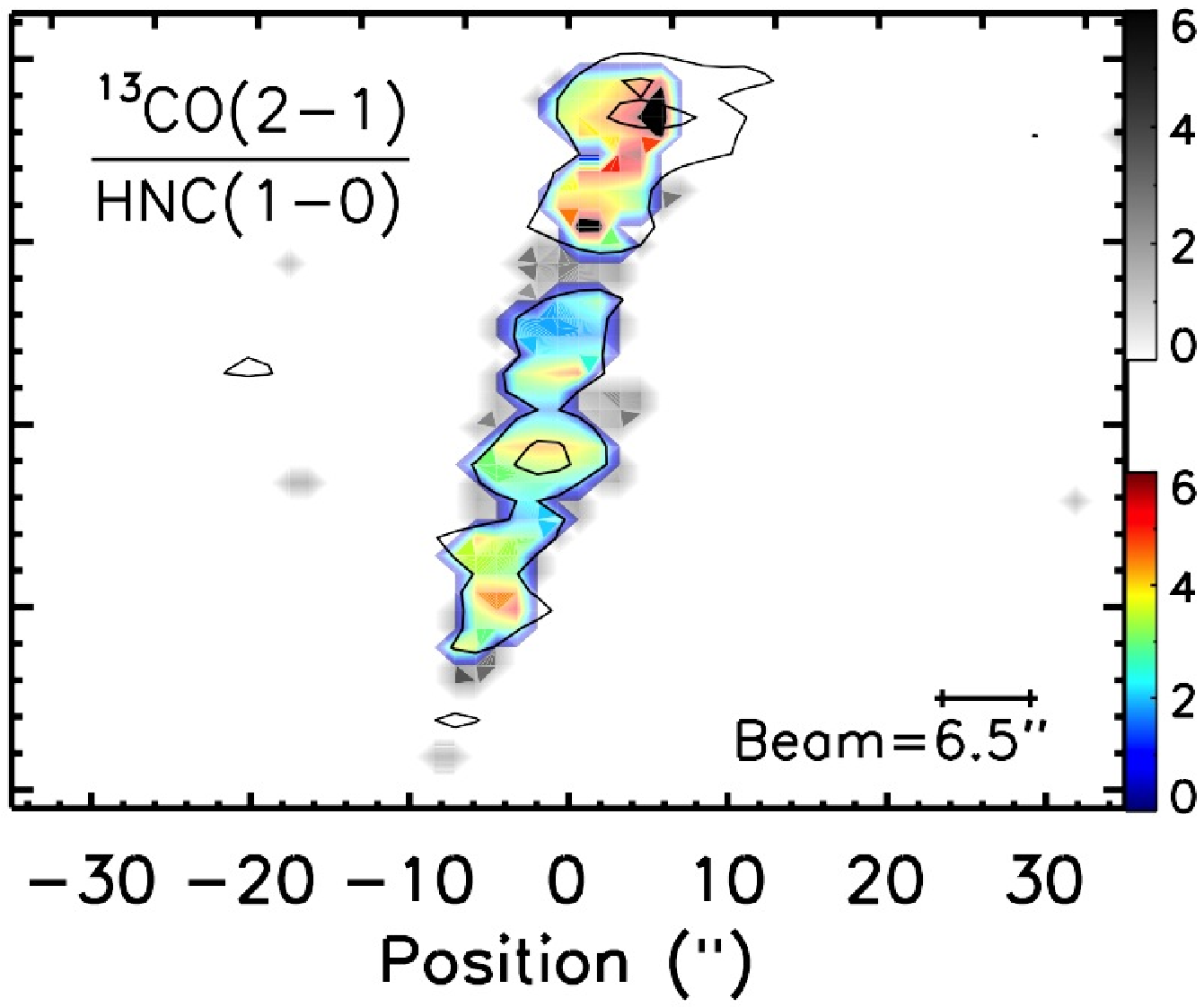}
  \hspace{-8pt}
  \includegraphics[width=4.0cm,clip=]{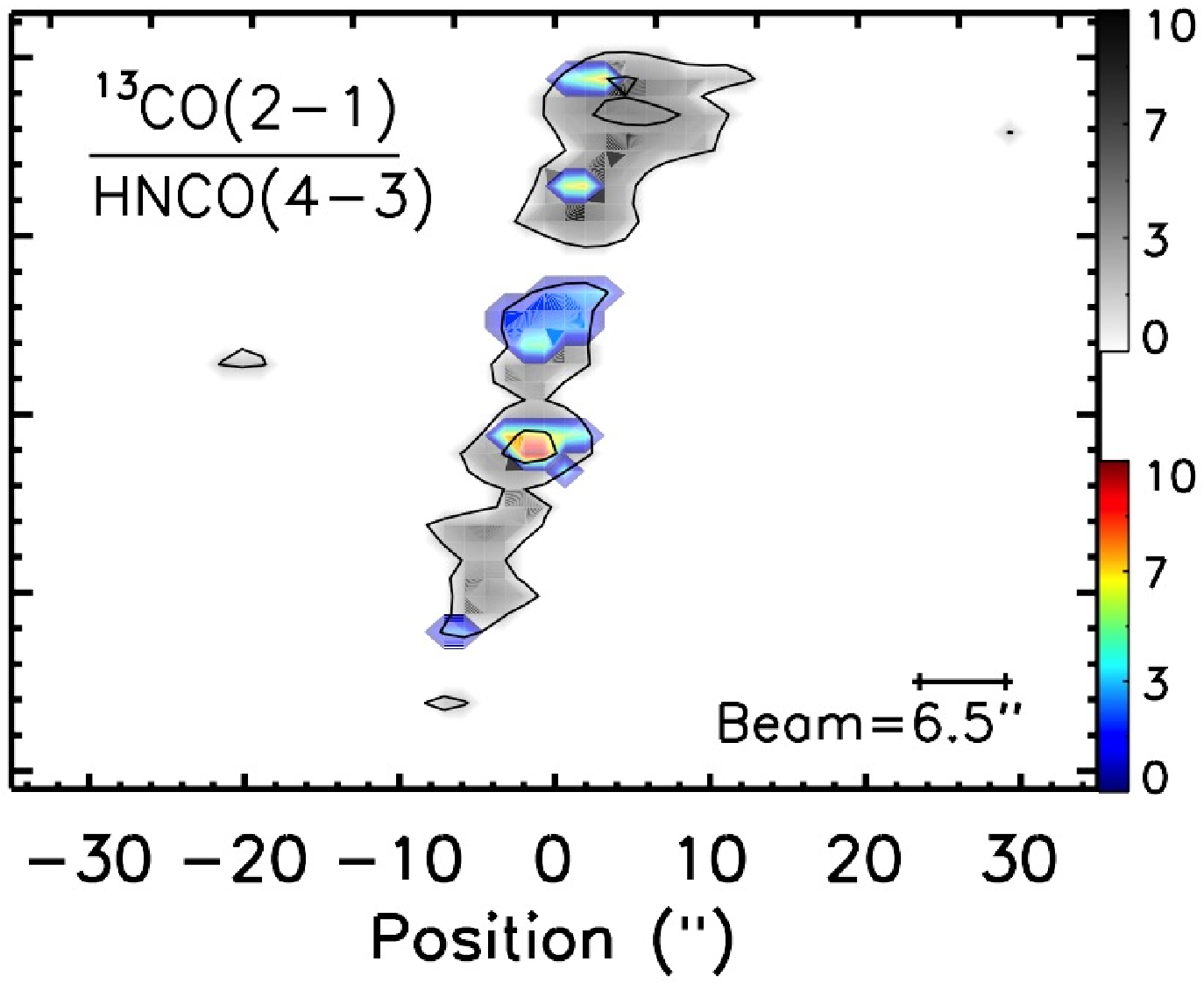}
  \caption{Same as Figure~\ref{fig:n4710ratioPVD1} but for the ratios
    of CO to dense gas tracer lines in NGC~4710.}
  \label{fig:n4710ratioPVD2}
\end{figure*}
%
%
\begin{figure*}
  \centering
  \hspace{1pt}
  \includegraphics[width=5.0cm,clip=]{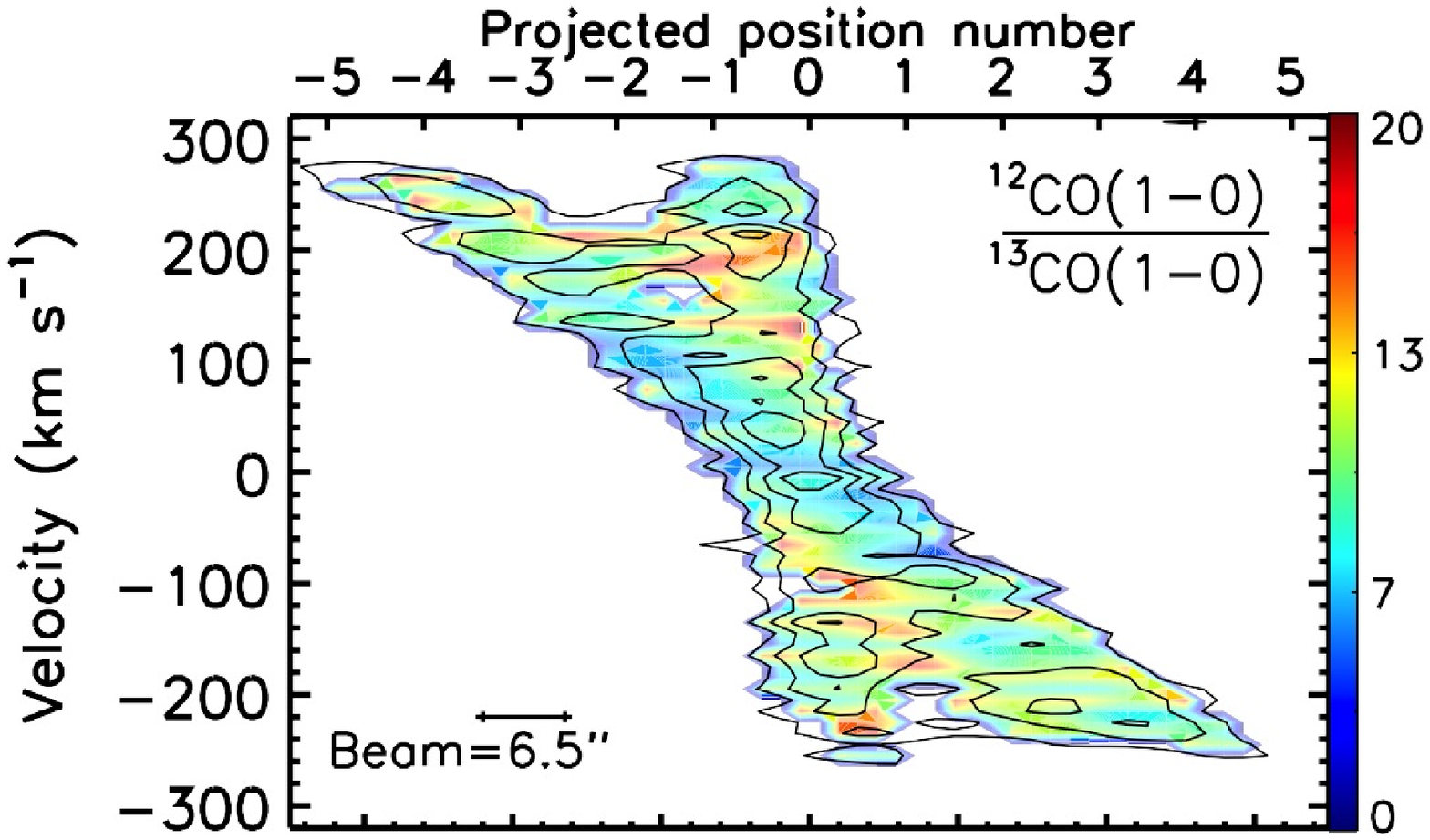}
  \hspace{-9pt}
  \includegraphics[width=4.2cm,clip=]{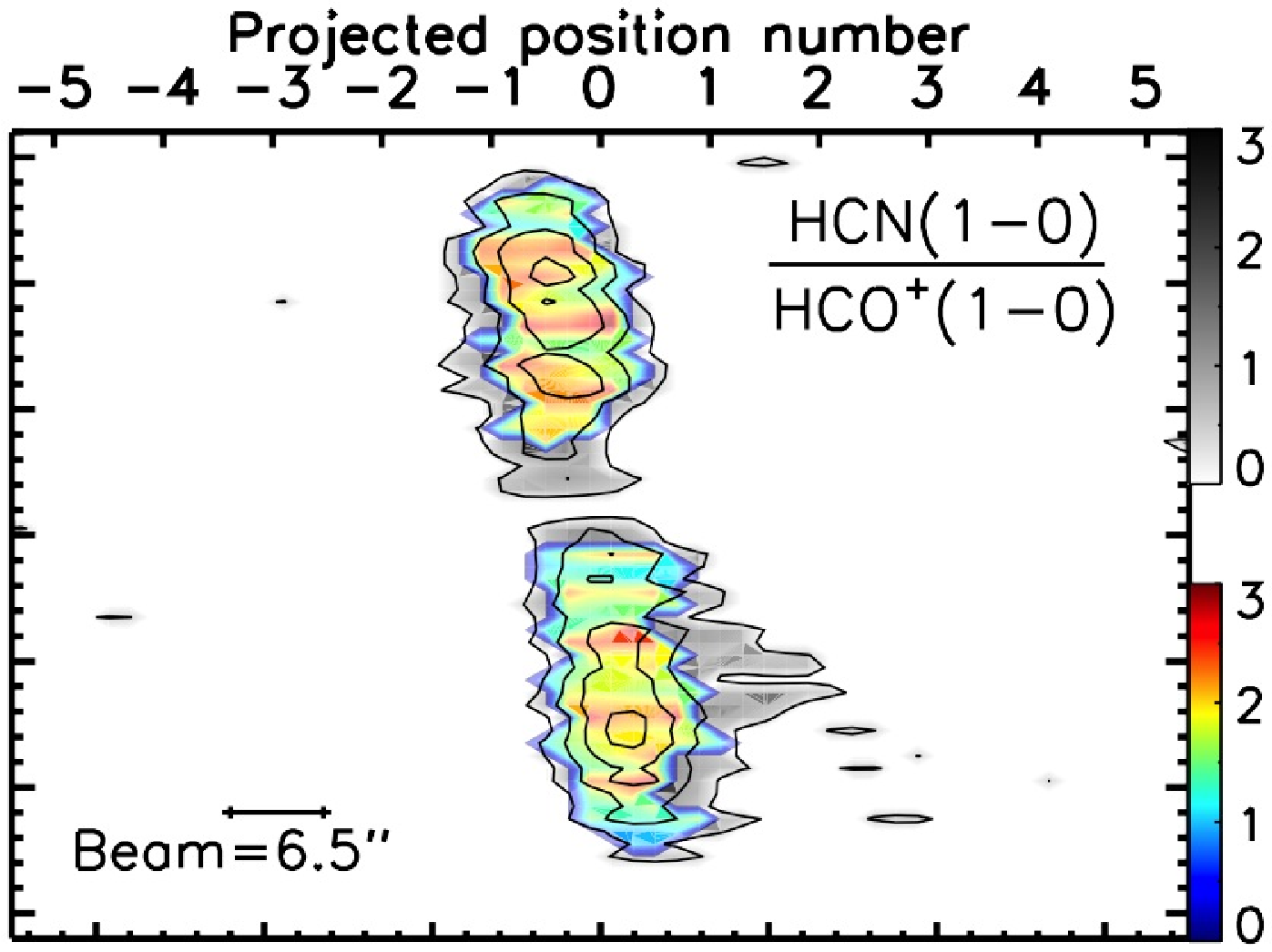}
  \hspace{-9pt}
  \includegraphics[width=4.2cm,clip=]{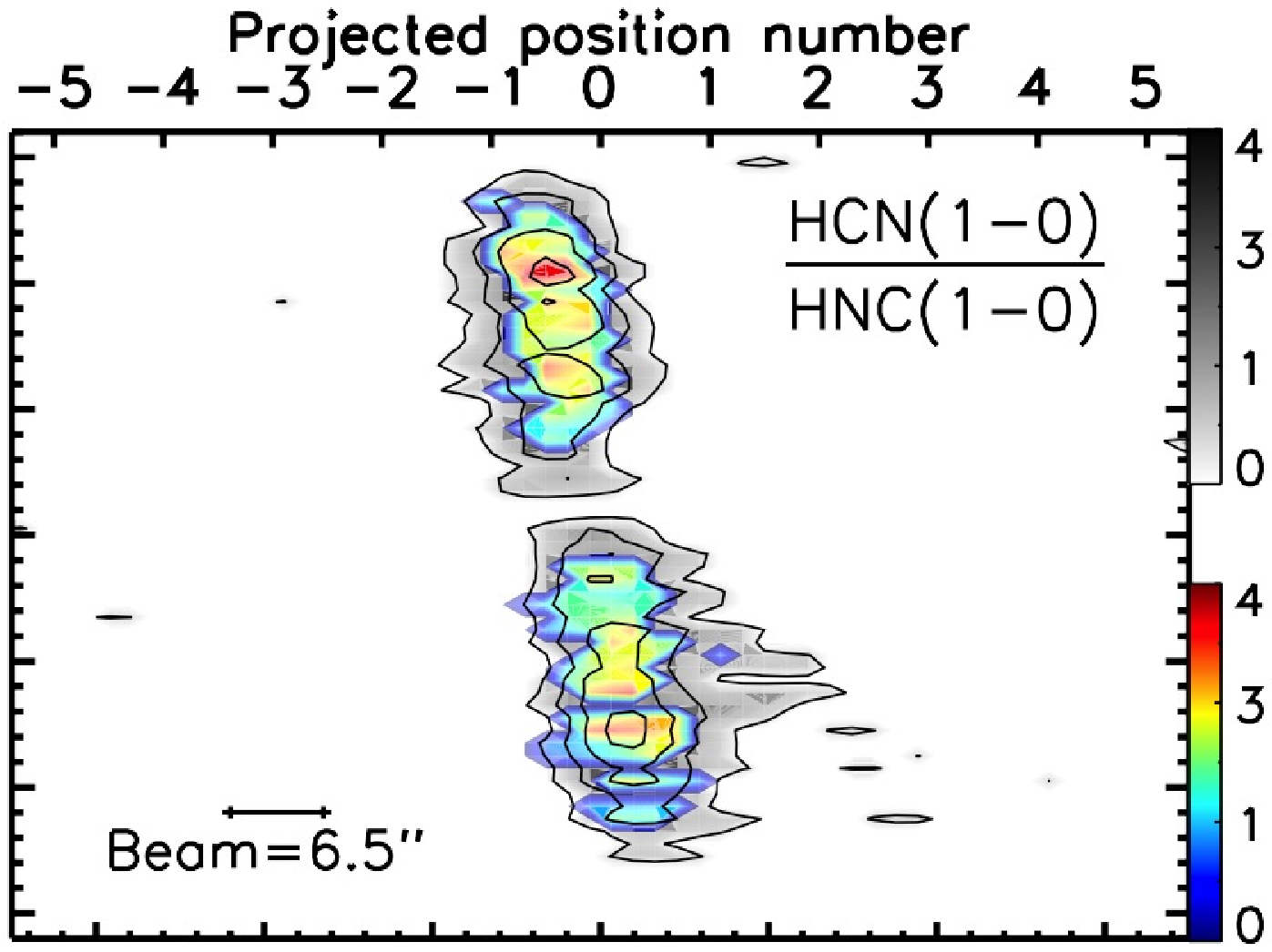}
  \hspace{-9pt}
  \includegraphics[width=4.2cm,clip=]{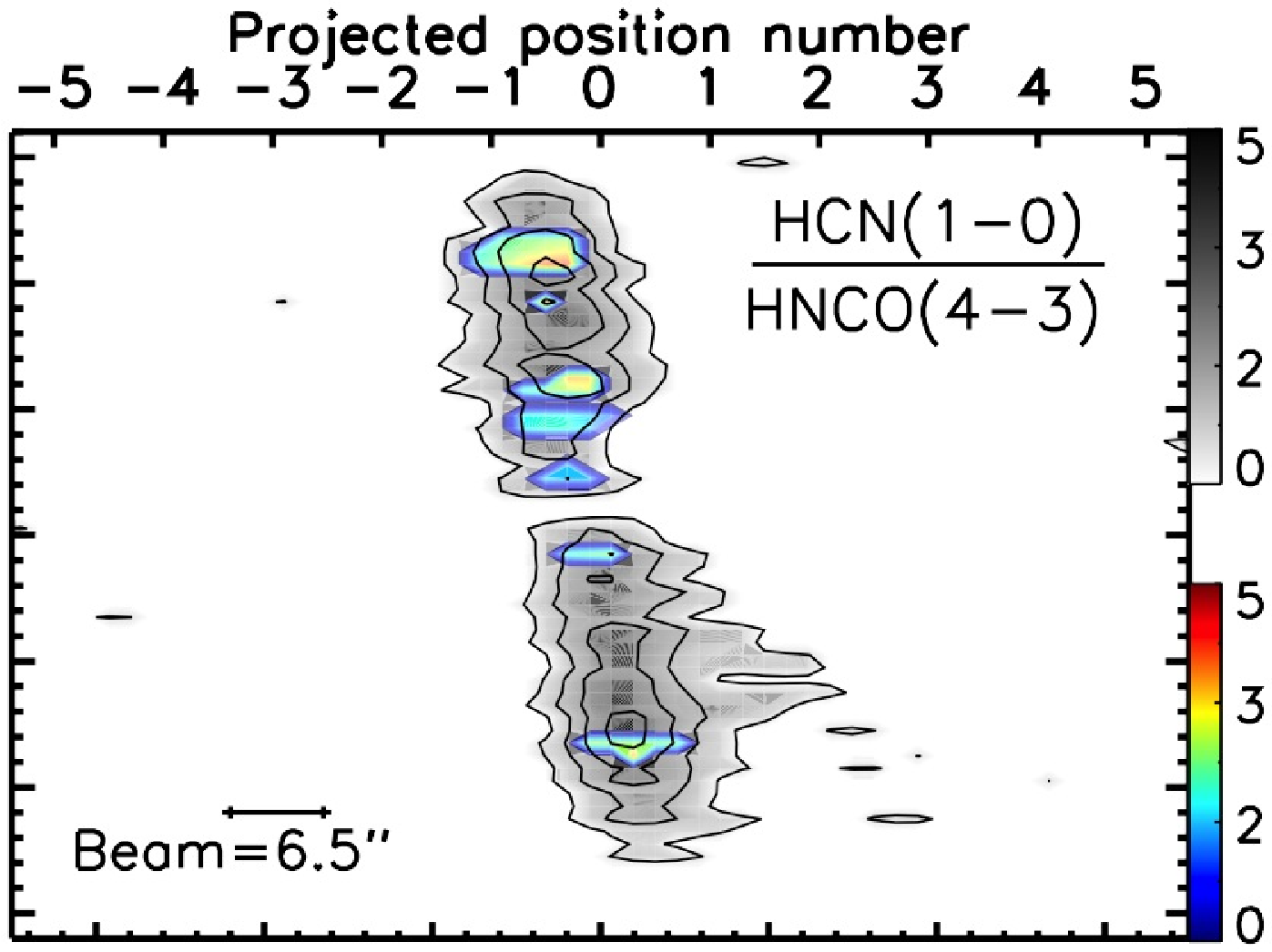}\\
  \vspace{-7pt}
  \includegraphics[width=5.0cm,clip=]{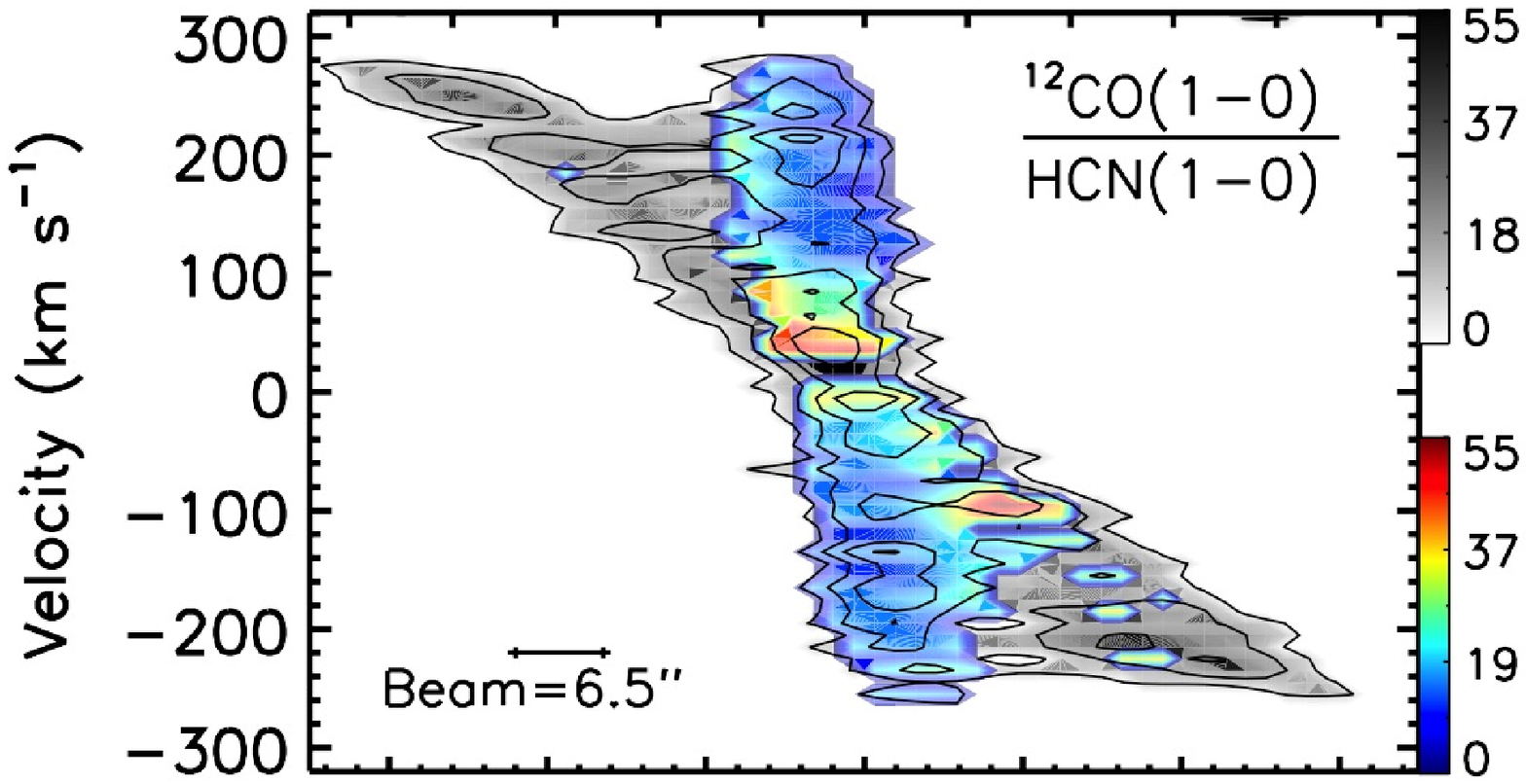}
  \hspace{-9pt}
  \includegraphics[width=4.2cm,clip=]{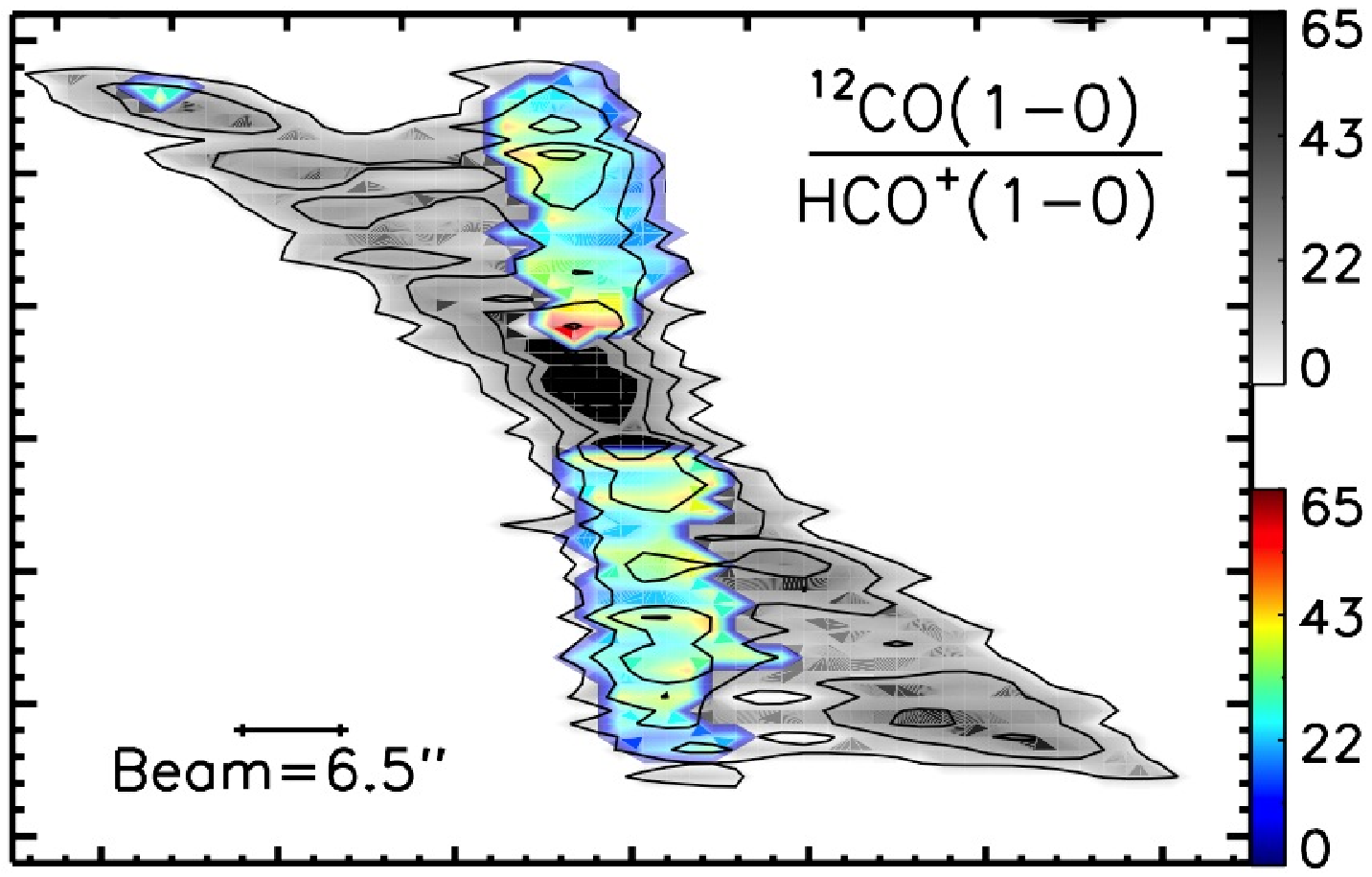}
  \hspace{-9pt}
  \includegraphics[width=4.2cm,clip=]{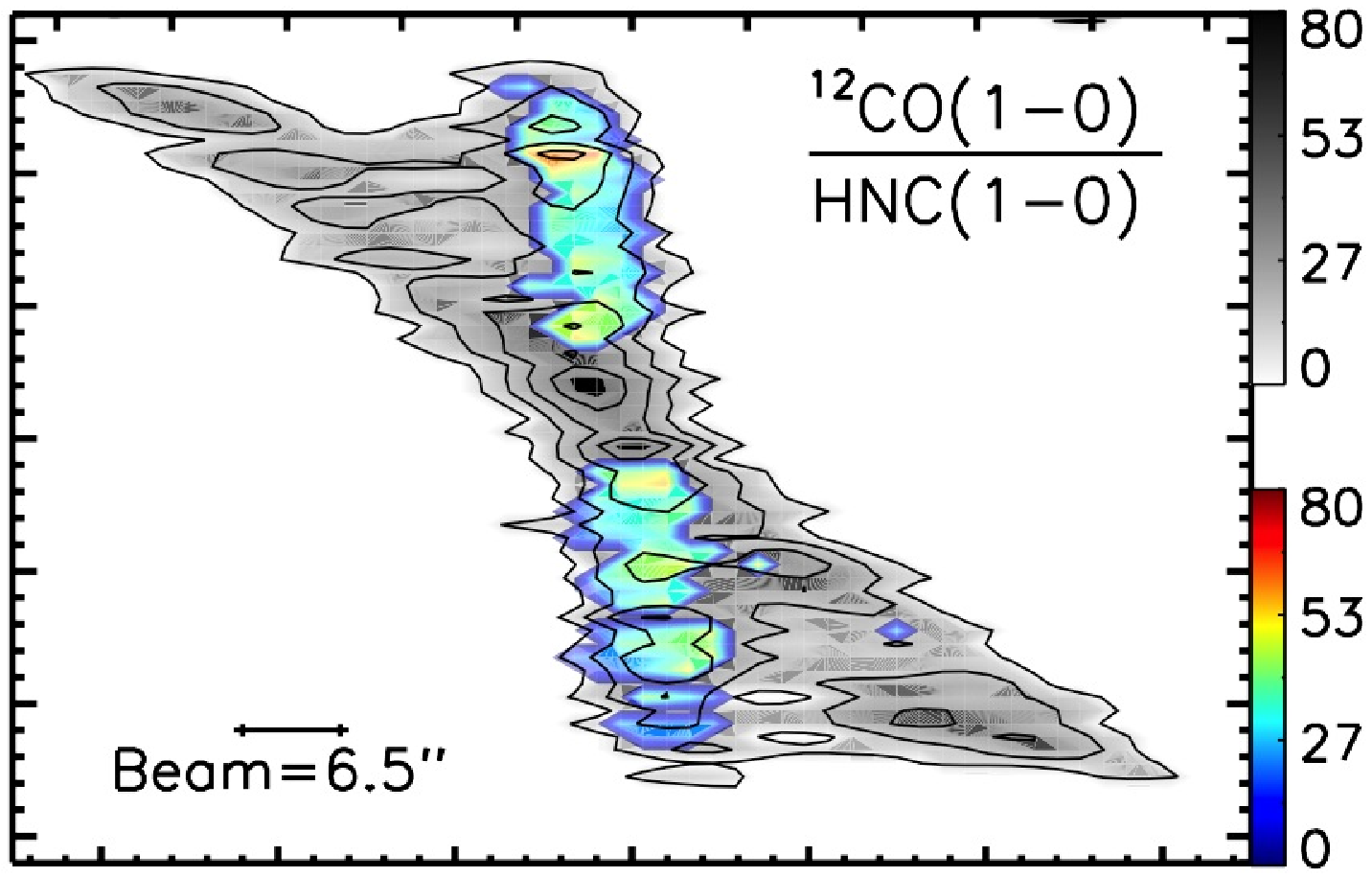}
  \hspace{-9pt}
  \includegraphics[width=4.2cm,clip=]{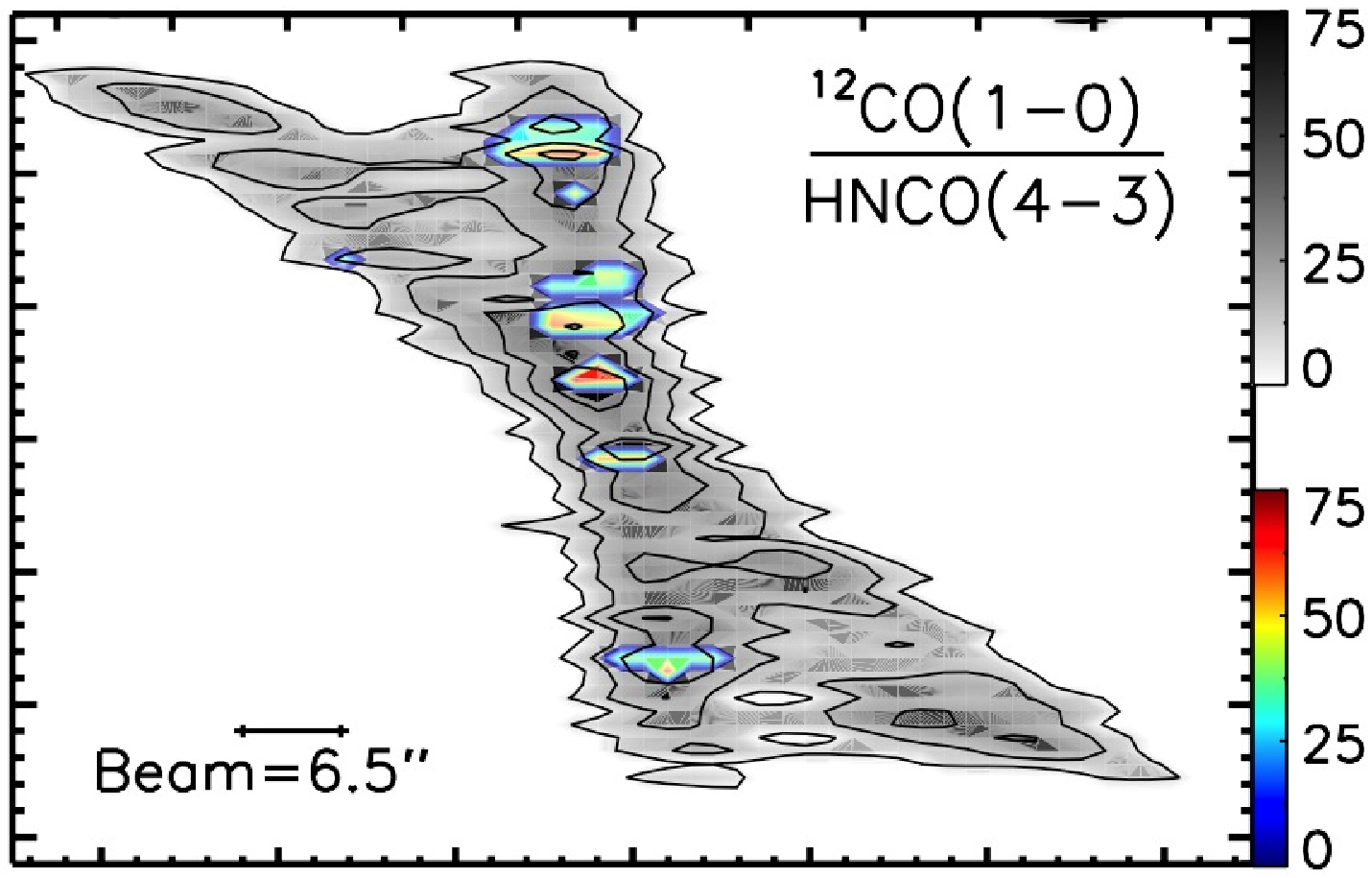}\\
  \vspace{-7pt}
  \includegraphics[width=5.0cm,clip=]{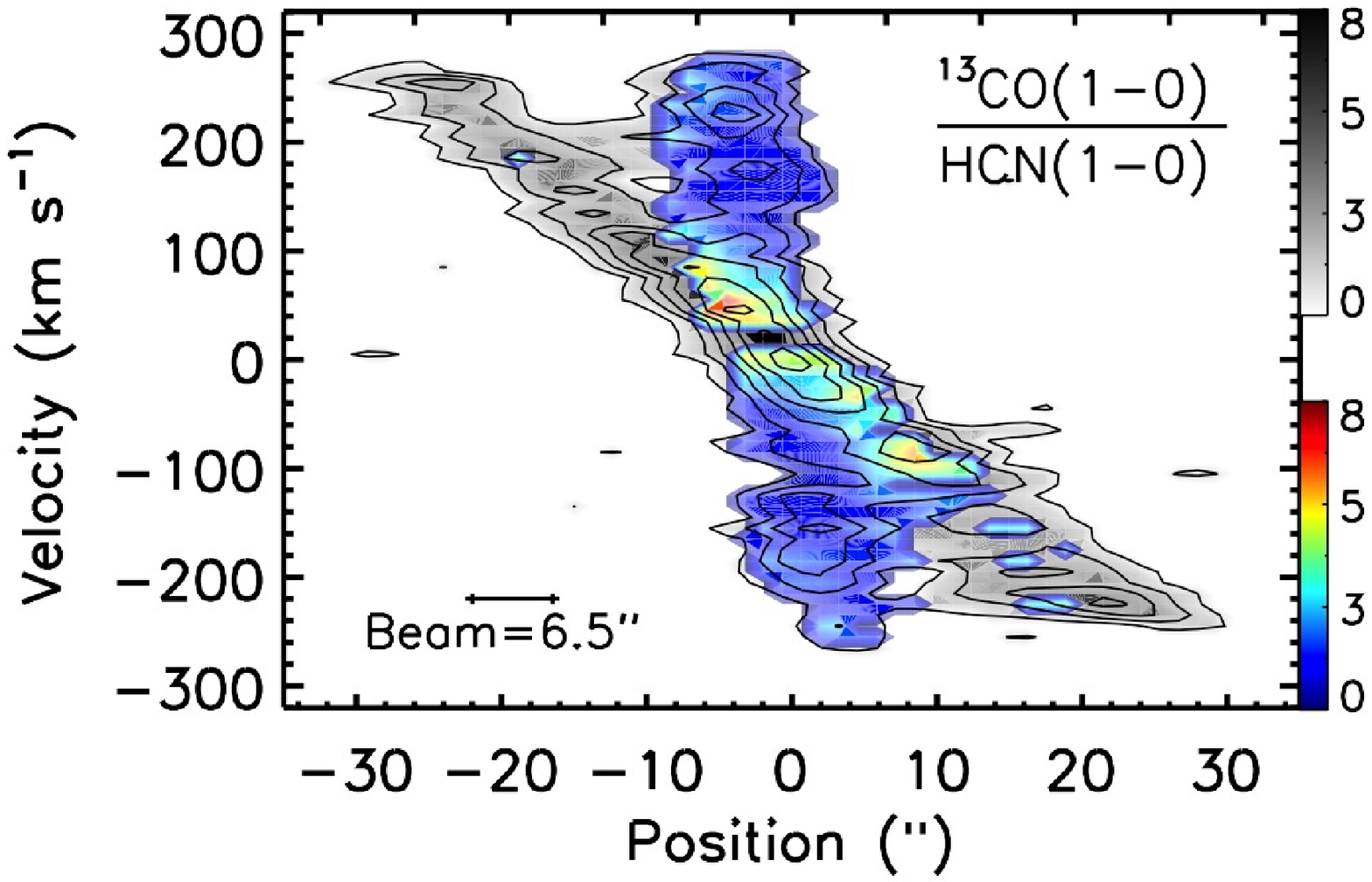}
  \hspace{-9pt}
  \includegraphics[width=4.2cm,clip=]{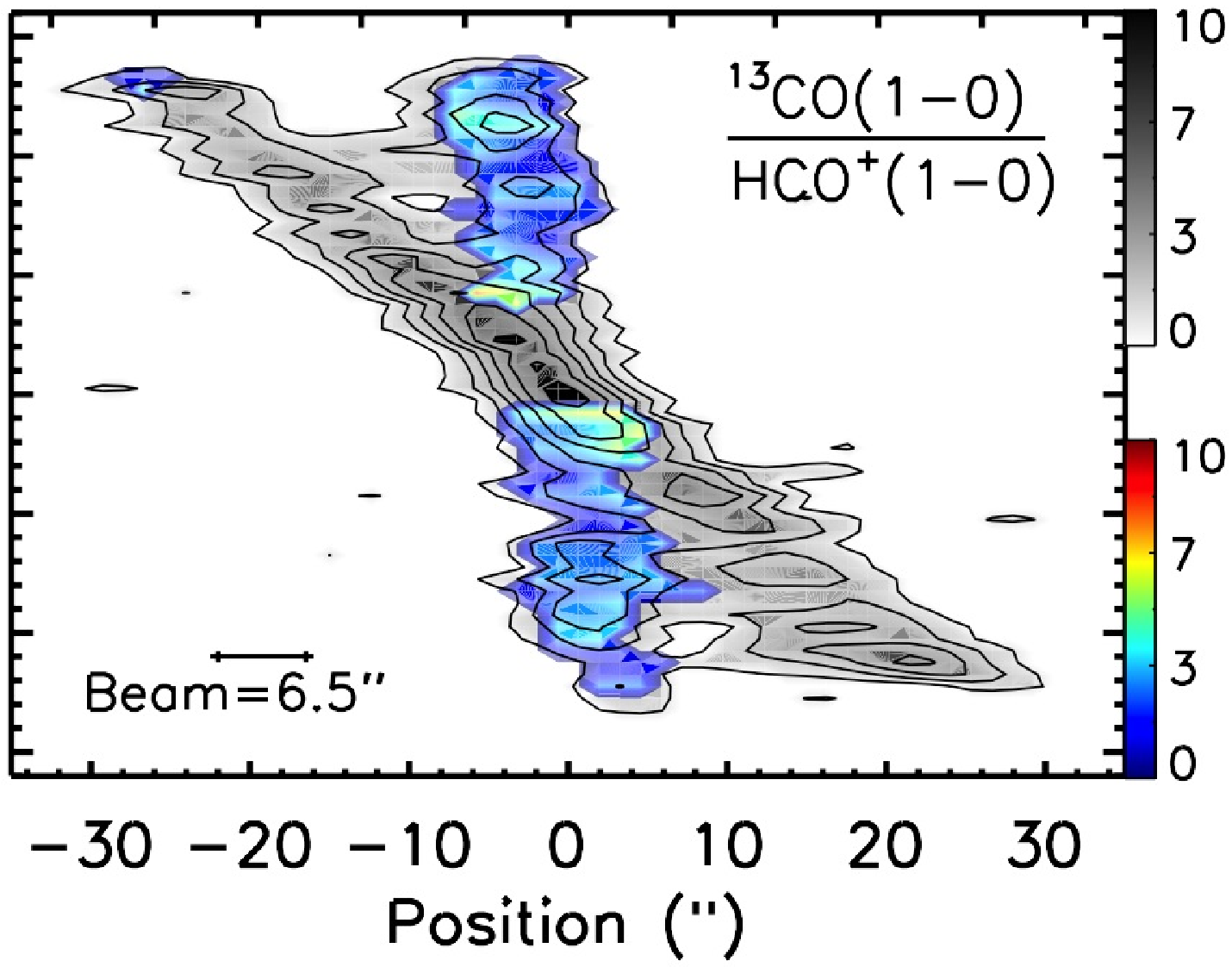}
  \hspace{-9pt}
  \includegraphics[width=4.2cm,clip=]{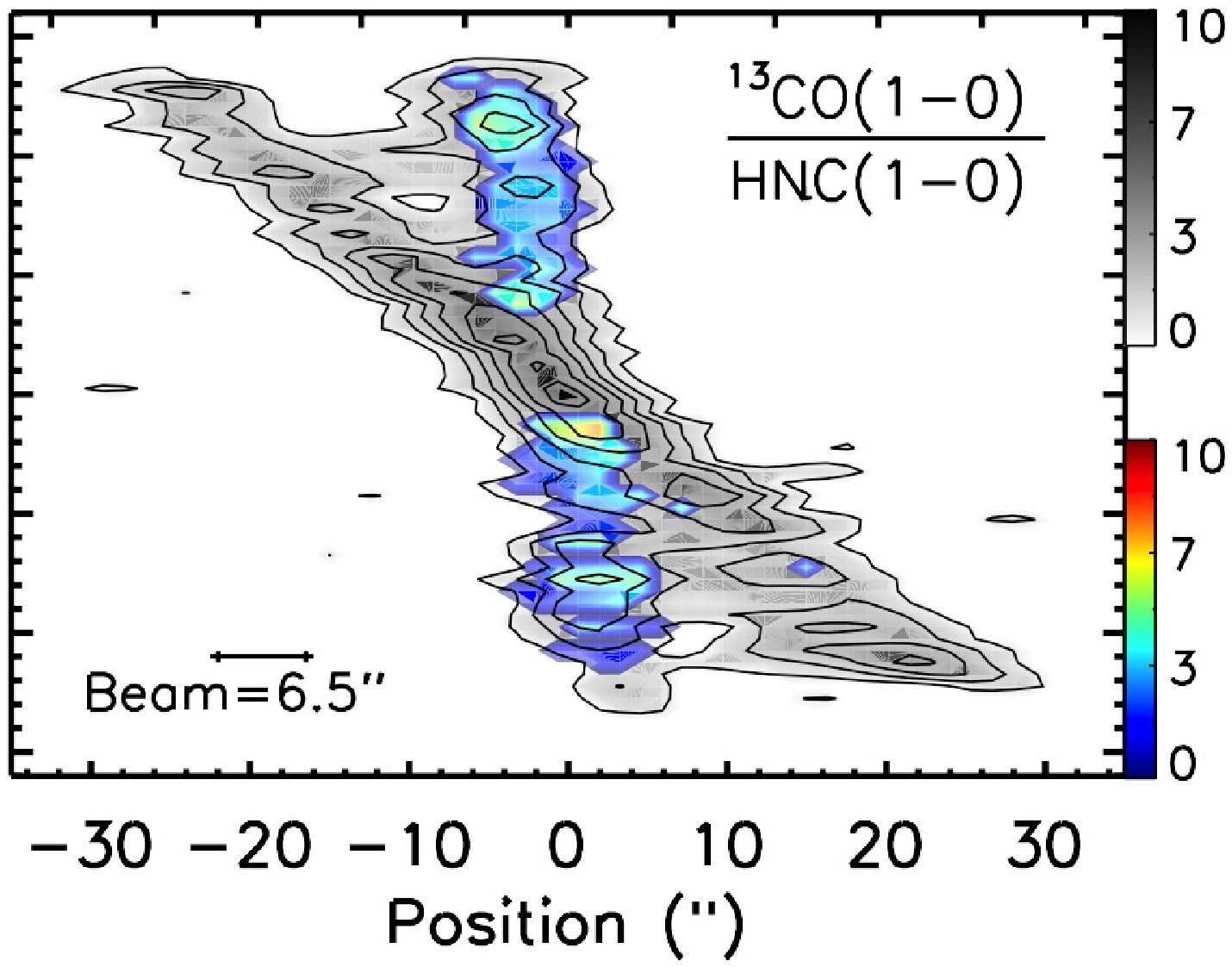}
  \hspace{-9pt}
  \includegraphics[width=4.2cm,clip=]{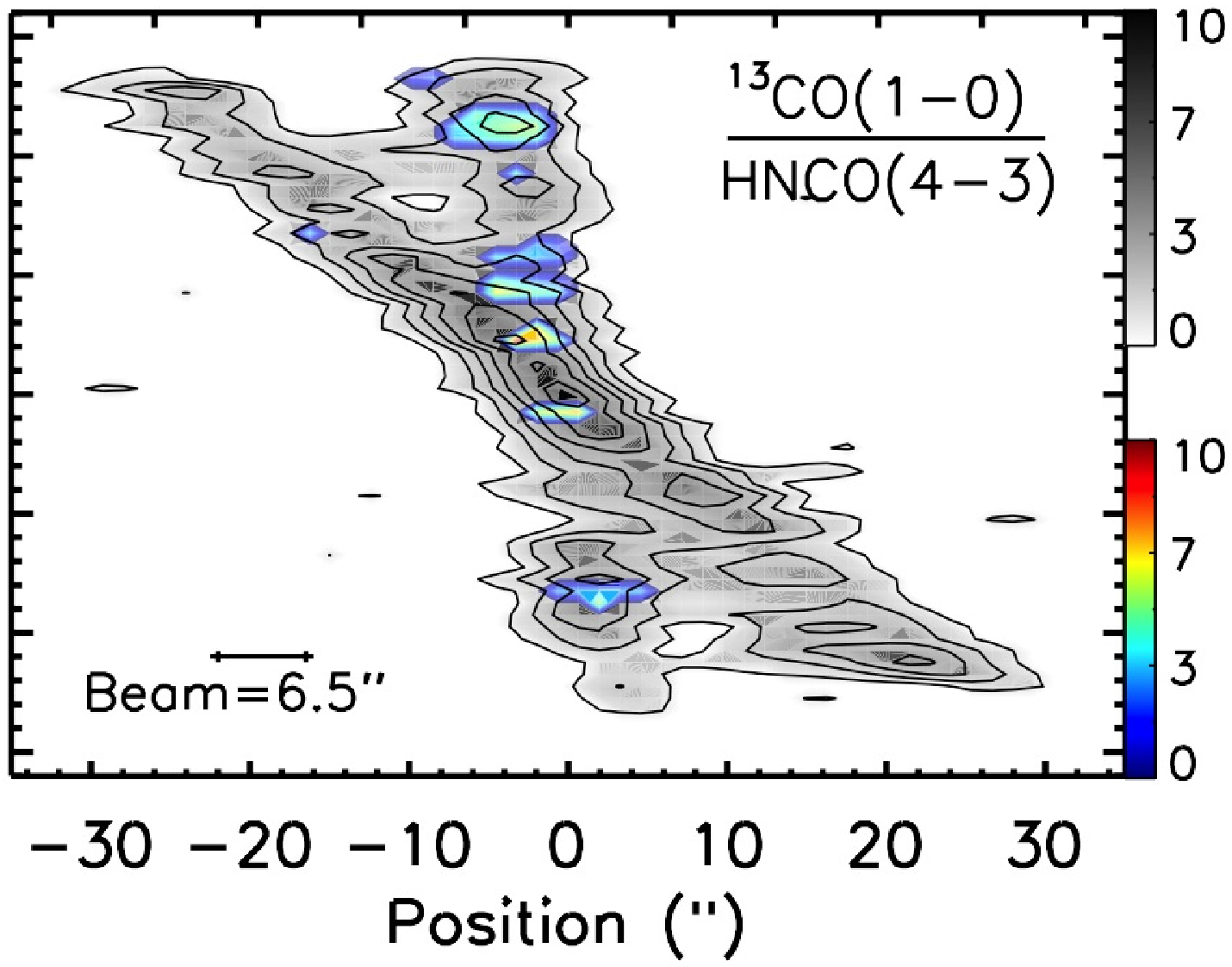}\\
  \caption{Same as Figures~\ref{fig:n4710ratioPVD1} and
    \ref{fig:n4710ratioPVD2} but for NGC~5866.}
  \label{fig:n5866ratioPVD}
\end{figure*}
%
%
\subsection{Integrated line intensity ratios as a function of projected radius}
\label{sec:extract}
\begin{figure*}
  \centering
  \includegraphics[width=7.0cm,clip=]{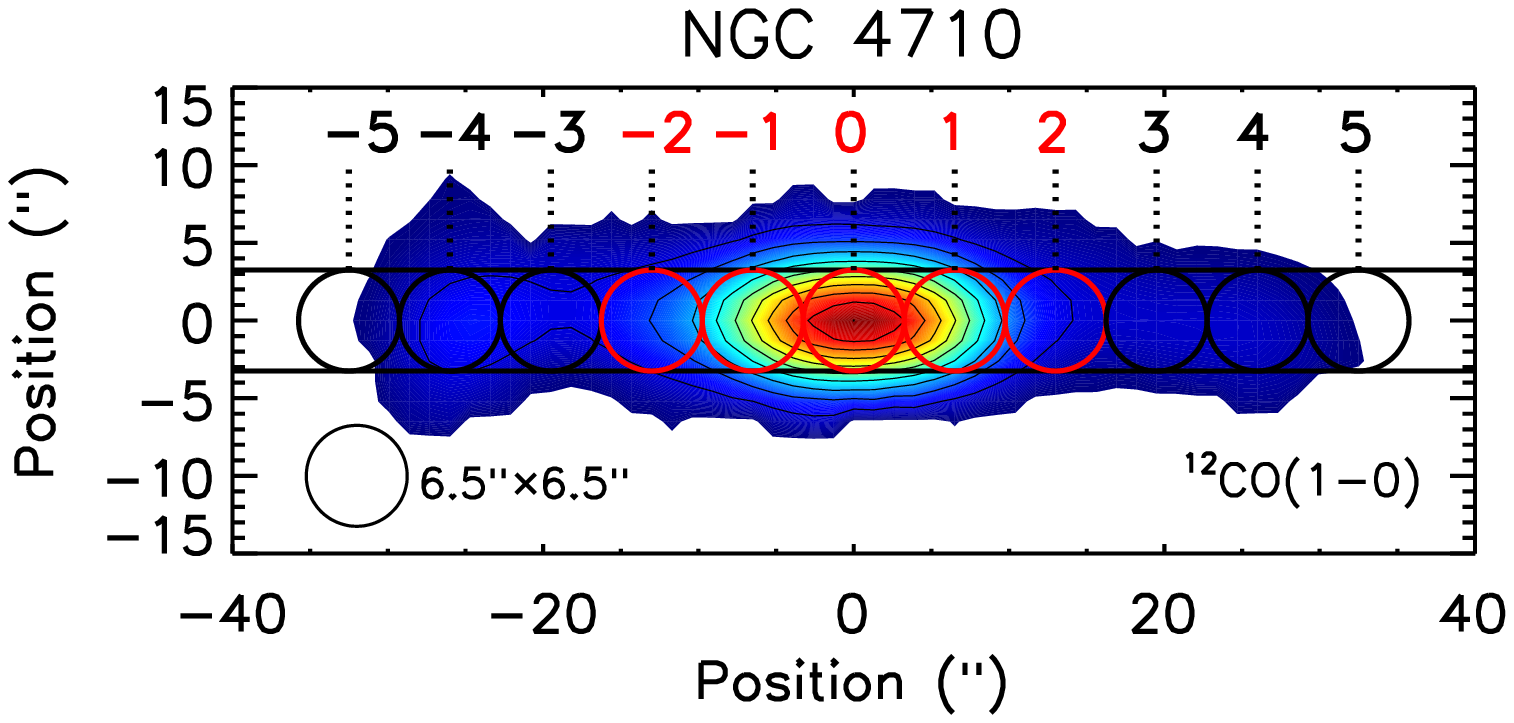}
  \hspace{-10pt}
  \includegraphics[width=7.0cm,clip=]{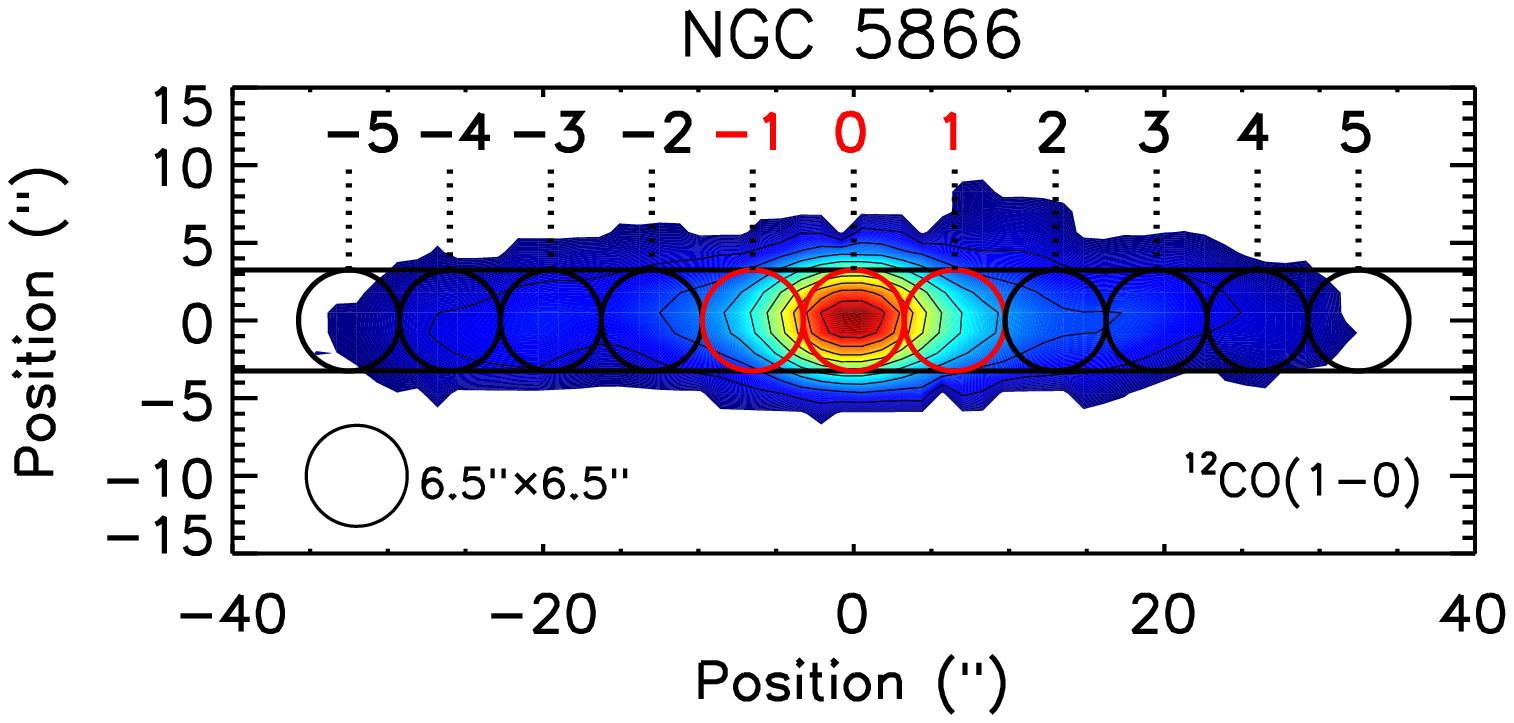} \\
  \includegraphics[width=7.0cm,clip=]{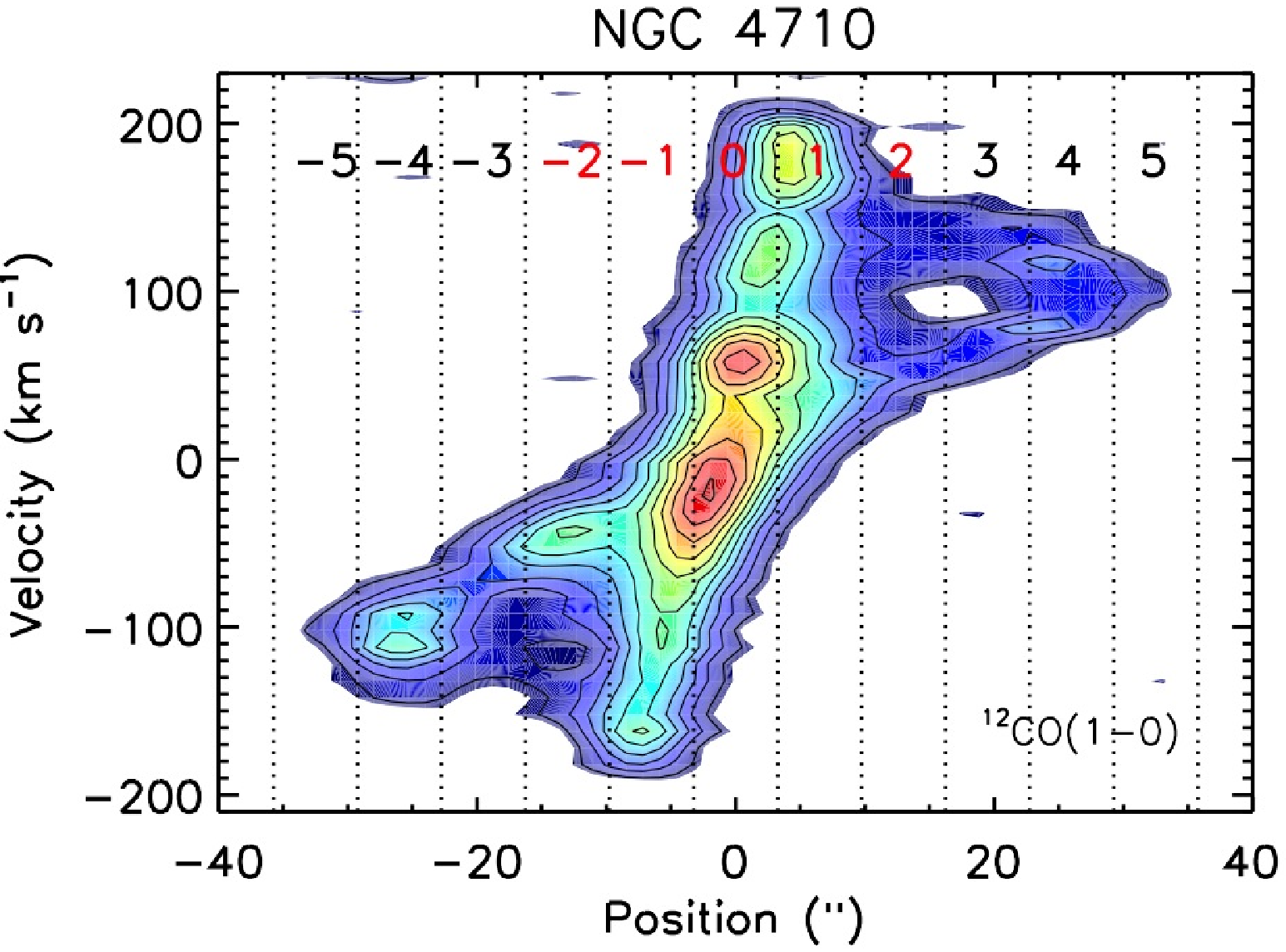}
  \hspace{-10pt}
  \includegraphics[width=7.0cm,clip=]{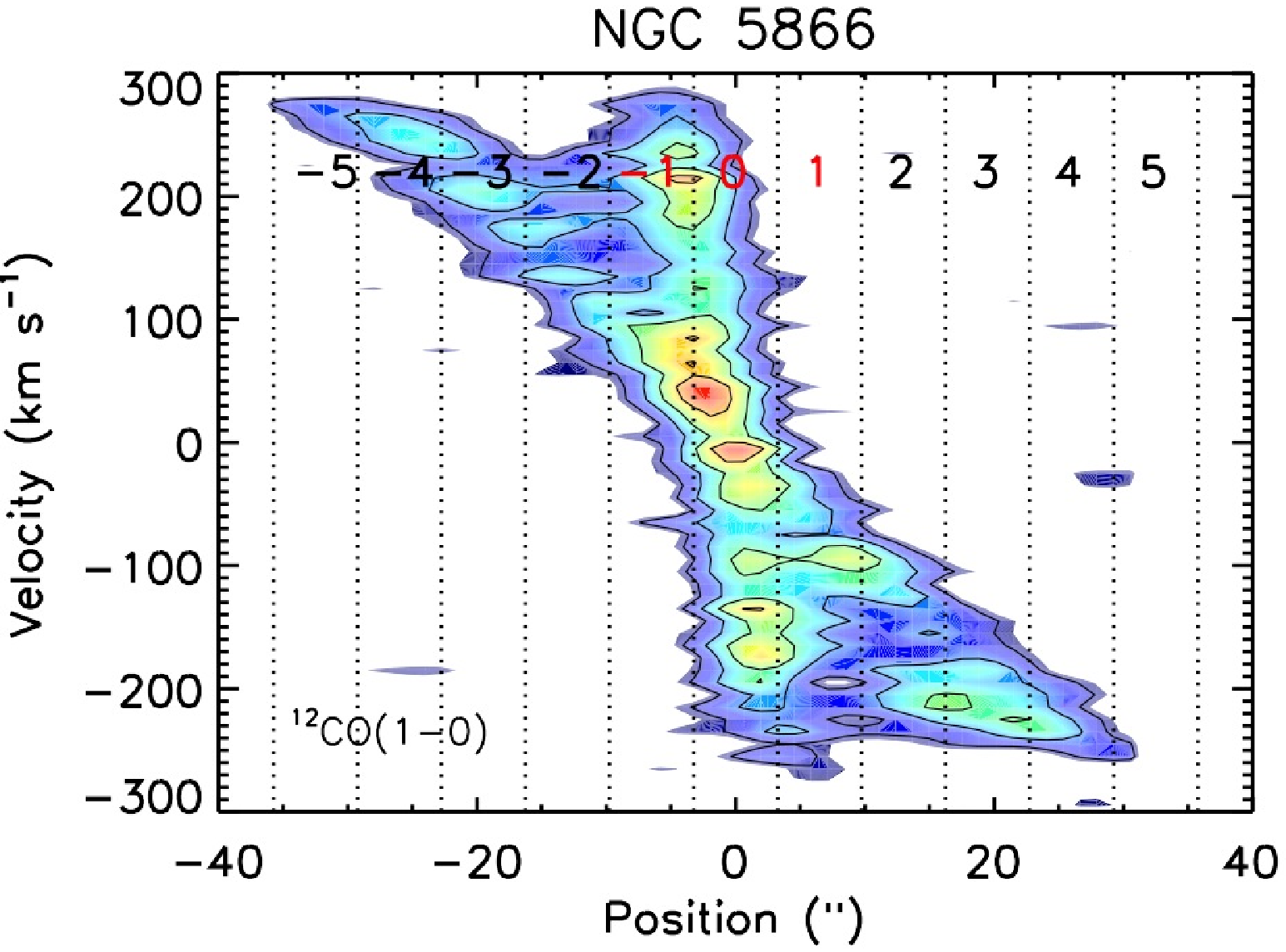}\\
  \caption{{\bf Top:} Illustration of the projected positions along
    the major-axis of NGC~4710 (left) and NGC~5866 (right), where the
    line ratios are extracted and studied. {\bf Bottom:} Corresponding
    $5$-pixel (i.e.\ one beam width) slices in the PVDs. Red circles
    and associated numbers indicate positions where both kinematic
    components (nuclear disc and inner ring) are present along the
    line of sight, requiring a double-Gaussian fit (see
    Fig.~\ref{fig:specext}, right). Black circles and associated
    numbers indicate positions where a single component (inner ring)
    is present, requiring a single Gaussian fit (see
    Fig.~\ref{fig:specext}, left). The circles are one beam width
    ($6\farcs5$) in diameter, equal to the separation
      between the projected positions.}
  \label{fig:pos}
\end{figure*}

\begin{figure*}
  \includegraphics[width=6.0cm,clip=,angle=90]{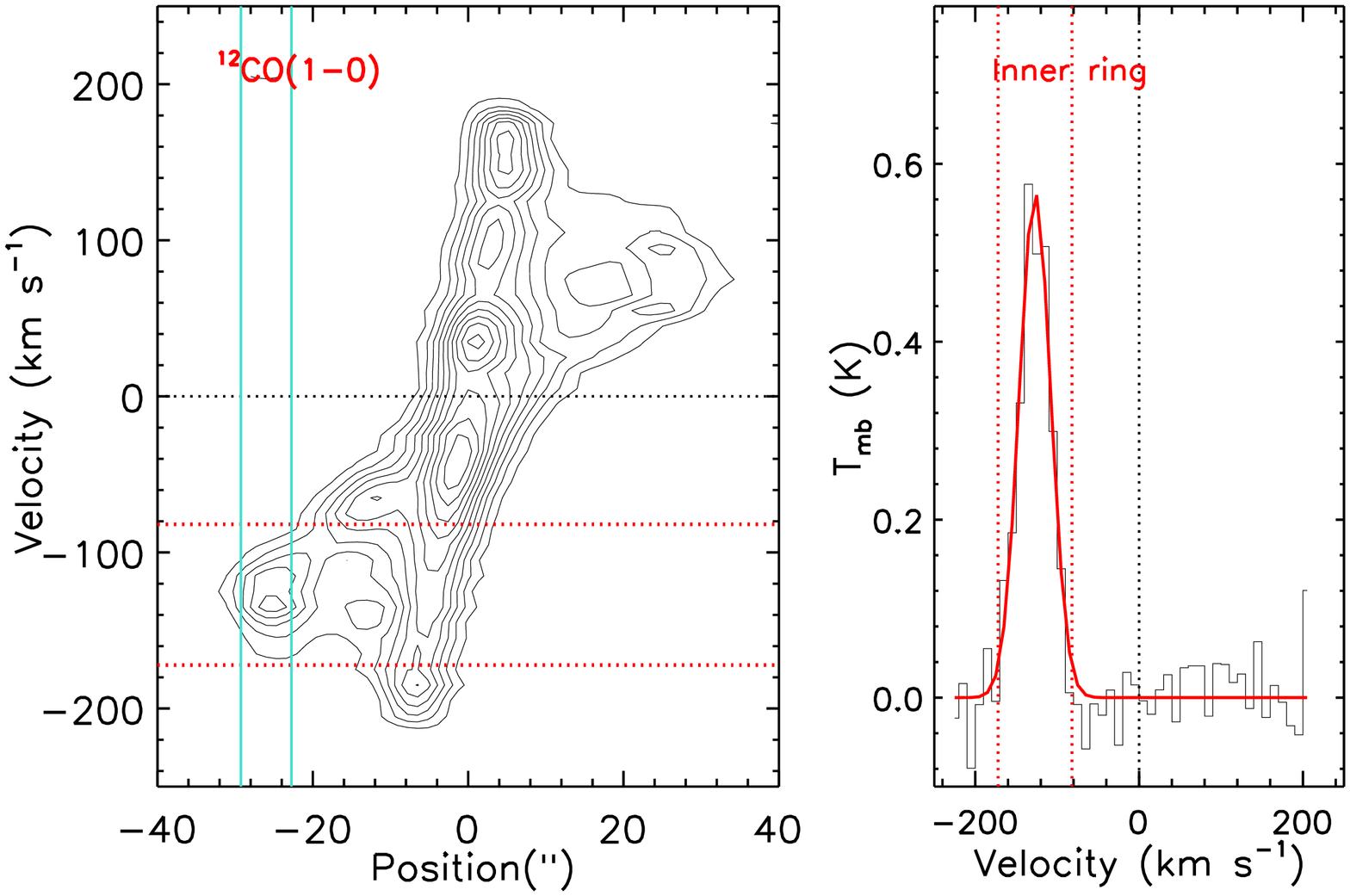}
  \hspace{10mm}
  \includegraphics[width=6.0cm,clip=,angle=90]{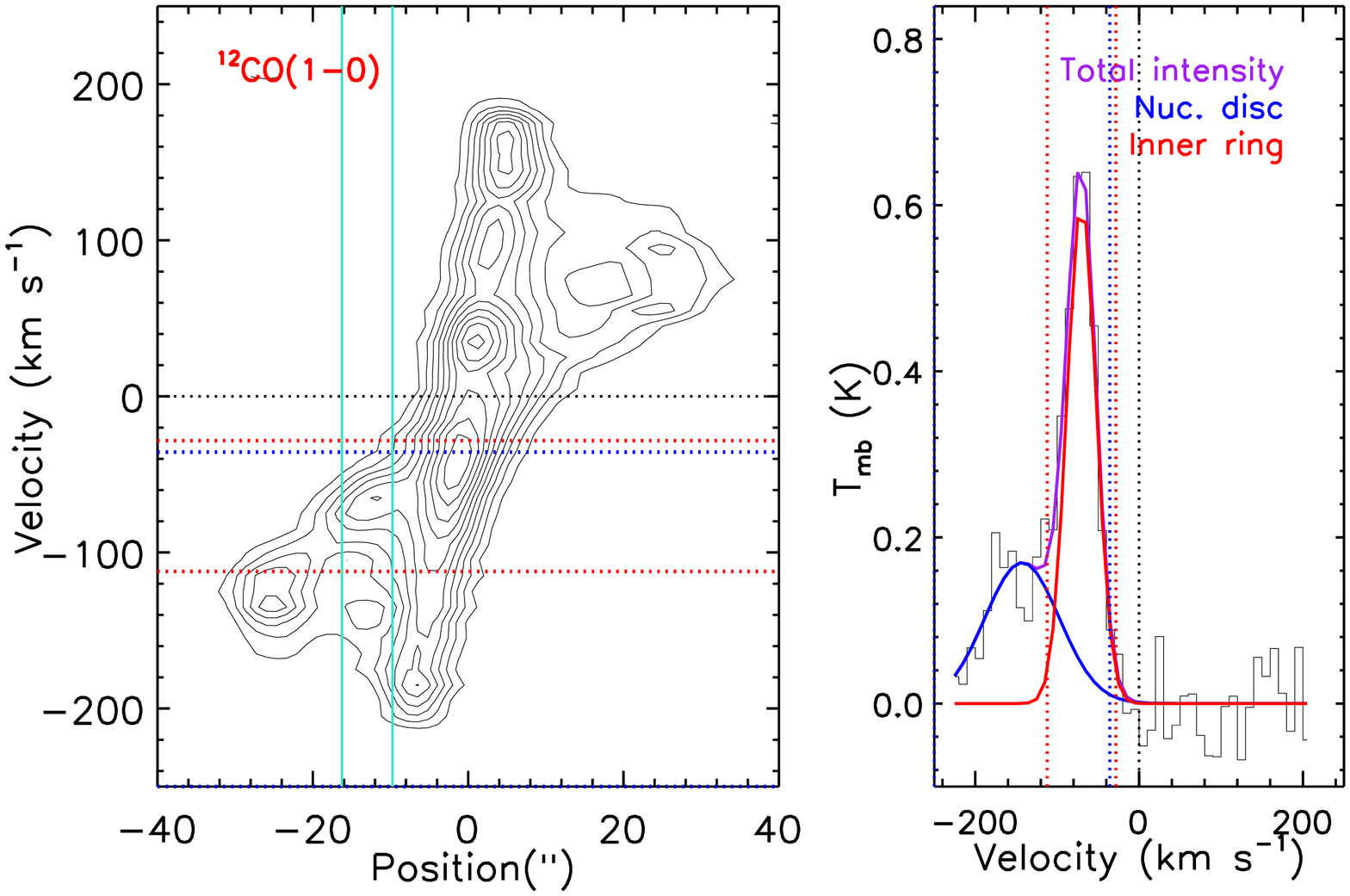}\\
  \includegraphics[width=6.0cm,clip=,angle=90]{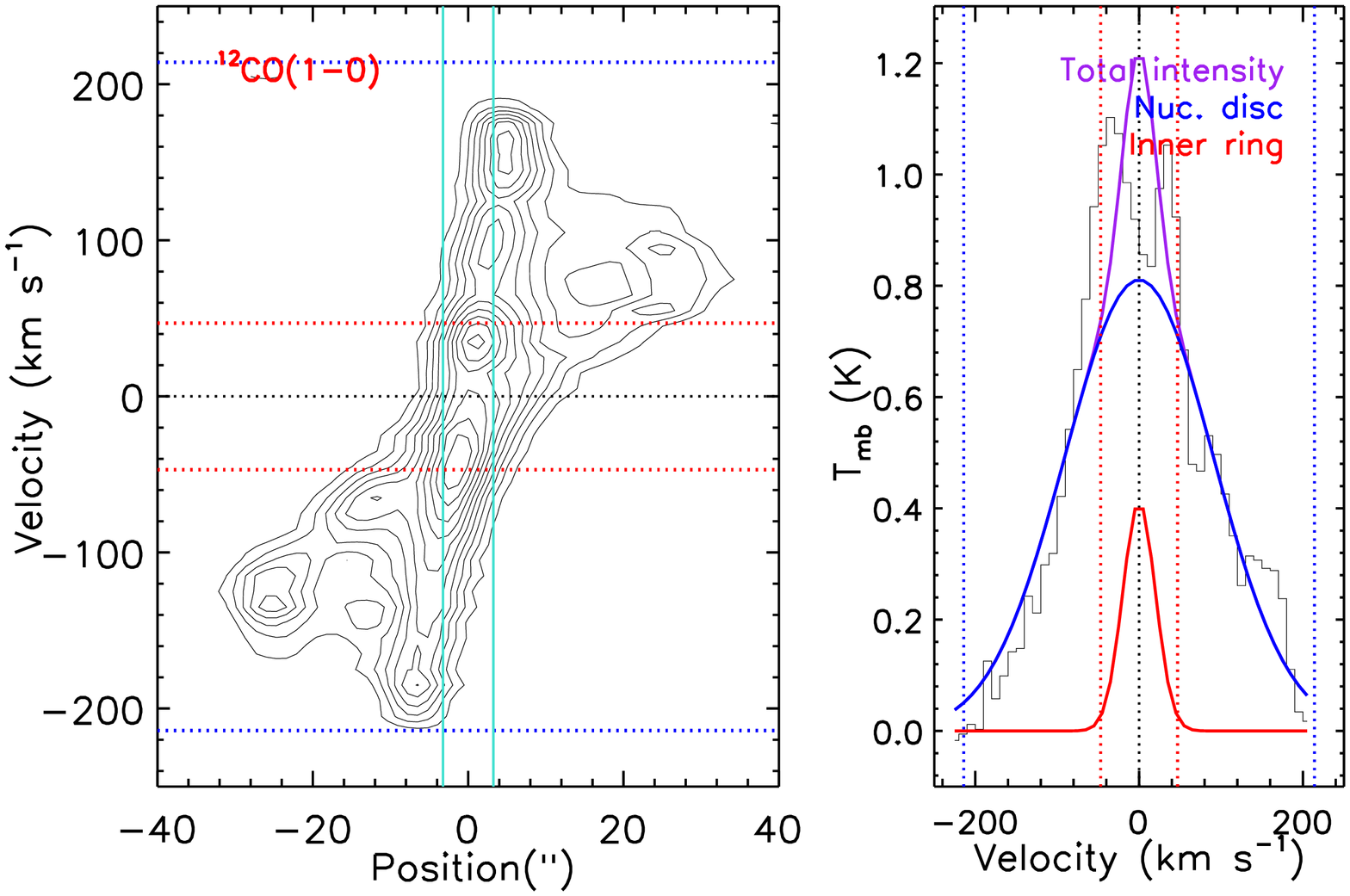}
  \caption{Illustration of the spectrum extraction process using the
    NGC~4710 $^{12}$CO(1-0) PVD. {\bf Top left:} $^{12}$CO(1-0) PVD
    and the spectrum extracted at position~$-4$, in the outskirt of the
    disc where a single kinematic component (inner ring) is present
    along the line of sight. {\bf Top right:} Same for position~$-2$,
    where both kinematic components are present (nuclear disc and
    inner ring). {\bf Bottom:} Same for position~$0$ (galaxy centre),
    where both kinematic components are again present. The red and
    blue solid lines overlaid on the spectra show the Gaussian
    profiles separately fitted to the emission of the inner ring and
    the nuclear disc, respectively, while the magenta solid lines show
    the sums of the multiple Gaussians. The red and blue dotted lines
    overlaid on the PVDs and spectra indicate velocities of $\pm$~FWHM
    with respect to the centre of the associated Gaussian (thus
    encompassing $\approx95\%$ of the total emission of each
    component). Black dotted lines overlaid on the PVDs and spectra
    indicate the galaxy heliocentric velocity. Turquoise solid lines
    overlaid on the PVDs show the (one beam) width of the spatial
    slice considered to extract the averaged spectrum at that
    position.}
  \label{fig:specext}
\end{figure*}

\begin{figure*}
  \centering
  \includegraphics[width=5.5cm,clip=]{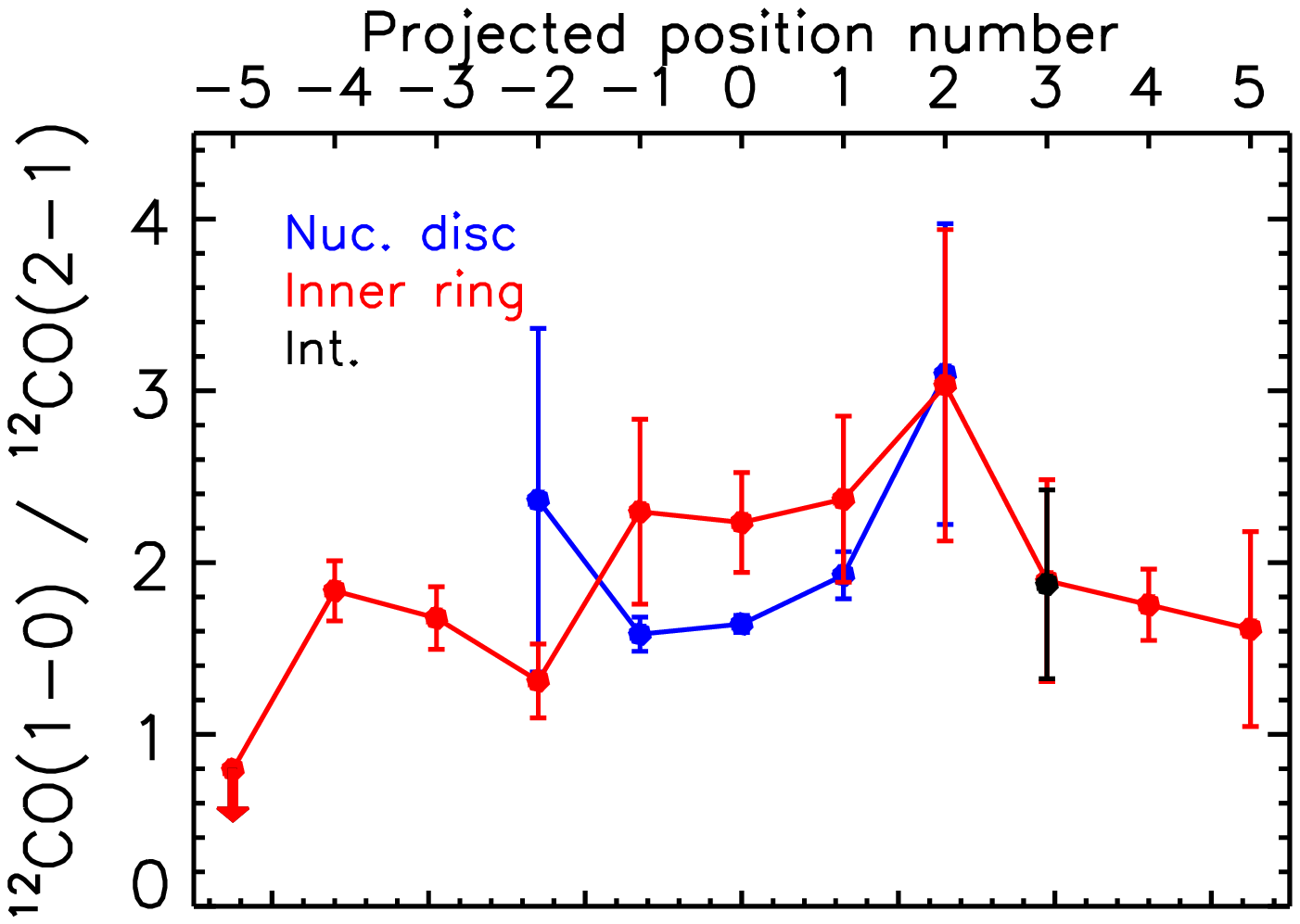}
  \includegraphics[width=5.5cm,clip=]{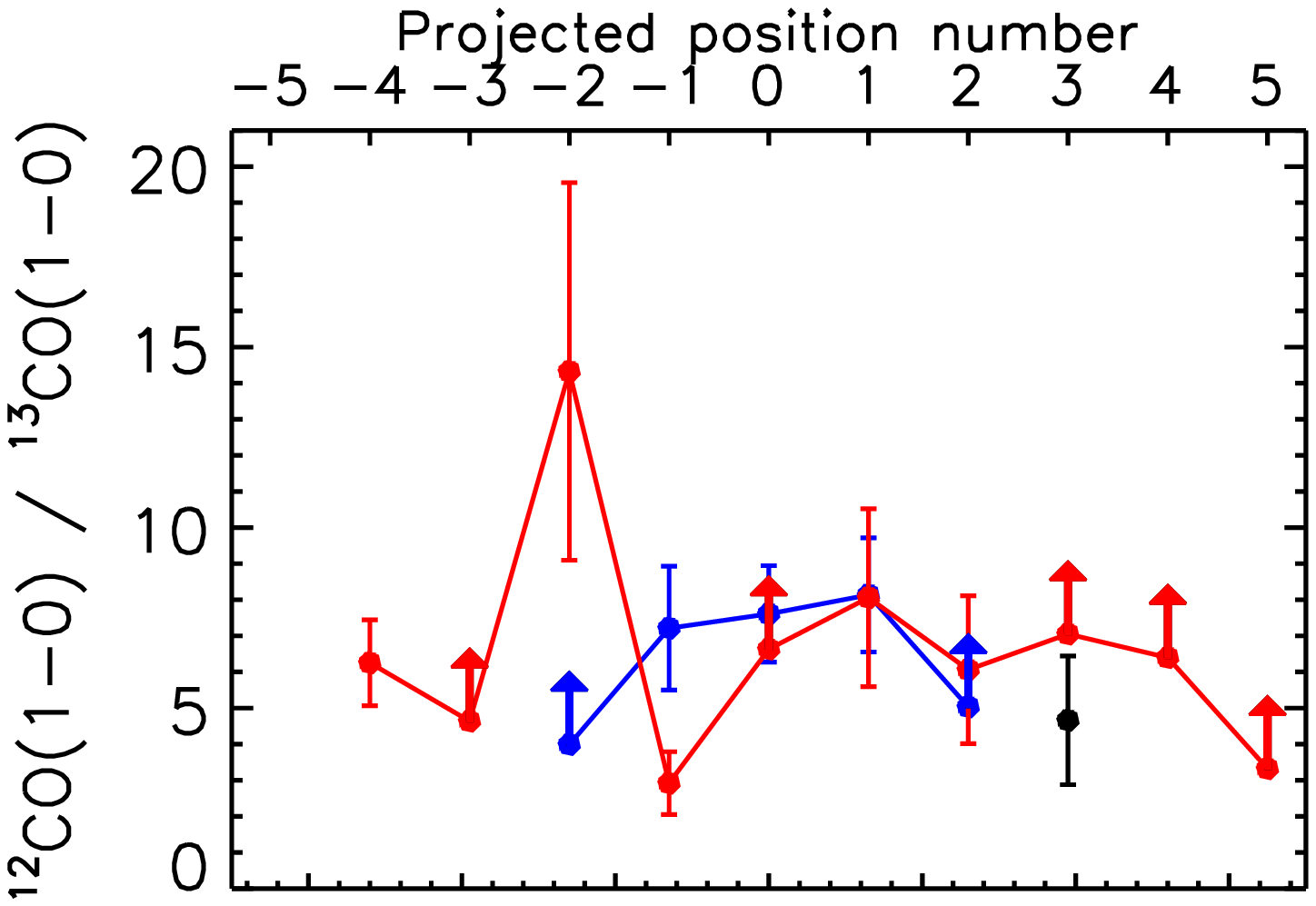}
  \includegraphics[width=5.5cm,clip=]{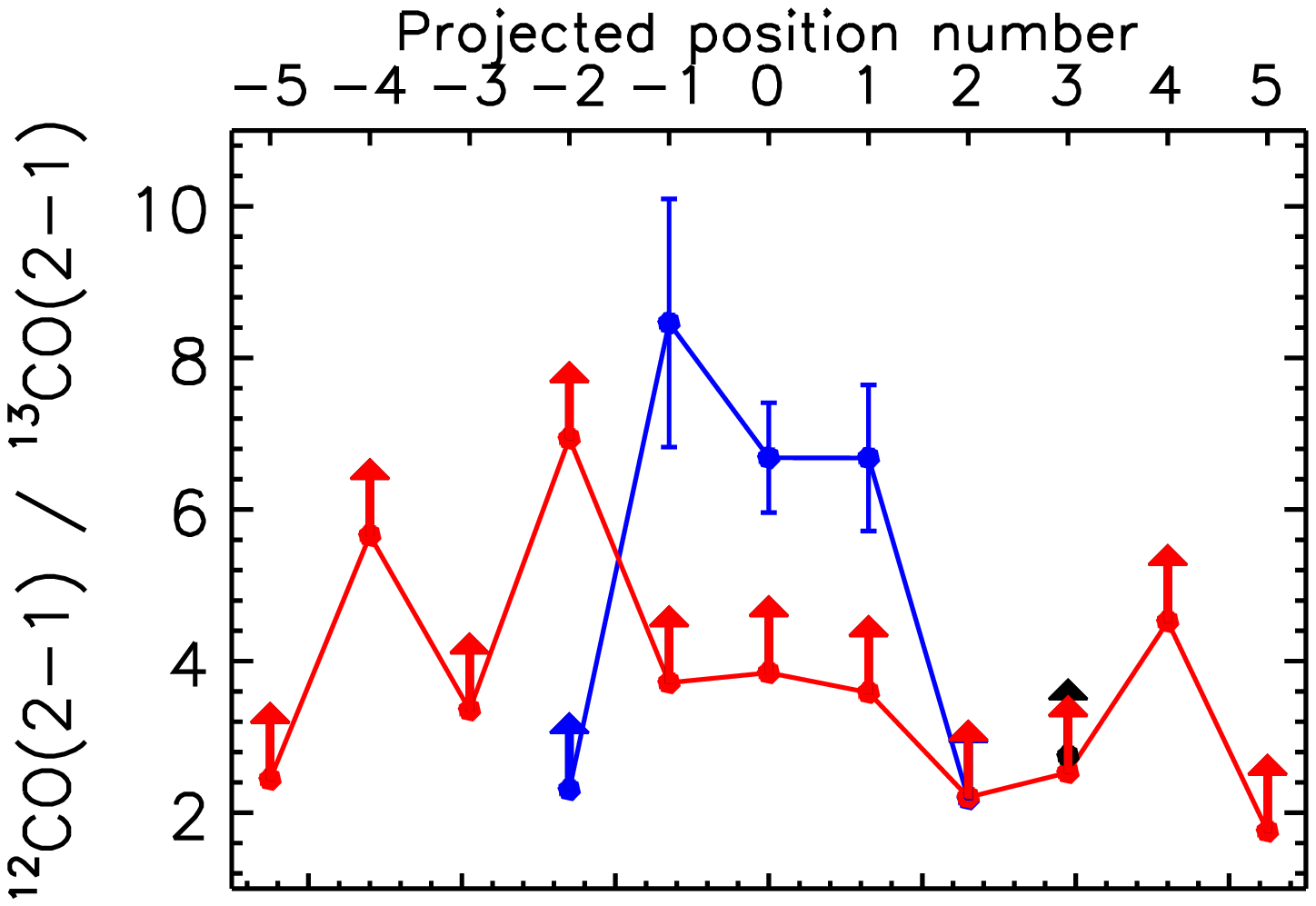}\\
  \vspace{-30pt}
  \includegraphics[width=5.5cm,clip=]{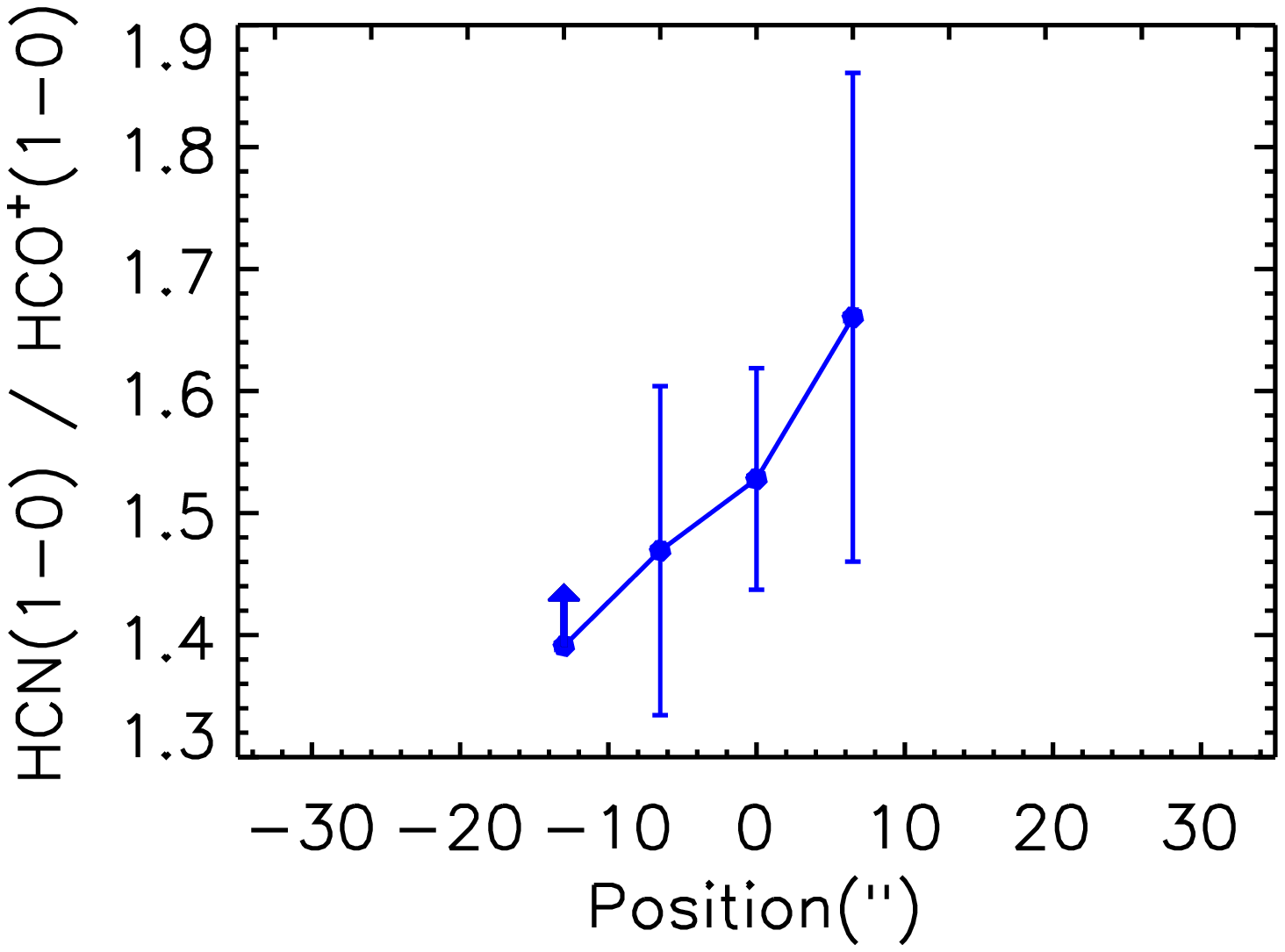}
  \includegraphics[width=5.5cm,clip=]{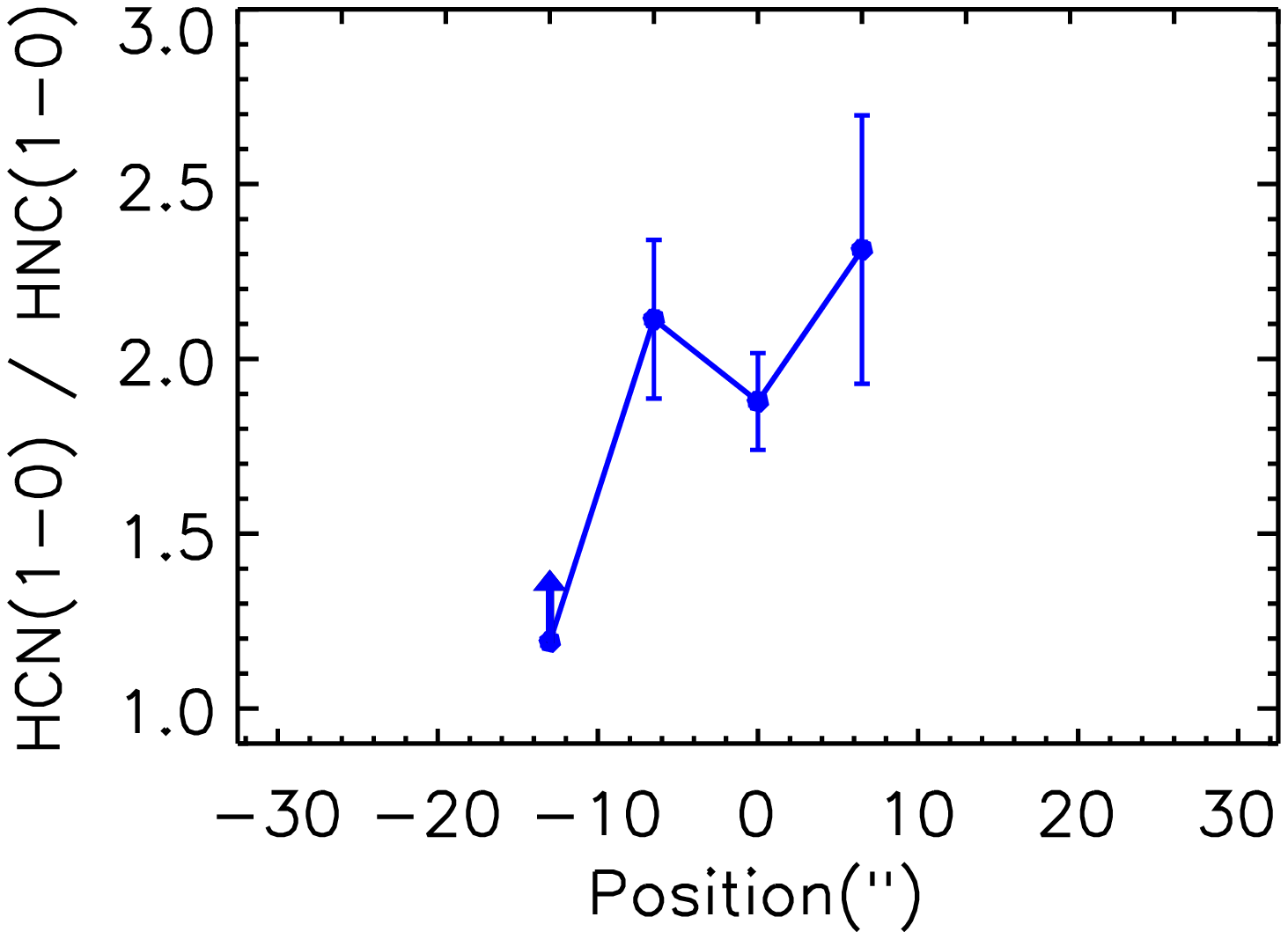}
  \includegraphics[width=5.5cm,clip=]{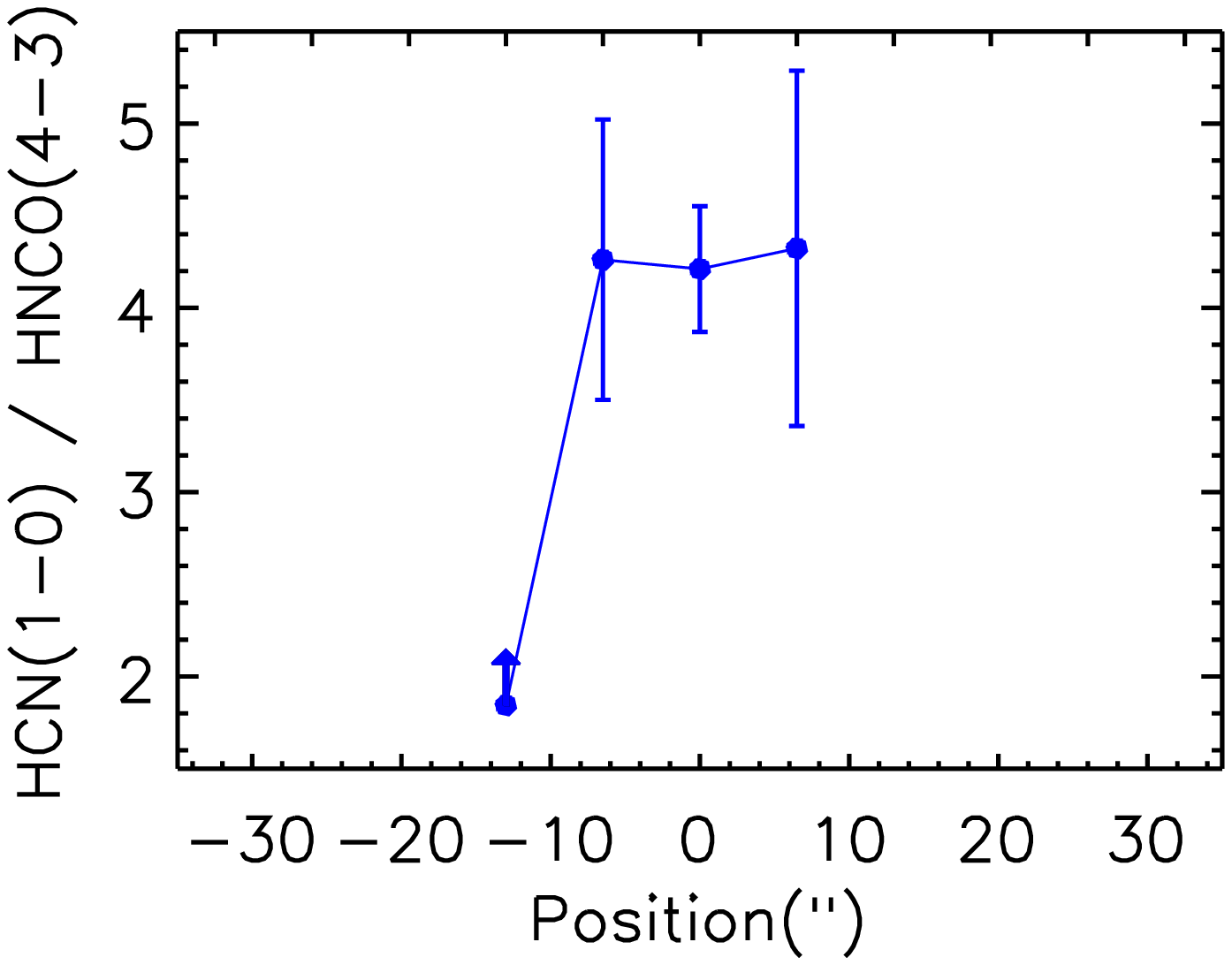}\\
  \caption{Ratios of CO lines only and dense gas tracer lines only as
    a function of projected radius in NGC~4710. {\bf Top:} Ratios of
    CO lines only, along the inner ring (red), nuclear disc (blue) and
    intermediate region (black; position~$3$ only). {\bf Bottom:} Ratios
    of dense gas tracer lines only, along the nuclear disc
    (blue). Upper and lower limits are indicated with arrows. The
    projected positions, as illustrated in Figure~\ref{fig:pos}, are
    indicated on the top axis of each panel.}
  \label{fig:n4710ratioPOS1}
\end{figure*}

Using the PVDs created from the identical data cubes, for our second
approach we attempted to disentangle the spectra of the two kinematic
components seen in the PVDs (nuclear disc and inner ring; see
\S~\ref{sec:posvel}), this as a function of projected position along
the disc. First, we extracted ``integrated'' spectra at a number of
positions along the discs, each separated by one beam width, by taking
averages of subsequent $5$-pixel (i.e.\ one beam width) slices, as
illustrated in Figure~\ref{fig:pos}. Second, we calculated the
integrated line intensity (i.e.\ $\int T_{\rm mb}\,{\rm d}v$ in
K~km~s$^{-1}$) of each kinematic component at each position by fitting
a single or double Gaussian to the extracted spectrum at each
position, depending on whether a single or both kinematic components
were present along the line of sight (see Fig.~\ref{fig:specext}). The
package $\emph{MPFIT}$ was used to optimize the fits
\citep{mark09}. When both components are detected, a single Gaussian
was normally sufficient in the outer parts (inner ring), while a
double Gaussian was normally required in the inner parts (nuclear disc
and inner ring). When a single component is detected in the inner
parts (nuclear disc), a single Gaussian is of course always
sufficient. The ratios of the integrated line intensities were then
calculated as a function of projected radius along the discs. The
corresponding line ratio profiles are shown in
Figures~\ref{fig:n4710ratioPOS1}\,--\,\ref{fig:n5866ratioPOS} and are
tabulated in Tables~\ref{tab:ratios}\,--\,\ref{tab:ratioir}.  Note
that while the PVD ratios shown in
Figures~\ref{fig:n4710ratioPVD1}\,--\,\ref{fig:n5866ratioPVD} are
ratios of fluxes (i.e.\ K), the ratios shown in
Figures~\ref{fig:n4710ratioPOS1}\,--\,\ref{fig:n5866ratioPOS} are
ratios of integrated line intensities (i.e.\ K~km~s$^{-1}$).

%
%
%
%
%
%
%
%
\begin{figure*}
  \hspace{-15pt}
  \includegraphics[width=4.7cm,clip=]{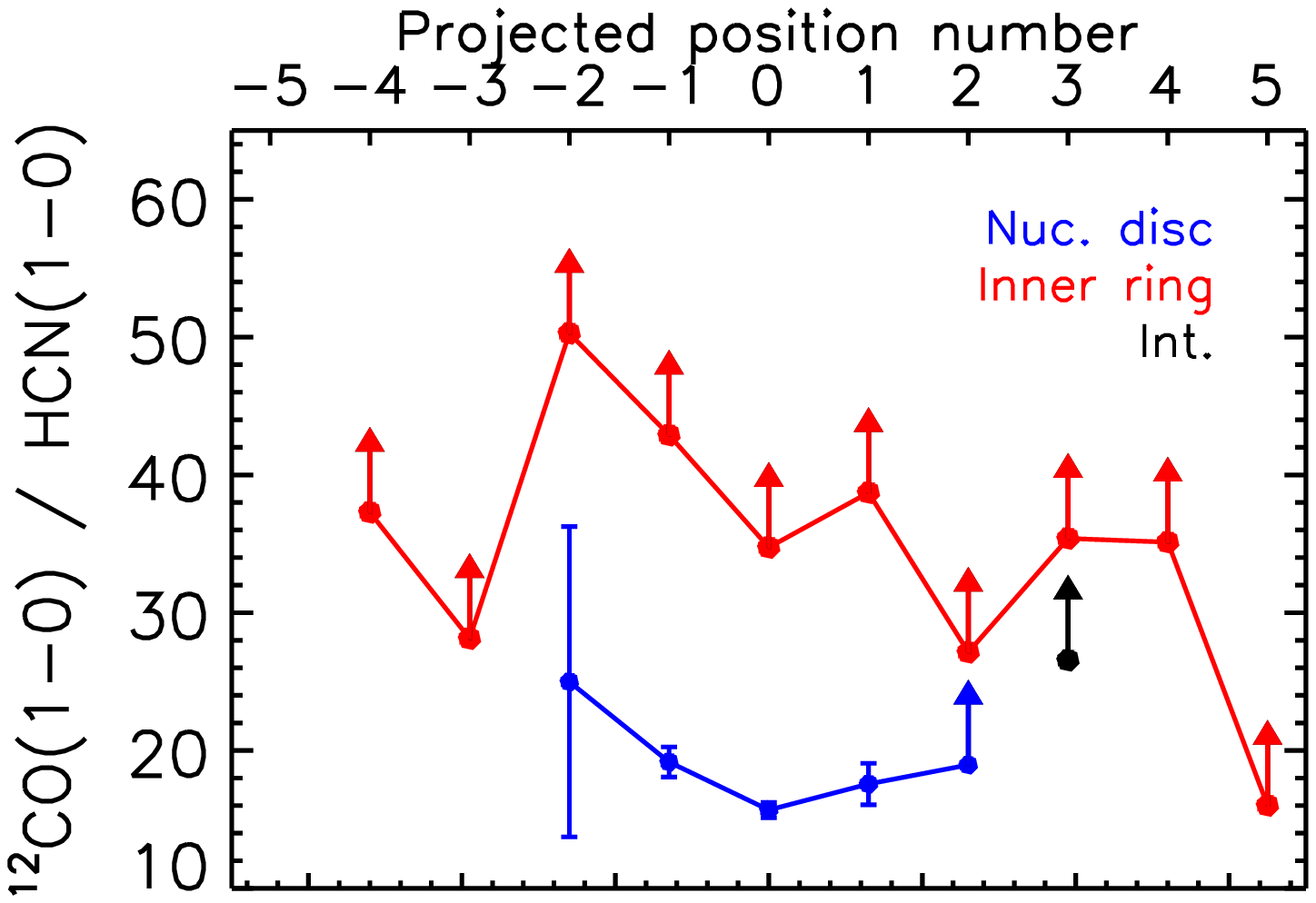}
  \hspace{-15pt}
  \includegraphics[width=4.7cm,clip=]{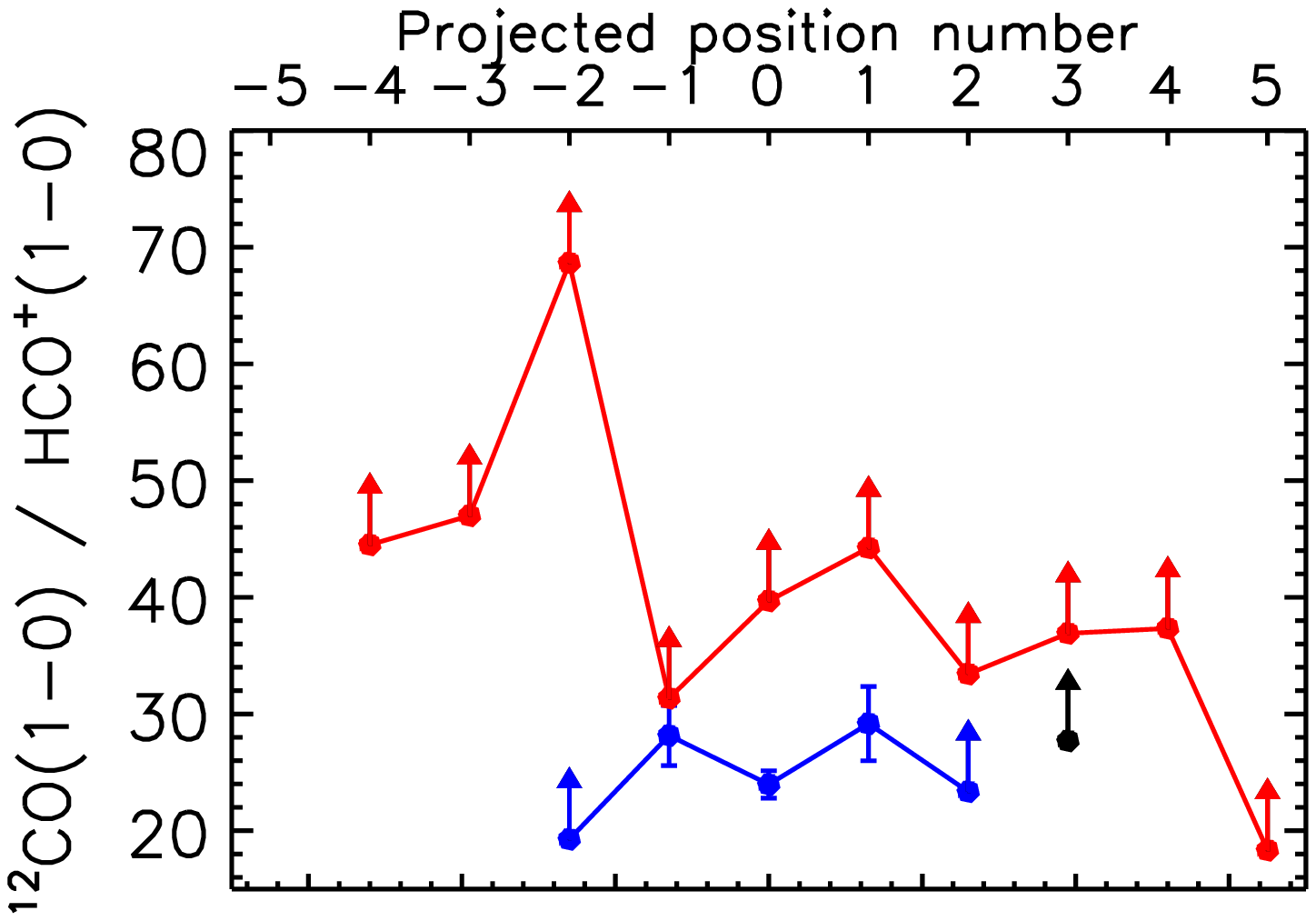}
  \hspace{-15pt}
  \includegraphics[width=4.7cm,clip=]{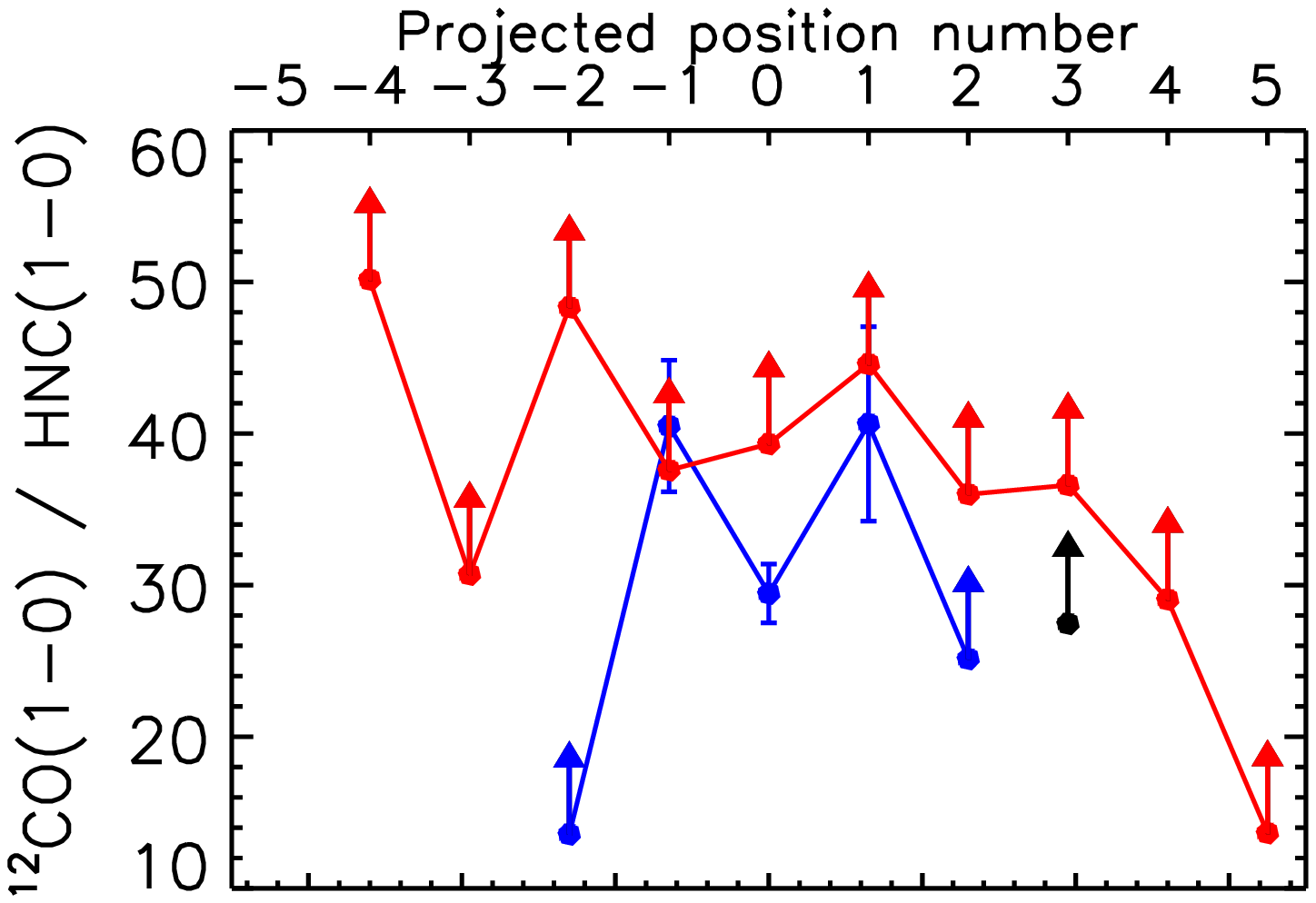}
  \hspace{-15pt}
  \includegraphics[width=4.7cm,clip=]{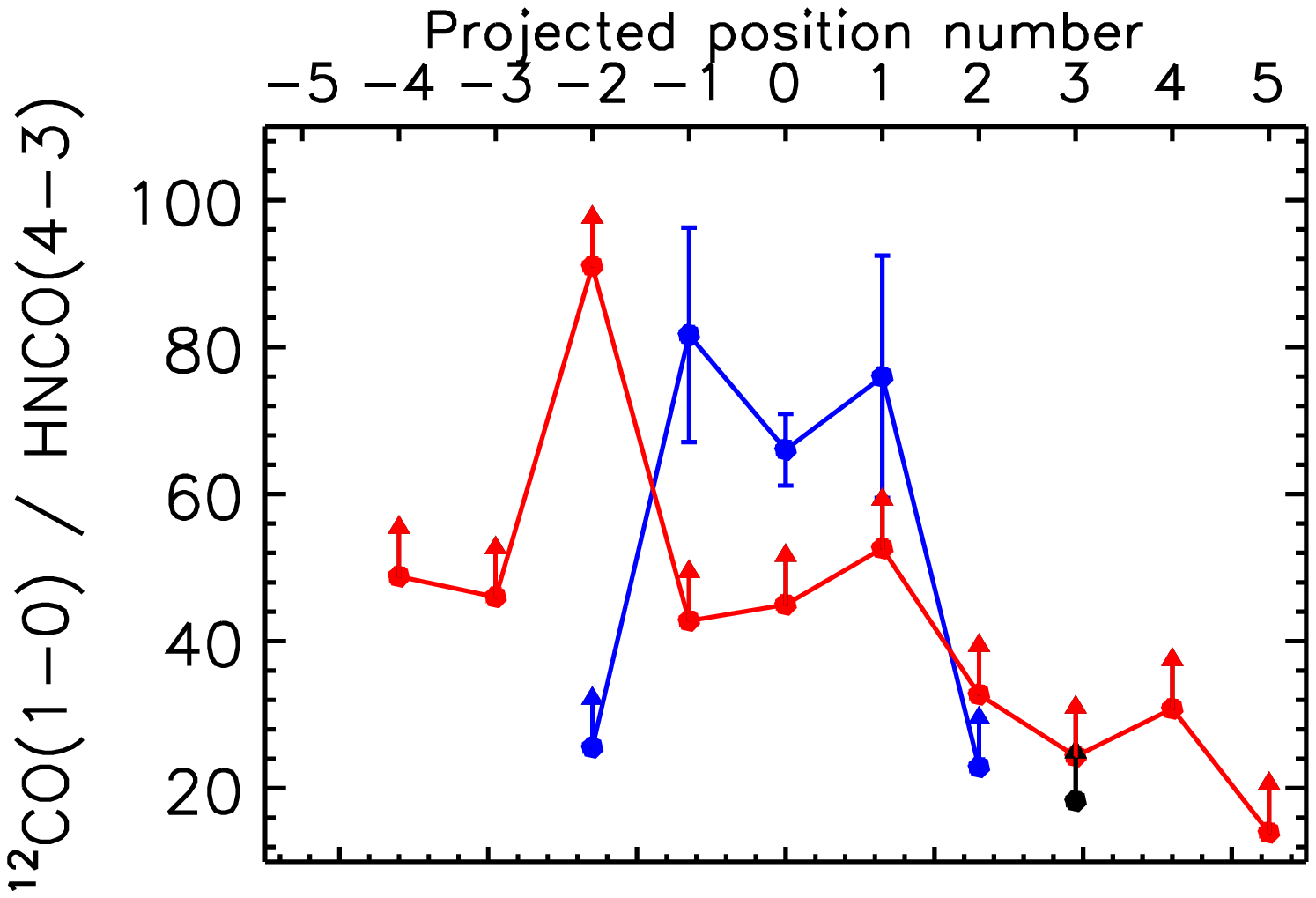}\\
  \vspace{-25pt}
  \hspace{-15pt}
  \includegraphics[width=4.7cm,clip=]{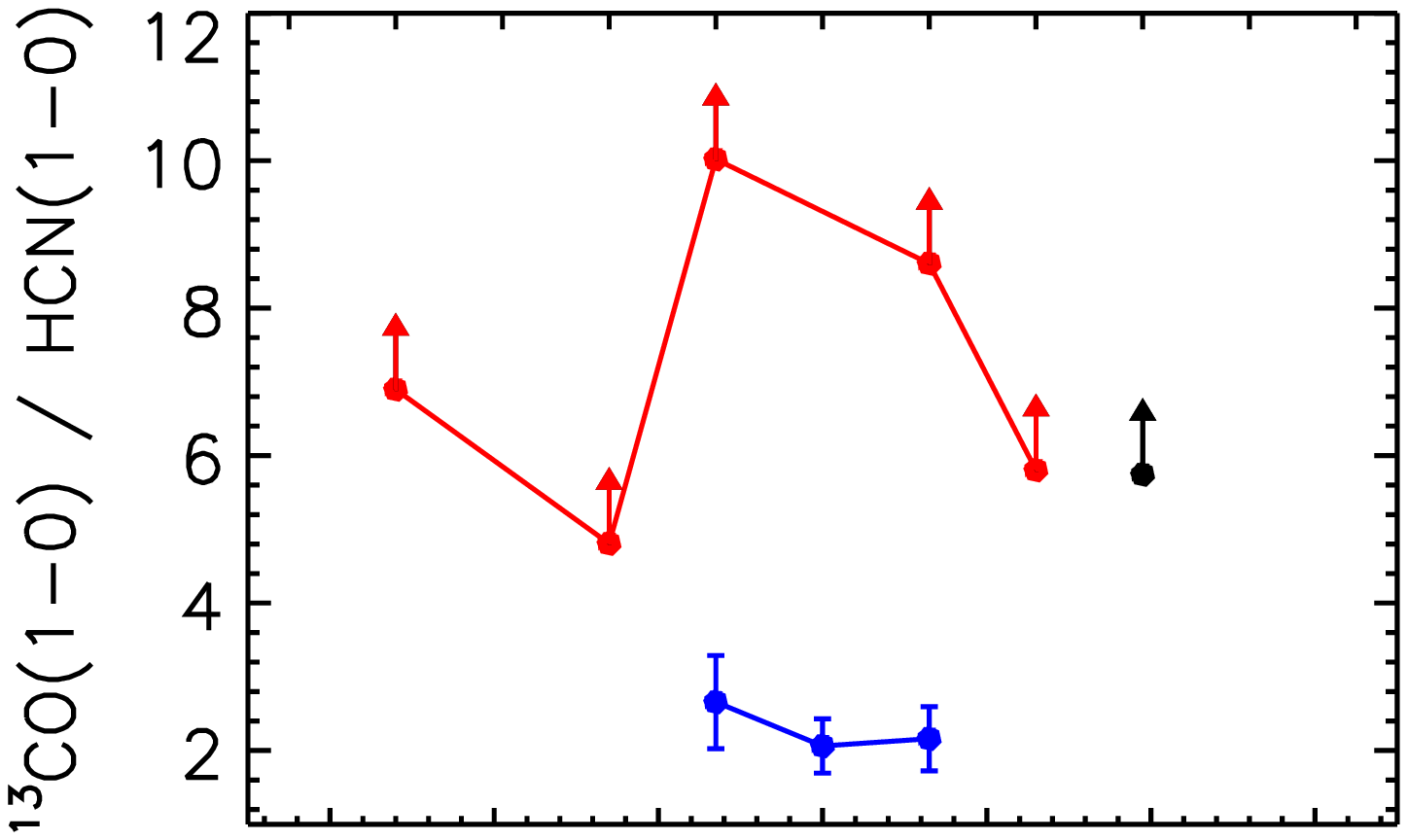}
  \hspace{-15pt}
  \includegraphics[width=4.7cm,clip=]{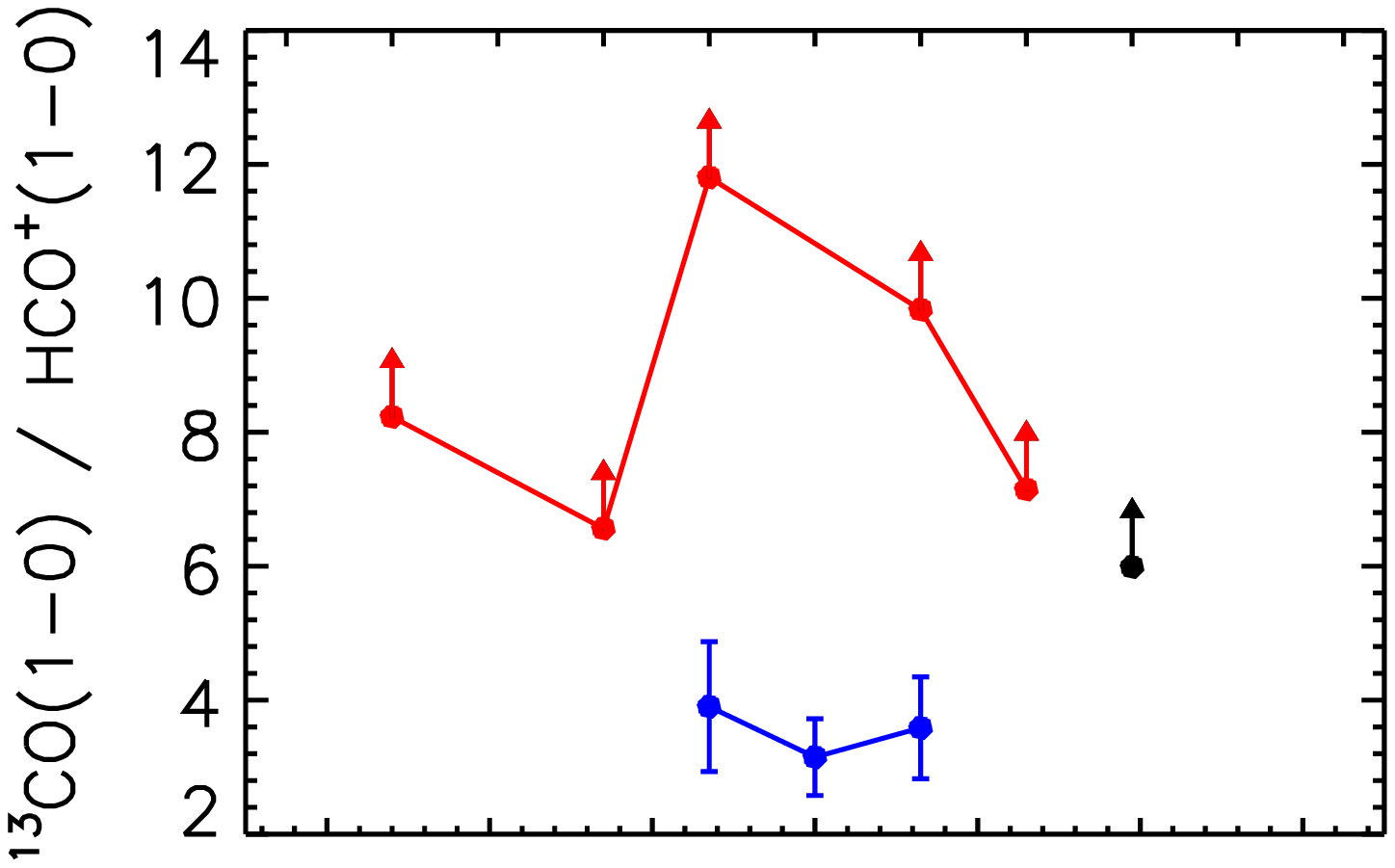}
  \hspace{-15pt}
  \includegraphics[width=4.7cm,clip=]{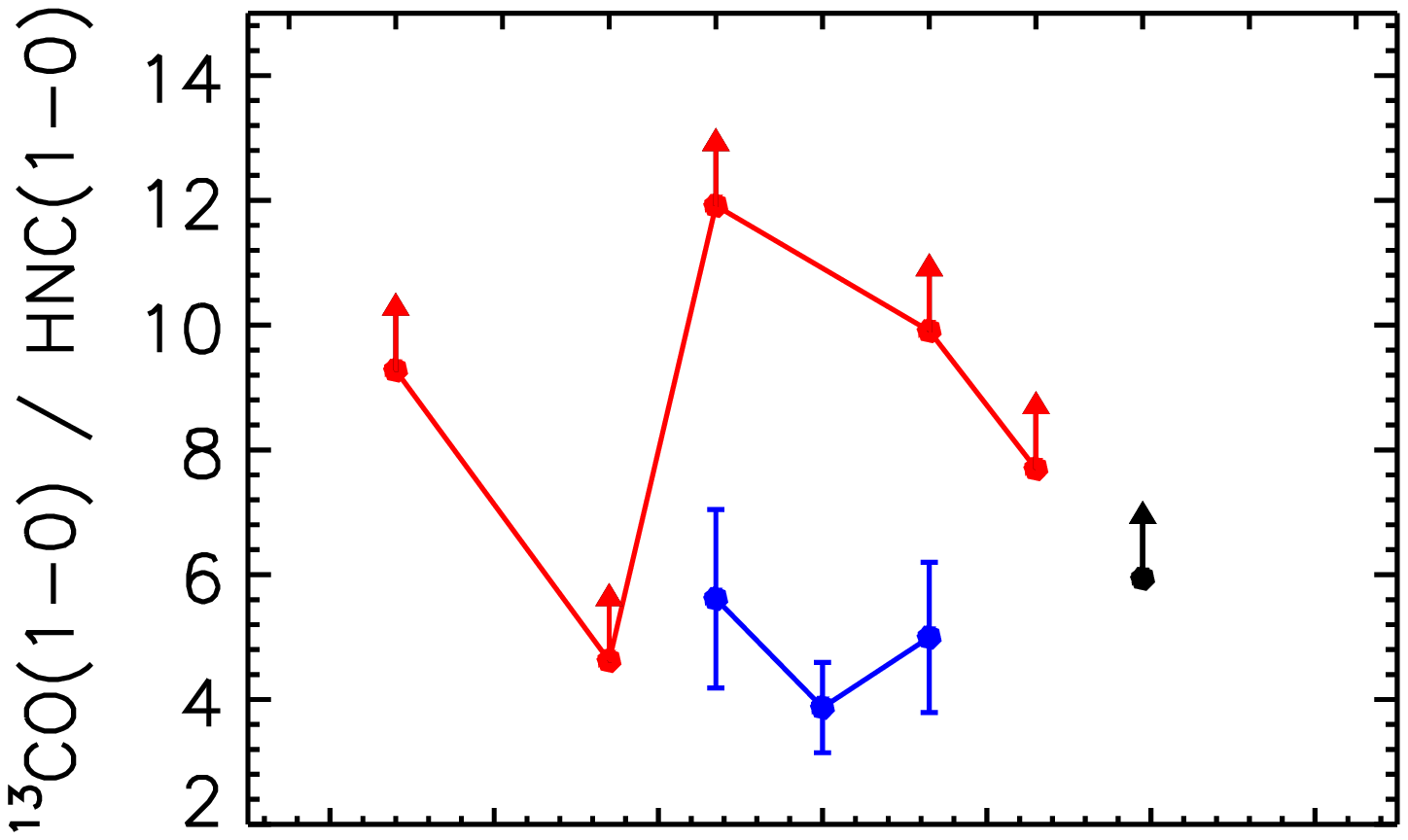}
  \hspace{-15pt}
  \includegraphics[width=4.7cm,clip=]{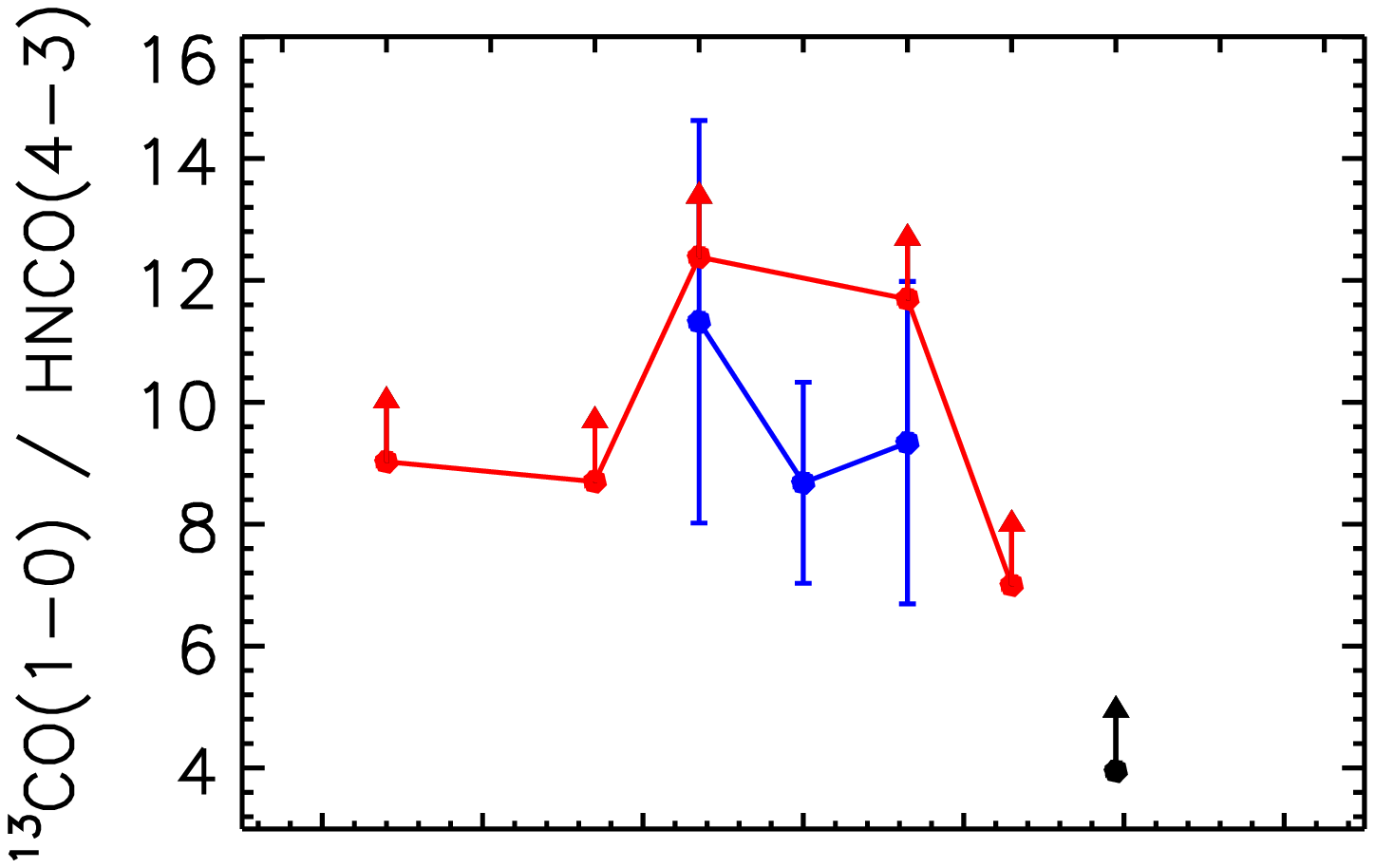}\\
  \vspace{-25pt}
  \hspace{-15pt}
  \includegraphics[width=4.7cm,clip=]{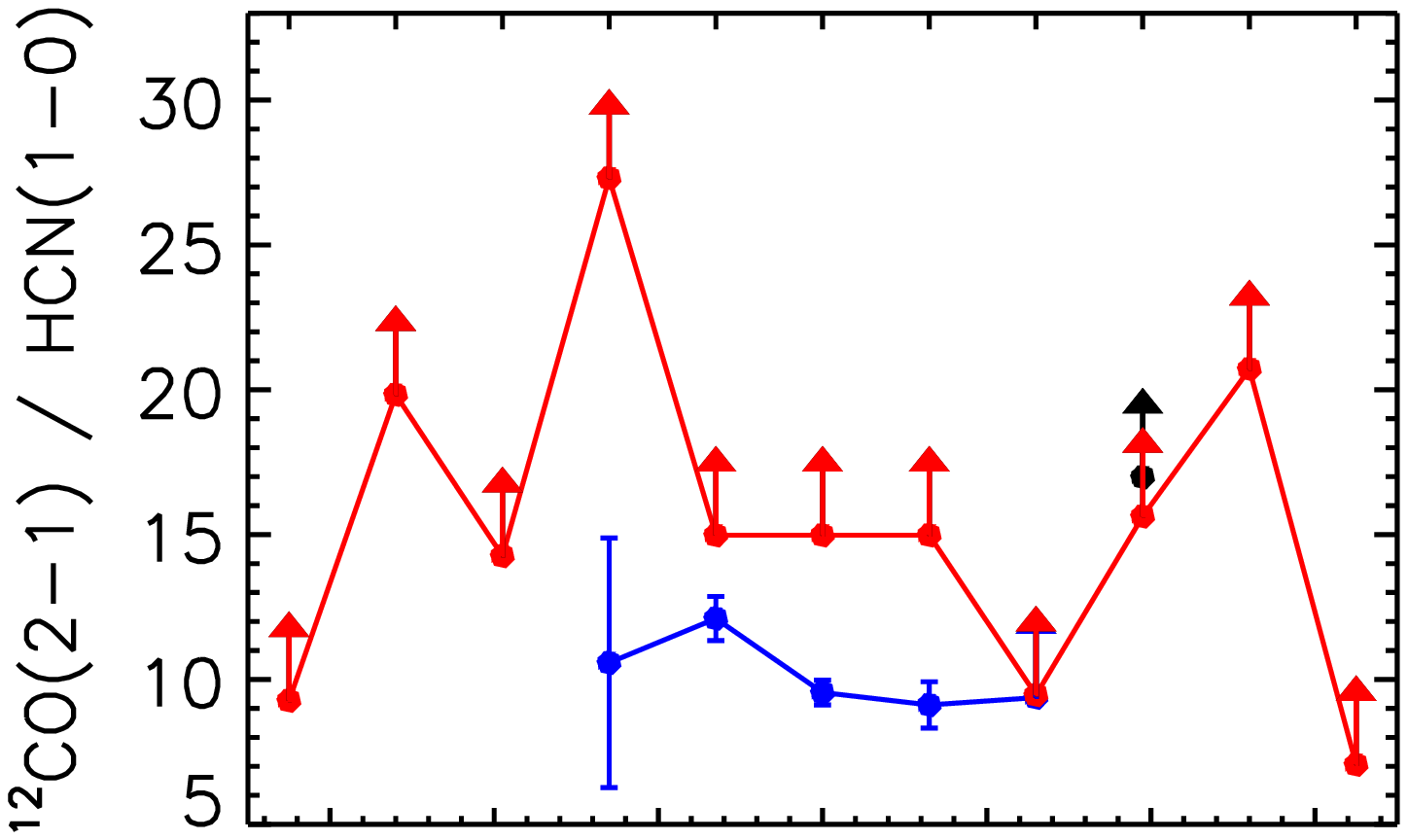}
  \hspace{-15pt}
  \includegraphics[width=4.7cm,clip=]{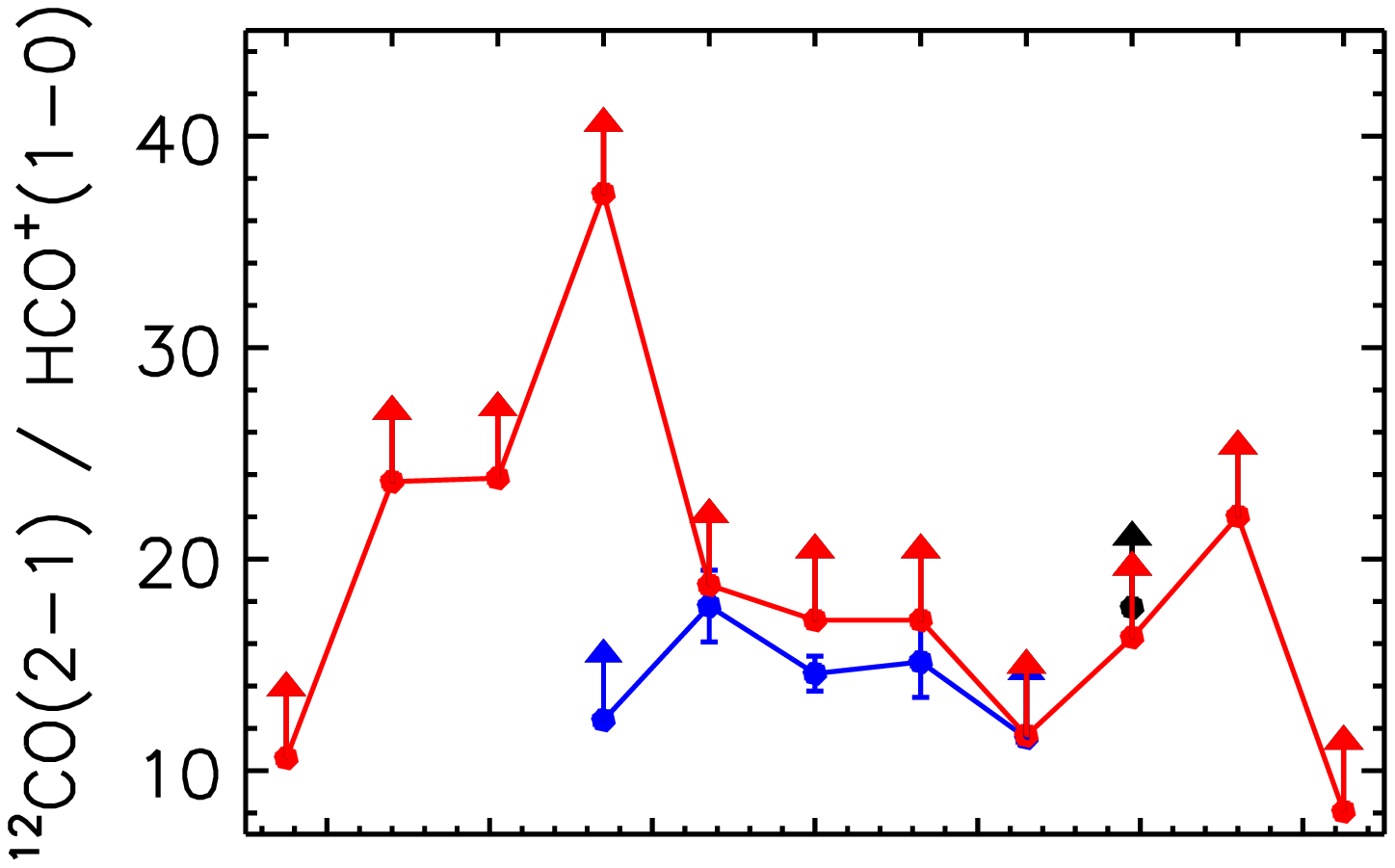}
  \hspace{-15pt}
  \includegraphics[width=4.7cm,clip=]{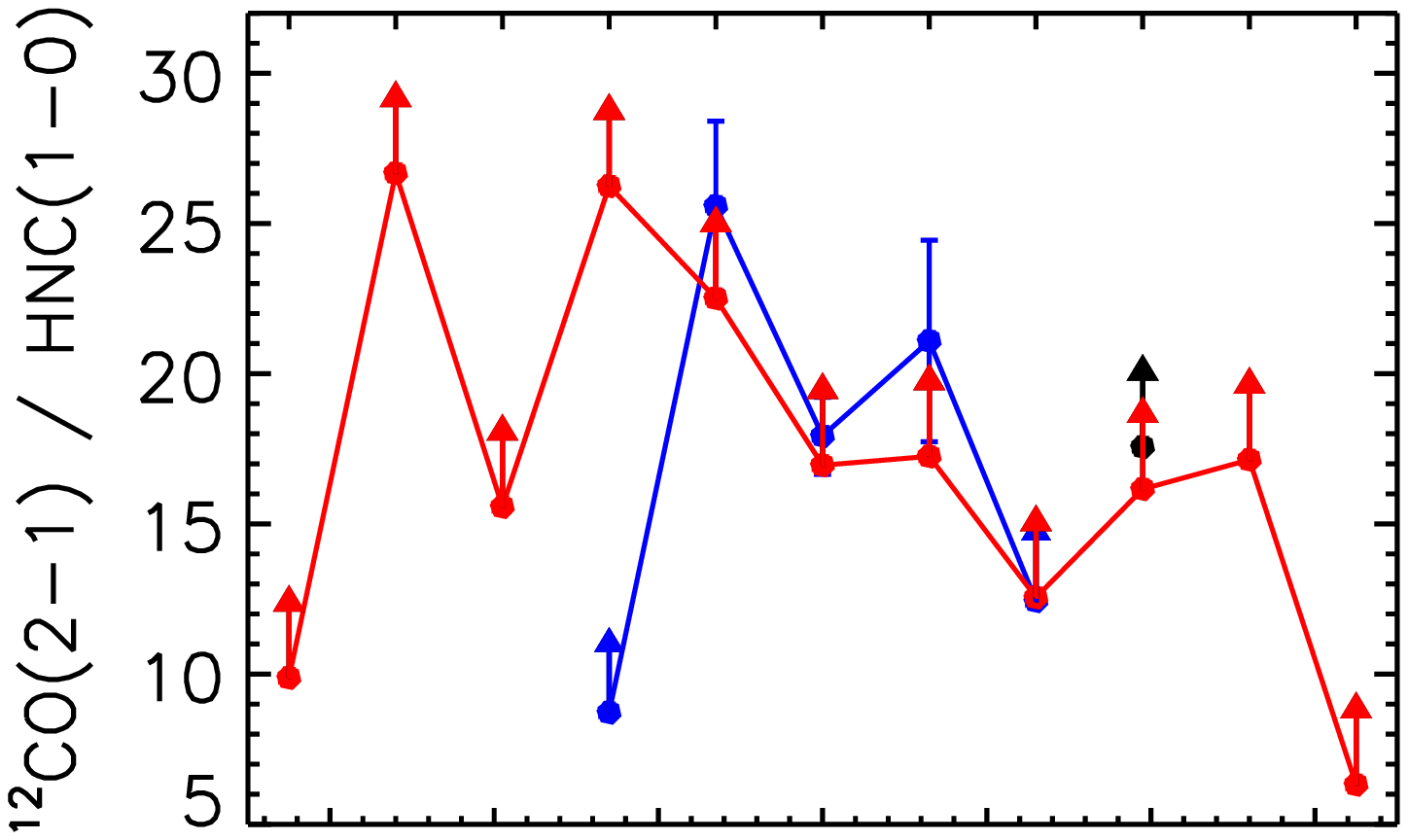}
  \hspace{-15pt}
  \includegraphics[width=4.7cm,clip=]{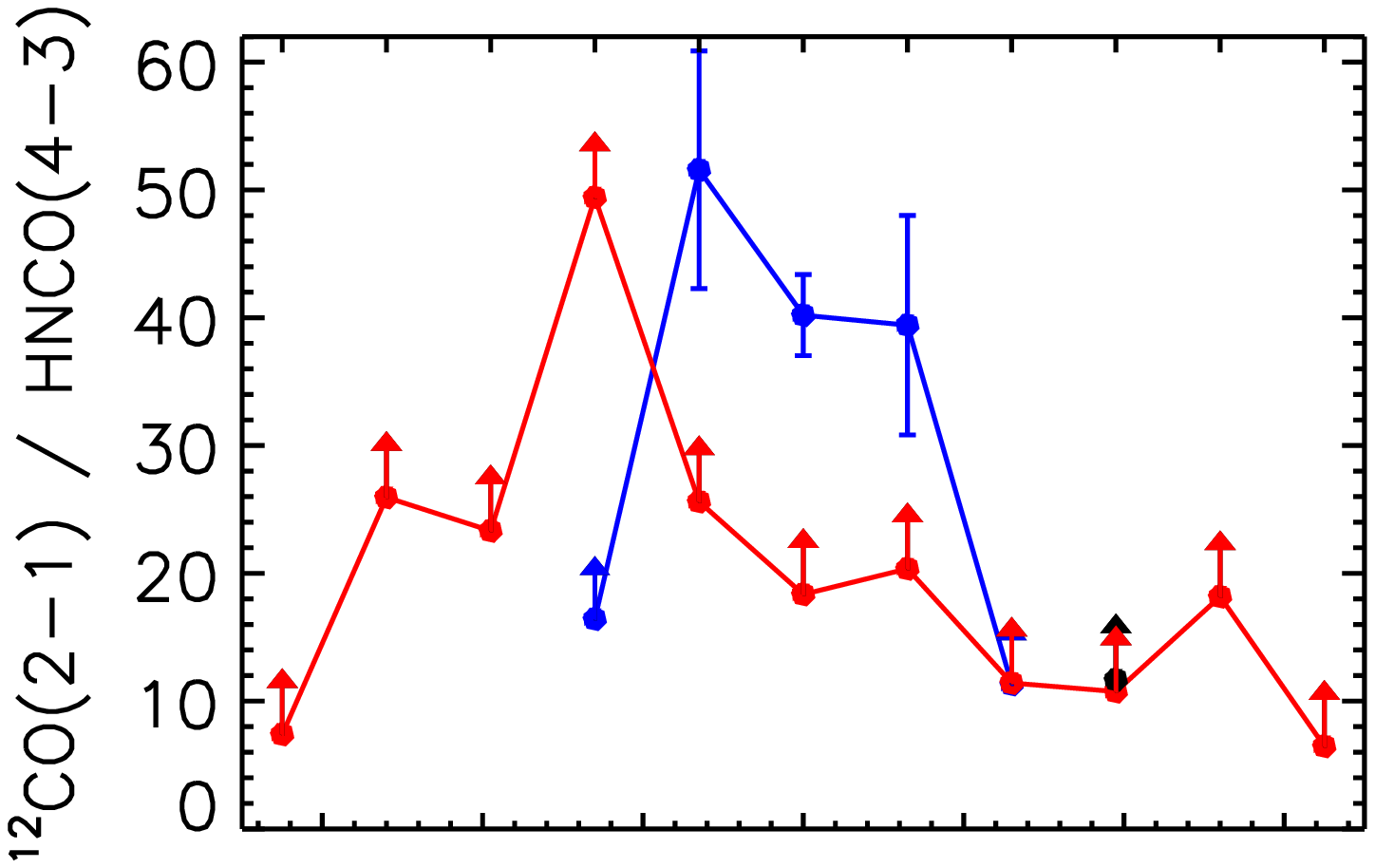}\\
  \vspace{-25pt}
  \hspace{-15pt}
  \includegraphics[width=4.7cm,clip=]{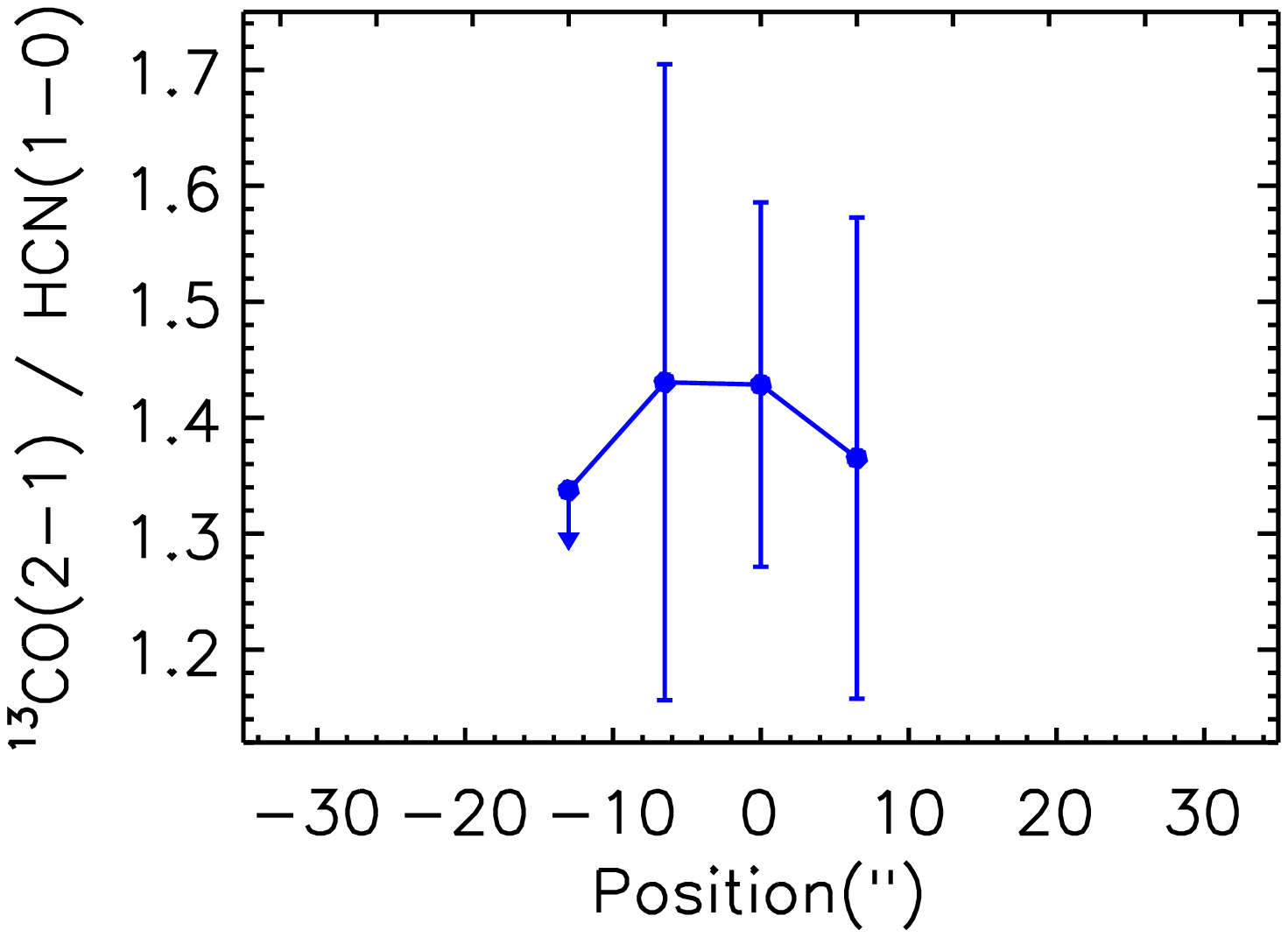}
  \hspace{-15pt}
  \includegraphics[width=4.7cm,clip=]{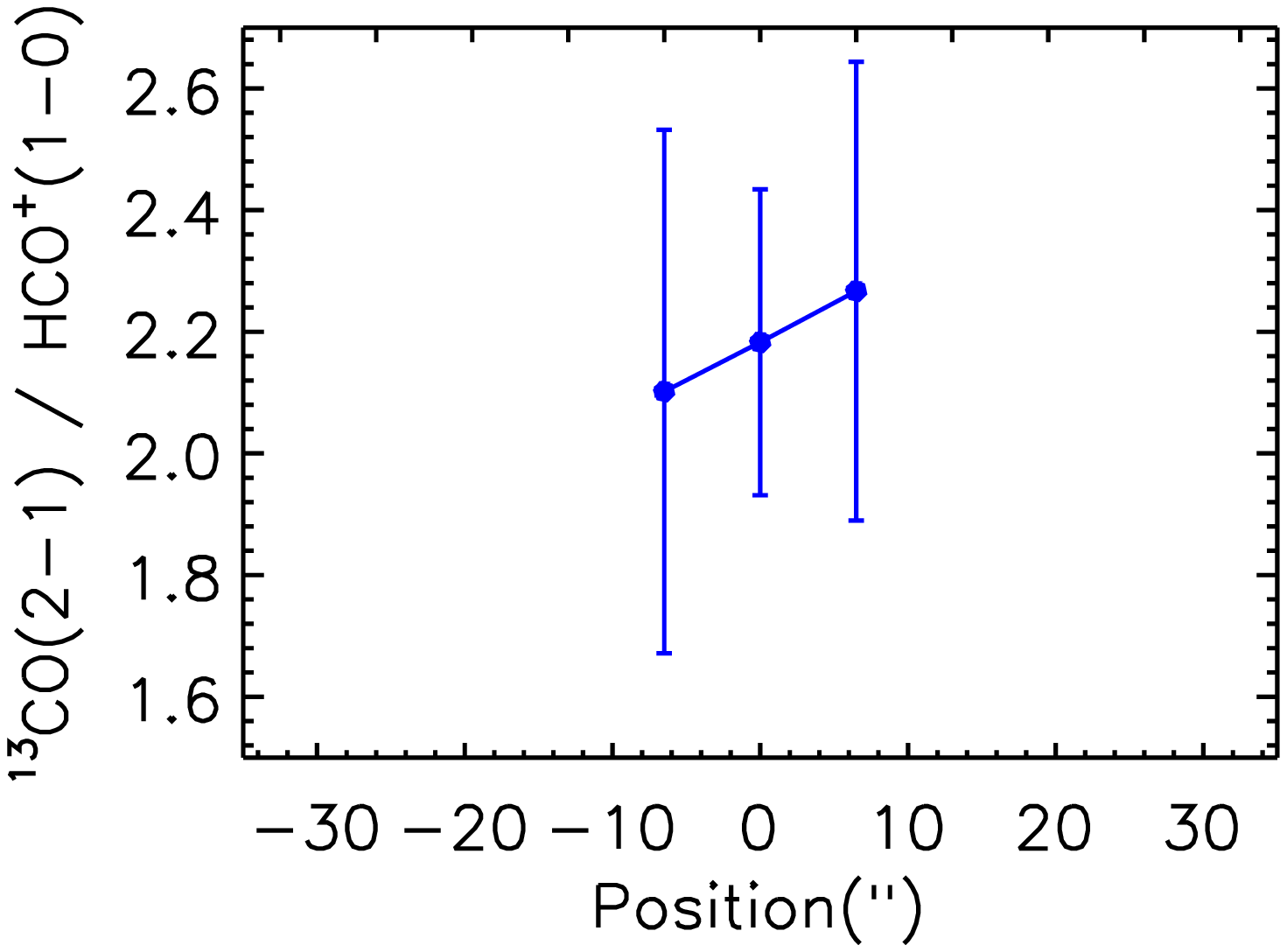}
  \hspace{-15pt}
  \includegraphics[width=4.7cm,clip=]{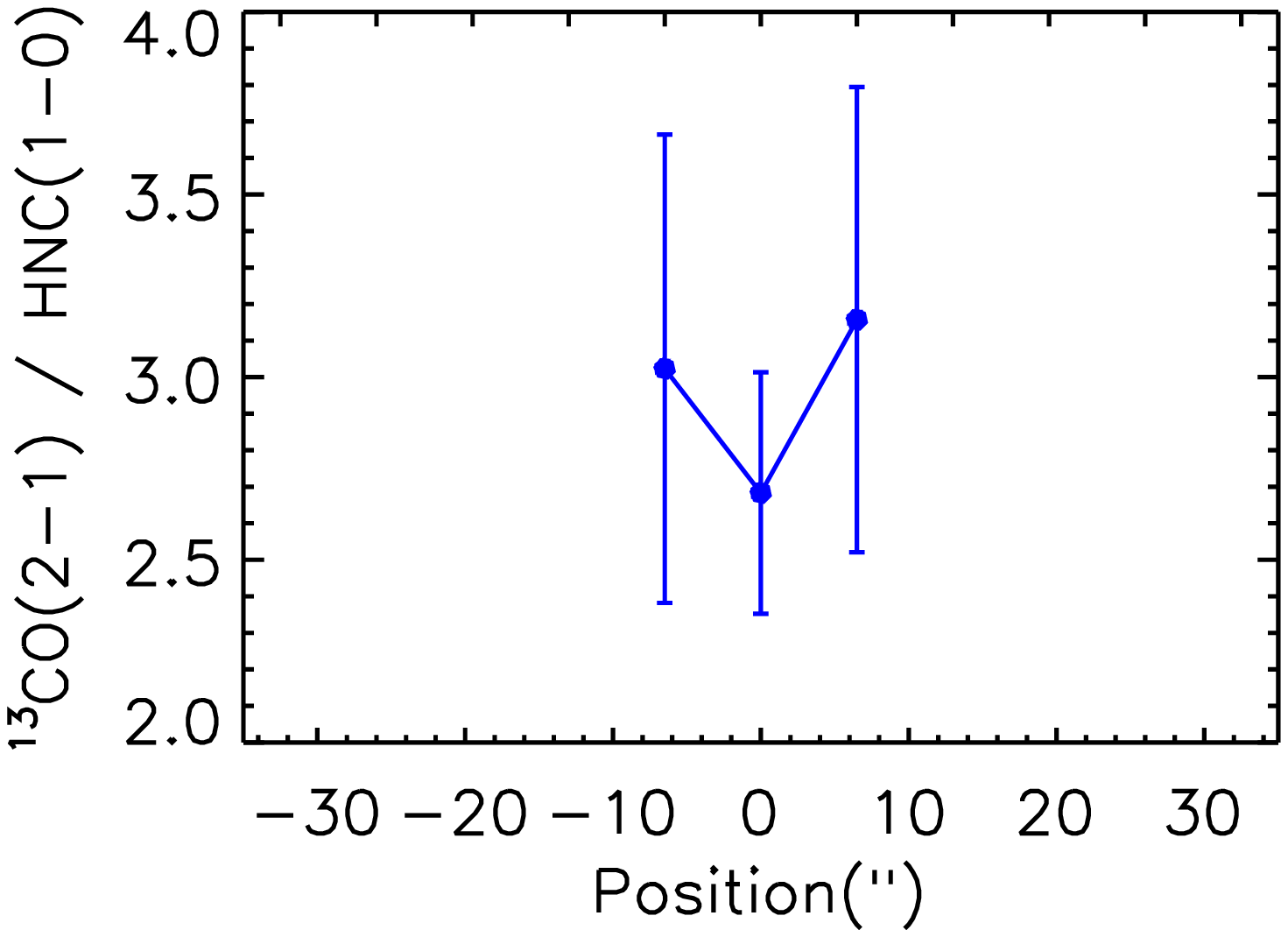}
  \hspace{-15pt}
  \includegraphics[width=4.7cm,clip=]{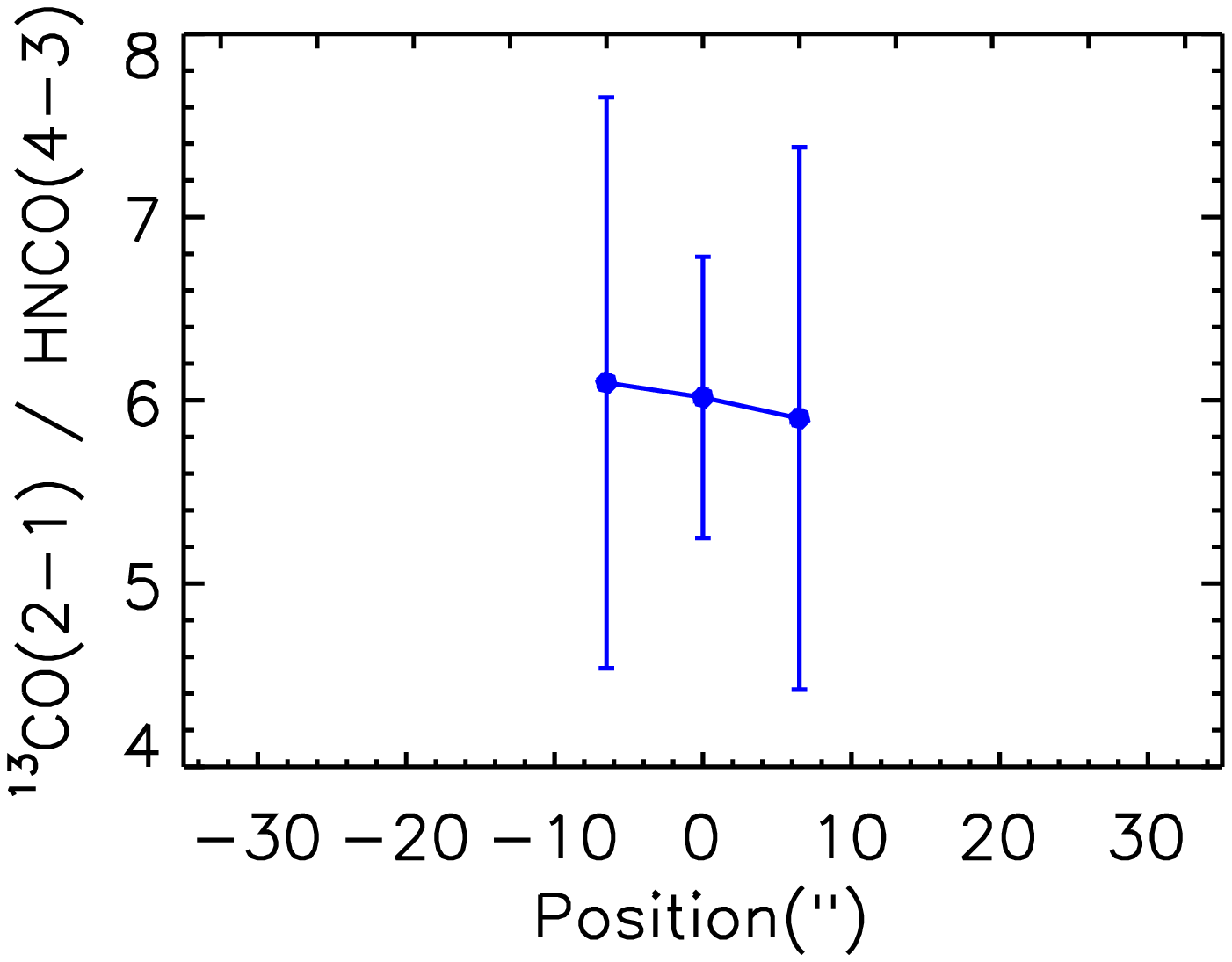}\\
  \caption{Same as Figure~\ref{fig:n4710ratioPOS1} but for the ratios
    of CO to dense gas tracer lines in NGC~4710.}
  \label{fig:n4710ratioPOS2}
\end{figure*}
%
%
\begin{figure*}
  \hspace{-15pt}
  \includegraphics[width=4.7cm,clip=]{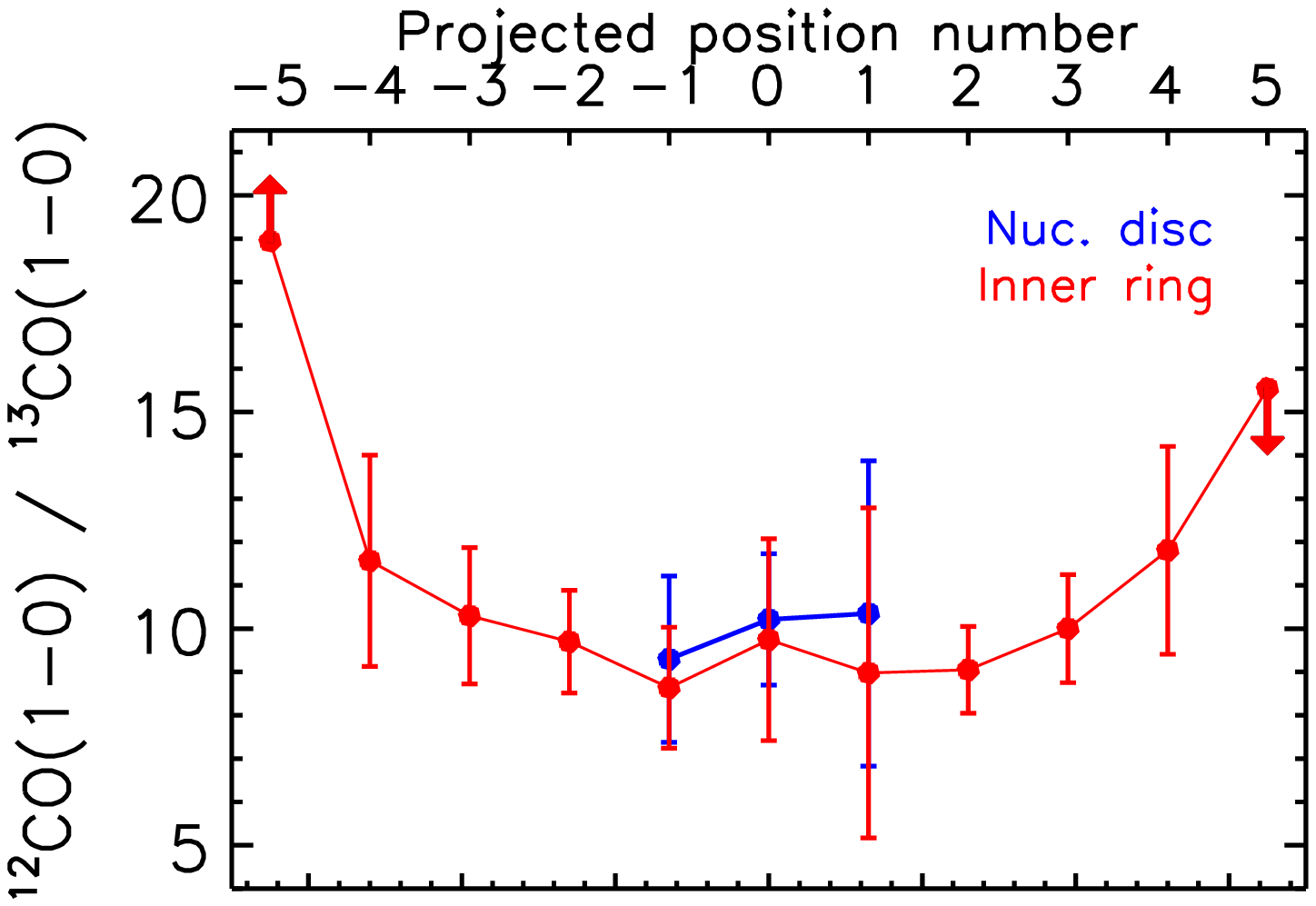}
  \hspace{-15pt}
  \includegraphics[width=4.7cm,clip=]{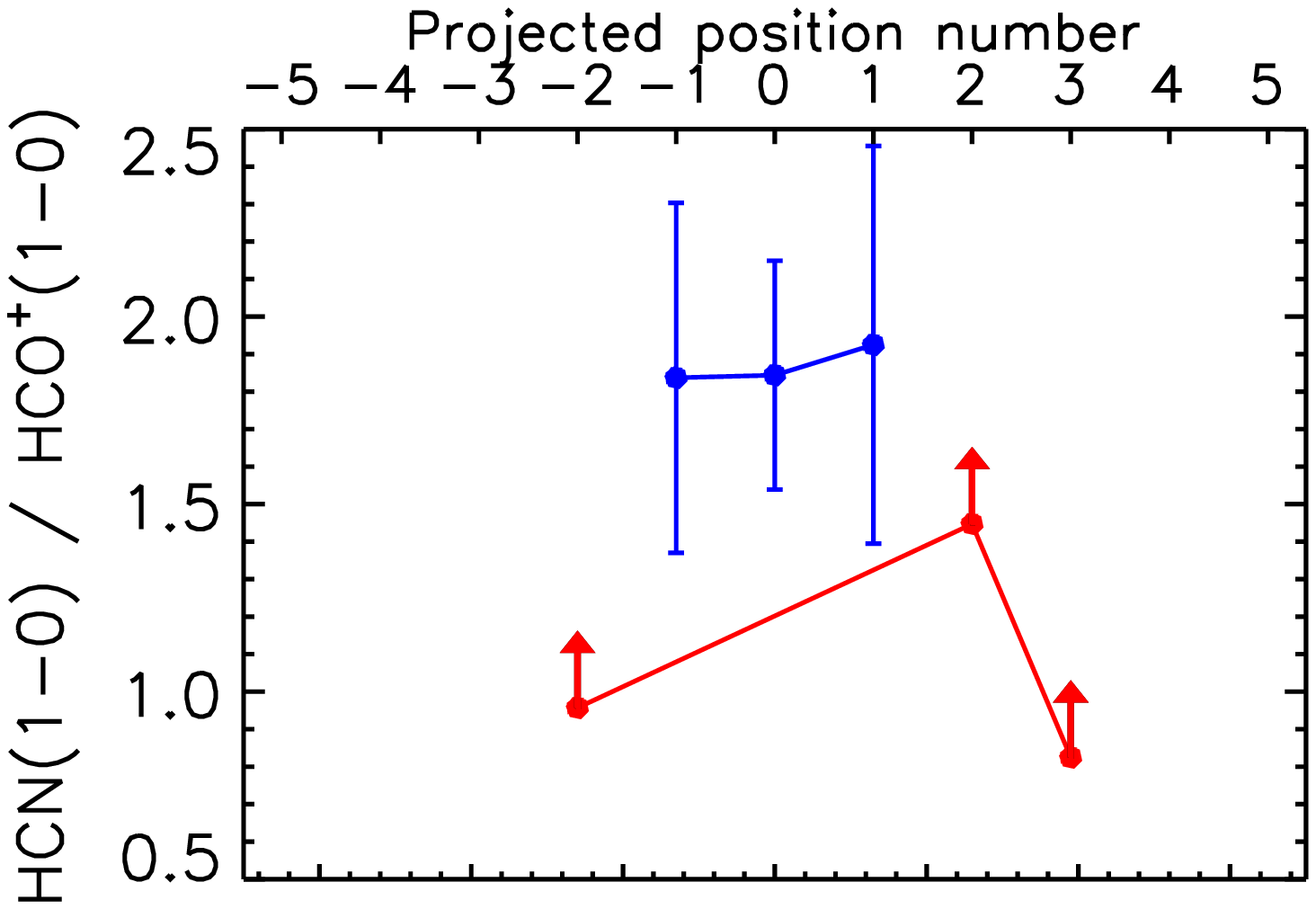}
  \hspace{-15pt}
  \includegraphics[width=4.7cm,clip=]{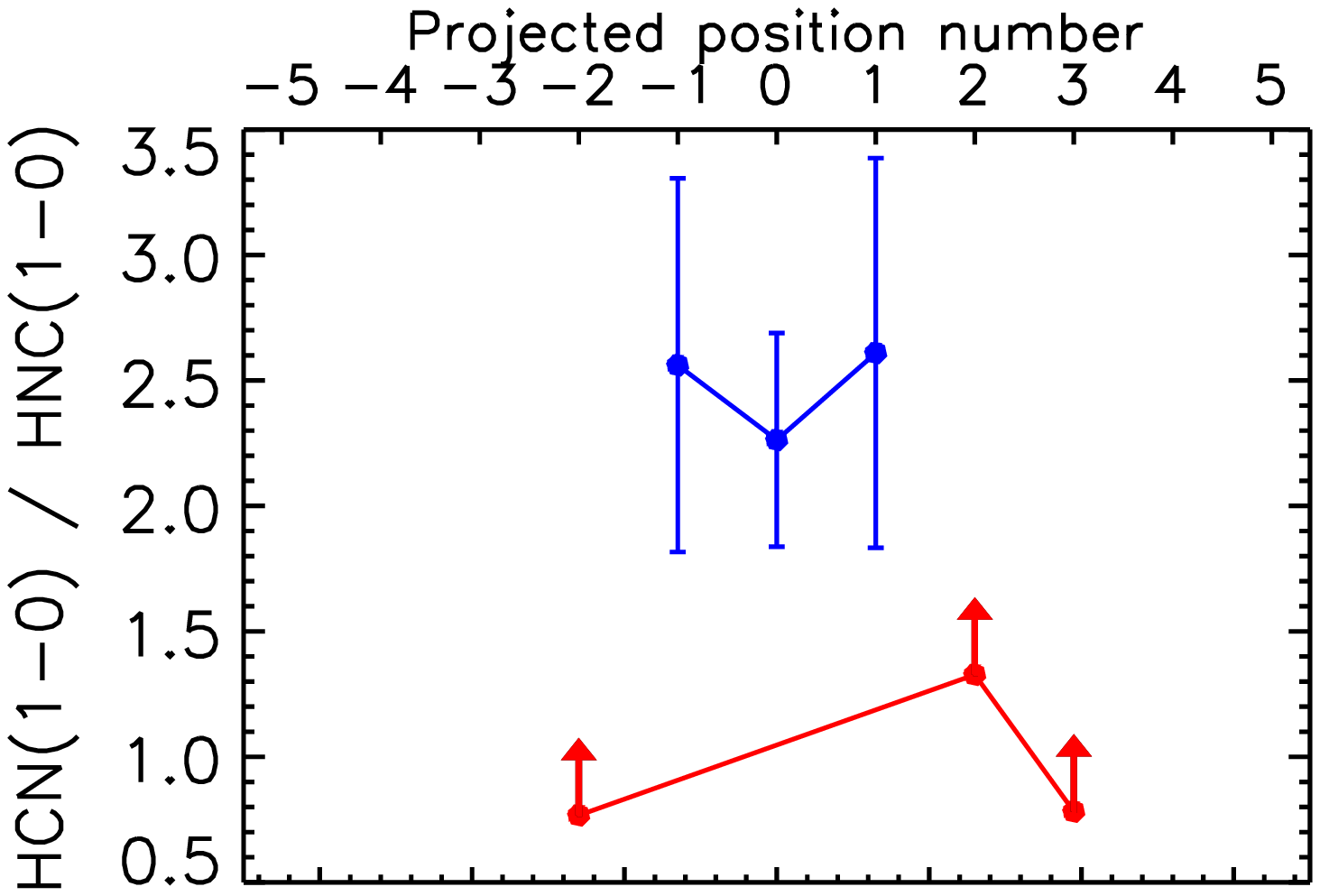}
  \hspace{-15pt}
  \includegraphics[width=4.7cm,clip=]{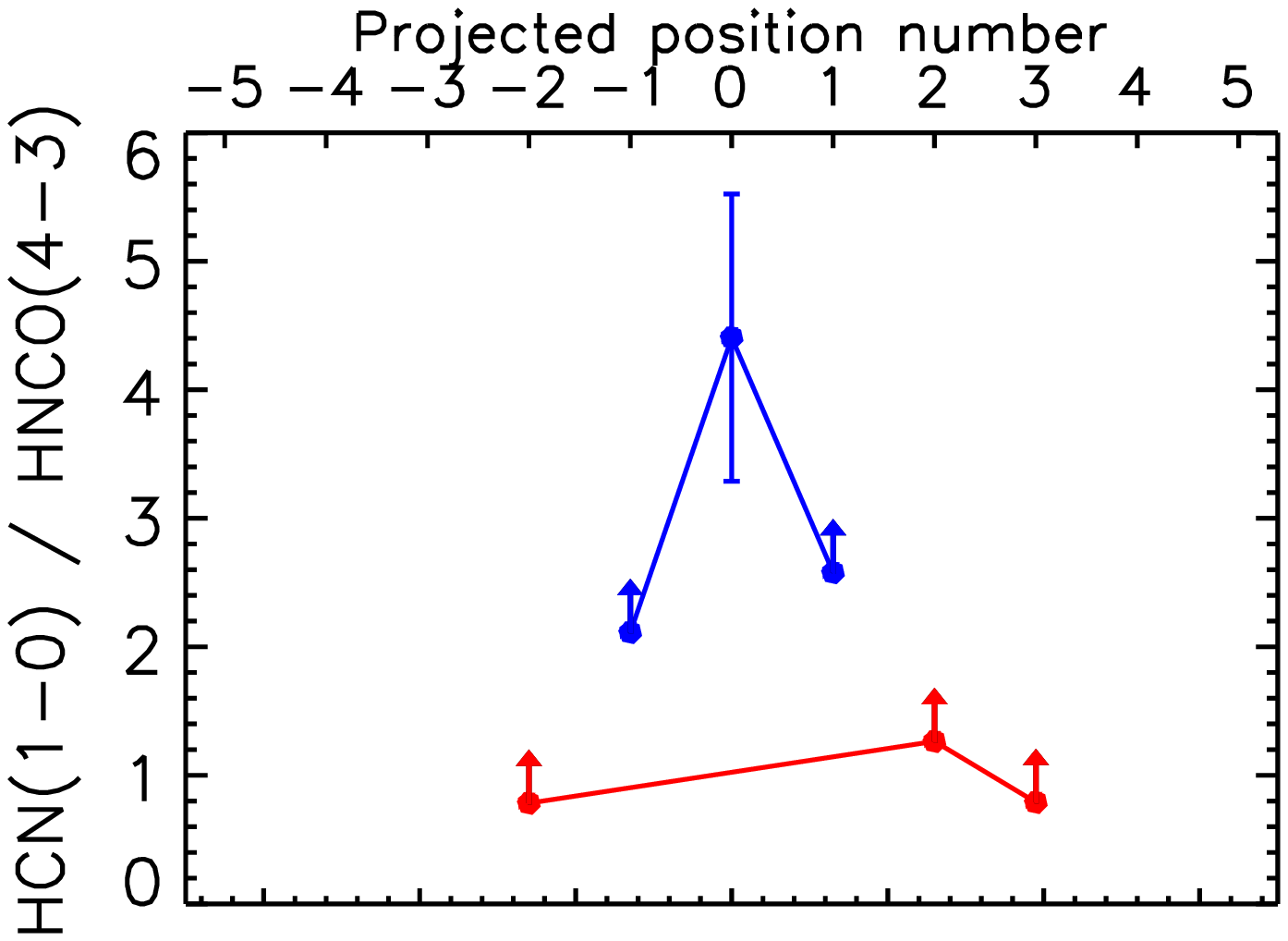}\\
  \vspace{-25pt}
  \hspace{-15pt}
  \includegraphics[width=4.7cm,clip=]{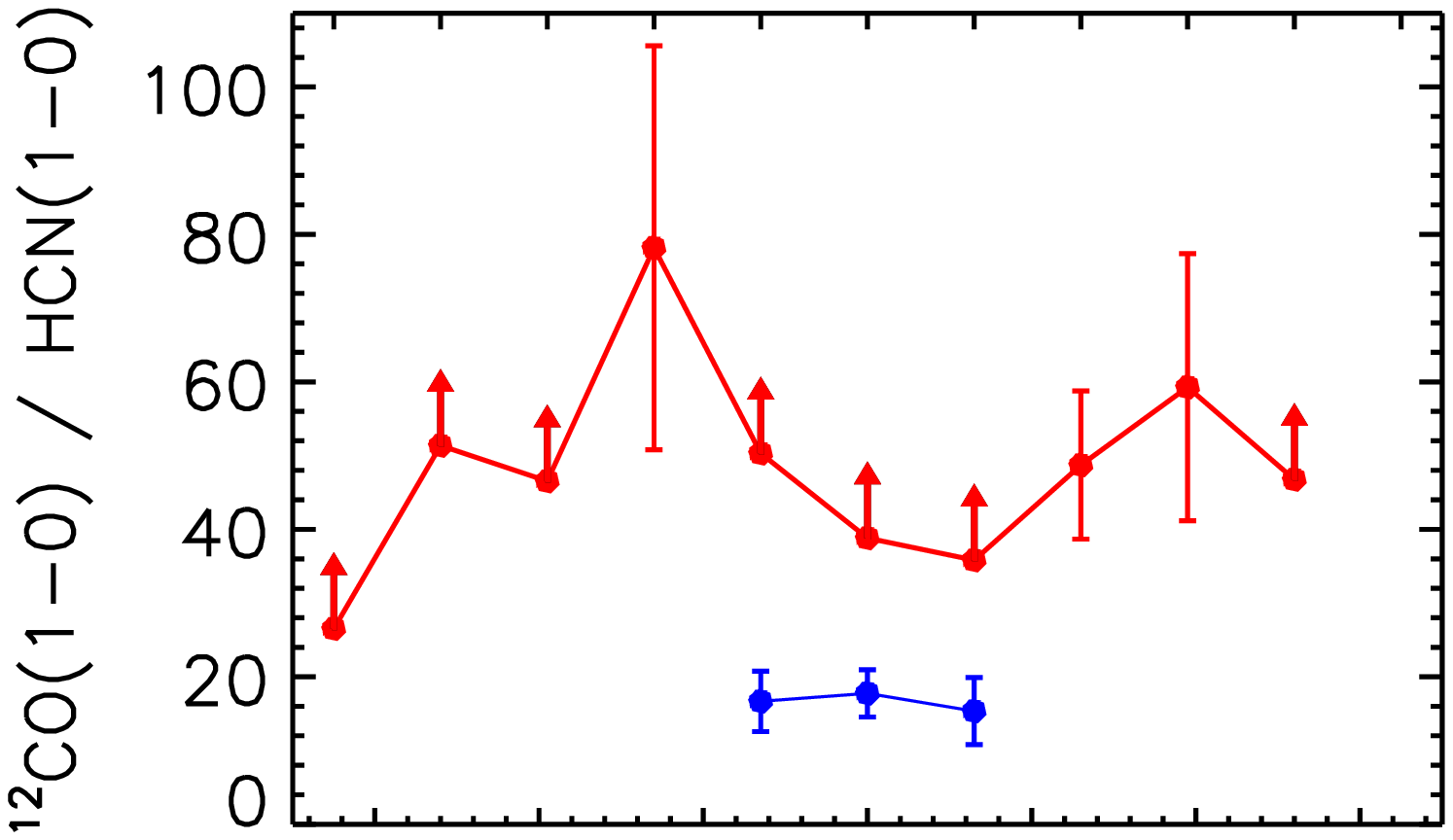}
  \hspace{-15pt}
  \includegraphics[width=4.7cm,clip=]{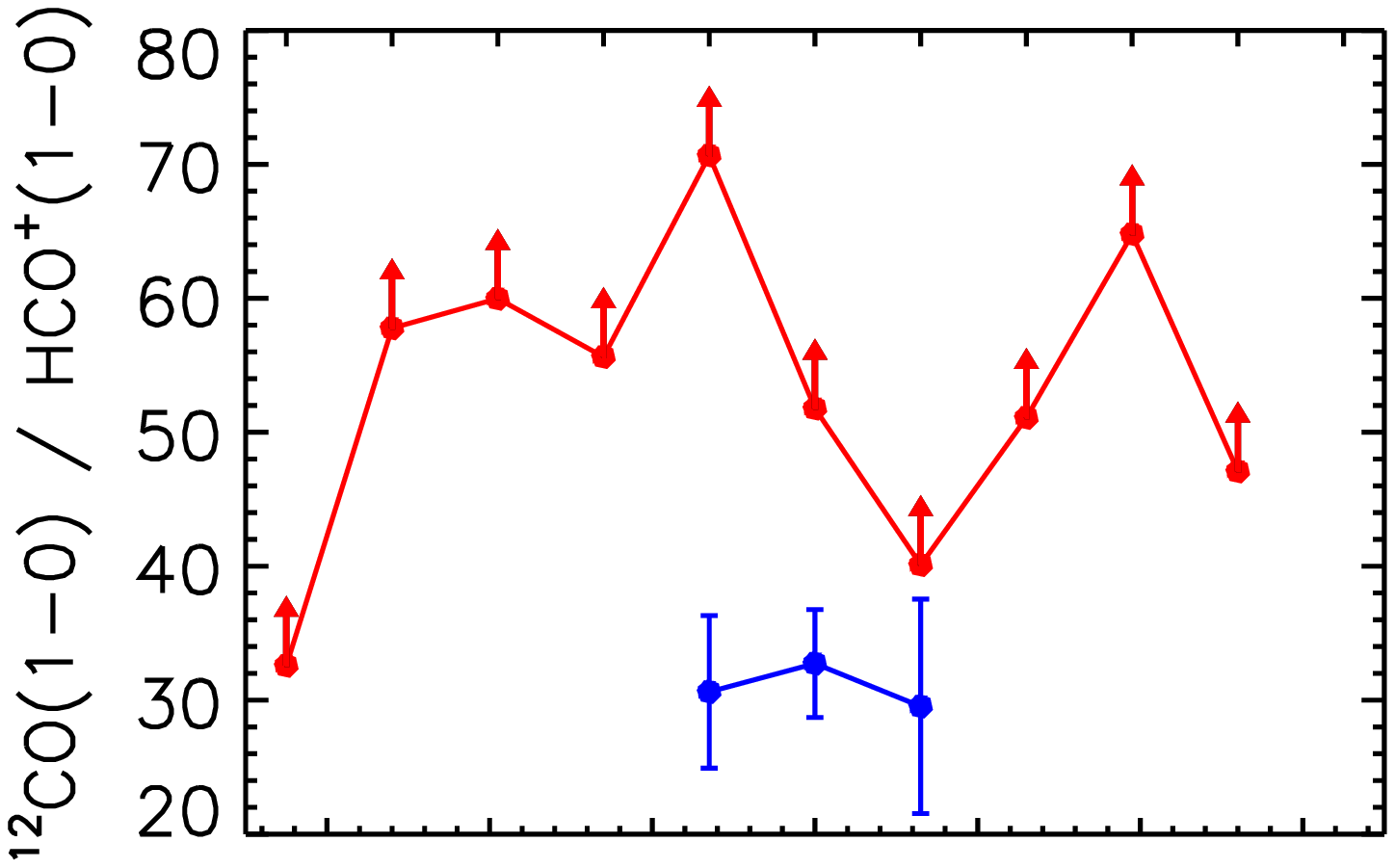}
  \hspace{-15pt}
  \includegraphics[width=4.7cm,clip=]{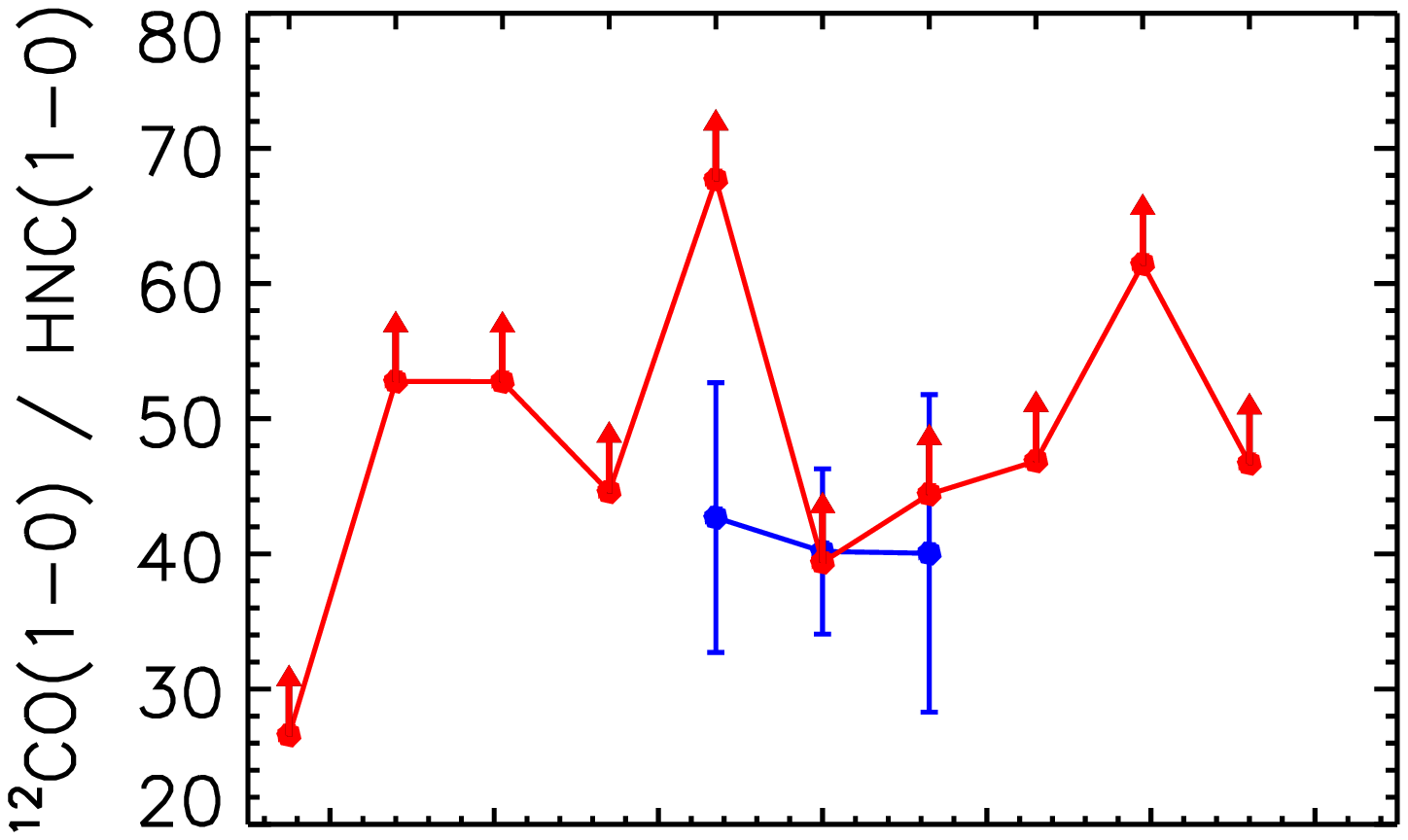}
  \hspace{-15pt}
  \includegraphics[width=4.7cm,clip=]{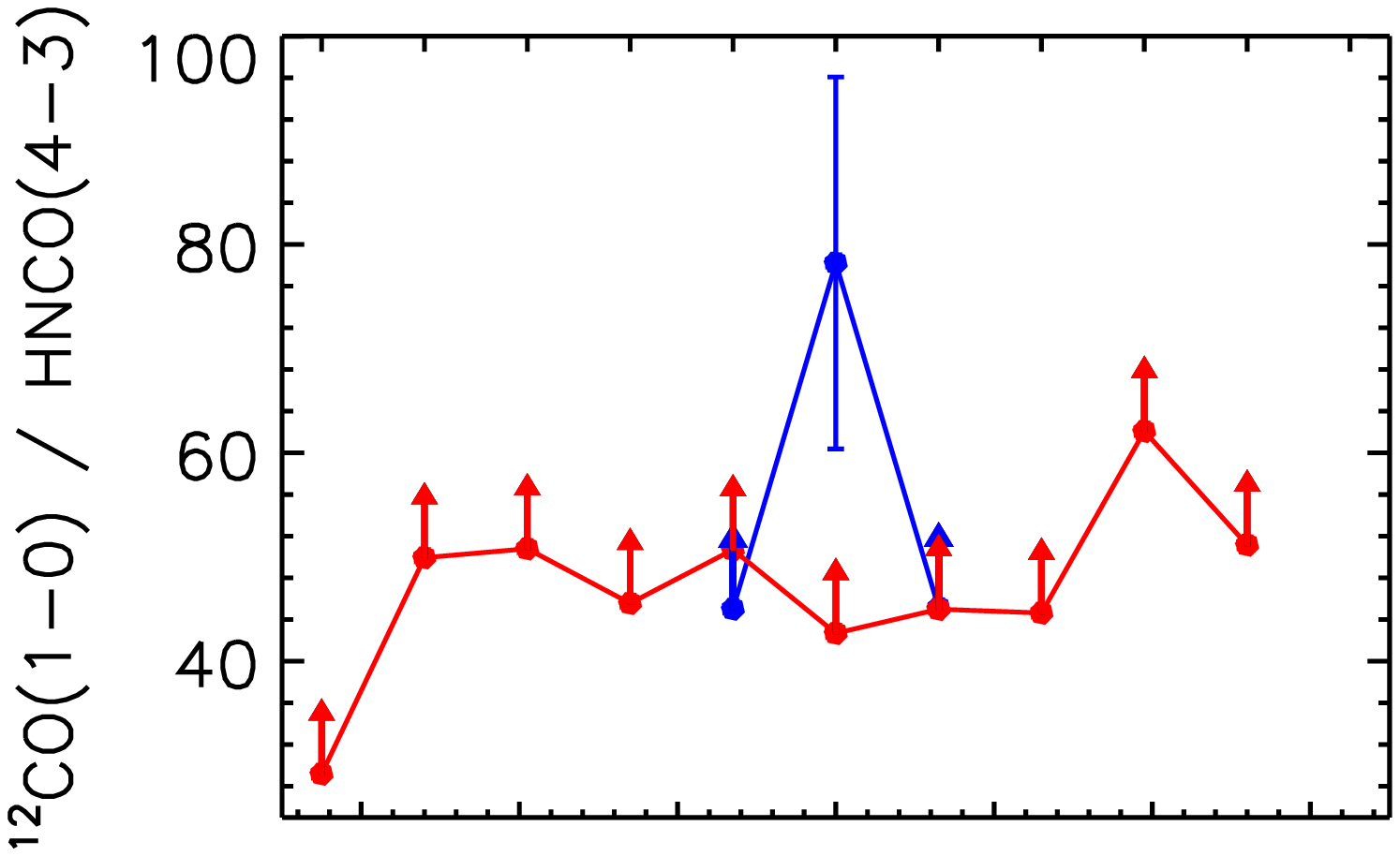}\\
  \vspace{-25pt}
  \hspace{-15pt}
  \includegraphics[width=4.7cm,clip=]{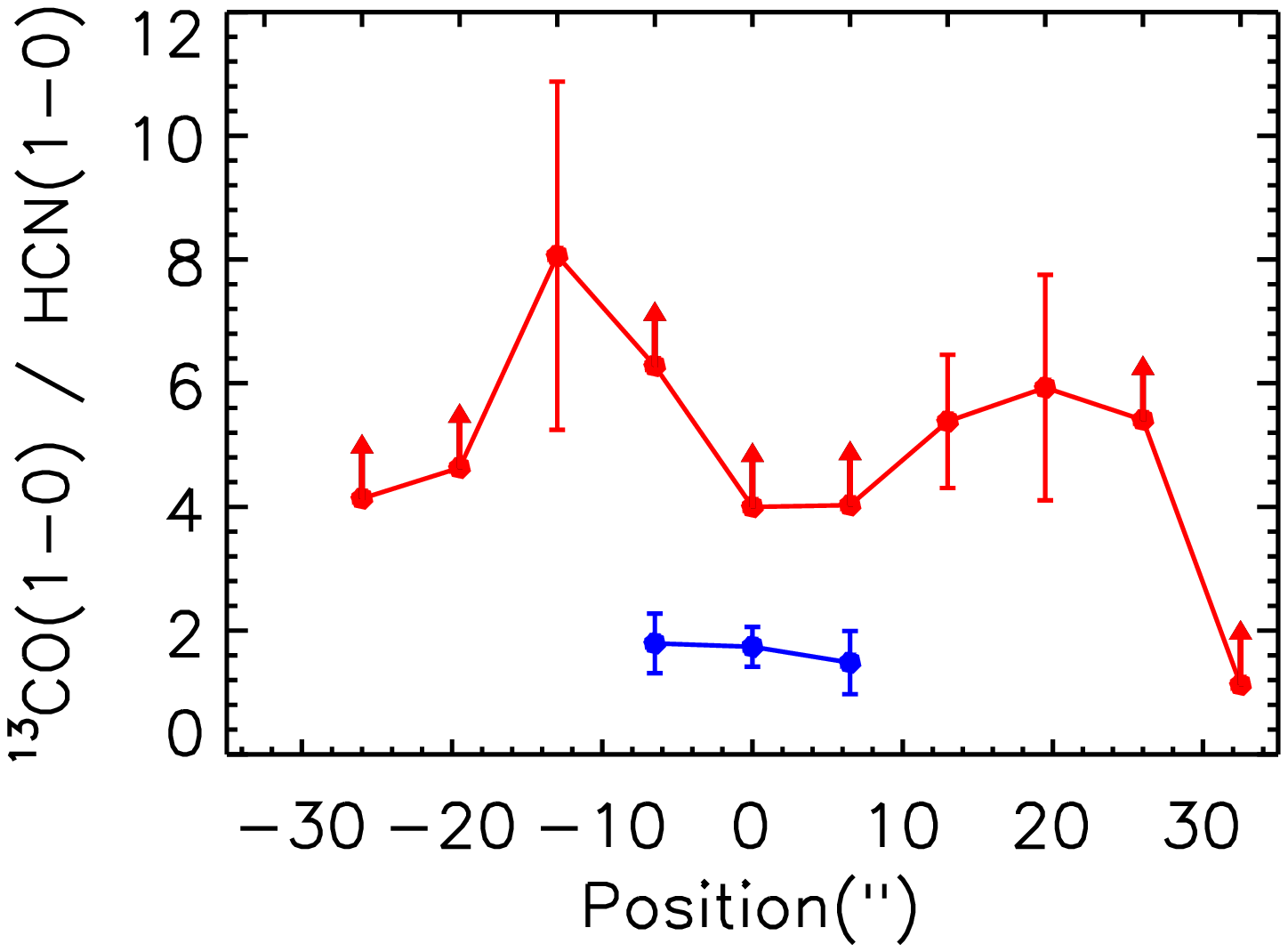}
  \hspace{-15pt}
  \includegraphics[width=4.7cm,clip=]{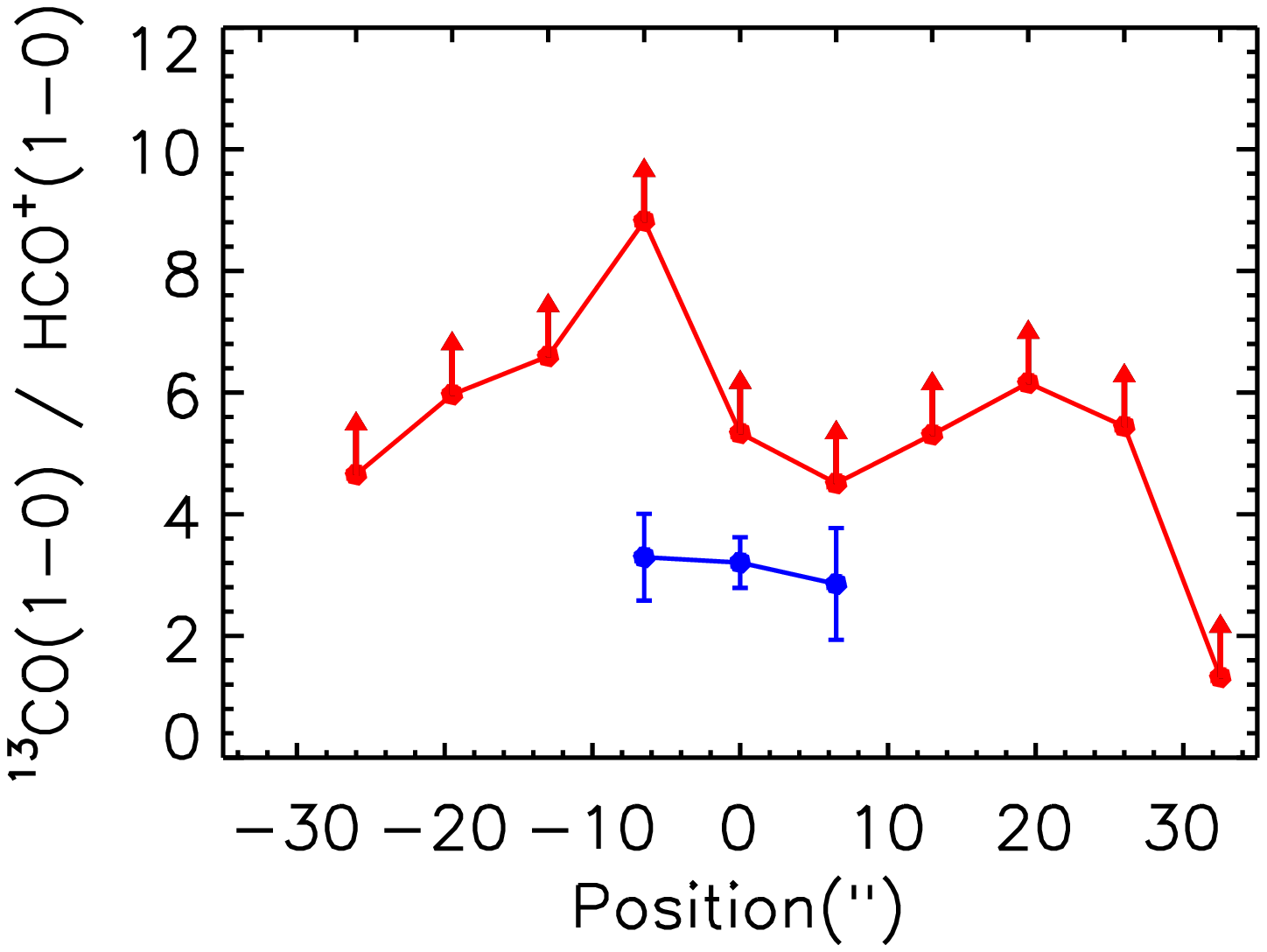}
  \hspace{-15pt}
  \includegraphics[width=4.7cm,clip=]{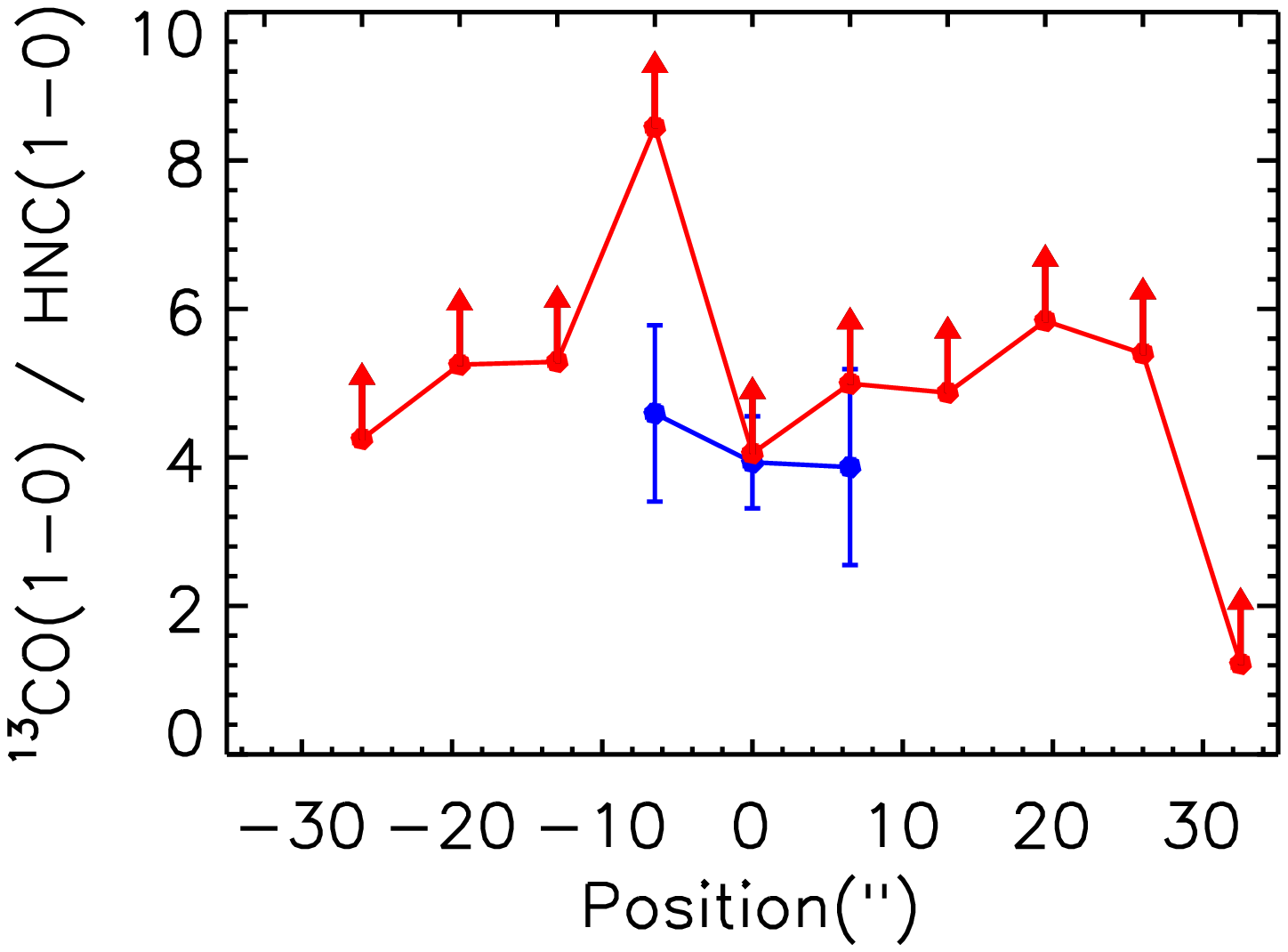}
  \hspace{-15pt}
  \includegraphics[width=4.7cm,clip=]{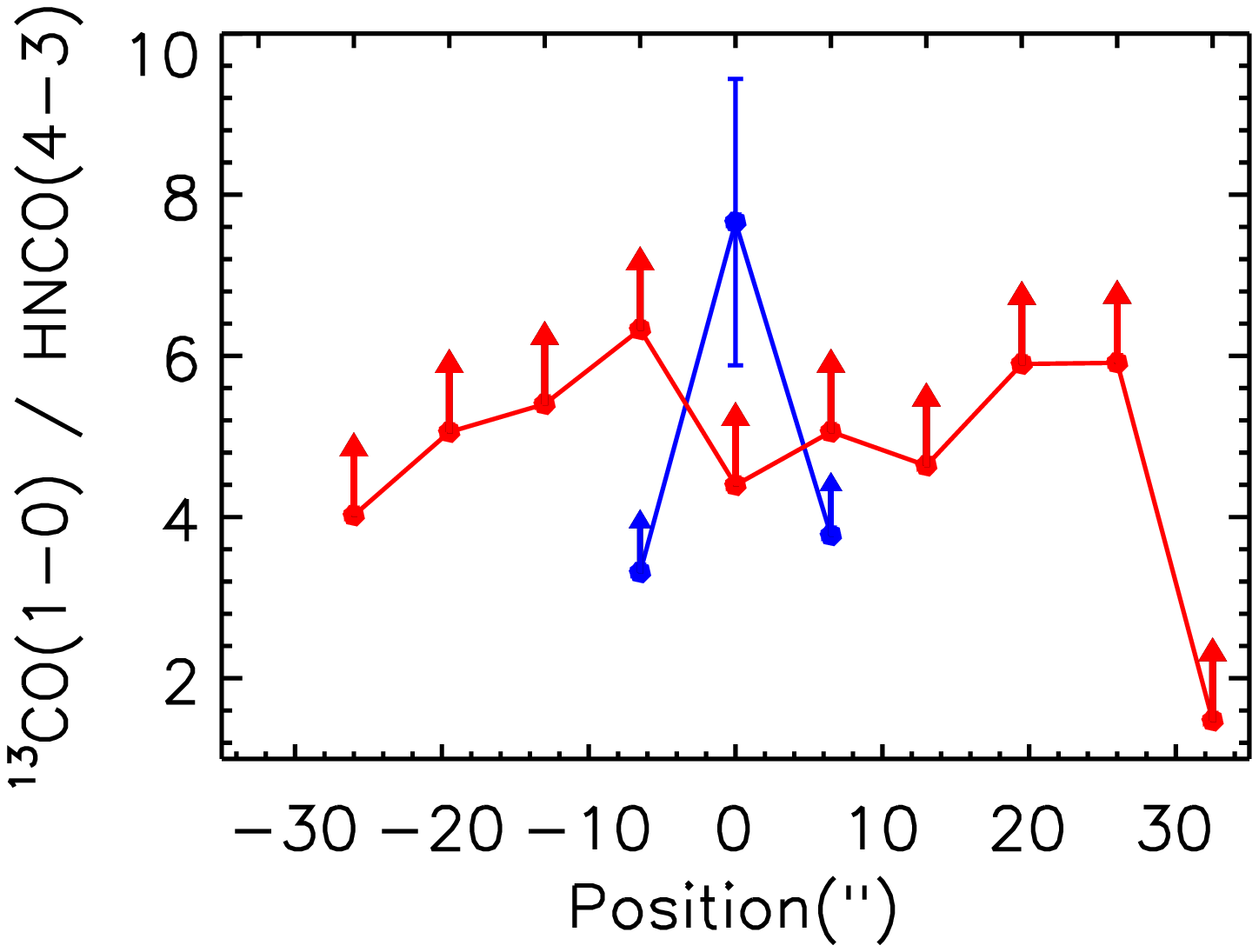}\\
  \caption{Same as Figures~\ref{fig:n4710ratioPOS1} and
    \ref{fig:n4710ratioPOS2} but for NGC~5866.}
  \label{fig:n5866ratioPOS}
\end{figure*}

At the projected positions where there is at least one undetected line
($\int T_{\rm mb}\,{\rm d}v<3\sigma$, where $\sigma$ is the
uncertainty in the integrated line intensity), we assigned an upper
limit to the integrated line intensity of
$3\sigma_{\rm rms}\times$FWHM, where $\sigma_{\rm rms}$ is the noise
in the spectrum of the undetected line and FWHM is the FWHM of the
(other) detected line used to define the line ratio at that
position. The error on this integrated line intensity upper limit was
estimated using the noise in the spectrum and the error on the
FWHM. Integrated line intensity ratios of particular interest are
$R_{\rm 12}\,\equiv\,^{12}$CO(1-0)/$^{12}$CO(2-1),
$R_{\rm 11}\,\equiv\,^{12}$CO(1-0)/$^{13}$CO(1-0) and
$R_{\rm 22}\,\equiv\,^{12}$CO(2-1)/$^{13}$CO(2-1) for the low-$J$ CO
lines only, and $R_{\rm D1}\,\equiv\,$HCN(1-0)/HCO$^+$(1-0),
$R_{\rm D2}\,\equiv\,$HCN(1-0)/HNC(1-0) and
$R_{\rm D3}\,\equiv\,$HCN(1-0)/HNCO(4-3) for the dense gas tracer
lines only.
%
%
\subsection{LVG Modelling}
\label{sec:model}
For our third approach, we probe the physical conditions of the
molecular gas quantitatively by modeling the observed line ratios
using the non-LTE radiative transfer code RADEX \citep{van07}. The
physics and assumptions adopted by the code are described at length in
Appendix~\ref{sec:parameters}. The three free parameters solved for
are the molecular gas kinetic temperature $T_{\rm K}$, H$_2$ volume
number density $n$(H$_2$) and species column number densities
$N$(mol). As the critical densities of the high density tracers are
$\approx3$ orders of magnitude larger than those of the low-$J$ CO
lines, we model the two sets of lines separately and hereafter refer
to the associated molecular gas as dense and tenuous,
respectively. Given the line ratios available (see
Tables~\ref{tab:ratios}\,--\,\ref{tab:ratioir}), four sets of models
are possible: tenuous and dense gas in the nuclear disc of NGC~4710,
tenuous gas in the inner ring of NGC~4710, and dense gas in the
nuclear disc of NGC~5866.

The models are characterised in two ways described in details in
Appendix~\ref{sec:bestlike}: first, the best-fit model in a $\chi^{2}$
sense (see Figs.~\ref{fig:n4710chifc}\,--\,\ref{fig:n4710chisc} and
Tables~\ref{tab:result1} and \ref{tab:result2}), and second, the most
likely model as determined from the probability distribution function
(PDF) of each model parameter marginalised over the others (see
Figs.~\ref{fig:n4710likefc}\,--\,\ref{fig:n4710likesc} and
Tables~\ref{tab:result1} and \ref{tab:result2}). The best-fit models
are generally consistent with the most likely models, but the
uncertainties on the latter are usually large due to the flatness of
the PDFs, and as expected results within a single kinematic component
are indistinguishable.

Looking at the results for NGC~4710 in more details, the H$_2$ volume
density $n$(H$_2$) is generally well constrained (with frequent
single-peaked PDFs) for both the tenuous and dense molecular gas
components, at least in the nuclear disc. Unsurprisingly, the model
results indicate that $n$(H$_2$) is larger and $N$(CO) smaller in the
dense gas component than in the tenuous gas component. The kinetic
temperature $T_{\rm K}$ of the tenuous gas component in the nuclear
disc always has a peak at $10$~K (at the low-temperature edge of the
model grid) in the PDFs, with a monotonic decrease at higher
temperatures, causing the large uncertainties seen in the most likely
model results. However, the most likely $T_{\rm K}$ model results in
the inner ring are just upper limits (i.e.\ all models are equally
likely). $T_{\rm K}$ for the dense gas component (nuclear discs only)
shows single-peaked PDFs with much higher temperatures, but again the
uncertainties are large. Overall, while there seems to be a clear
trend for the dense gas component to have a higher $T_{\rm K}$, higher
$n$(H$_2$) and smaller $N$(CO), the $T_{\rm K}$ trend is not
significant given the uncertainties (see Table~\ref{tab:result1}).

For NGC~5866, only the dense gas in the nuclear disc was modeled. At
the central projected position (i.e.\ position~$0$), the model results
indicate a similar dense gas volume density, column density and
temperature as those in the nuclear disc of NGC~4710. And again, the
uncertainties for $T_{\rm K}$ are proportionally higher than those for
$n$(H$_2$) and $N$(CO). However, as two positions in the nuclear disc
of NGC~5866 have at least one line ratio that is a lower limit, the
model results at those positions are just upper limits (and thus
unconstrained; see Table~\ref{tab:result2}).
%
%
\section{Results and Discussion}	
\label{sec:result}
In the previous sections, we quantified the physical conditions of a
two-component molecular ISM, i.e.\ tenuous and dense gas components,
along the discs of the edge-on early-type galaxies NGC~4710 and
NGC~5866. We achieved this by probing the variations of molecular line
ratios as a function of projected radius along the galaxy discs, and
by performing radiative transfer modeling of these multiple molecular
gas tracers.

In \S~\ref{sec:intro}, we also briefly discussed the properties of
NGC~4710 and NGC~5866, highlighting the fact that they are relatively
isolated, their barred nature, and their molecular gas richness
compared to other ETGs. Before putting our results in context,
however, we must also discuss their star formation.
%
%
\subsection{Star formation}
\label{sec:sf}
\cite{tim14} used $22\micron$ fluxes from the {\it Wide-field Infrared
  Survey Explorer} ({\it WISE}) catalogue to estimate the star
formation rates (SFRs) of molecule-rich ETGs in the ATLAS$^{\rm 3D}$
survey. While the total SFRs of NGC~4710 and NGC~5866 are average,
that of NGC~4710 is half that of NGC~5866 ($0.11$ vs.\
$0.21$~$M_\odot$~yr$^{-1}$; see Table~\ref{tab:gprop}). However, as
the star formation is concentrated into rings rather than being spread
across the entire discs, NGC~4710 and NGC~5866 have reasonably high
SFR surface densities, in the top $\approx30\%$ of the
ATLAS$^{\rm 3D}$ ETGs. \cite{tim14} also found that, like other ETGs,
NGC~4710 and NGC~5866 have lower star formation efficiencies (SFEs)
than those of spiral and starburst galaxies, with molecular gas
depletion times of $10$ and $4$~Gyr, respectively (indicating SFEs
respectively $\approx5$ and $2$ times lower than that of normal
spirals with $T_{\rm dep}=2$~Gyr; see, e.g.,
\citealt{ken98}). Nevertheless, the molecular line ratios we measured
along the discs of both galaxies (particularly in the nuclear discs)
show similarities to those found in the centres of some spirals and
starbursts (see \S~\ref{sec:ratiovar}). This suggests that the
molecular line ratios are not sensitive to whatever is suppressing
star formation in these systems.
%
%
\subsection{Moment maps}
\label{sec:mps}
As our interferometric maps show (see Figs.~\ref{fig:n4710mom} and
\ref{fig:n5866mom}), the tenuous molecular gas extends to
$\approx30^{\prime\prime}$ on either side of the centre in both
galaxies. The tenuous gas is brighter in the nuclear disc than in the
inner ring of NGC~4710, but while this is also the case in NGC~5866,
its inner ring is clearly more prominent (relative to the nuclear
disc). As expected, the tenuous gas is more extended than the dense
gas along the galaxy discs (at least given the roughly comparable
noise levels in most cubes), the latter being generally centrally
concentrated and restricted to the nuclear disc only (HCN(1-0)
emission in the outskirts of NGC~5866 again suggests a more prominent
inner ring). Already, the integrated molecular gas maps therefore
suggest that different physical conditions are likely to prevail in
the nuclear discs and inner rings.

In addition, the mean velocity maps reveal surprisingly complex
behaviour beyond the nuclear disc in both galaxies, already hinting at
multiple velocity components along the line of sight in these edge-on
discs, a characteristic confirmed by the galaxies' PVDs (see
\S~\ref{sec:pvds} below).
%
%
\subsection{PVDs}
\label{sec:pvds}
As discussed in \S~\ref{sec:posvel}, the PVDs of NGC~4710 and NGC~5866
show a characteristic X shape (see Figs~\ref{fig:n4710pvd} and
\ref{fig:n5866pvd}). This is easily understood in the context of
barred galaxy dynamics, whereby the central rapidly-rising PVD
component is associated with a nuclear disc within the ILR, while the
outer slowly-rising PVD component can be associated with an inner ring
near corotation.

In both NGC~4710 and NGC~5866, it is obvious that both $^{12}$CO and
$^{13}$CO (i.e.\ all the tenuous gas) have similar distributions and
kinematics, although as hinted from the moment~0 maps the inner ring
in NGC~5866 is more prominent. In fact, separating the nuclear disc
from the inner ring in the PVD, the integrated $^{12}$CO(1-0) (and
$^{13}$CO(1-0)) flux of the inner ring in NGC~5866 is larger than that
of the nuclear disc (opposite in NGC~4710). There does not seem to be
any molecular gas beyond the inner ring (i.e.\ beyond corotation) in
either galaxy, where the rotation curve (i.e.\ the high-velocity
envelope of the PVD) would be expected to be flat. The particularly
strong emission at the edges of the inner rings (especially in
NGC~4710) is simply due to edge brightening (as the rings are seen
edge-on). However, other local emission maxima are observed along the
nuclear discs and inner rings in the PVDs (see Figs~\ref{fig:n4710pvd}
and \ref{fig:n5866pvd}), indicating that the molecular gas in both of
these components is clumpy. The molecular gas peaks likely trace
individual giant molecular cloud complexes, and they should be
associated with regions of increased star formation.

The gap between the nuclear disc and inner ring in the PVDs is easily
understood by a lack of gas on $x_1$ orbits, as they are swept free of
gas by the shocks on the leading edge of the bar. However, the feature
seen at the intermediate region in $^{12}$CO(1-0) PVD (see
\S~\ref{sec:posvel} and Fig.~\ref{fig:pos}) is not fully understood,
but \cite{ab99} speculated based on hydrodynamical simulations that
this emission originates in secondary enhancements of gas in front of
the leading edges of particularly strong (and sharp) bars.

Overall, it is clear that the molecular gas in the barred edge-on
galaxies NGC~4710 and NGC~5866 is primarily concentrated in two
distinct but clumpy kinematic components, a nuclear disc and an inner
ring. As the dense gas tracers (HCN(1-0), HCO$^+$(1-0), HNC(1-0) and
HNCO(4-3)) are not detected outside the nuclear discs (except HCN(1-0)
in NGC~5866), their PVDs do not show the characteristic X shape. It is
thus natural to expect the ISM physical conditions to be similar
within each kinematic component (with possibly small variations
between clumps, especially if they are star-forming), but for the
conditions to be different across the two components. The clumpy
nature of the two kinematic components is confirmed by the observed
line ratios, as local emission maxima in the PVDs (i.e.\ clumps) seem
to correlate with local variations in the line ratios (particularly
the ratios of CO to dense gas tracer lines; see
Figs.~\ref{fig:n4710ratioPOS1}\,--\,\ref{fig:n5866ratioPOS}).

Lastly, we note that unlike the behaviour of the other lines in both
galaxies, there is almost no dense gas detected at the very centre of
the PVDs in NGC~5866. The same effect, if present at all, is much
weaker in NGC~4710 (see Figs.~\ref{fig:n4710pvd} and
\ref{fig:n5866pvd}). This suggests that a different excitation
mechanism (or possibly gas distribution) dominates in the very centre
of NGC~5866 (e.g.\ active galactic nucleus or nuclear starburst
activity), unless the local emission minimum is in fact due to
absorption against a nuclear continuum source. The latter possibility
however seems unlikely, since while there is a continuum source in
NGC~5866 (see Section~\ref{sec:contem}), no absorption is seen in the
spectra.
%
%
\subsection{Integrated spectra}
\label{sec:compiram}
Except for HNC(1-0) and HNCO(4-3) (shown here for the first time), all
the molecular lines discussed were previously observed with the
IRAM~30m telescope (see \citealt{y11} and \citealt{c12}). As
illustrated in the moment maps (Figs.~\ref{fig:n4710mom} and
\ref{fig:n5866mom}), where we overlaid the corresponding IRAM~30m beam
sizes, the CO extent (at all transitions) is covered entirely only by
the interferometric observations (with larger primary beams). The
single-dish beams systematically cover the nuclear discs only, thus
missing out much of the tenuous molecular gas (this is not a problem
for the dense gas tracers, that are not detected in the inner rings).

In Figures~\ref{fig:spec1} and \ref{fig:spec2}, the integrated CARMA
and PdBI spectra, simulated (i.e.\ integrated with a spatial Gaussian
weighting) IRAM 30m spectra, and observed IRAM 30m spectra are
overlaid with each other (see \S~\ref{sec:comp}). As expected given
the comments above, the integrated interferometric CO (i.e.\ tenuous
molecular gas) fluxes are systematically larger than the single-dish
fluxes (simulated and observed). However, again as expected given
their spatially compact emission (nuclear discs only), the integrated
and single-dish dense gas tracer spectra are consistent with each
other. Differences in the shapes of the spectra from line to line can
further indicate different gas physical conditions along the discs of
the galaxies (and thus a slightly different spatial distribution for
each tracer). Interestingly, only the dense gas tracers of NGC~5866
show clearly double-peaked integrated line profiles (and to a lesser
extent $^{12}$CO(1-0); see Figs.~\ref{fig:spec1} and \ref{fig:spec2}).

The only discordant note in relation to the integrated spectra is the
fact that the simulated IRAM 30m spectra of the tenuous molecular gas
in NGC~4710 are significantly brighter than the true IRAM 30m spectra
(see the top three panels of Fig.~\ref{fig:spec1}). The relatively
large offset between the simulated and observed spectra is thus most
likely due to a combination of single-dish pointing errors (supported
by the slight asymmetry of the true IRAM 30m spectra, not present in
the CARMA data) and flux calibration uncertainties ($20\%$ is standard
for milimetric observations flux-calibrated using planetary models;
see e.g.\ \citealt{al13}).
%
%
\subsection{Molecular line ratios}
\label{sec:ratiovar}
The ratios of the PVDs from different lines are shown in
Figures~\ref{fig:n4710ratioPVD1}\,--\,\ref{fig:n5866ratioPVD}, while
Figures~\ref{fig:n4710ratioPOS1}\,--\,\ref{fig:n5866ratioPOS} show the
ratios of the integrated line intensities as a function of projected
radius for each kinematic component separately (nuclear disc and inner
ring; the values are listed in
Tables~\ref{tab:ratios}\,--\,\ref{tab:ratioir}). Similarly, the LVG
modeling results for the tenuous and dense molecular gas components,
in the nuclear discs and inner rings, are listed in
Tables~\ref{tab:result1} and \ref{tab:result2}.

We recall here that for rings, the different projected positions
simply correspond to different azimuthal positions (i.e.\ angles)
within the rings, so we do not expect significant gradients with
projected position. The nuclear disc, as its name suggests, may well
however be filled in.
%
%
\subsubsection{$^{12}$CO(1-0) / $^{12}$CO(2-1) ratio and tenuous gas
  temperature} 
\label{sec:co12co}
In NGC~4710, the ratio of the PVDs of the $^{12}$CO(1-0) and
$^{12}$CO(2-1) lines shows that the ratio is greater than unity
everywhere, but is slightly smaller in the nuclear disc than the inner
ring (see Fig.~\ref{fig:n4710ratioPVD1}), suggesting a higher
temperature in the tenuous molecular gas of the nuclear disc. This is
confirmed by the ratio of the integrated line intensities along the
two kinematic components (see Fig.~\ref{fig:n4710ratioPOS1}), clearly
showing that $R_{\rm 12}\,\equiv\,^{12}$CO(1-0)/$^{12}$CO(2-1) at the
central three projected positions along the nuclear disc (i.e.\
projected positions~$-1$, $0$ and $1$) is smaller than that at the same
positions in the inner ring. This trend appears marginal at the edges
of the nuclear disc (projected positions~$-2$ and $2$), but this is
simply due to the very weak emission (and associated large
uncertainties) there.

Ideally, the temperature difference between the tenuous gas in the
nuclear disc and inner ring of NGC~4710 would be confirmed by our LVG
modeling, but the results are inconclusive. This is however not
totally surprising, as we have only two CO transitions
($J=1\rightarrow0$, $2\rightarrow1$) to constrain the temperature.
Within each ring, the best-fit temperatures cover nearly the entire
range allowed by the models. This is probably because the
$\Delta\chi_{\rm r}^2$ contours at each position are often shallow and
extended (see Figs.~\ref{fig:n4710chifc} and \ref{fig:n4710chisc}), so
that the best-fit model results should be taken with a grain of salt.
The likelihood results are better suited to such situations, and
suggest a low to intermediate temperature in the nuclear disc
($T_{\rm K}\approx10$\,--\,$60$~K, but perhaps much higher), but no
useful constraint can be derived for the inner ring (due to some line
ratio lower limits).

Although $^{12}$CO(2-1) and its isotopologue were not observed
interferometrically in NGC~5866, single-dish observations of these
lines in the central regions of NGC~5866 show that $R_{\rm 12}$ is
larger there than in the central regions of NGC~4710 \citep{c12}. This
suggests that the CO gas at the centre of NGC~5866 is colder than that
in NGC~4710. However, since the single-dish observations cannot
disentangle the emission from the nuclear disc and inner ring, this
should be verified with interferometric data.
%
%
\subsubsection{$^{12}$CO / $^{13}$CO ratio and tenuous gas opacity}
\label{sec:co13co}
As shown in Figure~\ref{fig:n4710ratioPVD1}, the ratio of $^{12}$CO to
its isotopologue $^{13}$CO is larger by a factor of
  $\approx2$ in the central regions of the nuclear disc of NGC~4710
(positions~$-1$, $0$ and $1$) than in all other
regions (external parts of the nuclear disc and the few positions in
the inner ring where a measurement is possible). This is true of both
the $1\rightarrow0$ and $2\rightarrow1$ transitions.

Looking at the ratios of the integrated line intensities of CO an its
isotopologue ($R_{\rm 11}\,\equiv\,^{12}$CO(1-0)/$^{13}$CO(1-0) and
$R_{\rm 22}\,\equiv\,^{12}$CO(2-1)/$^{13}$CO(2-1)) as a function of
projected radius (Fig.~\ref{fig:n4710ratioPOS1}), the above behaviour
is confirmed. Indeed, the $R_{\rm 22}$ ratios at positions~$-1$, $0$
and $1$ of the nuclear disc of NGC~4710 (the central three projected
positions) are twice those at the same projected positions in the
inner ring (so is $R_{\rm 11}$ at position $-1$). We must note however
that the $R_{\rm 22}$ ratios in the inner ring of NGC~4710 are all
lower limits (the same applies to some positions for $R_{\rm 11}$ and
to positions $-2$ and $2$ of the nuclear disc).

If confirmed (e.g.\ by the detection of $^{13}$CO(1-0) and
$^{13}$CO(2-1) at all the positions along the inner ring), the
behaviour described above would indicate optically thinner tenuous
molecular gas in the central regions of the nuclear disc (assuming
that all of the $^{13}$CO is optically thin but only some of the
$^{12}$CO). However, the emission in the central regions includes
contributions from the inner ring as well (in projection).

Interestingly, the $R_{\rm 11}$ ratio of the inner ring of NGC~4710
shows a significant maximum at projected position $-2$, that clearly
corresponds to a clump (see Figure~\ref{fig:n4710ratioPOS1} and
\ref{fig:pos}), while the weaker $R_{\rm 11}$ minimum at position~$-1$
has no obvious clump associated with it.

While a $\chi^2$ (and thus likelihood) analysis of the opacity is not
possible, since it is not a model parameter, each model computed does
return the optical depths of the associated lines. The optical depths
of the $^{12}$CO(1-0) and $^{12}$CO(2-1) lines for the best-fit models
are thus known. In addition, one would expect the CO column number
density to correlate with its optical depth. Looking at the model
results (Table~\ref{tab:result1}), the $^{12}$CO(1-0) optical depth,
$^{12}$CO(2-1) optical depth, and CO column number density in the
nuclear disc are generally slightly larger than those in the inner
ring, contrary to our expectations based on the empirical
$^{12}$CO/$^{13}$CO ratios. However, as the best-fit models are not
representative of the $\Delta\chi_{\rm r}^2$ geometry and most lie at
the edge of the model grid, the associated opacities and column number
densities are questionable.

Unlike NGC~4710, the $R_{\rm 11}$ ratio in NGC~5866 appears smaller in
the central regions by $\approx25\%$ (compared with
all other regions in the ratio PVD; see Fig.~\ref{fig:n5866ratioPVD}).
However, the effect is weak and seems to be associated with a
generally slightly smaller ratio in the inner ring compared to the
nuclear disc (if more marked in the central regions). Looking at the
ratio of the integrated line intensities ($R_{\rm 11}$ only), the
$R_{\rm 11}$ ratios of the nuclear disc and inner ring are consistent
in the central regions, although the ratios do increase at larger
projected radii in the inner ring (see
Figure~\ref{fig:n5866ratioPOS}). The effect is marginal at best
however, as nearly all $R_{\rm 11}$ ratios measured are consistent
within the uncertainties.

\citet{c12} also measured the $R_{\rm 11}$ and $R_{\rm 22}$ ratios in
the central regions of both galaxies (without disentangling the two
kinematic components). While our average values along the nuclear disc
and the inner ring of NGC~4710 agree with the ratios reported by
\citet{c12} (within the uncertainties), our average $R_{\rm 11}$ ratio
for both the nuclear disc and the inner ring of NGC~5866 is larger
than that found by \citet{c12}.

As hinted above, as $^{13}$CO is less abundant than its parent
molecule $^{12}$CO and is generally considered optically thin,
$^{12}$CO/$^{13}$CO variations largely reflect variations in the
$^{12}$CO optical depth. The larger the ratio, the thinner the
$^{12}$CO gas (and vice-versa). The general behaviour in NGC~4710 and
NGC~5866 (but more prominent in NGC~4710) is thus that the CO gas is
optically thinner in the nuclear disc than in the inner ring, although
the trend is weak and may be restricted to the inner parts of the
nuclear disc in NGC~5866. Variations within each kinematic component
(e.g.\ local maximum in the inner ring of NGC~4710 and projected
radius trend in NGC~5866) likely indicate azumithal variations of the
physical conditions (i.e.\ clumpiness) within them.

High $R_{\rm 11}$ and $R_{\rm 22}$ ratios tracing diffuse gas are
probably the result of stellar feedback, and therefore indicate more
active current and/or recent star formation. Radial $R_{\rm 11}$
gradients are indeed seen in spirals \citep{pag01}, and the ratio can
have local maxima in star-forming regions along spiral arms
\citep{tan11}. This thus suggests that star formation is more intense
in the nuclear discs of our galaxies than their inner rings (at least
in NGC~4710), a behaviour entirely consistent with the general
behaviour of barred disc galaxies, particularly early-type spirals
\citep[e.g.][]{kor04}. This also agrees with the higher temperatures
inferred for the molecular gas in the nuclear discs (compared to that
in the inner rings) in the previous sub-section (\S~\ref{sec:co12co}).
Higher $S/N$ $^{13}$CO observations would however help to strengthen
this result.
%
%
%
\subsubsection{HCN / HCO$^+$, HCN / HNC and HCN / HNCO ratios and
  dense gas excitation}
\label{sec:den}
The ratios of the PVDs of the dense gas tracers only are shown in
Figures~\ref{fig:n4710ratioPVD1} and \ref{fig:n5866ratioPVD} for
NGC~4710 and NGC~5866, respectively (where greyscales represent lower
limits, where HCN was detected but not the other lines). The ratios of
the integrated line intensities, namely
$R_{\rm D1}\,\equiv\,$HCN(1-0)/HCO$^+$(1-0),
$R_{\rm D2}\,\equiv\,$HCN(1-0)/HNC(1-0) and
$R_{\rm D3}\,\equiv\,$HCN(1-0)/HNCO(4-3) are shown as a function of
projected radius in Figures~\ref{fig:n4710ratioPOS1} and
\ref{fig:n5866ratioPOS}. Generally, the dense gas tracers are detected
only in the nuclear discs (typically projected positions~$-1$, $0$ and
$1$). The only exception to this is the detection of HCN(1-0) in the
inner ring of NGC~5866 (with significant detections at projected
positions $-2$, $2$ and $3$).

The $R_{\rm D1}$, $R_{\rm D2}$ and $R_{\rm D3}$ ratios are larger than
$1$ at all projected positions, and they do not show any evidence of a
gradient with projected position (and thus azimuthal angle).
$R_{\rm D1}\,<\,R_{\rm D2}\,<\,R_{\rm D3}$, indicating that HCN(1-0)
is the brightest line among the dense gas tracers, followed by
respectively HCO$^+$(1-0), HNC(1-0) and HNCO(4-3). \citet{c12} also
detected HCN(1-0) and HCO$^+$(1-0) in the central regions of both
galaxies, finding $R_{\rm D1}\approx1.5$, consistent with the values
found here for the nuclear discs.

UV radiation from young massive (O and B) stars, X-rays from active
galactic nuclei (AGN) and CRs from supernova explosions all have
distinct characteristic effects on the molecular gas physical
conditions, and each plays an important role in the dissociation and
ionisation of molecules, thus changing the chemistry of the ISM and
affecting the molecular line ratios observed. UV radiation primarily
affects the outermost layers of clouds (photon dissociation regions or
PDRs; e.g.\ \citealt{ti85b,ti85a,bl87}), while X-rays penetrate deeper
and form X-ray dissociation regions (XDRs;
\citealt[e.g.][]{le96,mal96}). Both UV and X-ray radiation can enhance
HCN, but since X-rays can affect the gas chemistry much deeper into
clouds, they can do so more efficiently (if present; e.g.\
\citealt{kr08}).

The theoretical results of \citet{mei07} suggest that if the volume
density exceeds a value $10^{5}$~cm$^{-3}$, then $R_{\rm D1}>1$ in
PDRs, while $R_{\rm D1}<1$ in XDRs. Our LVG modelling results for a
two-component molecular ISM indicate that for the dense gas component
$n$(H$_2$)$\ge10^{5}$~cm$^{-3}$ for both galaxies (see
Tables~\ref{tab:result1} and \ref{tab:result2}). Combined with the
fact that $R_{\rm D1}>1$ everywhere in both galaxies, this therefore
indicates that PDRs are most likely dictating the physical conditions
(and thus the observed line ratios) of the molecular ISM in the
nuclear discs of both galaxies. Reassuringly, those densities are also
much larger than those obtained for the tenuous molecular gas
component (typically by $3$ orders of magnitude).

Our results are also consistent with the absence of any substantial
AGN activity (and thus X-rays) in NGC~4710 and NGC~5866. However,
since $R_{\rm D1}>1$, supernova explosions are also unlikely to be
significant, and young massive OB stars are likely responsible for the
HCN enhancement. If supernova explosions were a dominant force in the
ISM, HCO$^+$ would have been enhanced by CRs (with respect to HCN),
leading to $R_{\rm D1}<1$ as seen in starbursts (although not all
starbursts and spiral galaxies have $R_{\rm D1}<1$; see
\S~\ref{sec:compg}). Overall, the high $R_{\rm D1}$ ratio observed
thus suggests a low-CR PDR-dominant environment in the nuclear disc of
both galaxies, but there could be some impact from supernova
explosions as seen in starbursts/spirals that have similar
$R_{\rm D1}$ ratios.

We note briefly that, assuming the $R_{\rm D1}$, $R_{\rm D2}$ and
$R_{\rm D3}$ upper limits in the inner ring of NGC~5866 are not too
far off the mark, these ratios are then larger in the nuclear disc
than in the inner ring, indicating different dissociation and
ionisation mechanisms in the two kinematic components. The HCN would
need to be suppressed or HCO$^+$, HNC and HNCO enhanced in the inner
ring. More sensitive observations of HCN, HCO$^+$, HNC and HNCO in the
inner ring of both galaxies are however necessary to clearly constrast
the physical conditions of their dense molecular ISM.
%
%
\subsubsection{Ratios of CO to HCN, HCO$^+$, HNC and HNCO and dense
  gas fraction}
\label{sec:covsden}
The ratios of CO PVDs to high density tracer PVDs (hereafter
$R_{\rm CD}$) are shown in Figures~\ref{fig:n4710ratioPVD2} and
\ref{fig:n5866ratioPVD} (where greyscales represent lower limits,
where CO was detected but not the high density lines), while the
ratios of the integrated line intensities as a function of projected
radius are shown in Figures~\ref{fig:n4710ratioPOS2} and
\ref{fig:n5866ratioPOS} (see also Tables~\ref{tab:ratiocd} and
~\ref{tab:ratioir}).

As seen from all those figures, there is a clear difference of the CO
to high density tracer ratios between the nuclear disc and inner ring,
for both NGC~4710 and NGC~5866. Indeed, the $R_{\rm CD}$ ratios are
significantly larger in the inner rings (clearest for the ratios
involving HCN and HCO$^+$), and the true differences are likely to be
even greater than that hinted by
Figures~\ref{fig:n4710ratioPVD2}\,--\,\ref{fig:n5866ratioPVD} and
\ref{fig:n4710ratioPOS2}\,--\,\ref{fig:n5866ratioPOS} as the
$R_{\rm CD}$ ratios in the inner rings are all lower limits (except
for a few positions in HCN in NGC~5866). This contrast is
incontrovertible for the HCN, HCO$^+$ and HNC lines, but may well also
hold true for HNCO (where some inner ring lower limits are smaller
than the corresponding nuclear disc ratios). For example, the ratios
of $^{12}$CO(1-0)/HCN(1-0) and $^{12}$CO(1-0)/HCO$^+$(1-0) in the
inner ring of NGC~4710 are at least $50\%$ higher than those in its
nuclear disc (and often more than twice). Similarly, the ratios of
$^{12}$CO(1-0)/HCN(1-0) in the inner ring of NGC~5866 are at least
twice those in its nuclear disc, whereas the ratios of
$^{12}$CO(1-0)/HCO$^+$(1-0) are at least $50\%$ higher, typically
more.

As the $R_{\rm CD}$ ratios essentially trace the fraction of dense
molecular gas (see below), these empirical results strongly suggest
that the fraction of dense gas is larger in the nuclear discs of the
galaxies than in their inner rings. This is consistent with dense gas
being generally centrally-concentrated in galaxies, and with the more
intense star formation activity (with hotter and optically thinner CO
gas) inferred in the previous sub-sections
(\S~\ref{sec:co12co}\,--\,\ref{sec:den}) for NGC~4710 and NGC~5866
specifically.

We note that the average $^{12}$CO(1-0)/HCN(1-0),
$^{12}$CO(1-0)/HCO$^+$(1-0), $^{13}$CO(1-0)/HCN(1-0) and
$^{13}$CO(1-0)/HCO$^+$(1-0) ratios along the nuclear disc of NGC~4710
are consistent with those found by \citet{c12} for the central regions
of the galaxy, while the average $^{12}$CO(2-1)/HCN(1-0),
$^{12}$CO(2-1)/HCO$^+$(1-0), $^{13}$CO(2-1)/HCO$^+$(1-0) and
$^{13}$CO(2-1)/HCO$^+$(1-0) ratios in the nuclear disc are slightly
smaller (by $20$--$30\%$). Similarly, the average
$^{12}$CO(1-0)/HCN(1-0), $^{12}$CO(1-0)/HCO$^+$(1-0) and
$^{13}$CO(1-0)/HCO$^+$(1-0) ratios along the nuclear disc of NGC~5866
are consistent with those found by \citet{c12}, but the average
$^{13}$CO(1-0)/HCN(1-0) ratio in the nuclear disc is slightly smaller
($\approx30\%$). See Table~\ref{tab:ratiocd} and
  Table~4 in \citet{c12}.

It is generally agreed that CO(1-0) traces the total molecular gas
content of galaxies, due to its low critical density
($n_{\rm crit}\approx10^3$~cm$^{-3}$). Some of the gas traced by
CO(1-0) may therefore not be involved in star formation. However,
transitions of more complex molecules such as HCN(1-0), HCO$^+$(1-0),
HNC(1-0) and HNCO(4-3) have higher critical densities (up to
$n_{\rm crit}\approx10^6$~cm$^{-3}$) because of their larger dipole
moments. These molecules are therefore generally taken as tracers of
high-density molecular gas, more closely related to star-forming
regions than CO itself \citep[e.g.][]{gao04b,s13}. As hinted above,
the ratios of CO to these dense gas tracers therefore trace the dense
gas fraction of the ISM (larger for smaller ratios), underlying our
statement that the nuclear discs have a larger fraction of dense gas
than the inner rings.

Nevertheless, another possible reason specifically for the low
CO/HCO$^+$ ratios observed in the nuclear discs is an enhancement of
HCO$^+$ via supernova explosions (as the CRs generated can ionize
H$_2$, producing H$^+_3$ that reacts with CO to form
HCO$^+$). However, as discussed in the previous sub-section
(\S~\ref{sec:den}), the CRs ionisation rates are not likely to be as
high as what is normally inferred in starbursts (enhancing HCO$^+$
with respect to HCN and other molecules), so the role of CRs in the
ISM ionisation and gas chemistry is probably limited.

In parallel to the discussion in the previous sub-section
(\S~\ref{sec:den}), the low $^{12}$CO(1-0)/HCN(1-0) ratios observed in
the nuclear discs can further be driven by the chemical enhancement of
HCN via UV radiation from young stars. \citet{mei07} claims that at
high densities ($n_{\rm crit}>10^5$~cm$^{-3}$), PDRs produce lower
CO(1-0)/HCN(1-0) ratios than XDRs, similar to the low ratios found in
the nuclear disc of both galaxies. Our empirical
  results thus suggest that a high dense gas fraction ISM harbouring
  PDRs with relatively few CRs but strong UV radiation is the most
  likely set of physical conditions explaining the low $R_{\rm CD}$
  ratios (and $R_{\rm D1}$, $R_{\rm D2}$ and $R_{\rm D3}$ ratios
  greater than unity) observed in the nuclear disc of both galaxies.

In the Milky Way, M31 and some other spirals, the $^{12}$CO/HCN ratio
increases with radius \citep{hb97,gao04b,br05}, in turn indicating a
decrease of the dense gas fraction with radius. Our results in
NGC~4710 and NGC~5866 are similar, as the $R_{\rm CD}$ ratios in their
inner rings (with radii of $\approx2.6$ and $\approx2.4$~kpc,
respectively) are larger than those in the nuclear discs (with radii
of $\approx1$ and $\approx0.6$~kpc). Given the geometries we have
argued for, however, we are only sampling possible dense gas fraction
gradients at two discrete locations in the galaxies (nuclear disc and
inner ring radii).

\citet{hb97} also argued that the CO/HCN ratio is directly related to
the hydrostatic pressure, the ratio decreasing as the pressure
increases. Our results thus suggest that the ambient pressure is
higher in the nuclear discs than in the inner rings. This is as
expected for the central regions of galaxies, as the nuclear discs of
NGC~4710 and NGC~5866 are located well within their bulges, and is
consistent with the idea that higher pressures lead to larger dense
gas fractions. In fact, the $^{12}$CO(1-0)/HCN and
$^{12}$CO(1-0)/HCO$^+$ ratios in the nuclear discs of both galaxies
are similar (likewise for $^{13}$CO(1-0)), indicating that the dense
gas fraction and/or the ionisation mechanisms of the gas in the
nuclear discs of both galaxies are similar.

Finally, we note that if the low $R_{\rm CD}$ ratios observed in the
nuclear discs of both galaxies were the result of chemical enhancement
of HCN via X-rays from an AGN, then that could also explain the
smaller CO/HCN ratios observed in the nuclear discs compared with the
inner rings. However, while the mm continuum sources in NGC~4710 and
NGC~5866 (see \S~\ref{sec:contem}) do suggest the presence of some AGN
activity, the dense gas tracer line ratios discussed above (see
\S~\ref{sec:den}) indicate that this activity is weak and is not a
dominant driver of the physical conditions and chemistry of the
ISM. It is therefore more likely that the other mechanisms highlighted
above are dominant.

Unfortunately, as we have considered the tenuous (CO) and dense (HCN,
HCO$^+$, HNC and HNCO) gas tracers separately for our LVG modeling,
this modeling does not inform on the dense gas fraction.
%
%
\subsubsection{All ratios and  gas physical conditions}
\label{sec:allratios}
Overall, considering all line ratios and associated diagnoses, our
results seem to suggest that while the nuclear discs and inner rings
of NGC~4710 and NGC~5866 have similar molecular gas physical
conditions, the nuclear discs have a slightly larger dense gas
fraction with hotter and optically thinner molecular gas than their
inner rings. Physically, this in turn suggests that compared to the
inner rings, the nuclear discs have a more inhomogeneous ISM, with
more dense clumps bathed in a hotter and more diffuse molecular
medium. Conversely, the inner rings have a more homogeneous ISM, with
fewer clumps immersed in colder and denser molecular gas. This is
consistent with a dominant PDR-like environment with few CRs but
intense UV radiation in the nuclear discs.
%
%
\subsection{Line ratios and galaxy type/morphology}
\label{sec:compg}
\begin{figure*}
  \includegraphics[width=9.0cm,clip=]{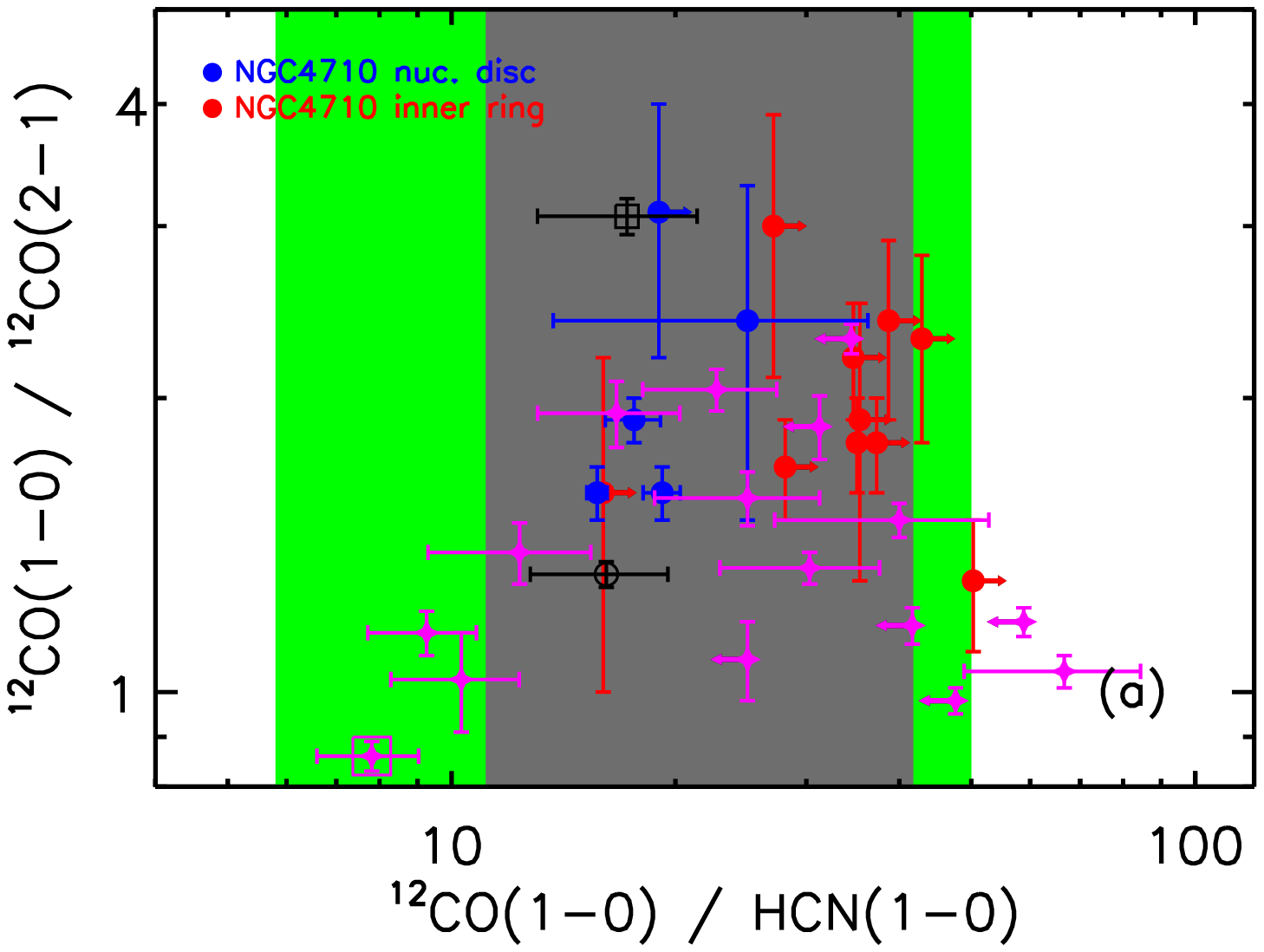}
  \hspace*{-5mm}
  \includegraphics[width=9.0cm,clip=]{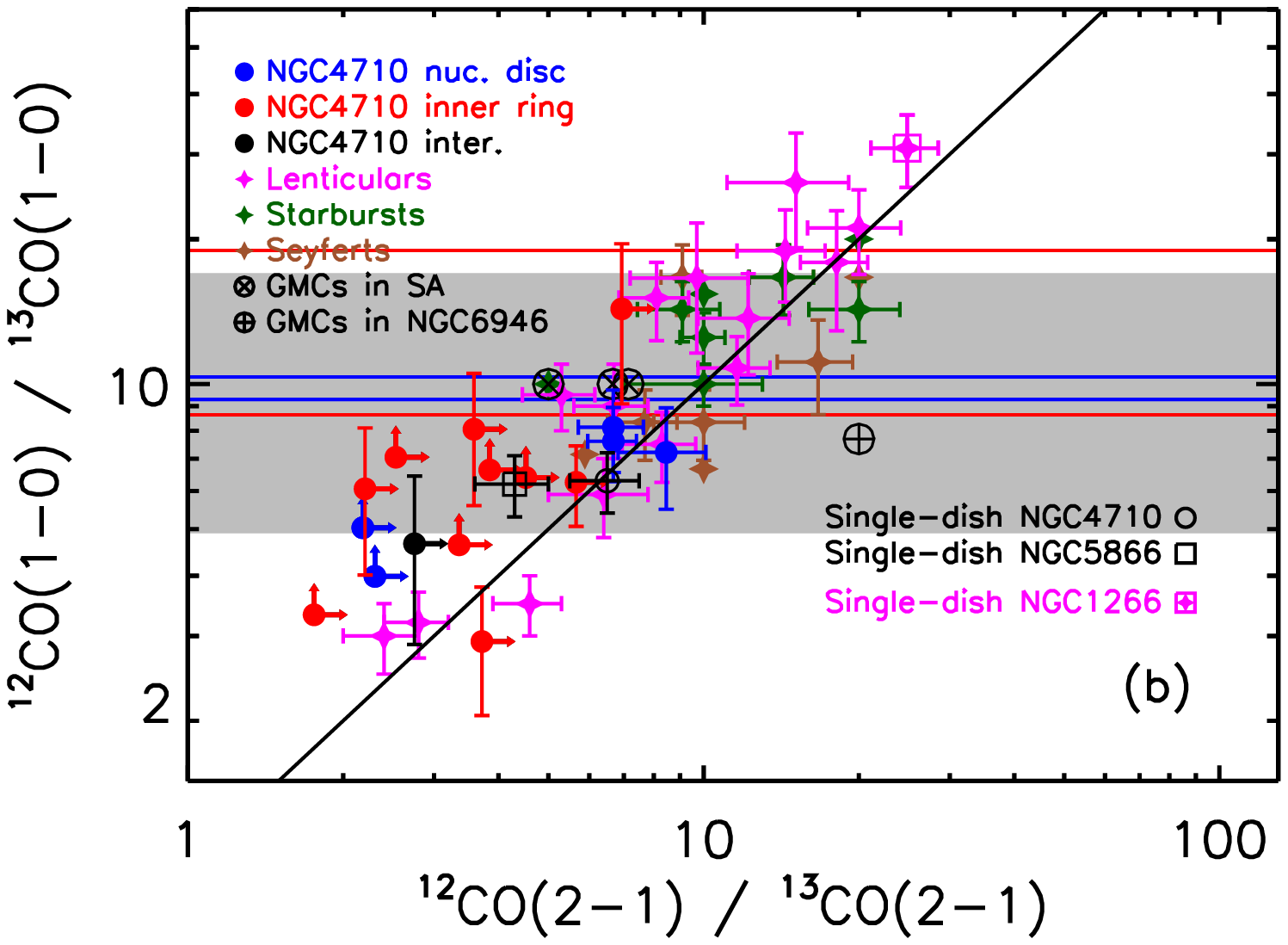}\\
  \vspace*{-5mm}
  \includegraphics[width=9.0cm,clip=]{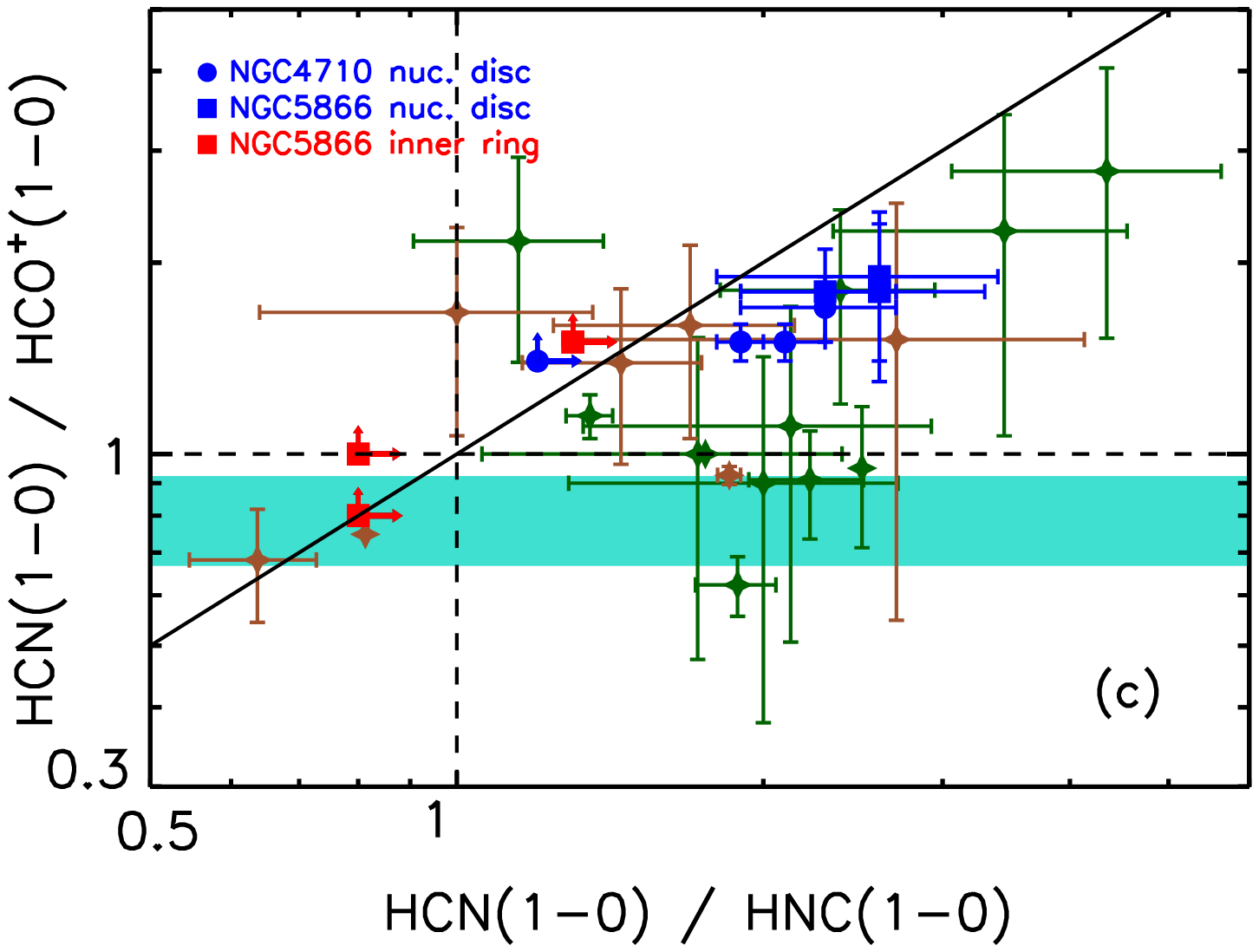}
  \hspace*{-5mm}
  \includegraphics[width=9.0cm,clip=]{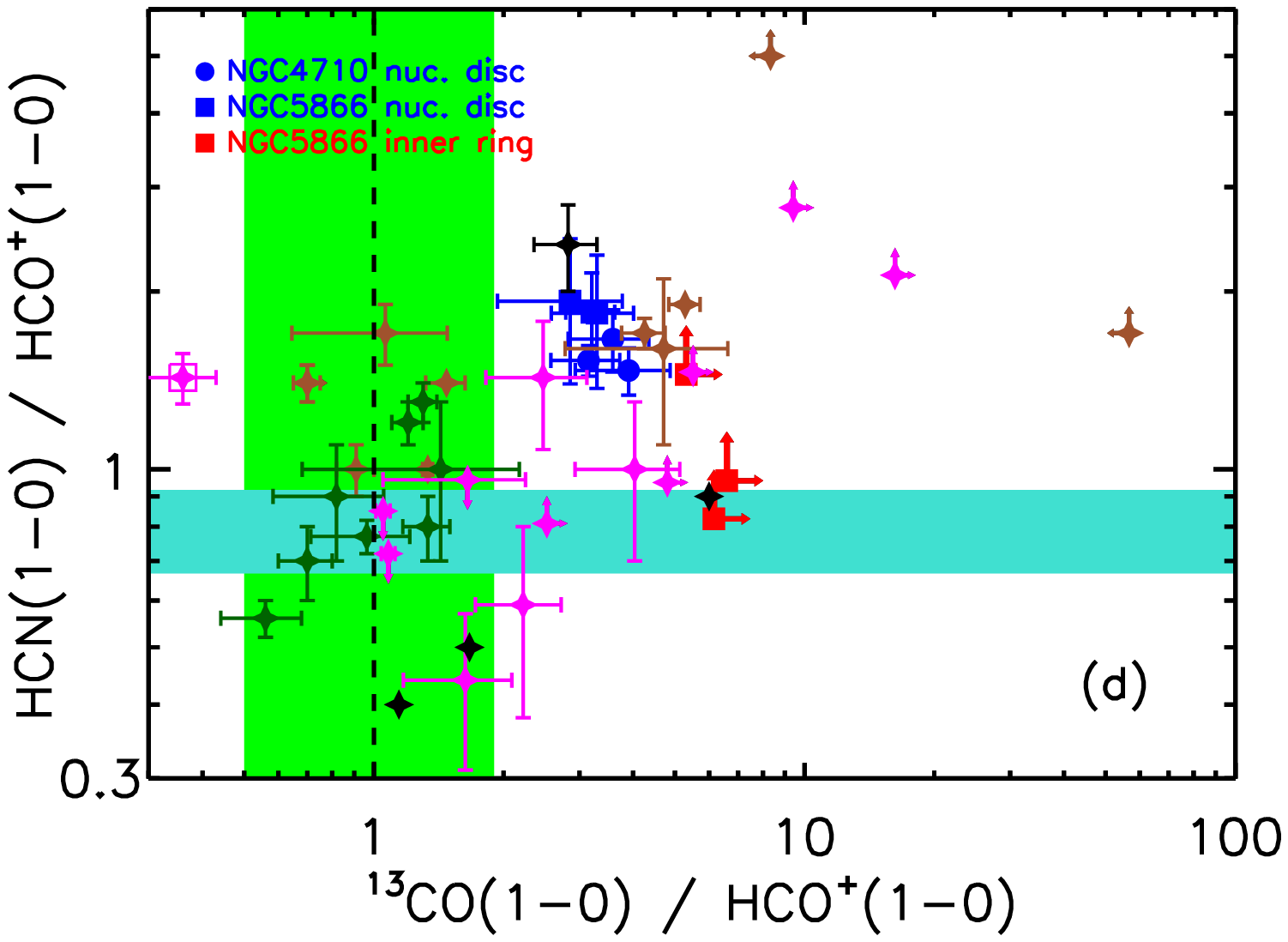}\\
  \vspace*{-5mm}
  \includegraphics[width=9.0cm,clip=]{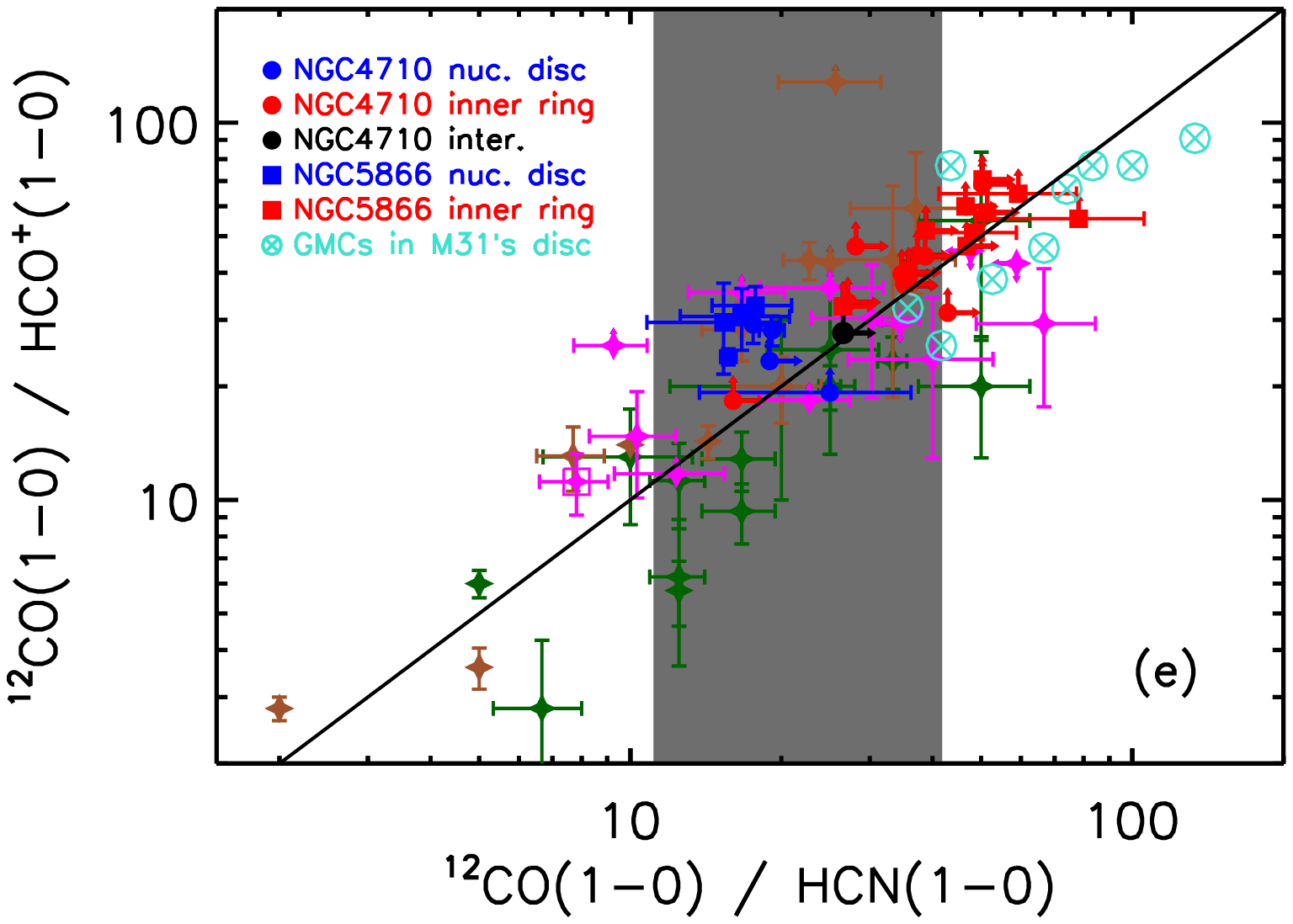}
  \hspace*{-5mm}
  \includegraphics[width=9.0cm,clip=]{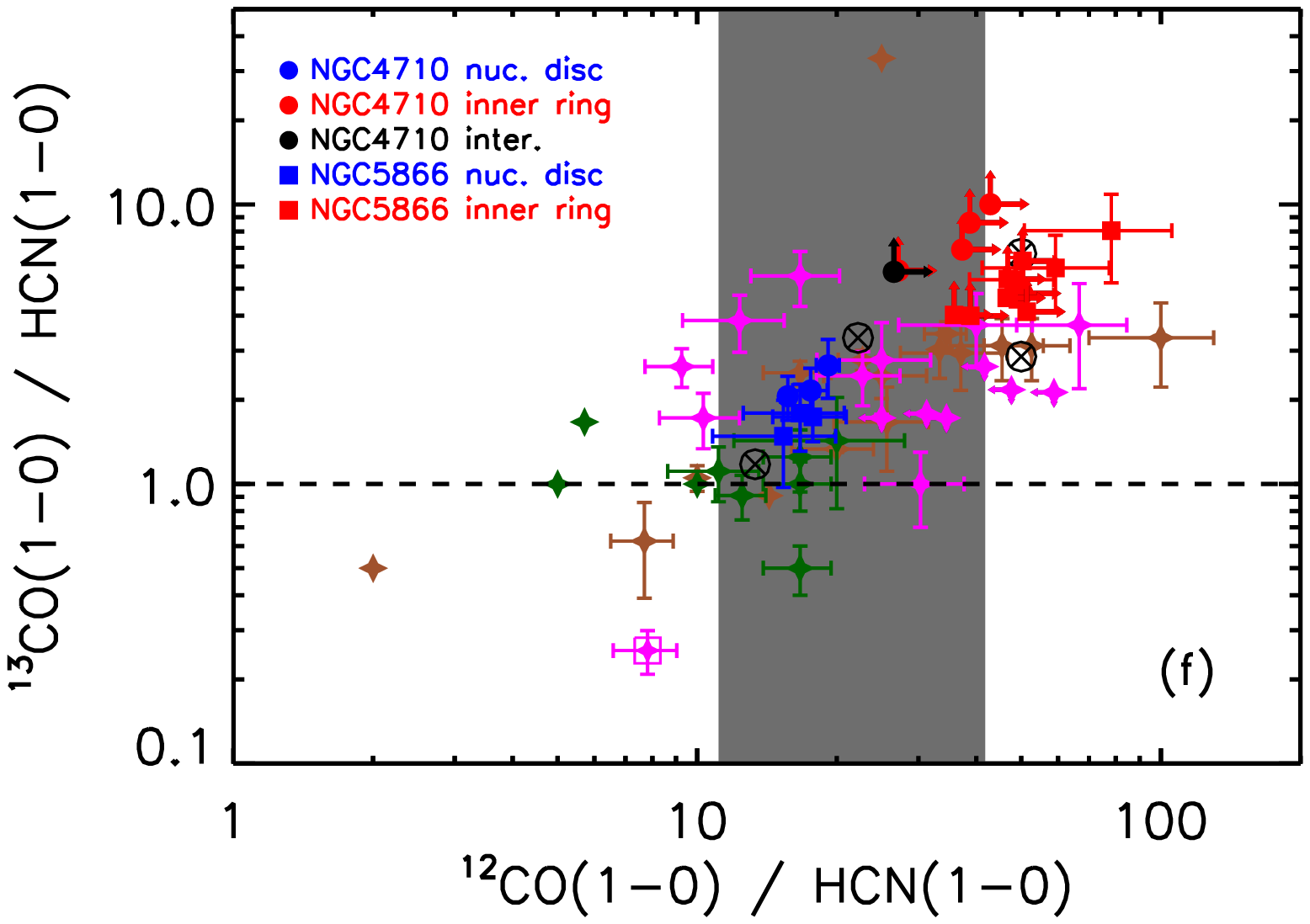}
  \caption{Molecular line ratio diagrams for NGC~4710, NGC~5866 and a
    variety of other galaxies. Our data for NGC~4710 and NGC~5866 are
    shown as filled circles and squares, respectively, while our data
    for the nuclear discs and inner rings are shown in blue and red,
    respectively (black for the intermediate region). Upper and lower
    limits are represented by arrows. Other lenticular galaxies are
    indicated by magenta filled stars \citep{ka10, c12}, starburst
    nuclei by dark green filled stars, Seyferts by brown filled stars,
    spiral arm GMCs by black circles with an X (see \citealt{baa08};
    Table~3 in \citealt{ka10} and references therein), NGC~6946
    (starburst) GMCs by black circles with a cross \citep{to14}, and
    M31 GMCs by turquoise circles with an X \citep{br05}. The data for
    NGC~1266 (a lenticular galaxy with a molecular outflow) are shown
    by magenta squares with a filled star \citep{al11}. The green
    shaded region in panel {\em a} indicates the typical range of
    $^{12}$CO(1-0)/HCN(1-0) ratios in starbursts with
      $L_{\textrm{FIR}}>10^{11}$~$L_{\odot}$ (see Table~B2 in
    \citealt{baa08}). The range of $R_{\rm 11}$ ratios in the nuclear
    disc and inner ring of NGC~5866 (this work) is indicated by
    respectively the blue and red horizontal lines in panel {\em b},
    while the typical range in spirals with
      $L_{\textrm{FIR}}<10^{11}$~$L_{\odot}$ \citep{pag01} is
    indicated by the pale grey shaded region. In panels {\em a} and
    {\em b}, \citeauthor{c12}'s (\citeyear{c12}) single-dish
    observations of NGC~4710 and NGC~5866 are shown as an open black
    circle and an open black square, respectively (see Table~4 of
    \citealt{c12}). The HCN(1-0)/HCO$^+$ ratios for M31 GMCs
    \citep{br05} are indicated by the turquoise shaded region in
    panels {\em c} and {\em d}. The green shaded region in panel {\em
      d} indicates the typical range of $^{13}$CO(1-0)/HCO$^+$(1-0)
    ratios in the disc of M82 (starburst; \citealt{tan11}). The
    $^{12}$CO(1-0)/HCN(1-0) ratios in spirals \citep{gao04a} are
    indicated by the dark grey shaded region in panels {\em a}, {\em
      e} and {\em f}, respectively. The black solid lines in a number
    of panels show the $1:1$ relation and are there to guide the
    eye. Similarly, the black dashed lines show a ratio of $1$ in
    panels {\em c}, {\em d} and {\em f}.}
  \label{fig:complit}
\end{figure*}

In Figure~\ref{fig:complit}, we compare the ratios of the integrated
line intensities obtained along the equatorial plane of NGC~4710 and
NGC~5866 with those obtained in the central regions of a variety of
galaxies (lenticulars, spirals, starbursts, Seyferts) as well as
spatially-resolved spiral galaxy GMCs. The figure caption describes
all the symbols and overlaid lines in detail.

%
%
%

In Figure~\ref{fig:complit}{\em a}, the $R_{\rm 12}$ ratio is shown as
a function of $^{12}$CO(1-0)/HCN(1-0) for the nuclear disc and inner
ring of NGC~4710 only, as $^{12}$CO(2-1) was not observed in NGC~5866.
However, the single-dish data for the centre of NGC~5866 and NGC~4710
are also shown for comparison. While the range of
$^{12}$CO(1-0)/HCN(1-0) ratios in lenticulars and particularly
starbursts (green shaded region in Figure~\ref{fig:complit}{\em
  a}) is larger than that in NGC~4710, the striking feature of this
figure is that the $R_{\rm 12}$ ratio in NGC~4710 appears much larger
than that in most lenticulars, suggesting that the molecular gas
temperature in most lenticulars is higher than that in NGC~4710.

As seen in Figure~\ref{fig:complit}{\em b}, the $R_{\rm 11}$ ratio is
larger than $R_{\rm 22}$ in a majority of sources of all kinds,
including NGC~4710 and NGC~5866. However, the range of $R_{\rm 11}$
ratios in NGC~5866 ($9.3\le R_{\rm 11}\le10.3$ for the
  nuclear disc and $8.6\le R_{\rm 11}<18.9$ for the inner ring, as
  shown by respectively the blue and red horizontal lines) is smaller
than that in lenticulars ($3\le R_{\rm 11}\le30$;
  \citealt{ka10,c12}) and spirals
($5\le R_{\rm 11}\le17$; pale grey shaded region;
  \citealt{pag01}), and it is located toward the upper end of the
distribution, similarly to starbursts. The $R_{\rm 11}$ ratios in
NGC~4710 are smaller (mostly $3\le R_{\rm 11}\le8$) and occupy the
lower end of the spiral range. This indicates that the CO gas is
optically thinner in NGC~5866 and starbursts than in NGC~4710,
supporting the idea that strong star formation feedback in starbursts
(both radiative via UV light from OB stars and mechanical via
supernova explosions) leads to diffuse gas. The difference between
NGC~4710 and NGC~5866 is however consistent with the wide range of
$R_{\rm 11}$ and $R_{\rm 22}$ ratios observed in lenticular galaxies,
analogous to the significant variations observed in their CO to dense
gas tracer ratios (Figs.~\ref{fig:complit}{\em a}, {\em d}\,--\,{\em
  f}).

As seen from Figure~\ref{fig:complit}{\em c}, the HCN(1-0)/HNC(1-0)
ratios in the nuclear discs of NGC~4710 and NGC~5866 are greater than
$1$ and similar to those in starbursts, while some Seyferts have much
smaller ratios. The behaviour is similar for the HCN(1-0)/HCO$^+$(1-0)
ratio, although many starbursts also then have smaller ratios. As
hinted by the $1:1$ line, the HCO$^+$/HNC ratio is essentially always
larger than $1$ in the nuclear discs of NGC~4710 and NGC~5866, as for
all starbursts (except one), while many Seyferts cluster around the
$1:1$ relation. The ratios may be different in the inner rings of
NGC~4710 and NGC~5866, but this is unclear as all ratios are lower
limits.

As shown in Figure~\ref{fig:complit}{\em d}, there is a clear
difference between the $^{13}$CO(1-0)/HCO$^+$ and
HCN(1-0)/HCO$^+$(1-0) ratios of NGC~4710 and NGC~5866 and those of
starbursts (including M82 shown as a green shaded
  region) in particular, while both Seyferts and other lenticulars
have broad ranges of ratios encompassing those of NGC~4710 and
NGC~5866. As discussed in \S~\ref{sec:covsden}, the
$^{13}$CO(1-0)/HCO$^+$ ratio is smaller in the nuclear discs than the
inner rings of NGC~4710 and NGC~5866, but now it also appears that
this ratio is smaller in all starbursts than in either component,
suggesting that starbursts have a larger fraction of dense molecular
gas than either the nuclear discs or inner rings of NGC~4710 and
NGC~5866 (in that order). As usual, NGC~1266 stands out, here with an
exceptionally high dense gas fraction. The HCN(1-0)/HCO$^+$(1-0)
ratios in the nuclear discs of NGC~4710 and NGC~5866 are also larger
than those in starbursts or the GMCs of M31, indicating less HCO$^+$
and thus fewer CRs from supernova explosions in NGC~4710 and
NGC~5866. Observations of both $^{13}$CO and high density tracers in
the outskirts of our galaxies and normal star-forming (i.e.\ not
star-bursting) spirals would help to establish a better comparison
sample for all galaxy types.

As can be seen from Figure~\ref{fig:complit}{\em e}, while the
$^{12}$CO(1-0)/HCN(1-0) ratios in the inner rings of NGC~4710 and
NGC~5866 are at the upper end of the range for spirals (dark grey
shaded region), and are entirely consistent with the ratios in M31's
GMCs, the $^{12}$CO(1-0)/HCN(1-0) ratios in their nuclear discs are
smaller and in the middle of the spiral range. While overlapping
slightly with the nuclear discs, the ratios for starbursts are
generally even smaller, suggesting again a sequence of dense gas
fraction (increasing from the GMCs of M31, to the inner rings and
nuclear discs of NGC~4710 and NGC~5866, to starbursts). And indeed,
the $^{12}$CO(1-0)/HCO$^+$(1-0) ratios in NGC~4710 and NGC~5866 are
larger than those in starbursts, this both for the nuclear discs and
inner rings (although more so for the latter and M31's GMCs). While
M31's GMCs and most starbursts lie below the $1:1$ lines, indicating
an HCO$^+$ enhancement, the opposite is true in both the nuclear discs
and inner rings of NGC~4710 and NGC~5866 (see also
Figs.~\ref{fig:complit}{\em c} and {\em d}). This indicates that while
the aforementioned dense gas fraction sequence may dominate both
ratios, the enhancement of HCO$^+$ via CRs is strongest in starbursts,
less so in M31's GMCs, and is weakest in NGC~5866 and NGC~4710.

The behaviour shown in Figure~\ref{fig:complit}{\em f} largely mimics
that in Figure~\ref{fig:complit}{\em e}, although being unaffected by
optical depth effects the $^{13}$CO(1-0) line is arguably a better
tracer of the total molecular gas content, thus introducing additional
scatter. NGC~1266 remains the only lenticular galaxy with
$^{13}$CO(1-0)/HCN(1-0) significantly smaller than $1$.
%
%
\section{Conclusions}
\label{sec:conc}
Interferometric observations of tenuous ($^{12}$CO(1-0),
$^{12}$CO(2-1), $^{13}$CO(1-0) and $^{13}$CO2-1)) and dense (HCN(1-0),
HCO$^+$(1-0), HNC(1-0) and HNCO(4-3)) molecular gas tracers were
presented for the edge-on lenticular galaxies NGC~4710 and
NGC~5866. Our main conclusions are:
\begin{enumerate}

\item The PVDs of the CO lines are X-shaped and reveal that the gas is
  constrained to two bar-driven kinematic components in both galaxies,
  a nuclear disc (contained within the inner Lindblad resonance) and
  an inner ring (around corotation). Although brighter in the nuclear
  discs, the tenuous molecular gas is clearly detected in both
  kinematic components and is radially more extended than the dense
  gas, the latter being generally detected in the
  nuclear discs only. However, as suggested by the HCN(1-0) detection
  in the inner ring of NGC~5866, it is likely that the inner rings
  also contain dense gas below our detection thresholds. Both
  components appear clumpy, and no molecular gas is detected beyond
  the inner rings.

\item A comparison of our interferometric data with published
  single-dish data (with much smaller primary beams) reveals that the
  latter were missing significant flux associated with the CO lines in
  the radially-extended inner rings.

\item Molecular line ratios were probed empirically by studying the
  ratios of the PVDs of CO lines only, dense gas tracer lines only,
  and CO to dense gas tracer lines, as well as by extracting the
  integrated line intensity ratios of these same lines as a function
  of projected position along the galaxy discs, this for each
  kinematic component separately. The CO(1-0)/CO(2-1) ratios are
  smaller in the nuclear discs than the inner rings, suggesting that
  the nuclear discs have higher tenuous molecular gas
  temperatures. The $^{12}$CO/$^{13}$CO ratios are slightly larger in
  the nuclear discs, suggesting that the tenuous gas there has
  slightly smaller optical depths and column densities (at the very
  least in NGC~4710). The line ratios of the dense gas tracers only
  (detected only in the nuclear discs) are all larger than $1$ and
  HCN(1-0)/HCO$^+$(1-0) $<$ HCN(1-0)/HNC(1-0) $<$ HCN(1-0)/HNCO(4-3),
  suggesting that the environment is similar to PDRs, with a chemical
  enhancement of HCN via UV radiation from young massive OB stars and
  relatively few (supernova explosion-related) CRs. The ratios of CO
  to dense gas tracers (e.g.\ $^{12}$CO/HCN(1-0) or
  $^{13}$CO/HCN(1-0)) are significantly lower in the nuclear discs
  than in the inner rings, suggesting a higher fraction of dense gas
  there, possibly linked to a higher ambient pressure. Overall,
  the picture that emerges from these empirical line
    ratio diagnostics is that of nuclear discs that have a more
  inhomogeneous ISM, with more dense clumps immersed in a hotter and
  optically thinner molecular gas medium, consistent with a more
  intense star formation activity (conversely for the inner rings).

\item LVG (RADEX) modeling was also carried out, considering a
  two-component molecular ISM traced by the CO lines only (tenuous
  component) and dense gas tracer lines only (dense component). The
  results are however inconclusive. The best-fit models within a
  single kinematic component often cover the entire range of
  parameters allowed by the models and are often driven to the edge of
  the model grid, presumably as a result of the shallowness and extent
  of the $\chi^2$ contours. The most likely model parameters have very
  large uncertainties (particularly for $T_{\rm K}$), due to the shape
  of the marginalised probability distribution functions. As such,
  they do not reveal clear differences between the inner rings and
  nuclear discs. This is exacerbated by the facts that most model
  results are upper limits in the inner ring of NGC~4710, and that we
  are unable to model the tenuous gas component in the disc of
  NGC~5866.

\item We further compared the line ratios measured in NGC~4710 and
  NGC~5866 with those obtained in other galaxy types, revealing
  interesting contrasts. The CO(1-0)/CO(2-1) ratio is larger and thus
  the tenuous molecular gas temperature lower in NGC~4710 than in most
  other lenticular galaxies. The $^{12}$CO/$^{13}$CO ratios in
  NGC~5866 are larger than those in NGC~4710 and similar to those in
  starbursts, suggesting an optically thinner tenuous component
  similar to that in starbursts. The range of ratios in other
  lenticulars and Seyferts is however much larger than that in either
  NGC~4710 or NGC~5866. While the $^{12}$CO(1-0)/HCN(1-0) ratios in
  the nuclear discs of NGC~4710 and NGC~5866 are similar to those of
  starbursts near the lower end of the range for spirals, the ratios
  in the inner rings are larger and rather similar to those observed
  in some Seyferts and lenticulars near the upper end of the spiral
  range (where M31's GMCs are found). The $^{13}$CO(1-0)/HCN(1-0)
  ratios of both the inner rings and nuclear discs are however larger
  than those of starbursts (in that order), indicating smaller dense
  gas fractions. The HCN(1-0)/HCO$^+$(1-0) ratio is greater than unity
  everywhere in NGC~4710 and NGC~5866, as in most starbursts, while it
  is smaller than unity in spatially-resolved GMCs. High
  $^{13}$CO(1-0)/HCO$^+$(1-0) ratios in NGC~4710 and NGC~5866 further
  indicate relatively low HCO$^+$ enhancement (few CRs from supernova
  explosions) compared to that seen in starbursts and GMCs. Overall,
  the molecular line ratios in the nuclear discs of NGC~4710 and
  NGC~5866 thus suggest that the physical conditions of the molecular
  gas are intermediate between those of spiral galaxies and
  starbursts, with intense but not extreme star-formation activity,
  while the inner rings host even milder star formation.
  Interestingly, the star formation efficiency of both galaxies is
  much lower than that of normal spirals, suggesting that the line
  ratios are not sensitive to whatever is suppressing star formation.
\end{enumerate}

In summary, based on empirical line ratios, star formation feedback is
likely to be stronger in the nuclear discs of NGC~4710 and NGC~5866
than in their inner rings, leading to hotter and optically thinner CO
gas with a higher fraction of dense gas clumps.  However, due to their
large uncertainties, the most likely model results are
  unable to confirm this apparent dichotomy. As the resolution of our
observations ($\approx300$~pc for the CO lines) is not enough to
resolve individual GMCs ($<80$~pc), the physical conditions estimated
either empirically or via LVG modeling are only averages over GMC
associations. Higher angular resolution observations of high-$J$ CO
lines and dense gas tracers with ALMA will ultimately resolve the
GMCs, and therefore allow us to verify and expand these statements
with much greater accuracy.
%
%
\section*{Acknowledgements}
ST was supported by the Republic of Turkey, Ministry of National
Education, and The Philip Wetton Graduate Scholarship at Christ
Church. MB was supported by the rolling grants ``Astrophysics at
Oxford'' ST/H002456/1 and ST/K00106X/1 from the UK Research Councils.
TAD acknowledges the support provided by an STFC Ernest Rutherford
Fellowship. This research also made use of the NASA/IPAC Extragalactic
Database (NED), which is operated by the Jet Propulsion Laboratory,
California Institute of Technology, under contract with the National
Aeronautics and Space Administration. The research leading to these
results has received funding from the European Community's Seventh
Framework Programme (/FP7/2007-2013/) under grant agreement No 229517.
%
%
\bibliographystyle{mn2e}
\bibliography{reference}
%
%
\appendix
%
%
\section{Molecular line ratios}
\label{sec:rcd}
Tables of molecular line ratios along the discs of NGC~4710 and
NGC~5866, for CO lines only and dense gas tracer lines only
(Table~\ref{tab:ratios}) as well as CO to dense gas tracer lines
(Tables~\ref{tab:ratiocd} and \ref{tab:ratioir}), as described in
\S~\ref{sec:extract}.

%
%
\begin{table*}
  \begin{center}
    \caption{Ratios of CO lines only and dense gas tracer lines only,
      in both the nuclear disc and inner ring of NGC~4710 and
      NGC~5866.}
    \label{tab:ratios}
    \begin{tabular}{cccccccccc}
      Galaxy & Position & Ratio & Nuclear disc & Inner ring & Galaxy &
      Position & Ratio & Nuclear disc & Inner ring \\ 
      (1) & (2) & (3) & (4) & (5) & (6) & (7) & (8) & (9) & (10) \\ 
      \hline
      NGC~4710 &$-5$&$R_{\rm 12}$& $\phantom{0}$- & $\le0.8\,\pm\,0.3$ & 
      \hspace*{10mm} NGC~5866 &$-5$&$R_{\rm 11}$& $\phantom{0}$- &$\ge18.9\,\pm\,7.8$\\
      &$(-32\farcs5)$&$R_{\rm 22}$& $\phantom{0}$- & $\ge2.4\,\pm\,1.1$ &&$(-32\farcs5)$&&&\\ \\
      &$-4$&$R_{\rm 12}$ & $\phantom{0}$- & $\phantom{00}1.8\,\pm\,0.2$&
      &$-4$&$R_{\rm 11}$ & $\phantom{0}$- &$\phantom{0}11.6\,\pm\,2.4$\\
      &  $(-26\farcs0)$     &$R_{\rm 11}$ & $\phantom{0}$- & $\phantom{00}6.3\,\pm\,1.2$&&$(-26\farcs0)$&&&\\
      &        &$R_{\rm 22}$ & $\phantom{0}$- & $\ge5.7\,\pm\,1.9$&&&&&\\ \\
      &$-3$&$R_{\rm 12}$ & $\phantom{0}$- &$\phantom{00}1.7\,\pm\,0.2$&
      &$-3$&$R_{\rm 11}$ &$\phantom{0}$-&$\phantom{0}10.3\,\pm\,1.6$ \\
      &  $(-19\farcs5)$      &$R_{\rm 11}$ &$\phantom{0}$-&$\ge4.6\,\pm\,1.6$&&$(-19\farcs5)$ &&& \\
      &       &$R_{\rm 22}$ & $\phantom{0}$- &$\ge3.4\,\pm\,1.2$&&&&&\\ \\
      &$-2$&$R_{\rm 12}$ &$\phantom{00}2.4\,\pm\,0.9$&$\phantom{00}1.3\,\pm\,0.2$&
      &$-2$&$R_{\rm 11}$ & $\phantom{0}$- &$\phantom{00}9.7\,\pm\,1.2$\\
      &   $(-13\farcs0)$     &$R_{\rm 11}$ &$\ge4.0\,\pm\,2.2$&$\phantom{0}14.3\,\pm\,5.2$&
      &$(-13\farcs0)$ &$R_{\rm D1}$& $\phantom{0}$- &$\ge1.0\,\pm\,0.5$\\
      &       &$R_{\rm 22}$ &$\ge2.3\,\pm\,1.1$&$\ge6.9\,\pm\,2.5$&
      &&$R_{\rm D2}$& $\phantom{0}$- &$\ge0.8\,\pm\,0.4$\\
      &       &$R_{\rm D1}$ &$\ge1.4\,\pm\,0.7$& $\phantom{0}$- &
      &&$R_{\rm D3}$&$\phantom{0}$-&$\ge0.8\,\pm\,0.4$\\
      &       &$R_{\rm D2}$ &$\ge1.2\,\pm\,0.6$& $\phantom{0}$- &&&&&\\
      &       &$R_{\rm D3}$&$\ge1.8\,\pm\,0.9$& $\phantom{0}$- &&&&&\\ \\
      &$-1$&$R_{\rm 12}$&$\phantom{00}1.6\,\pm\,0.1$&$\phantom{00}2.3\,\pm\,0.5$&
      &$-1$&$R_{\rm 11}$&$\phantom{00}9.3\,\pm\,1.9$&$\phantom{00}8.6\,\pm\,1.4$\\
      &  $(-6\farcs5)$     &$R_{\rm 11}$&$\phantom{00}7.2\,\pm\,1.7$&$\phantom{00}2.9\,\pm\,0.9$&
      &$(-6\farcs5)$&$R_{\rm D1}$&$\phantom{00}1.8\,\pm\,0.5$& $\phantom{0}$- \\
      &       &$R_{\rm 22}$ &$\phantom{00}8.5\,\pm\,1.6$&$\ge3.7\,\pm\,1.5$&
      &&$R_{\rm D2}$&$\phantom{00}2.6\,\pm\,0.7$& $\phantom{0}$-\\
      &       &$R_{\rm D1}$&$\phantom{00}1.5\,\pm\,0.1$& $\phantom{0}$- &
      &&$R_{\rm D3}$&$\ge2.1\,\pm\,0.9$& $\phantom{0}$-\\
      &       &$R_{\rm D2}$&$\phantom{00}2.1\,\pm\,0.2$& $\phantom{0}$- &&&&&\\
      &       &$R_{\rm D3}$&$\phantom{00}4.3\,\pm\,0.8$& $\phantom{0}$- &&&&&\\ \\
      &$0$&$R_{\rm 12}$ &$\phantom{00}1.6\,\pm\,0.1$&$\phantom{00}2.2\,\pm\,0.3$&
      &$0$&$R_{\rm 11}$&$10.2\,\pm\,1.5$&$\phantom{00}9.7\,\pm\,2.3$\\
      &  $(0^{\prime\prime})$     &$R_{\rm 11}$&$\phantom{00}7.6\,\pm\,1.3$&$\ge6.6\,\pm\,2.2$&
      &$(0^{\prime\prime})$&$R_{\rm D1}$&$\phantom{0}1.8\,\pm\,0.3$& $\phantom{0}$-\\
      &       &$R_{\rm 22}$&$\phantom{00}6.7\,\pm\,0.7$&$\ge3.8\,\pm\,1.4$&
      &&$R_{\rm D2}$&$\phantom{0}2.3\,\pm\,0.4$&$\phantom{0}$-\\
      &       &$R_{\rm D1}$&$\phantom{00}1.5\,\pm\,0.1$& $\phantom{0}$- &
      &&$R_{\rm D3}$&$\phantom{0}4.4\,\pm\,1.1$& $\phantom{0}$-\\
      &       &$R_{\rm D2}$&$\phantom{00}1.9\,\pm\,0.1$& $\phantom{0}$- &&&&&\\
      &       &$R_{\rm D3}$&$\phantom{00}4.2\,\pm\,0.3$& $\phantom{0}$- &&&&&\\ \\
      &$1$&$R_{\rm 12}$&$\phantom{00}1.9\,\pm\,0.1$&$\phantom{00}2.4\,\pm\,0.5$&
      &$1$&$R_{\rm 11}$&$\phantom{0}10.4\,\pm\,3.5$&$\phantom{00}8.9\,\pm\,3.8$\\
      &    $(+6\farcs5)$   &$R_{\rm 11}$&$\phantom{00}8.1\,\pm\,1.6$&$\phantom{00}8.1\,\pm\,2.5$&
      &$(+6\farcs5)$&$R_{\rm D1}$&$\phantom{00}1.9\,\pm\,0.5$& $\phantom{0}$-\\
      &       &$R_{\rm 22}$ &$\phantom{00}6.7\,\pm\,0.9$&$\ge3.6\,\pm\,1.4$&
      &&$R_{\rm D2}$&$\phantom{00}2.6\,\pm\,0.8$& $\phantom{0}$-\\
      &       &$R_{\rm D1}$&$\phantom{00}1.7\,\pm\,0.2$& $\phantom{0}$- &
      &&$R_{\rm D3}$&$\ge2.6\,\pm\,1.7$&$\phantom{0}$-\\
      &       &$R_{\rm D2}$&$\phantom{00}2.3\,\pm\,0.4$& $\phantom{0}$- &&&&&\\
      &       &$R_{\rm D3}$&$\phantom{00}4.3\,\pm\,0.9$& $\phantom{0}$- &&&&&\\ \\
      &$2$&$R_{\rm 12}$&$\phantom{00}3.1\,\pm\,0.9$&$\phantom{00}3.0\,\pm\,0.9$&
      &$2$&$R_{\rm 11}$& $\phantom{0}$- &$\phantom{00}9.0\,\pm\,1.0$\\
      &  $(+13\farcs0)$     &$R_{\rm 11}$&$\ge5.0\,\pm\,2.1$&$\phantom{00}6.1\,\pm\,2.1$&
      &$(+13\farcs0)$&$R_{\rm D1}$& $\phantom{0}$- &$\ge1.5\,\pm\,0.6$\\
      &       &$R_{\rm 22}$& $\ge2.2\,\pm\,1.0$&$\ge2.2\,\pm\,1.0$&
      &&$R_{\rm D2}$& $\phantom{0}$- &$\ge1.3\,\pm\,0.5$\\ 
      &       &&&&&&$R_{\rm D3}$& $\phantom{0}$- &$\ge1.3\,\pm\,0.5$\\ \\
      &$3^{\ast}$&$R_{\rm 12}$& $\phantom{0}$- &$\phantom{00}1.9\,\pm\,0.6$&
      &$3$&$R_{\rm 11}$& $\phantom{0}$- &$\phantom{0}10.0\,\pm\,1.3$\\
      &  $(+19\farcs5)$     &$R_{\rm 11}$& $\phantom{0}$- &$\ge7.1\,\pm\,2.8$&
      &$(+19\farcs5)$&$R_{\rm D1}$& $\phantom{0}$- &$\ge0.8\,\pm\,0.4$\\
      &       &$R_{\rm 22}$& $\phantom{0}$- &$\ge2.5\,\pm\,1.2$&
      &&$R_{\rm D2}$& $\phantom{0}$- &$\ge0.8\,\pm\,0.4$\\
      &       &&&&&&$R_{\rm D3}$& $\phantom{0}$- &$\ge0.8\,\pm\,0.4$\\\\
      &$4$&$R_{\rm 12}$& $\phantom{0}$- &$\phantom{00}1.8\,\pm\,0.2$&
      &$4$&$R_{\rm 11}$& $\phantom{0}$- &$\phantom{0}11.8\,\pm\,2.4$\\
      &   $(+26\farcs0)$       &$R_{\rm 11}$& $\phantom{0}$- &$\ge6.4\,\pm\,2.2$&&$(+26\farcs0)$&&&\\
      &          &$R_{\rm 22}$& $\phantom{0}$- &$\ge4.5\,\pm\,1.6$&&&&&\\ \\
      &$5$&$R_{\rm 12}$ & $\phantom{0}$- &$\phantom{00}1.6\,\pm\,0.6$&
      &$5$&$R_{\rm 11}$& $\phantom{0}$- &$\phantom{0}15.5\,\pm\,6.5$\\
      &  $(+32\farcs5)$       &$R_{\rm 11}$ & $\phantom{0}$- &$\ge3.3\,\pm\,1.5$&&$(+32\farcs5)$&&&\\
      &         &$R_{\rm 22}$ & $\phantom{0}$- &$\ge1.8\,\pm\,0.8$&&&&&\\	
      \hline
    \end{tabular}
  \end{center}
  Notes: $^{\ast}$ The ratios of CO lines only were also measured for
  the intermediate region at position~$3$ in NGC~4710 (see
  Fig.~\ref{fig:pos}). These ratios are $R_{\rm 12}=1.9\,\pm\,0.6$,
  $R_{\rm 11}=4.7\,\pm\,1.8$ and $R_{\rm
    22}\ge2.7\,\pm\,1.3$. The offset along the major
    axis (with respect to the galaxy centre) of each projected
    position is listed under the projected position number in
    column~2. The uncertainties on the line ratios are calculated
  from the uncertainties on the integrated line intensities using the
  standard error propagation formula.
\end{table*}
%

%
%
\begin{table*}
  \begin{center}
    \setlength{\tabcolsep}{1.0pt}
    \caption{Ratios of CO to dense gas tracer lines, in both the
      nuclear disc and inner ring of NGC~4710 and NGC~5866.}
    \label{tab:ratiocd}
    \begin{tabular}{cccccccccc}
      Galaxy & Position & Ratio & Nuclear disc & Inner ring & Galaxy &
      Position & Ratio & Nuclear disc & Inner ring \\ 
      (1) & (2) & (3) & (4) & (5) & (6) & (7) & (8) & (9) & (10) \\ 
      \hline
      NGC~4710 &$-5$&$^{12}$CO(2-1) / HCN(1-0) & - & $\,\,\,\ge\phantom{0}9.3\,\pm\,\,\,\,4.1$ & 
      NGC~5866 & $-5$ & $^{12}$CO(1-0) / HCN(1-0) & - &$\ge26.5\,\pm\,13.9$\\	
      &$(-32\farcs5)$&\,\,\,\,$^{12}$CO(2-1) / HCO$^+$(1-0) & - & $\,\,\,\ge10.6\,\pm\,\,\,\,4.6$ &
      &$(-32\farcs5)$&\,\,\,\,$^{12}$CO(1-0) / HCO$^+$(1-0)& - &$\ge32.6\,\pm\,17.1$\\ 
      &&$^{12}$CO(2-1) / HNC(1-0)& - & $\,\,\,\ge\phantom{0}9.9\,\pm\,\,\,\,4.3$ &
      &&$^{12}$CO(1-0) / HNC(1-0)& - &$\ge26.6\,\pm\,14.0$\\
      &&\,\,\,\,$^{12}$CO(2-1) / HNCO(4-3)& - & $\,\,\,\ge\phantom{0}7.4\,\pm\,\,\,\,3.2$ &
      &&\,\,\,\,$^{12}$CO(1-0) / HNCO(4-3)& - &$\ge29.2\,\pm\,15.3$\\\\
      &$-4$&$^{12}$CO(1-0) / HCN(1-0) & - &$\,\,\,\ge37.3\,\pm\,12.8$&
      &$-4$&$^{12}$CO(1-0) / HCN(1-0) & - &$\ge51.4\,\pm\,19.5$\\
      &$(-26\farcs0)$&\phantom{0}$^{12}$CO(1-0) / HCO$^+$(1-0) & - & $\,\,\,\ge44.5\,\pm\,15.3$&
      &$(-26\farcs0)$&\phantom{0}$^{12}$CO(1-0) / HCO$^+$(1-0)& - &$\ge57.8\,\pm\,21.9$\\
      &&$^{12}$CO(1-0) / HNC(1-0)& - &$\,\,\,\ge50.2\,\pm\,17.3$&
      &&$^{12}$CO(1-0) / HNC(1-0)& - &$\ge52.8\,\pm\,20.1$\\
      &&\phantom{0}$^{12}$CO(1-0) / HNCO(4-3) & - & $\,\,\,\ge48.8\,\pm\,16.8$&
      &&\phantom{0}$^{12}$CO(1-0) / HNCO(4-3)& - &$\ge49.9\,\pm\,18.9$\\
      &&$^{13}$CO(1-0) / HCN(1-0) & - &$\,\,\,\ge\phantom{0}6.9\,\pm\,\,\,\,2.8$&
      &&$^{13}$CO(1-0) / HCN(1-0) & - &$\ge\phantom{0}4.1\,\pm\,\,\,\,1.6$\\
      &&\phantom{0}$^{13}$CO(1-0) / HCO$^+$(1-0) & - & $\,\,\,\ge\phantom{0}8.2\,\pm\,\,\,\,3.3$&
      &&\phantom{0}$^{13}$CO(1-0) / HCO$^+$(1-0) & - &$\ge\phantom{0}4.7\,\pm\,\,\,\,1.8$\\
      & &$^{13}$CO(1-0) / HNC(1-0)& - &$\,\,\,\ge\phantom{0}9.3\,\pm\,\,\,\,3.7$&
      &&$^{13}$CO(1-0) / HNC(1-0) & - &$\ge\phantom{0}4.3\,\pm\,\,\,\,1.6$\\
      &&\phantom{0}$^{13}$CO(1-0) / HNCO(4-3) & - & $\,\,\,\ge\phantom{0}9.0\,\pm\,\,\,\,3.6$&
      &&\phantom{0}$^{13}$CO(1-0) / HNCO(4-3) & - &$\ge\phantom{0}4.0\,\pm\,\,\,\,1.5$\\
      &&$^{12}$CO(2-1) / HCN(1-0) & - &$\,\,\,\ge19.8\,\pm\,\,\,\,6.8$&
      &&&&\\
      &&\phantom{0}$^{12}$CO(2-1) / HCO$^+$(1-0) & - &$\,\,\,\ge23.7\,\pm\,\,\,\,8.1$&&&&&\\
      &&$^{12}$CO(2-1) / HNC(1-0)& - &$\,\,\,\ge26.7\,\pm\,\,\,\,9.2$&&&&&\\ 
      &&\phantom{0}$^{12}$CO(2-1) / HNCO(4-3)& - & $\,\,\,\ge26.0\,\pm\,\,\,\,8.9$&&&&&\\ \\
      &$-3$&$^{12}$CO(1-0) / HCN(1-0) & - &$\,\,\,\ge28.1\,\pm\,\,\,\,9.7$&
      &$-3$&$^{12}$CO(1-0) / HCN(1-0) & - &$\ge46.5\,\pm\,16.8$\\
      &$(-19\farcs5)$&\phantom{0}$^{12}$CO(1-0) / HCO$^+$(1-0) & - & $\,\,\,\ge47.0\,\pm\,16.2$&
      &$(-19\farcs5)$&\phantom{0}$^{12}$CO(1-0) / HCO$^+$(1-0)& - &$\ge60.0\,\pm\,21.7$\\
      &&$^{12}$CO(1-0) / HNC(1-0) & - & $\,\,\,\ge30.7\,\pm\,10.5$&
      &&$^{12}$CO(1-0) / HNC(1-0)& - &$\ge52.8\,\pm\,19.1$\\
      &&\phantom{0}$^{12}$CO(1-0) / HNCO(4-3) & - & $\,\,\,\ge46.0\,\pm\,15.8$&
      &&\phantom{0}$^{12}$CO(1-0) / HNCO(4-3)& - &$\ge50.8\,\pm\,18.3$\\
      &&$^{12}$CO(2-1) / HCN(1-0) & - &$\,\,\,\ge14.3\,\pm\,\,\,\,5.0$&
      &&$^{13}$CO(1-0) / HCN(1-0) & - &$\ge\,\,\,4.6\,\pm\,\,\,\,1.7$\\
      &&\phantom{0}$^{12}$CO(2-1) / HCO$^+$(1-0)& - &$\,\,\,\ge23.8\,\pm\,\,\,\,8.4$&
      &&\phantom{0}$^{13}$CO(1-0) / HCO$^+$(1-0) & - &$\ge\,\,\,6.0\,\pm\,\,\,\,2.1$\\
      &&$^{12}$CO(2-1) / HNC(1-0)& - &$\,\,\,\ge15.6\,\pm\,\,\,\,5.5$&
      &&$^{13}$CO(1-0) / HNC(1-0) & - &$\ge\,\,\,5.3\,\pm\,\,\,\,1.9$\\
      &&\phantom{0}$^{12}$CO(2-1) / HNCO(4-3)& - &$\,\,\,\ge23.3\,\pm\,\,\,\,8.2$&
      &&\phantom{0}$^{13}$CO(1-0) / HNCO(4-3) & - &$\ge\,\,\,5.1\,\pm\,\,\,\,1.8$\\ \\	
      &$-2$&$^{12}$CO(1-0) / HCN(1-0) &$\phantom{000}25.0\,\pm\,11.3$&$\,\,\,\ge50.3\,\pm\,18.7$&
      &$-2$&$^{12}$CO(1-0) / HCN(1-0) & - &$\phantom{00}78.2\,\pm\,27.4$\\
      &$(-13\farcs0)$&\phantom{0}$^{12}$CO(1-0) / HCO$^+$(1-0) &$\,\,\,\ge19.3\,\pm\,10.8$&$\,\,\,\ge68.6\,\pm\,25.6$&
      &$(-13\farcs0)$&\phantom{0}$^{12}$CO(1-0) / HCO$^+$(1-0)& - &$\ge55.6\,\pm\,19.6$\\
      &&$^{12}$CO(1-0) / HNC(1-0) &$\,\,\,\ge13.6\,\pm\,\,\,\,7.6$&$\,\,\,\ge48.3\,\pm\,18.0$&
      &&$^{12}$CO(1-0) / HNC(1-0)& - &$\ge44.6\,\pm\,15.7$\\
      &&\phantom{0}$^{12}$CO(1-0) / HNCO(4-3) &$\,\,\,\ge25.5\,\pm\,14.3$&$\,\,\,\ge91.0\,\pm\,33.9$&
      &&\phantom{0}$^{12}$CO(1-0) / HNCO(4-3)& - &$\ge45.6\,\pm\,16.0$\\
      &&$^{13}$CO(1-0) / HCN(1-0) & - &$\,\,\,\ge\phantom{0}4.8\,\pm\,\,\,\,2.6$&
      &&$^{13}$CO(1-0) / HCN(1-0) & - &$\,\,\,\phantom{00}8.1\,\pm\,\,\,\,2.8$\\
      &&\phantom{0}$^{13}$CO(1-0) / HCO$^+$(1-0) & - & $\,\,\,\ge\phantom{0}6.6\,\pm\,\,\,\,3.6$&
      &&\phantom{0}$^{13}$CO(1-0) / HCO$^+$(1-0) & - &$\ge\,\,\,6.6\,\pm\,\,\,\,2.3$\\
      &&$^{13}$CO(1-0) / HNC(1-0) & - & $\,\,\,\ge\phantom{0}4.6\,\pm\,\,\,\,2.5$&
      &&$^{13}$CO(1-0) / HNC(1-0) & - &$\ge\,\,\,5.3\,\pm\,\,\,\,1.9$\\
      &&\phantom{0}$^{13}$CO(1-0) / HNCO(4-3) & - & $\,\,\,\ge\phantom{0}8.7\,\pm\,\,\,\,4.7$&
      &&\phantom{0}$^{13}$CO(1-0) / HNCO(4-3) & - &$\ge\,\,\,5.4\,\pm\,\,\,\,1.9$\\
      &&$^{12}$CO(2-1) / HCN(1-0) &$\,\,\,\phantom{00}10.6\,\pm\,\,\,\,4.3$&$\,\,\,\ge27.3\,\pm\,\,\,\,9.7$&&&&&\\
      &&\phantom{0}$^{12}$CO(2-1) / HCO$^+$(1-0) &$\,\,\,\ge12.4\,\pm\,\,\,\,6.0$&$\,\,\,\ge37.3\,\pm\,13.2$&&&&&\\
      &&$^{12}$CO(2-1) / HNC(1-0) &$\,\,\,\ge\phantom{0}8.7\,\pm\,\,\,\,4.2$&$\,\,\,\ge26.3\,\pm\,\,\,\,9.3$&&&&&\\
      &&\phantom{0}$^{12}$CO(2-1) / HNCO(4-3)&$\,\,\,\ge16.4\,\pm\,\,\,\,8.0$&$\,\,\,\ge49.4\,\pm\,17.5$&&&&&\\
      &&$^{13}$CO(2-1) / HCN(1-0) &$\,\,\,\le\phantom{0}1.3\,\pm\,\,\,\,1.0$& - &&&&&\\ \\
      &$-1$&$^{12}$CO(1-0) / HCN(1-0)&$\,\,\,\phantom{00}19.2\,\pm\,\,\,\,1.1$&$\,\,\,\ge42.9\,\pm\,16.3$&
      &$-1$&$^{12}$CO(1-0) / HCN(1-0)&$\,\,\,\phantom{00}16.7\,\pm\,\,\,\,4.1$&$\ge50.3\,\pm\,18.4$\\
      &$(-6\farcs5)$&\phantom{0}$^{12}$CO(1-0) / HCO$^+$(1-0)&$\,\,\,\phantom{00}28.2\,\pm\,\,\,\,2.6$&$\,\,\,\ge31.4\,\pm\,11.9$&
      &$(-6\farcs5)$&\phantom{0}$^{12}$CO(1-0) / HCO$^+$(1-0)&$\,\,\,\phantom{00}30.6\,\pm\,\,\,\,5.7$&$\ge70.7\,\pm\,25.9$\\
      &&$^{12}$CO(1-0) / HNC(1-0)&$\,\,\,\phantom{00}40.5\,\pm\,\,\,\,4.3$&$\,\,\,\ge37.6\,\pm\,14.3$&
      &&$^{12}$CO(1-0) / HNC(1-0)&$\phantom{000}42.7\,\pm\,10.0$&$\ge67.7\,\pm\,24.8$\\
      &&\phantom{0}$^{12}$CO(1-0) / HNCO(4-3)&$\,\,\,\phantom{00}81.7\,\pm\,14.6$&$\,\,\,\ge42.8\,\pm\,16.2$&
      &&\phantom{0}$^{12}$CO(1-0) / HNCO(4-3)&$\,\,\,\ge45.0\,\pm\,\,\,\,7.4$&$\ge50.7\,\pm\,18.6$\\
      &&$^{13}$CO(1-0) / HCN(1-0)&$\,\,\,\phantom{000}2.7\,\pm\,\,\,\,0.6$&$\,\,\,\ge10.0\,\pm\,\,\,\,4.7$&
      &&$^{13}$CO(1-0) / HCN(1-0) &$\,\,\,\phantom{000}1.8\,\pm\,\,\,\,0.5$&$\ge\,\,\,6.3\,\pm\,\,\,\,2.3$\\
      &&\phantom{0}$^{13}$CO(1-0) / HCO$^+$(1-0)&$\,\,\,\phantom{000}3.9\,\pm\,\,\,\,1.0$&$\,\,\,\ge11.8\,\pm\,\,\,\,5.5$&
      &&\phantom{0}$^{13}$CO(1-0) / HCO$^+$(1-0)&$\,\,\,\phantom{000}3.3\,\pm\,\,\,\,0.7$&$\ge\,\,\,8.8\,\pm\,\,\,\,3.2$\\
      &&$^{13}$CO(1-0) / HNC(1-0)&$\,\,\,\phantom{000}5.6\,\pm\,\,\,\,1.4$&$\,\,\,\ge11.9\,\pm\,\,\,\,5.5$&
      &&$^{13}$CO(1-0) / HNC(1-0)&$\,\,\,\phantom{000}4.6\,\pm\,\,\,\,1.2$&$\ge\,\,\,8.4\,\pm\,\,\,\,3.1$\\
      &&\phantom{0}$^{13}$CO(1-0) / HNCO(4-3)&$\,\,\,\phantom{00}11.3\,\pm\,\,\,\,3.3$&$\,\,\,\ge12.4\,\pm\,\,\,\,5.8$&
      &&\phantom{0}$^{13}$CO(1-0) / HNCO(4-3)&$\,\,\,\ge\phantom{0}3.3\,\pm\,\,\,\,0.7$&$\ge\,\,\,6.3\,\pm\,\,\,\,2.3$\\
      &&$^{12}$CO(2-1) / HCN(1-0)&$\,\,\,\phantom{00}12.1\,\pm\,\,\,\,0.8$&$\,\,\,\ge14.9\,\pm\,16.1$&
      &&&&\\
      &&\phantom{0}$^{12}$CO(2-1) / HCO$^+$(1-0) &$\,\,\,\phantom{00}17.8\,\pm\,\,\,\,1.7$&$\,\,\,\ge18.8\,\pm\,\,\,\,7.7$&
      &&&&\\
      &&$^{12}$CO(2-1) / HNC(1-0)&$\,\,\,\phantom{00}25.6\,\pm\,\,\,\,2.8$&$\,\,\,\ge22.5\,\pm\,\,\,\,9.2$&
      &&&&\\
      &&\phantom{0}$^{12}$CO(2-1) / HNCO(4-3)&$\,\,\,\phantom{00}51.6\,\pm\,\,\,\,9.3$&$\,\,\,\ge25.6\,\pm\,10.5$&
      &&&&\\
      &&$^{13}$CO(2-1) / HCN(1-0)&$\,\,\,\phantom{000}1.4\,\pm\,\,\,\,0.3$& - &
      &&&&\\ 
      &&\phantom{0}$^{13}$CO(2-1) / HCO$^+$(1-0)&$\,\,\,\phantom{000}2.1\,\pm\,\,\,\,0.4$& - &
      &&&&\\
      &&$^{13}$CO(2-1) / HNC(1-0)&$\,\,\,\phantom{000}3.0\,\pm\,\,\,\,0.6$& - &
      &&&&\\
      &&\phantom{0}$^{13}$CO(2-1) / HNCO(4-3)&$\,\,\,\phantom{000}6.1\,\pm\,\,\,\,1.6$& - &
      &&&&\\
      \hline
    \end{tabular}
  \end{center}
  Notes: The offset along the major axis (with respect
    to the galaxy centre) of each projected position is listed under
    the projected position number in column~2. The uncertainties on
  the line ratios are calculated from the uncertainties on the
  integrated line intensities using the standard error propagation
  formula.
\end{table*}
%
%
\addtocounter{table}{-1}
\begin{table*}
  \begin{center}
    \setlength{\tabcolsep}{1pt}
    \caption{\em Continued.}
    \begin{tabular}{cccccccccc}
      Galaxy & Position & Ratio & Nuclear disc & Inner ring & Galaxy &
      Position & Ratio & Nuclear disc & Inner ring \\ 
      (1) & (2) & (3) & (4) & (5) & (6) & (7) & (8) & (9) & (10) \\ 
      \hline
      NGC~4710&$0$&$^{12}$CO(1-0) / HCN(1-0) &$15.7\,\pm\,0.5$&$\ge34.7\,\pm\,11.6$&
      NGC~5866&$0$&$^{12}$CO(1-0) / HCN(1-0)&$\,\,\,17.8\,\pm\,\,\,\,3.2$&$\,\,\,\ge38.8\,\pm\,15.3$\\
      &$(0^{\prime\prime})$&\phantom{0}$^{12}$CO(1-0) / HCO$^+$(1-0)&$24.0\,\pm\,1.2$&$\ge39.7\,\pm\,13.3$&
      &$(0^{\prime\prime})$&\phantom{0}$^{12}$CO(1-0) / HCO$^+$(1-0)&$\,\,\,32.7\,\pm\,\,\,\,4.0$&$\,\,\,\ge51.8\,\pm\,20.4$\\
      &&$^{12}$CO(1-0) / HNC(1-0)&$29.5\,\pm\,1.9$&$\ge39.3\,\pm\,13.2$&
      &&$^{12}$CO(1-0) / HNC(1-0)&$\,\,\,40.2\,\pm\,\,\,\,6.1$&$\,\,\,\ge39.4\,\pm\,15.5$\\
      &&\phantom{0}$^{12}$CO(1-0) / HNCO(4-3)&$66.0\,\pm\,4.9$&$\ge45.0\,\pm\,15.1$&
      &&\phantom{0}$^{12}$CO(1-0) / HNCO(4-3)&$\,\,\,78.2\,\pm\,17.9$&$\,\,\,\ge42.7\,\pm\,16.8$\\
      &&$^{13}$CO(1-0) / HCN(1-0)&$\phantom{0}2.1\,\pm\,0.4$& - &
      &&$^{13}$CO(1-0) / HCN(1-0) &$\,\,\,\phantom{0}1.7\,\pm\,\,\,\,0.3$&$\,\,\,\ge\phantom{0}4.0\,\pm\,\,\,\,1.6$\\
      &&\phantom{0}$^{13}$CO(1-0) / HCO$^+$(1-0)&$\phantom{0}3.2\,\pm\,0.6$& - &
      &&\phantom{0}$^{13}$CO(1-0) / HCO$^+$(1-0)&$\,\,\,\phantom{0}3.2\,\pm\,\,\,\,0.4$&$\,\,\,\ge\phantom{0}5.3\,\pm\,\,\,\,2.1$\\
      &&$^{13}$CO(1-0) / HNC(1-0)&$\phantom{0}3.9\,\pm\,0.7$& - &
      &&$^{13}$CO(1-0) / HNC(1-0)&$\,\,\,\phantom{0}3.9\,\pm\,\,\,\,0.6$&$\,\,\,\ge\phantom{0}4.1\,\pm\,\,\,\,1.6$\\
      &&\phantom{0}$^{13}$CO(1-0) / HNCO(4-3)&$\phantom{0}8.7\,\pm\,1.7$& - &
      &&\phantom{0}$^{13}$CO(1-0) / HNCO(4-3)&$\,\,\,\phantom{0}7.7\,\pm\,\,\,\,1.8$&$\,\,\,\ge\phantom{0}4.4\,\pm\,\,\,\,1.7$\\
      &&$^{12}$CO(2-1) / HCN(1-0) &$\phantom{0}9.6\,\pm\,0.4$&$\,\,\,\ge14.9\,\pm\,\,\,\,5.5$&&&&&\\
      &&\phantom{0}$^{12}$CO(2-1) / HCO$^+$(1-0) &$14.6\,\pm\,0.8$&$\,\,\,\ge17.1\,\pm\,\,\,\,6.3$&&&&&\\
      &&$^{12}$CO(2-1) / HNC(1-0)&$17.9\,\pm\,1.3$&$\,\,\,\ge17.0\,\pm\,\,\,\,6.3$&&&&&\\
      &&\phantom{0}$^{12}$CO(2-1) / HNCO(4-3)&$40.2\,\pm\,3.2$& $\,\,\,\ge18.4\,\pm\,\,\,\,6.8$&&&&&\\
      &&$^{13}$CO(2-1) / HCN(1-0) &$\phantom{0}1.4\,\pm\,0.2$& - &&&&&\\ 
      &&\phantom{0}$^{13}$CO(2-1) / HCO$^+$(1-0)&$\phantom{0}2.2\,\pm\,0.3$& - &&&&&\\
      &&$^{13}$CO(2-1) / HNC(1-0)&$\phantom{0}2.7\,\pm\,0.3$& - &&&&&\\
      &&\phantom{0}$^{13}$CO(2-1) / HNCO(4-3)&$\phantom{0}6.0\,\pm\,0.8$& - &&&&&\\ \\
      &$1$&$^{12}$CO(1-0) / HCN(1-0)&$\,\,\,17.6\,\pm\,\,\,\,1.5$&$\,\,\,\ge38.7\,\pm\,13.8$&
      &$1$&$^{12}$CO(1-0) / HCN(1-0)&$\,\,\,\phantom{0}15.4\,\pm\,\,\,\,4.6$&$\,\,\,\ge35.8\,\pm\,16.5$\\
      &$(+6\farcs5)$&\phantom{0}$^{12}$CO(1-0) / HCO$^+$(1-0)&$\,\,\,29.2\,\pm\,\,\,\,3.2$&$\,\,\,\ge44.2\,\pm\,15.7$&
      &$(+6\farcs5)$&\phantom{0}$^{12}$CO(1-0) / HCO$^+$(1-0)&$\,\,\,\phantom{0}29.5\,\pm\,\,\,\,8.0$&$\,\,\,\ge40.1\,\pm\,18.4$\\
      &&$^{12}$CO(1-0) / HNC(1-0)&$\,\,\,40.6\,\pm\,\,\,\,6.4$&$\,\,\,\ge44.6\,\pm\,15.9$&
      &&$^{12}$CO(1-0) / HNC(1-0)&$\phantom{00}40.0\,\pm\,11.8$&$\,\,\,\ge44.4\,\pm\,20.4$\\
      &&\phantom{0}$^{12}$CO(1-0) / HNCO(4-3)&$\,\,\,76.0\,\pm\,16.5$&$\,\,\,\ge52.6\,\pm\,18.7$&
      &&\phantom{0}$^{12}$CO(1-0) / HNCO(4-3)&$\ge45.1\,\pm\,12.6$&$\,\,\,\ge45.0\,\pm\,20.7$\\
      &&$^{13}$CO(1-0) / HCN(1-0)&$\,\,\,\phantom{0}2.2\,\pm\,\,\,\,0.4$&$\,\,\,\ge\phantom{0}8.6\,\pm\,\,\,\,4.3$&
      &&$^{13}$CO(1-0) / HCN(1-0) &$\,\,\,\phantom{00}1.5\,\pm\,\,\,\,0.5$&$\,\,\,\ge\phantom{0}4.0\,\pm\,\,\,\,2.0$\\
      &&\phantom{0}$^{13}$CO(1-0) / HCO$^+$(1-0)&$\,\,\,\phantom{0}3.6\,\pm\,\,\,\,0.8$&$\,\,\,\ge\phantom{0}9.8\,\pm\,\,\,\,4.9$&
      &&\phantom{0}$^{13}$CO(1-0) / HCO$^+$(1-0)&$\,\,\,\phantom{00}2.9\,\pm\,\,\,\,0.9$&$\,\,\,\ge\phantom{0}4.5\,\pm\,\,\,\,2.2$\\
      &&$^{13}$CO(1-0) / HNC(1-0)&$\,\,\,\phantom{0}5.0\,\pm\,\,\,\,1.2$&$\,\,\,\ge\phantom{0}9.9\,\pm\,\,\,\,4.9$&
      &&$^{13}$CO(1-0) / HNC(1-0)&$\,\,\,\phantom{00}3.9\,\pm\,\,\,\,1.3$&$\,\,\,\ge\phantom{0}5.0\,\pm\,\,\,\,2.4$\\
      &&\phantom{0}$^{13}$CO(1-0) / HNCO(4-3)&$\,\,\,\phantom{0}9.3\,\pm\,\,\,\,2.6$&$\,\,\,\ge11.7\,\pm\,\,\,\,5.8$&
      &&\phantom{0}$^{13}$CO(1-0) / HNCO(4-3)&$\,\,\,\ge3.8\,\pm\,\,\,\,1.4$&$\,\,\,\ge\phantom{0}5.1\,\pm\,\,\,\,2.5$\\
      &&$^{12}$CO(2-1) / HCN(1-0)&$\,\,\,\phantom{0}9.1\,\pm\,\,\,\,0.8$&$\,\,\,\ge14.9\,\pm\,\,\,\,6.0$&&&&&\\
      &&\phantom{0}$^{12}$CO(2-1) / HCO$^+$(1-0) &$\,\,\,15.1\,\pm\,\,\,\,1.7$&$\,\,\,\ge17.1\,\pm\,\,\,\,6.8$&&&&&\\
      &&$^{12}$CO(2-1) / HNC(1-0) &$\,\,\,21.1\,\pm\,\,\,\,3.4$&$\,\,\,\ge17.2\,\pm\,\,\,\,6.9$&&&&&\\
      &&\phantom{0}$^{12}$CO(2-1) / HNCO(4-3)&$\,\,\,39.4\,\pm\,\,\,\,8.6$&$\,\,\,\ge20.4\,\pm\,\,\,\,8.1$&&&&&\\
      &&$^{13}$CO(2-1) / HCN(1-0)&$\,\,\,\phantom{0}1.4\,\pm\,\,\,\,0.2$& - &&&&&\\ 
      &&\phantom{0}$^{13}$CO(2-1) / HCO$^+$(1-0)&$\,\,\,\phantom{0}2.3\,\pm\,\,\,\,0.4$& - &&&&&\\
      &&$^{13}$CO(2-1) / HNC(1-0)&$\,\,\,\phantom{0}3.2\,\pm\,\,\,\,0.6$& - &&&&&\\
      &&\phantom{0}$^{13}$CO(2-1) / HNCO(4-3)&$\,\,\,\phantom{0}5.9\,\pm\,\,\,\,1.5$& - &&&&&\\ \\
      &$2$&$^{12}$CO(1-0) / HCN(1-0)&$\,\,\,\ge19.0\,\pm\,\,\,\,7.7$&$\,\,\,\ge27.1\,\pm\,10.8$&
      &$2$&$^{12}$CO(1-0) / HCN(1-0)& - &$\phantom{00}48.7\,\pm\,10.1$\\
      &$(+13\farcs0)$&\phantom{0}$^{12}$CO(1-0) / HCO$^+$(1-0) &$\,\,\,\ge23.4\,\pm\,\,\,\,9.6$&$\,\,\,\ge33.4\,\pm\,13.3$&
      &$(+13\farcs0)$&\phantom{0}$^{12}$CO(1-0) / HCO$^+$(1-0)& - &$\ge51.1\,\pm\,17.9$\\
      &&$^{12}$CO(1-0) / HNC(1-0) &$\,\,\,\ge25.1\,\pm\,10.3$&$\,\,\,\ge36.0\,\pm\,14.3$&
      &&$^{12}$CO(1-0) / HNC(1-0)& - &$\ge46.9\,\pm\,16.4$\\
      &&\phantom{0}$^{12}$CO(1-0) / HNCO(4-3) &$\,\,\,\ge22.9\,\pm\,\,\,\,9.4$&$\,\,\,\ge32.7\,\pm\,13.0$&
      &&\phantom{0}$^{12}$CO(1-0) / HNCO(4-3) & - &$\ge44.6\,\pm\,15.6$\\
      &&$^{13}$CO(1-0) / HCN(1-0) & - &$\,\,\,\ge\phantom{0}5.8\,\pm\,\,\,\,2.9$&
      &&$^{13}$CO(1-0) / HCN(1-0) &- &$\,\,\,\phantom{00}5.4\,\pm\,\,\,\,1.1$\\
      &&\phantom{0}$^{13}$CO(1-0) / HCO$^+$(1-0) & - &$\,\,\,\ge\phantom{0}7.2\,\pm\,\,\,\,3.6$&
      &&\phantom{0}$^{13}$CO(1-0) / HCO$^+$(1-0) & - &$\ge\,\,\,5.3\,\pm\,\,\,\,1.8$\\
      &&$^{13}$CO(1-0) / HNC(1-0) & - &$\,\,\,\ge\phantom{0}7.7\,\pm\,\,\,\,3.8$&
      &&$^{13}$CO(1-0) / HNC(1-0)& - &$\ge\,\,\,4.9\,\pm\,\,\,\,1.7$\\
      &&\phantom{0}$^{13}$CO(1-0) / HNCO(4-3) & - &$\,\,\,\ge\phantom{0}7.0\,\pm\,\,\,\,3.5$&
      &&\phantom{0}$^{13}$CO(1-0) / HNCO(4-3)& - &$\ge\,\,\,4.6\,\pm\,\,\,\,1.6$\\
      &&$^{12}$CO(2-1) / HCN(1-0)&$\,\,\,\ge\phantom{0}9.4\,\pm\,\,\,\,4.1$&$\,\,\,\ge\phantom{0}9.5\,\pm\,\,\,\,4.4$&&&&&\\
      &&\phantom{0}$^{12}$CO(2-1) / HCO$^+$(1-0) &$\,\,\,\ge11.6\,\pm\,\,\,\,5.1$&$\,\,\,\ge11.7\,\pm\,\,\,\,5.4$&&&&&\\
      &&$^{12}$CO(2-1) / HNC(1-0)& $\,\,\,\ge12.4\,\pm\,\,\,\,5.5$&$\,\,\,\ge12.6\,\pm\,\,\,\,5.8$&&&&&\\ 
      &&\phantom{0}$^{12}$CO(2-1) / HNCO(4-3)&$\,\,\,\ge11.3\,\pm\,\,\,\,5.0$&$\,\,\,\ge11.4\,\pm\,\,\,\,5.3$&&&&&\\ \\
      &$3$&$^{12}$CO(1-0) / HCN(1-0)& - &$\,\,\,\ge35.4\,\pm\,13.9$&
      &$3$&$^{12}$CO(1-0) / HCN(1-0)& - &$\phantom{00}59.3\,\pm\,18.1$\\
      &$(+19\farcs5)$&\phantom{0}$^{12}$CO(1-0) / HCO$^+$(1-0)& - &$\,\,\,\ge36.9\,\pm\,14.5$&
      &$(+19\farcs5)$&\phantom{0}$^{12}$CO(1-0) / HCO$^+$(1-0)& - &$\ge64.8\,\pm\,22.7$\\
      &&$^{12}$CO(1-0) / HNC(1-0)& - &$\,\,\,\ge36.6\,\pm\,14.4$&
      &&$^{12}$CO(1-0) / HNC(1-0)& - &$\ge61.4\,\pm\,21.5$\\
      &&\phantom{0}$^{12}$CO(1-0) / HNCO(4-3)& - &$\,\,\,\ge24.3\,\pm\,\,\,\,9.6$&
      &&\phantom{0}$^{12}$CO(1-0) / HNCO(4-3) & - &$\ge62.1\,\pm\,21.7$\\
      &&$^{12}$CO(2-1) / HCN(1-0)& - &$\,\,\,\ge15.6\,\pm\,\,\,\,7.5$&
      &&$^{13}$CO(1-0) / HCN(1-0) & - &$\,\,\,\phantom{00}5.9\,\pm\,\,\,\,1.8$\\
      &&\phantom{0}$^{12}$CO(2-1) / HCO$^+$(1-0) & - &$\,\,\,\ge16.3\,\pm\,\,\,\,7.8$&
      &&\phantom{0}$^{13}$CO(1-0) / HCO$^+$(1-0) & - &$\ge\,\,\,6.2\,\pm\,\,\,\,2.2$\\
      &&$^{12}$CO(2-1) / HNC(1-0)& - &$\,\,\,\ge16.2\,\pm\,\,\,\,7.8$&
      &&$^{13}$CO(1-0) / HNC(1-0)& - &$\ge\,\,\,5.8\,\pm\,\,\,\,2.1$\\
      &&\phantom{0}$^{12}$CO(2-1) / HNCO(4-3)& - &$\,\,\,\ge10.7\,\pm\,\,\,\,5.2$&
      &&\phantom{0}$^{13}$CO(1-0) / HNCO(4-3)& - &$\ge\,\,\,5.9\,\pm\,\,\,\,2.1$\\
      \hline
    \end{tabular}
  \end{center}
\end{table*}
%
%
\addtocounter{table}{-1}
\begin{table*}
  \begin{center}
    \setlength{\tabcolsep}{2pt}
    \caption{\em Continued.}
    \begin{tabular}{cccccccccc}
      Galaxy & Position & Ratio & Nuclear disc & Nuclear ring & Galaxy &
      Position & Ratio & Nuclear disc & Inner ring \\ 
      (1) & (2) & (3) & (4) & (5) & (6) & (7) & (8) & (9) & (10) \\ 
      \hline
      NGC~4710&$4$&$^{12}$CO(1-0) / HCN(1-0)& - &$\,\,\,\ge35.1\,\pm\,12.2$&
      NGC~5866&$4$&$^{12}$CO(1-0) / HCN(1-0)& - &$\,\,\,\ge46.8\,\pm\,17.9$\\
      &$(+26\farcs0)$&\phantom{0}$^{12}$CO(1-0) / HCO$^+$(1-0)& - &$\,\,\,\ge37.3\,\pm\,13.0$&
      &$(+26\farcs0)$&\phantom{0}$^{12}$CO(1-0) / HCO$^+$(1-0)& - &$\,\,\,\ge47.1\,\pm\,18.1$\\
      &&$^{12}$CO(1-0) / HNC(1-0)& - &$\,\,\,\ge29.0\,\pm\,10.1$&
      &&$^{12}$CO(1-0) / HNC(1-0)& - &$\,\,\,\ge46.7\,\pm\,17.9$\\
      &&\phantom{0}$^{12}$CO(1-0) / HNCO(4-3)& - &$\,\,\,\ge30.8\,\pm\,10.7$&
      &&\phantom{0}$^{12}$CO(1-0) / HNCO(4-3) & - &$\,\,\,\ge51.2\,\pm\,19.6$\\
      &&$^{12}$CO(2-1) / HCN(1-0)& - &$\,\,\,\ge20.7\,\pm\,\,\,\,7.3$&
      &&$^{13}$CO(1-0) / HCN(1-0) & - &$\,\,\,\ge\phantom{0}5.4\,\pm\,\,\,\,2.0$\\
      &&\phantom{0}$^{12}$CO(2-1) / HCO$^+$(1-0) & - &$\,\,\,\ge22.0\,\pm\,\,\,\,7.7$&
      &&\phantom{0}$^{13}$CO(1-0) / HCO$^+$(1-0) & - &$\,\,\,\ge\phantom{0}5.4\,\pm\,\,\,\,2.0$\\
      &&$^{12}$CO(2-1) / HNC(1-0)& - &$\,\,\,\ge17.1\,\pm\,\,\,\,6.0$&
      &&$^{13}$CO(1-0) / HNC(1-0)& - &$\,\,\,\ge\phantom{0}5.4\,\pm\,\,\,\,2.0$\\
      &&\phantom{0}$^{12}$CO(2-1) / HNCO(4-3)& - &$\,\,\,\ge18.2\,\pm\,\,\,\,6.4$&
      &&\phantom{0}$^{13}$CO(1-0) / HNCO(4-3)& - &$\,\,\,\ge\phantom{0}5.9\,\pm\,\,\,\,2.2$\\ \\
      &$5$&$^{12}$CO(1-0) / HCN(1-0) & - &$\ge16.0\,\pm\,7.1$&
      &$5$&$^{13}$CO(1-0) / HCN(1-0) & - &$\ge\phantom{0}1.1\,\pm\,0.6$\\	
      &$(+32\farcs5)$&\phantom{0}$^{12}$CO(1-0) / HCO$^+$(1-0)& - &$\ge18.3\,\pm\,8.2$&
      &$(+32\farcs5)$&\phantom{0}$^{13}$CO(1-0) / HCO$^+$(1-0) & - &$\ge\phantom{0}1.3\,\pm\,0.7$\\
      &&$^{12}$CO(1-0) / HNC(1-0)& - &$\ge13.6\,\pm\,6.1$&
      &&$^{13}$CO(1-0) / HNC(1-0)& - &$\ge\phantom{0}1.2\,\pm\,0.7$\\
      &&\phantom{0}$^{12}$CO(1-0) / HNCO(4-3)& - &$\ge14.0\,\pm\,6.2$&
      &&\phantom{0}$^{13}$CO(1-0) / HNCO(4-3)& - &$\ge\phantom{0}1.5\,\pm\,0.8$\\ 
      &&$^{12}$CO(2-1) / HCN(1-0) & - &$\ge\phantom{0}7.1\,\pm\,3.3$&
      &&&&\\	
      &&\phantom{0}$^{12}$CO(2-1) / HCO$^+$(1-0) & - &$\ge\phantom{0}8.1\,\pm\,3.8$&
      &&&&\\
      &&$^{12}$CO(2-1) / HNC(1-0)& - &$\ge\phantom{0}6.3\,\pm\,3.0$&
      &&&&\\
      &&\phantom{0}$^{12}$CO(2-1) / HNCO(4-3)& - &$\ge\phantom{0}6.5\,\pm\,3.0$&
      &&&&\\
      \hline
    \end{tabular}
  \end{center}
\end{table*}
%

%
%
\begin{table}
  \begin{center}
    \setlength{\tabcolsep}{1pt}
    \caption{Ratios of CO to dense gas tracer lines in the
      intermediate region of NGC~4710.}
    \label{tab:ratioir}
    \begin{tabular}{cccc}
      Galaxy & Position & Ratio & Value \\ 
      \hline
      NGC~4710 & $3$&$^{12}$CO(1-0) / HCN(1-0)&$\,\,\,\ge26.6\,\pm\,10.8$ \\
      & $(+19\farcs5)$&\phantom{0}$^{12}$CO(1-0) / HCO$^+$(1-0)&$\,\,\,\ge27.7\,\pm\,11.2$\\
      & &$^{12}$CO(1-0) / HNC(1-0)&$\,\,\,\ge27.5\,\pm\,11.1$\\
      & &\phantom{0}$^{12}$CO(1-0) / HNCO(4-3) &$\,\,\,\ge18.3\,\pm\,\,\,\,7.4$\\
      & &$^{13}$CO(1-0) / HCN(1-0) &$\,\,\,\ge\,\,\,5.7\,\pm\,\,\,\,3.1$\\
      & &\phantom{0}$^{13}$CO(1-0) / HCO$^+$(1-0) &$\,\,\,\ge\,\,\,6.0\,\pm\,\,\,\,3.3$\\
      & &$^{13}$CO(1-0) / HNC(1-0)&$\,\,\,\ge\,\,\,5.9\,\pm\,\,\,\,3.3$\\
      & &\phantom{0}$^{13}$CO(1-0) / HNCO(4-3)&$\,\,\,\ge\,\,\,3.9\,\pm\,\,\,\,2.2$\\
      & &$^{12}$CO(2-1) / HCN(1-0) &$\,\,\,\ge17.0\,\pm\,\,\,\,7.7$\\
      & &\phantom{0}$^{12}$CO(2-1) / HCO$^+$(1-0) &$\,\,\,\ge17.7\,\pm\,\,\,\,8.0$\\
      & &$^{12}$CO(2-1) / HNC(1-0)&$\,\,\,\ge17.6\,\pm\,\,\,\,8.0$\\
      & &\phantom{0}$^{12}$CO(2-1) / HNCO(4-3)&$\,\,\,\ge11.7\,\pm\,\,\,\,5.3$\\
    \end{tabular}
  \end{center}
  Notes: The offset along the major axis (with respect
    to the galaxy centre) of each projected position is listed under
    the projected position number in column~2. The uncertainties on
  the line ratios are calculated from the uncertainties on the
  integrated line intensities using the standard error propagation
  formula.
\end{table} 
%
%
\section{LVG Models}
\label{sec:LVG}
%
%
\subsection{Model Parameters}
\label{sec:parameters}
The line ratio models discussed in \S~\ref{sec:model} exploit the
non-LTE radiative transfer code RADEX \citep{van07}, that uses the
large velocity gradient (LVG) approximation
\citep{sob60,cas70,gol74,jon75} and calculates the intensities of
molecular lines based on a statistical equilibrium involving
collisional processes (excitation and de-excitation through collisions
with H$_2$), radiative processes, and the cosmic microwave background
radiation ($T_{\rm CMB}=2.7$~K). The escape probability depends on the
optical depths \citep[see][]{sob60} and is related to the intensity
within the medium. RADEX offers three different medium geometries
(sphere, uniformly expanding shell and parallel slab), although the
differences between these are generally small \citep{van07}. We adopt
here the expanding shell geometry.

The main input parameters to RADEX are the molecular gas kinetic
temperature $T_{\rm K}$, H$_2$ volume number density $n$(H$_2$),
species column number densities $N$(mol), species line widths
$\Delta v$ (i.e.\ the FWHMs of the Gaussians fitted to the spectra),
and species intrinsic abundance ratios (e.g.\ [$^{12}$C] / [$^{13}$C],
[HCN] / [HCO$^+$], [HCN] / [HNC] and [HCN] / [HNCO]). By modeling the
dense gas tracer lines and low-$J$ CO lines separately, we effectively
adopt a two-component molecular ISM (at each projected position and
for each kinematic component).

We created model grids as follows. $T_{\rm K}$, $n$(H$_2$), and
$N$(mol) are kept as free parameters to be fit for. Our $T_{\rm K}$
grid ranges from $10$ to $250$~K in steps of $5$~K, $n$(H$_2$) ranges
from $10^2$ to $10^7$~cm$^{-3}$ in steps of $0.25$~dex, and $N$(mol)
ranges from $10^{13}$ to $10^{21}$~cm$^{-2}$ in steps of
$0.25$~dex. As the widths of all the lines are similar, we adopt a
single (average) line width for each kinematic component and each
group of lines (rather than, e.g., using the exact width of each line
in each kinematic component at each position). The line widths used
for the modeling were thus respectively $150$ and $180$~km~s$^{-1}$
for the CO and dense gas lines in the nuclear disc of NGC~4710,
$50$~km~s$^{-1}$ for the CO gas lines in the inner ring of NGC~4710,
and $300$~km~s$^{-1}$ for the dense gas lines in the nuclear disc of
NGC~5866. Other object--line--kinematic component combinations could
not be modeled (see below).

The intrinsic abundance ratios of different molecules are also
important inputs to the model calculations and must therefore be
chosen carefully. \citet{ro11} state that the ratios of
velocity-integrated line intensities are not good proxies for
intrinsic abundance ratios, while the ratios of column densities are
better. We therefore take here the ratios of the column densities as
the ratios of the intrinsic abundances. This indicates that changes in
the abundance ratios affect the column densities linearly, i.e.\ a
$50\%$ change in an intrinsic abundance ratio will change the
associated column densities by $50\%$. The [$^{12}$C] / [$^{13}$C]
ratio is $\approx20$ at the centre of the Milky Way, $\approx50$ in
the $4$~kpc molecular ring, $\approx70$ in the local ISM and
$\approx90$ in the Solar System \citep{ag89,wr94}. In the active
nuclear regions of nearby starburst galaxies, the [$^{12}$C] /
[$^{13}$C] ratio is $\geq40$ \citep{hm93a,hm93b,mar10}. There is thus
much variation in the intrinsic [$^{12}$C] / [$^{13}$C] ratio. The
situation is even less satisfactory for the abundance ratios of high
density tracers, since they have been much less studied in external
galaxies. The [HCN] / [HCO$^+$] and [HCN] / [HNC] ratios vary widely
from source to source \citep{woo78,gold81,gold86,ro11} and there is a
lack of information in the literature about the [HCN] / [HNCO]
ratio. These ratios also vary from region to region within galaxies,
as molecular clouds located in different regions of galactic discs may
well have different physical conditions and processes at play. For
example, HCO$^+$ is enhanced by shocks, and the [HCN] / [HNC] ratio is
known to vary with temperature (it increases with increasing
temperature, although the abundance of each species decreases with
increasing temperature and density;
\citealt{gold81,gold86,sc92}). \citet{ro11} suggest that [HCN] / [HNC]
and [HCN] / [HCO$^+$] range from $0.2$ to $100$ and $50$ to $0.02$,
respectively (at $n$(H$_2$) = $10^{6}$~cm$^{-3}$ and
$T_{\rm K}=10$--$200$~K). However, both abundance ratios are about $1$
at $\approx30$~K.  In this work, we therefore assume an intrinsic
ratio of $70$ for [$^{12}$C] / [$^{13}$C] and $1$ for the dense gas
tracers (i.e.\ [HCN] / [HCO$^+$], [HCN] / [HNC] and [HCN] / [HNCO]).

Overall, this thus yields $4$ possible sets of models for each
position in each galaxy: models for the tenuous molecular gas only
(i.e.\ $^{12}$CO(1-0), $^{12}$CO(2-1), $^{13}$CO(1-0) and
$^{13}$CO(2-1)) and for the dense molecular gas only (i.e.\ HCN,
HCO$^+$, HNC and HNCO), this for both the nuclear disc and the inner
ring separately (adopting an average $\Delta v$ for each group of
lines separately).

The positions along the disc of the galaxies where we apply the
radiative transfer modelling are the same as those discussed in
\S~\ref{sec:extract} and illustrated in Figure~\ref{fig:pos}. However,
since only $2$ low-$J$ CO lines were observed in NGC~5866, these sets
of models for the tenuous molecular gas would be unconstrained and are
thus not computed. Similarly, the dense gas tracer lines are generally
not detected in the inner ring, so these sets of models are not
computed either. Overall, we thus need models for the tenuous and
dense molecular gas in the nuclear disc of NGC~4710, for the tenuous
gas in the inner ring of NGC~4710, and for the dense gas in the
nuclear disc of NGC~5866.
%
%
\subsection{Best-fitting and most likely models}
\label{sec:bestlike}
Following \citet{to14}, we use two methods to characterise the models
(i.e.\ the physical parameters $T_{\rm K}$, $n$(H$_2$) and $N$(mol) of
a two-component molecular ISM) best representing the data at each of
the positions illustrated in Figure~\ref{fig:pos}. First we use a
$\chi^{2}$ method identifying the best-fitting model, and second a
likelihood method identifying the most likely model.

For each set of model parameters, the $\chi^{2}$ is defined as
\begin{equation}
  \chi^{2}\equiv\sum\limits_{i}\bigg(\frac{R_{i,{\rm
      mod}}-R_{i,{\rm obs}}}{\Delta R_{i,{\rm obs}}}\bigg)^2\,, 
  \label{eq:chi2}
\end{equation}
where $R_{\rm mod}$ is the modeled line ratio, $R_{\rm obs}$ is the
observed line ratio with uncertainty $\Delta R_{\rm obs}$, and the
summation is over all independent line ratios $i$ at the given
position (one fewer than the number of lines available at that
position).

For positions where $4$ lines are detected, so $3$ line ratios are
available, the $\chi^{2}$ is well-defined and the models are well
constrained. However, for positions where at least one observed line
ratio is a lower limit, we calculate the $\chi^{2}$ following one of
two procedures. (1) For models with a line ratio larger than or equal
to the observed lower limit, the $\chi^{2}$ is taken as $1$, thus
ensuring that all models meeting this criterion are equally
likely. (2) For models with a line ratio smaller than the observed
lower limit, the $\chi^{2}$ is calculated in the usual way, but the
resultant reduced $\chi^{2}$ value then indicates a lower limit only,
leading to an upper limit on the likelihood (see below).

\begin{figure*}
  \includegraphics[width=7.3cm,clip=]{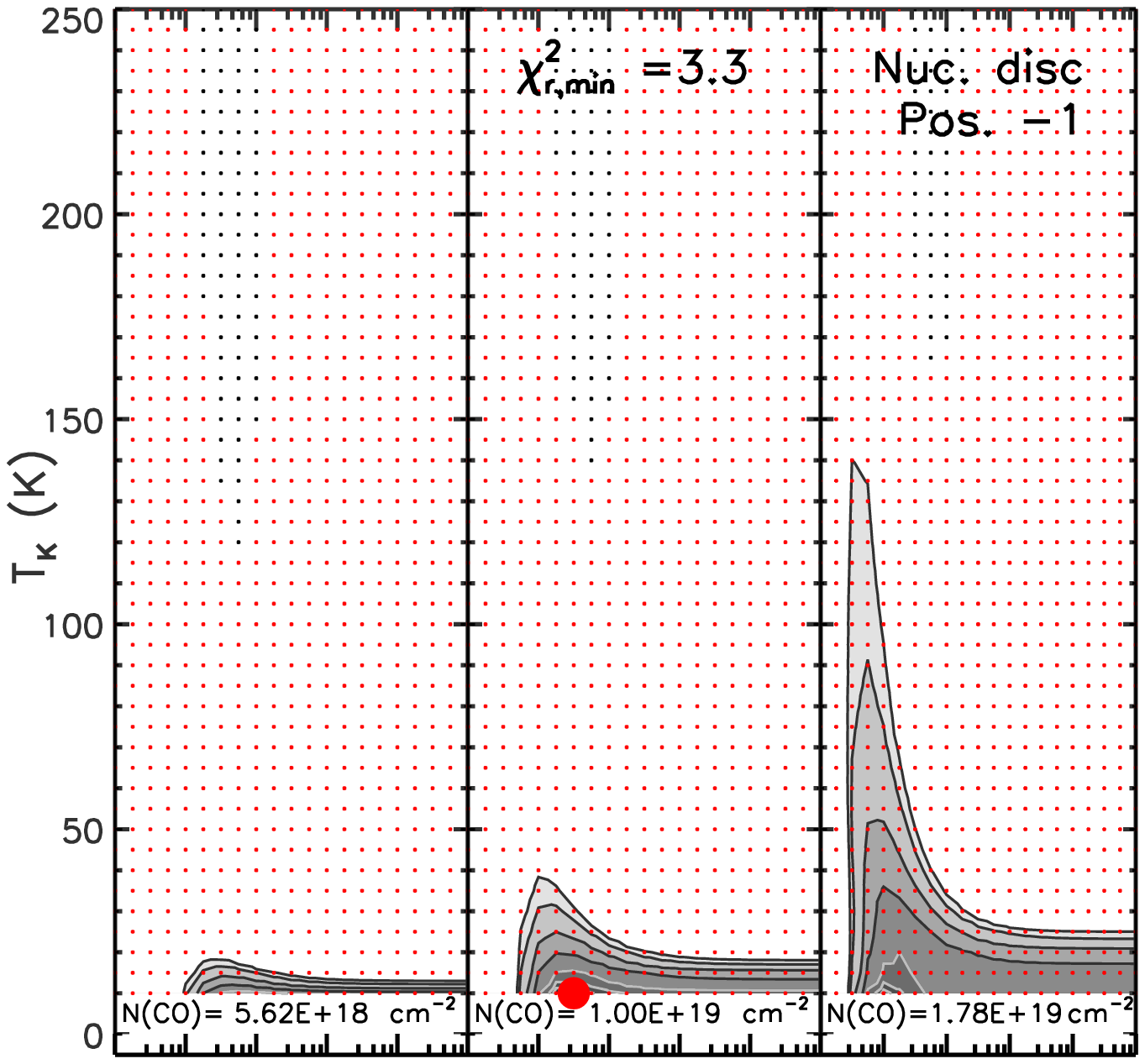}
  \hspace{-25pt}
  \includegraphics[width=7.3cm,clip=]{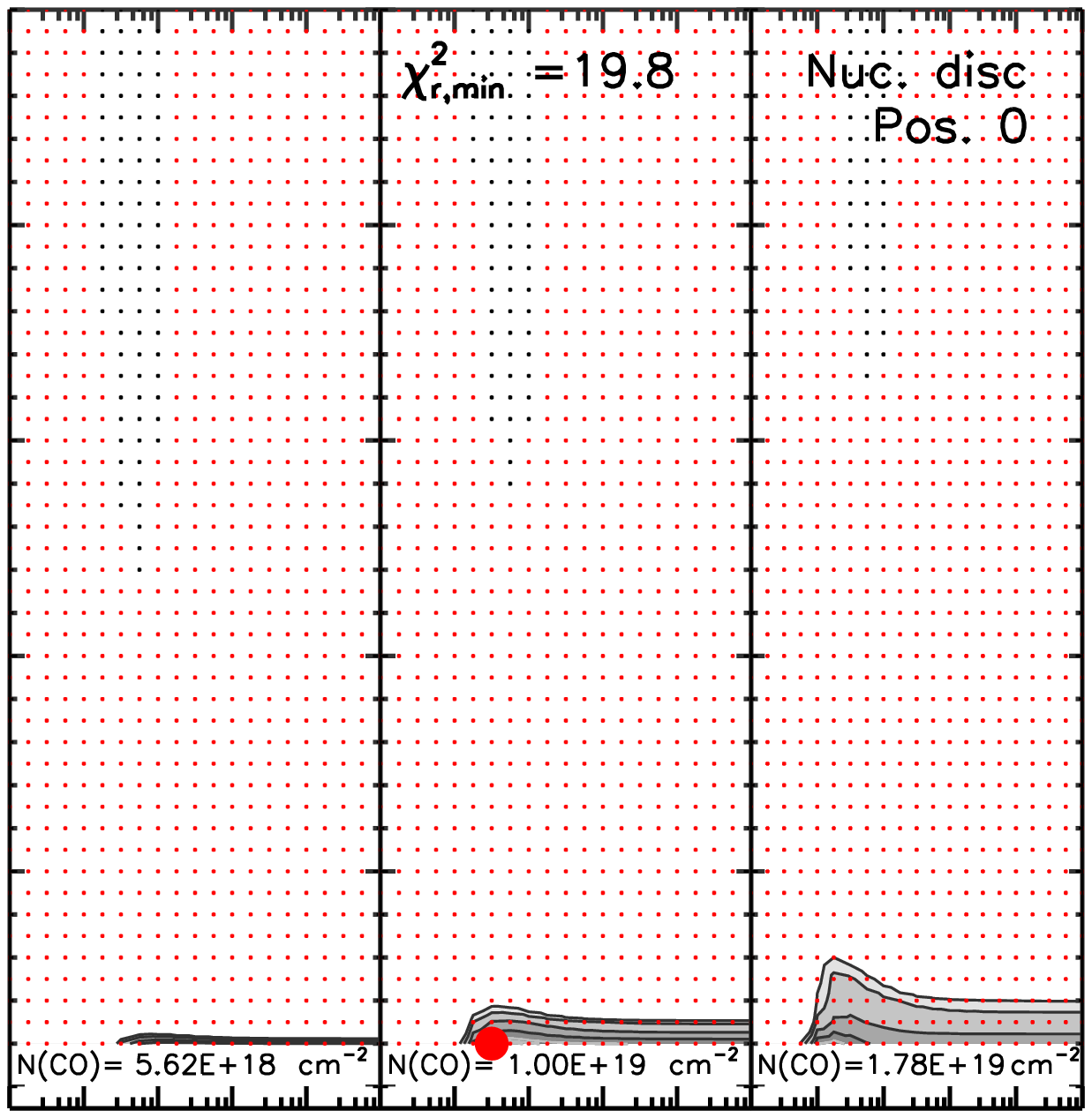}\\
  \vspace{-18pt}
  \includegraphics[width=7.3cm,clip=]{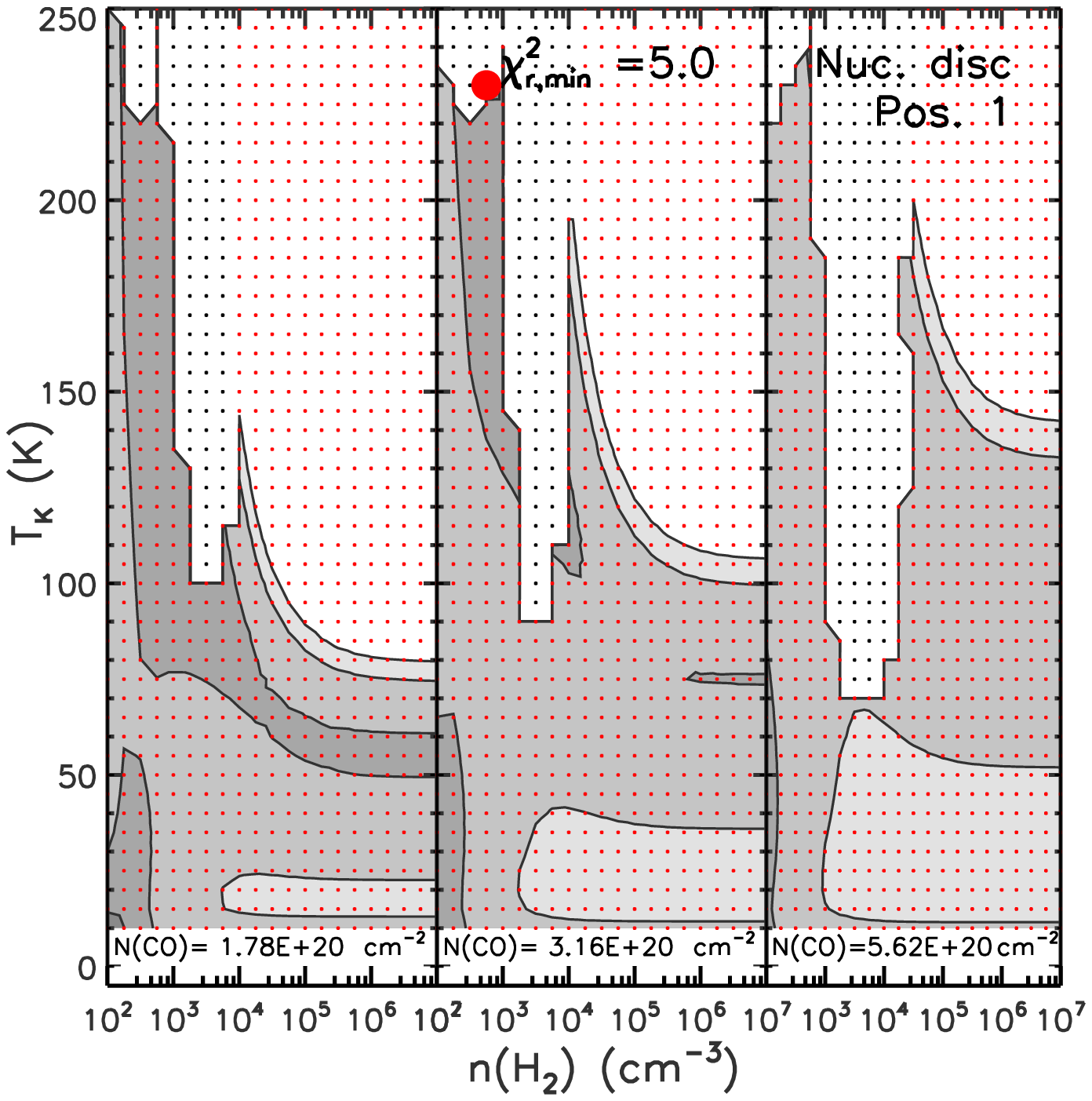}
  \hspace{-25pt}
  \includegraphics[width=7.3cm,clip=]{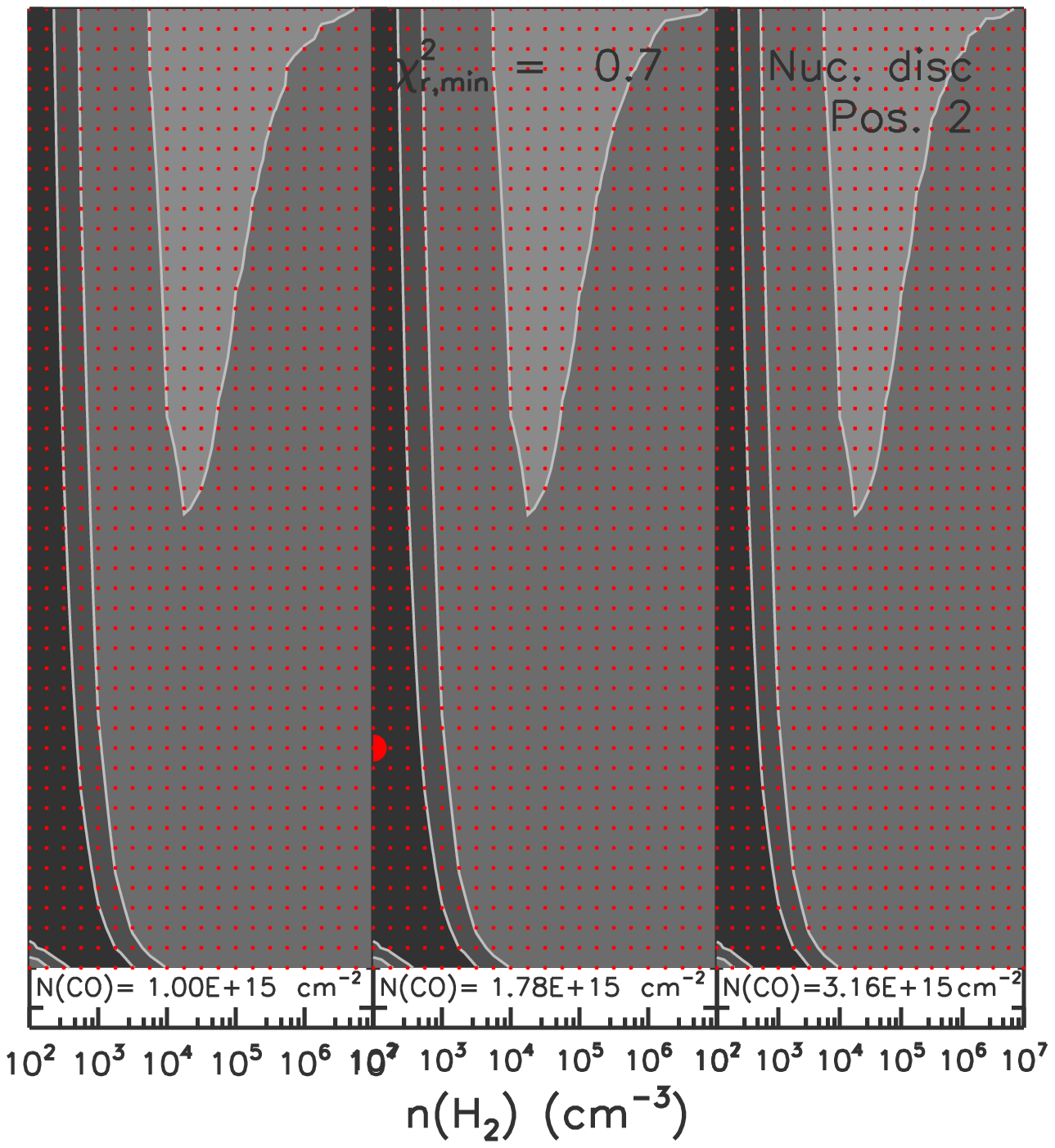}\\
  \caption{$\Delta\chi_{\rm r}^{2}\equiv\chi_{\rm r}^2-\chi_{\rm r,
      min}^2$ maps for the tenuous molecular gas in the nuclear disc of NGC~4710
    (positions $-1$, $0$, $1$ and $2$). For each region,
    $\Delta\chi_{\rm r}^{2}$ is shown as a function of $T_{\rm K}$ and
    $n$(H$_{\rm 2}$) for three values of $N$(CO) centred around the
    best-fit and indicated at the bottom of each panel. The models
    computed are indicated by red dots and the best-fit model with a
    red filled circle. Black dots represent bad models (e.g.\
    unacceptably low opacity; see \citealt{van07}). The
    $\Delta\chi_{\rm r}^2$ contours show the $0.2\sigma$ ($16\%$
    probability that the appropriate model is enclosed; darkest
    greyscale), $0.5\sigma$ ($38\%$), and $1\sigma$ ($68\%$) to
    $5\sigma$ ($99.9\%$; lightest greyscale) confidence levels in
    steps of $1\sigma$ for $3$ degrees of freedom ($3$ line
    ratios). The actual $\Delta\chi_{\rm r}^{2}$ levels from
    $0.2\sigma$ to $5\sigma$ are $0.8$, $1.8$, $3.5$, $8.0$, $14.2$,
    $22.1$ and $28.0$, respectively. The confidence levels from
    $2\sigma$ to $5\sigma$ are separated by black lines, those from
    $0.2\sigma$ to $1\sigma$ by grey lines. The area containing models
    with $\leq1\sigma$ confidence levels is much smaller than the
    best-fit model symbol at positions $-1$, $0$ and $1$. The
    $\chi_{\rm r,min}^2$ value, kinematic component and position along
    the galaxy disc are also indicated in each panel.}
  \label{fig:n4710chifc}
\end{figure*}

Contours and grey scales of reduced $\chi^{2}$ (formally
$\Delta\chi_{\rm r}^2\equiv\chi_{\rm r}^2-\chi_{\rm r,min}^2$, where
$\chi_{\rm r,min}^2$ is the minimum reduced $\chi^{2}$) in
$T_{\rm K}$, $n$(H$_2$) and $N$(CO) space are shown in
Figures~\ref{fig:n4710chifc}\,--\,\ref{fig:n4710chisc}. These
illustrate the uncertainties of the best-fit model parameters and
exhibit the usual degeneracies between physical parameters. The
best-fit models ($\chi_{\rm r,min}^2$) are listed in
Tables~\ref{tab:result1} and \ref{tab:result2}. For positions with a
line ratio lower limit where a model meets the criterion for case (2)
above, the contours and greyscales are instead shown in colour,
indicating a $\Delta\chi_{\rm r}^2$ lower limit. If
$\chi_{\rm r,min}^{2}$ lies in such a region of parameter space, the
best-fit model is then ill defined.

%
%
%

%
%
\begin{figure*}
  \includegraphics[width=7.3cm,clip=]{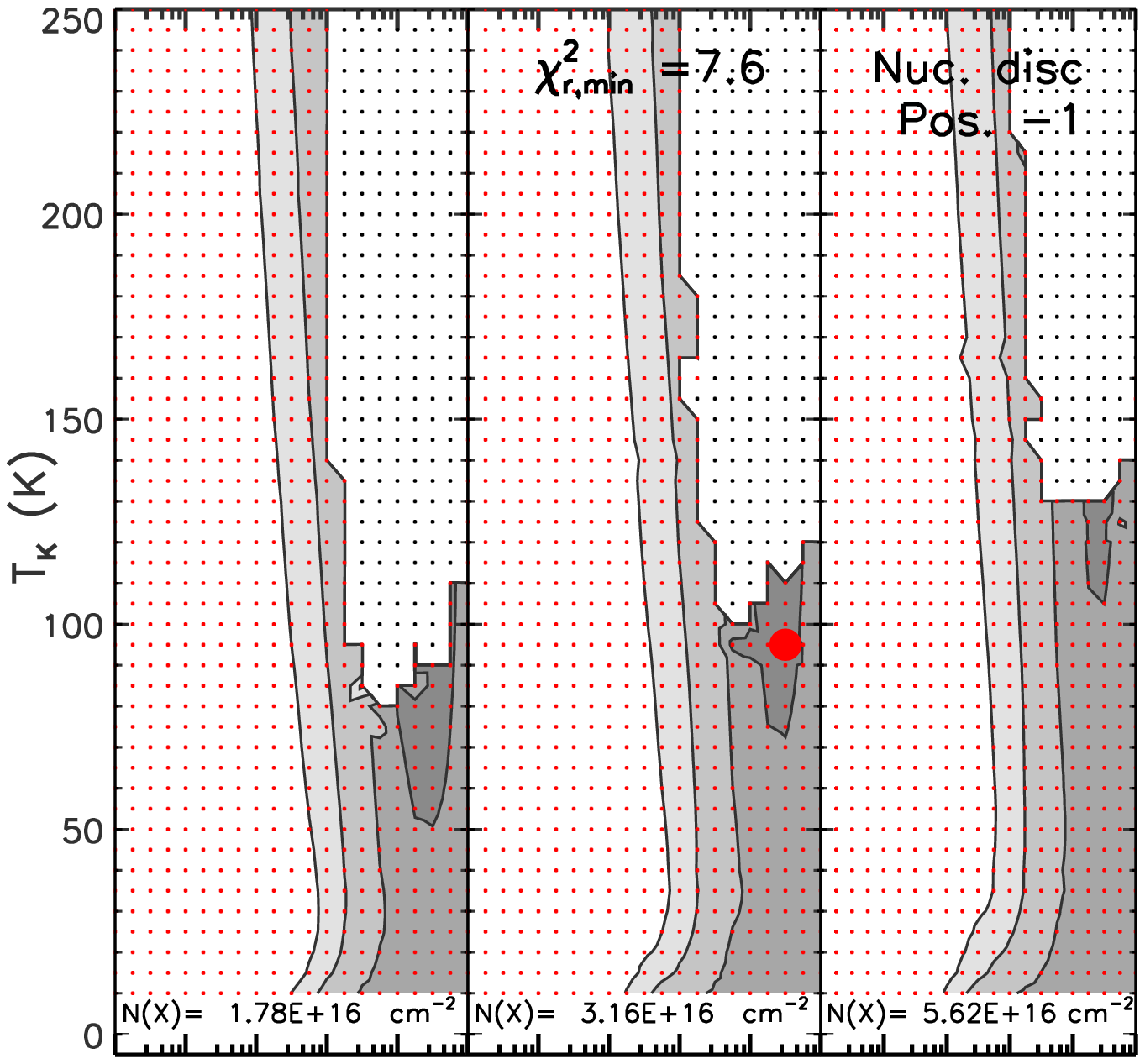}
  \hspace{-25pt}
  \includegraphics[width=7.3cm,clip=]{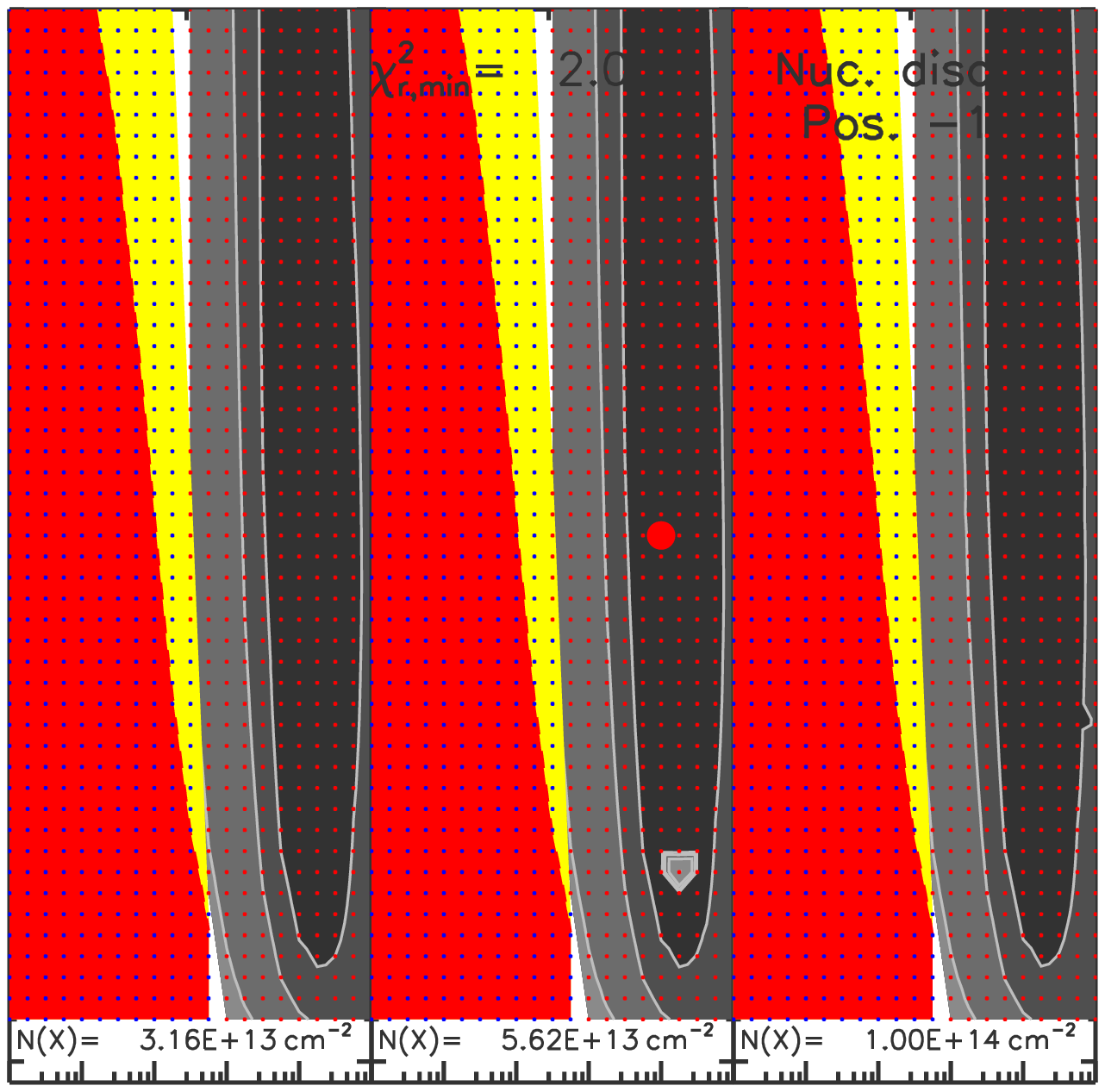}\\
  \vspace{-18pt}
  \includegraphics[width=7.3cm,clip=]{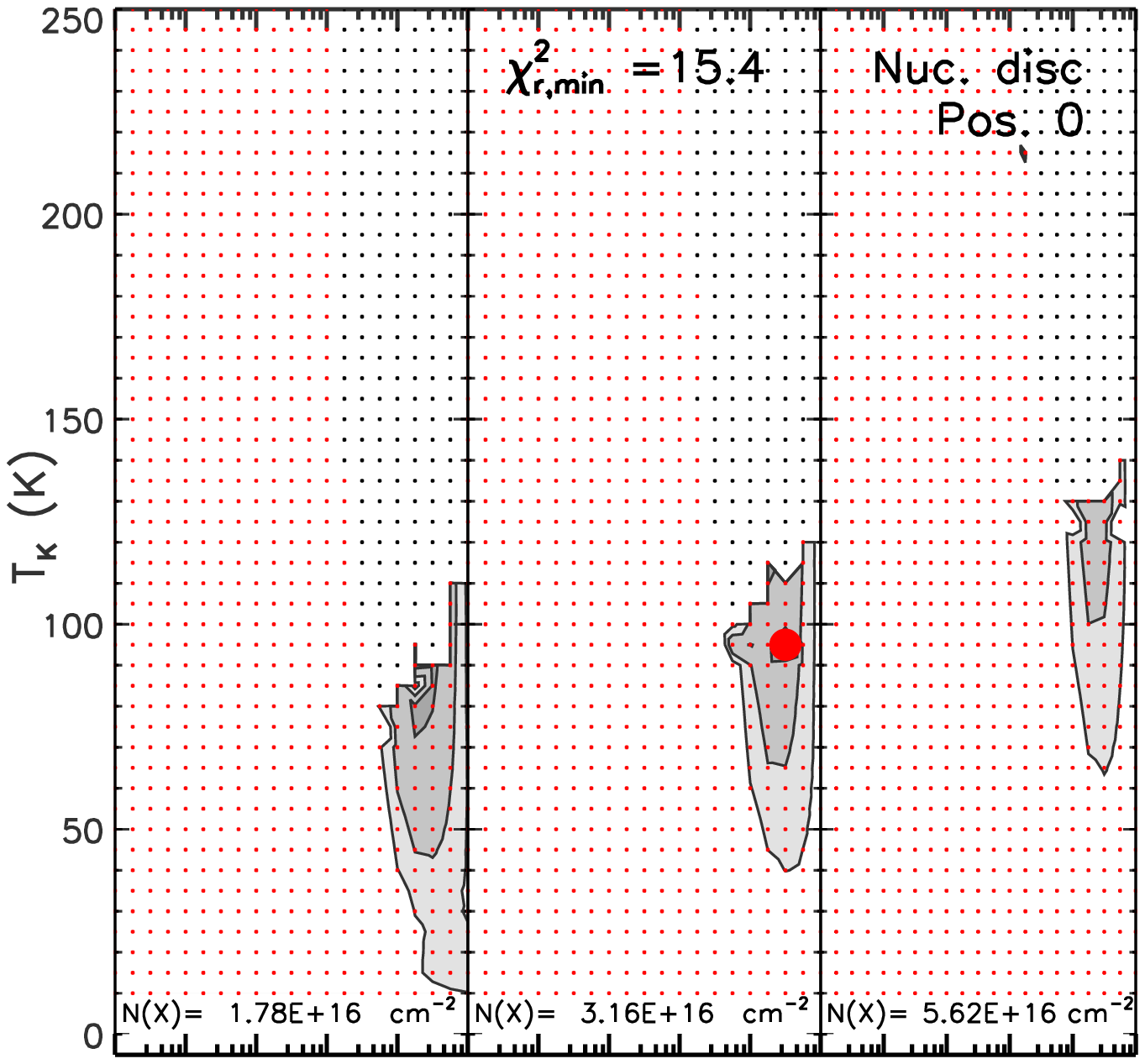}
  \hspace{-25pt}
  \includegraphics[width=7.3cm,clip=]{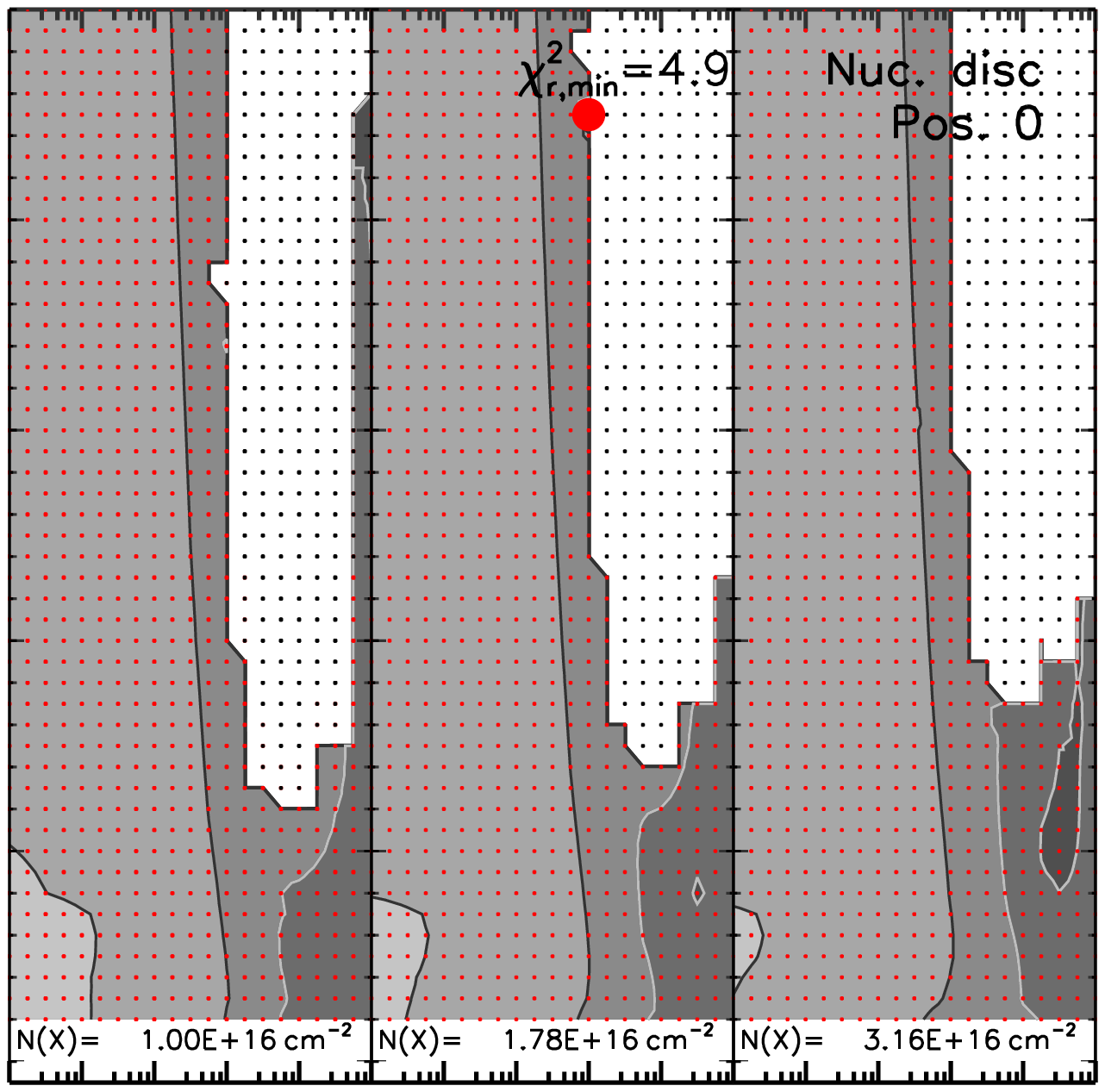}\\
  \vspace{-18pt}
  \includegraphics[width=7.3cm,clip=]{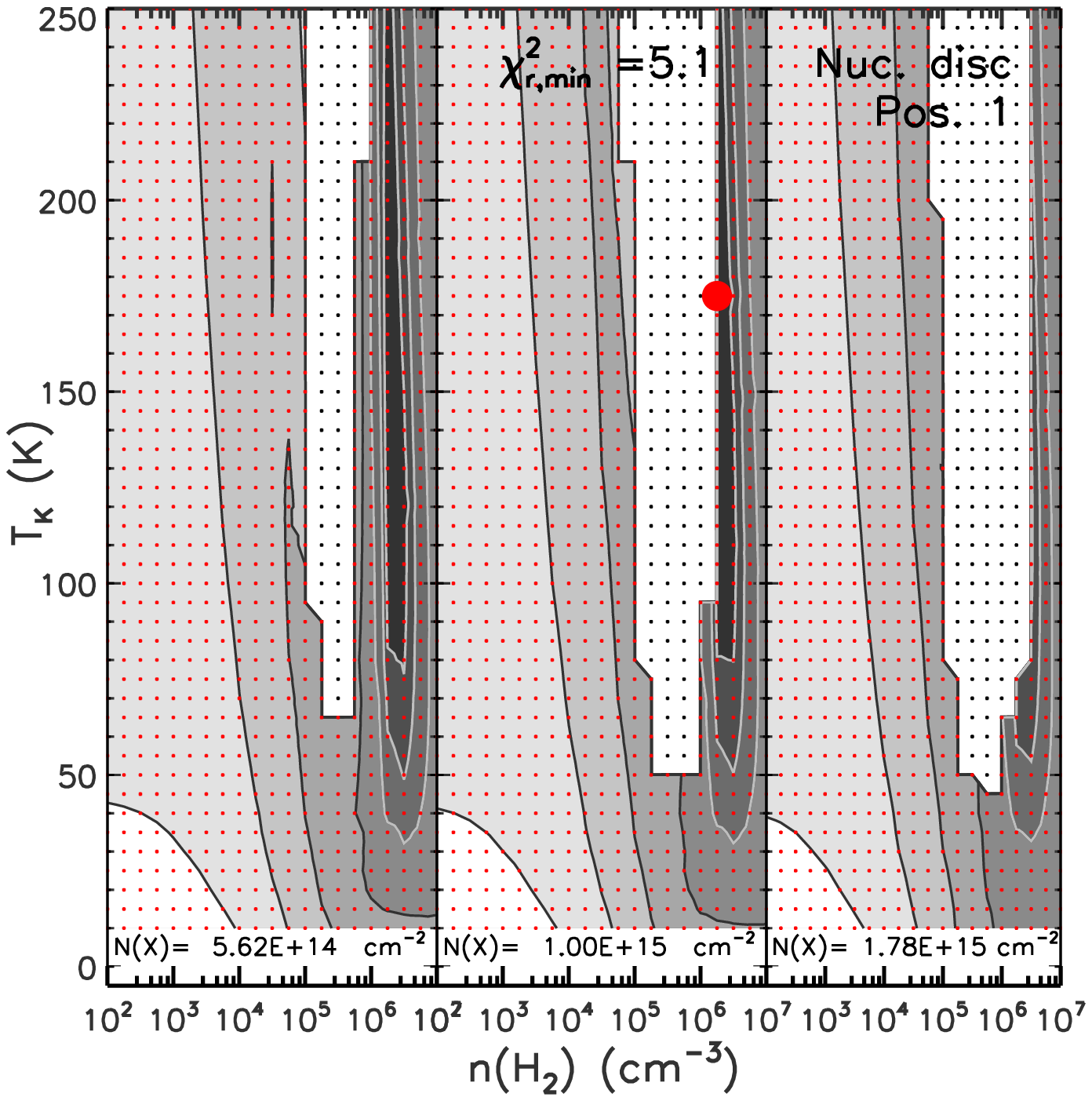}
  \hspace{-25pt}
  \includegraphics[width=7.3cm,clip=]{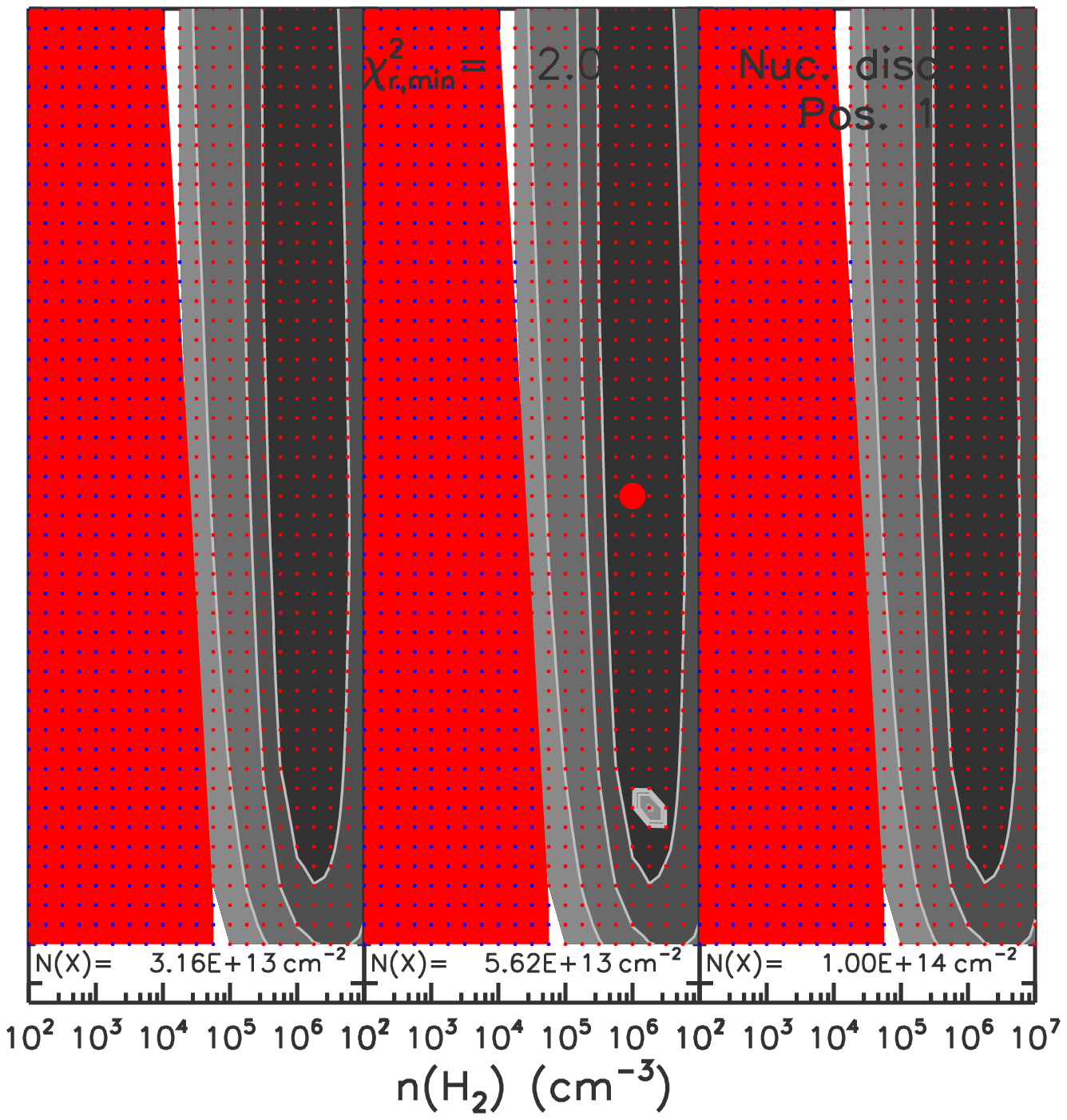} \\
  \caption{Same as Figure~\ref{fig:n4710chifc} but for the dense
    molecular gas in the nuclear disc of NGC~4710 (left) and NGC~5866
    (right) (positions $-1$, $0$ and $1$). $N$(X) stands for the column
    number density of all four high density tracers. The area
    containing models with $\leq1\sigma$ confidence levels is much
    smaller than the best-fit model symbol at positions $-1$ and $0$ in
    NGC~4710 and at position~$0$ in NGC~5866. For positions with a
    line ratio lower limit, blue dots indicate models meeting the
    criterion for case (2) described in the text
    (Appendix~\ref{sec:bestlike}). Associated $\Delta\chi_{\rm r}^{2}$
    contours (confidence levels) are shown in colour: yellow
    ($1\sigma$), red ($2\sigma$), orange ($3\sigma$), green
    ($4\sigma$) and brown ($5\sigma$).}
  \label{fig:n4758chifd}
\end{figure*}
%

%
%
\begin{figure*}
  \includegraphics[width=7.3cm,clip=]{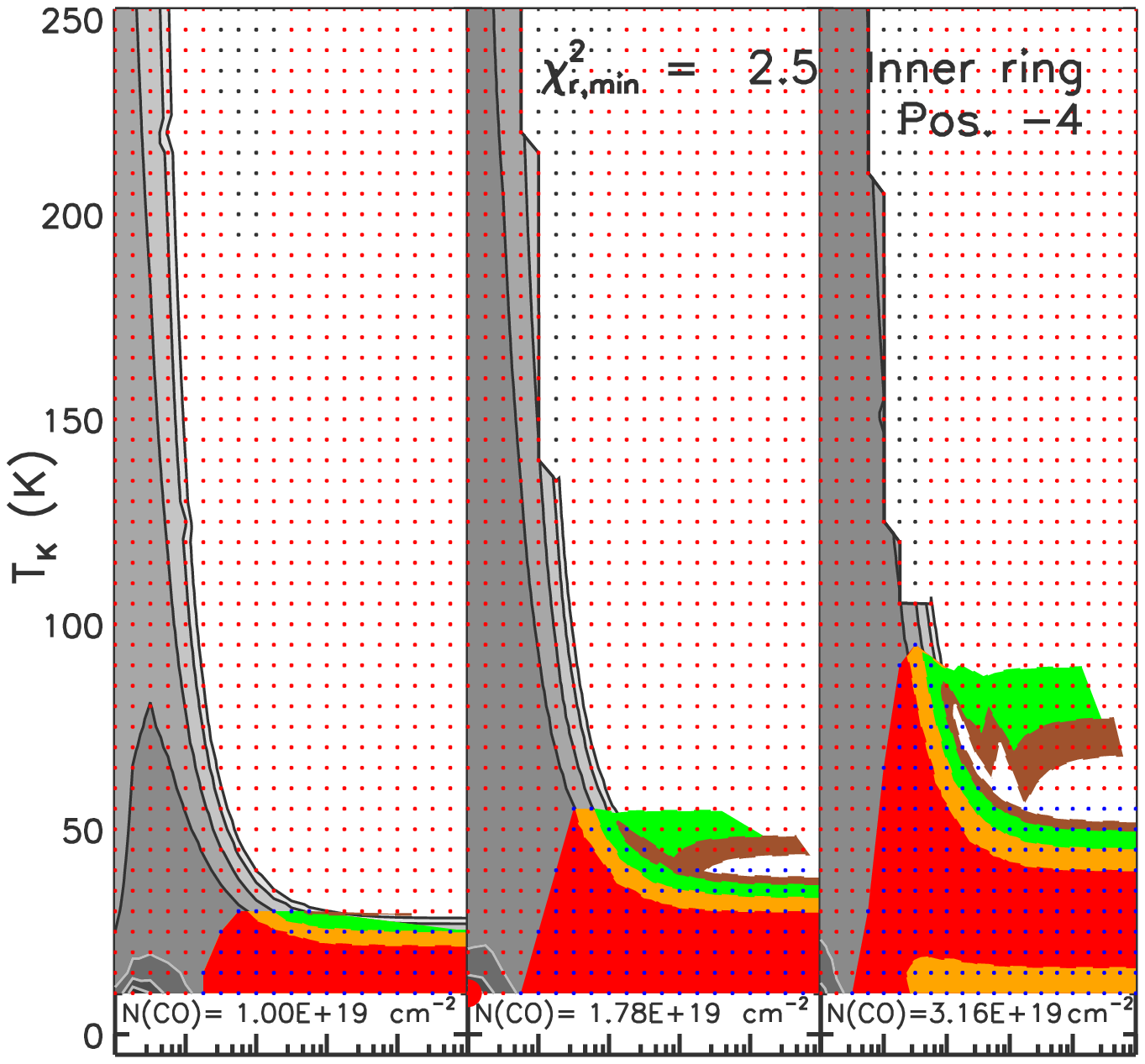}
  \hspace{-25pt}
  \includegraphics[width=7.3cm,clip=]{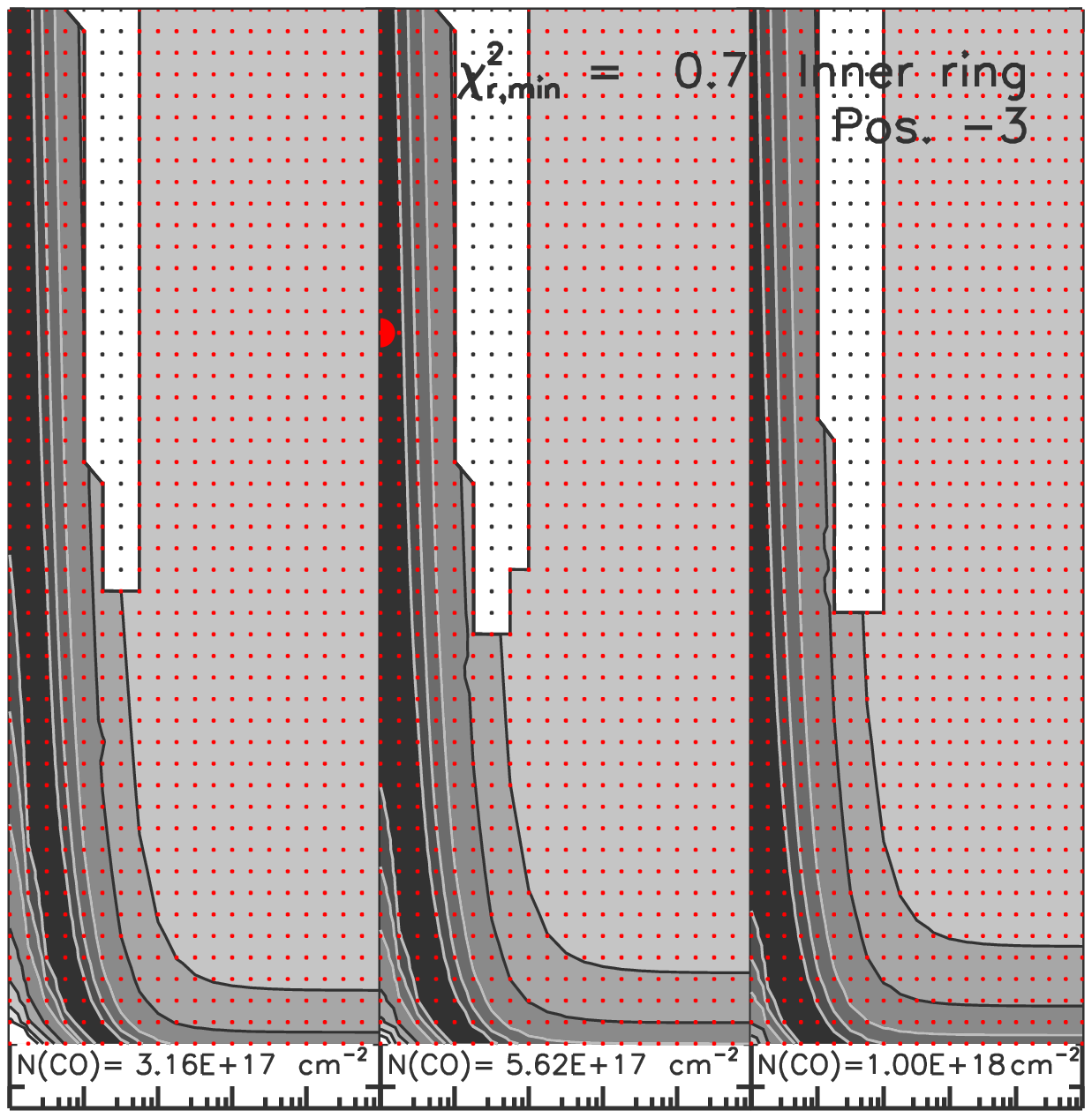}\\
  \vspace{-18pt}
  \includegraphics[width=7.3cm,clip=]{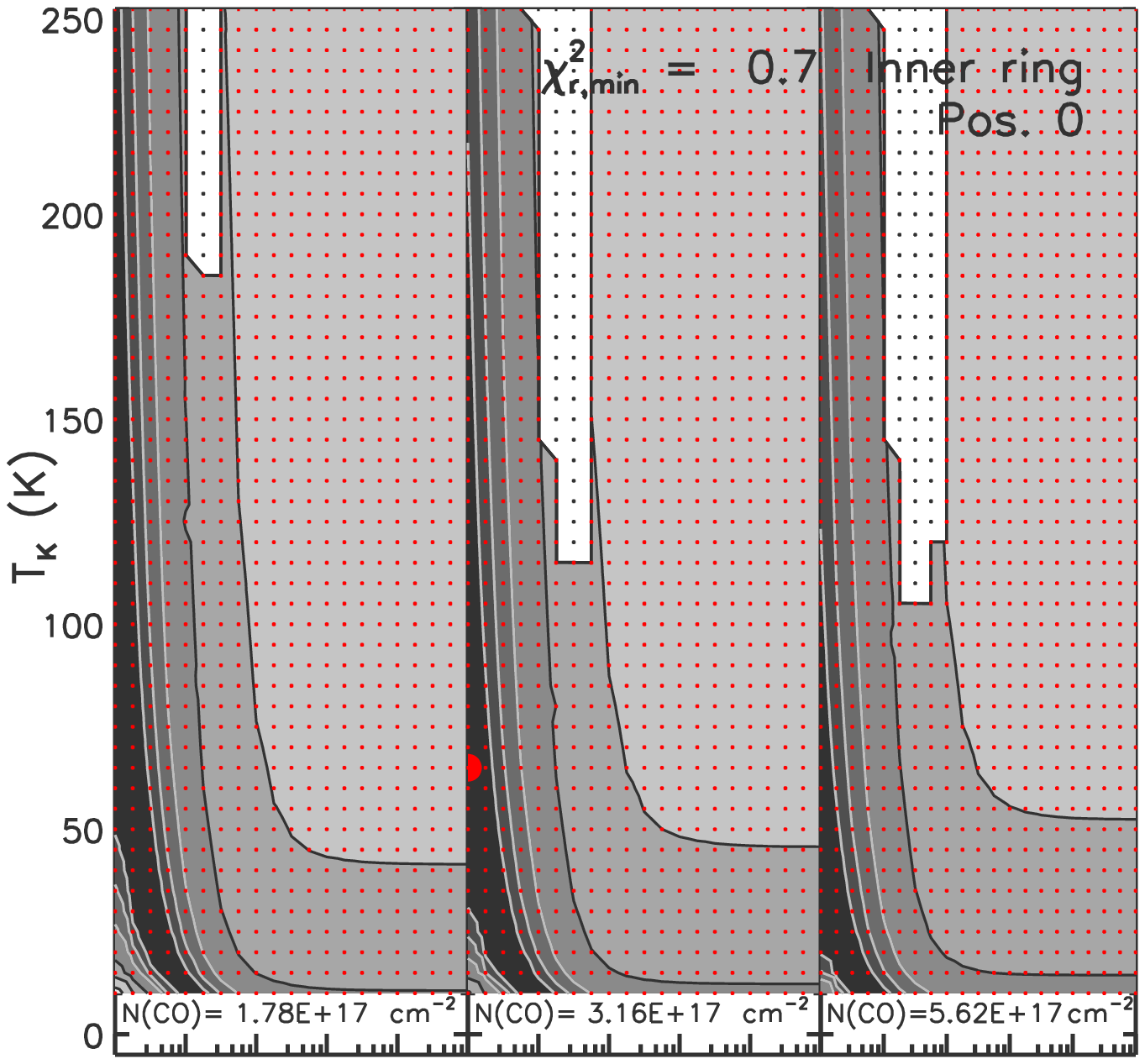}
  \hspace{-25pt}
  \includegraphics[width=7.3cm,clip=]{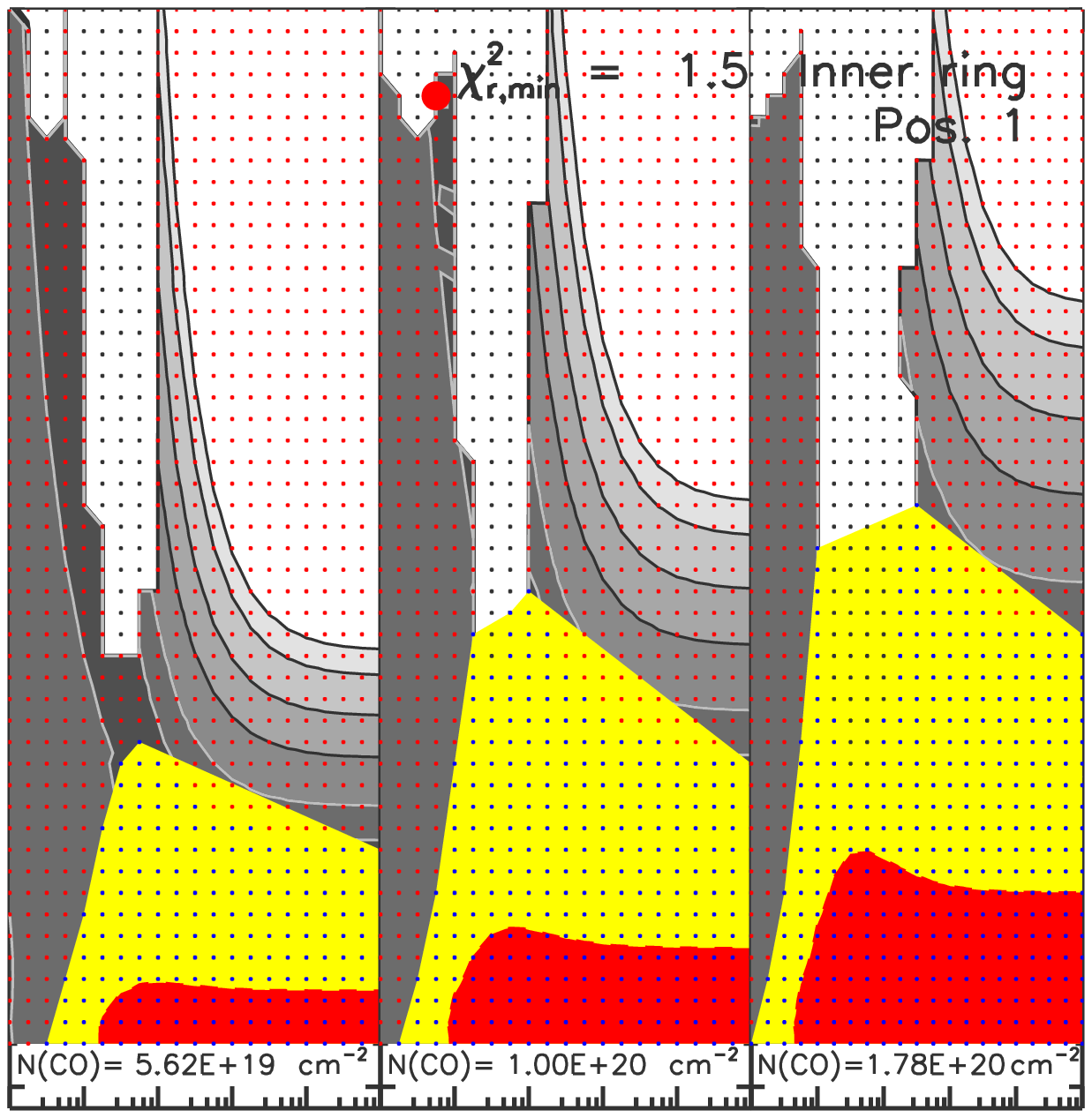}\\
  \vspace{-18pt}
  \includegraphics[width=7.3cm,clip=]{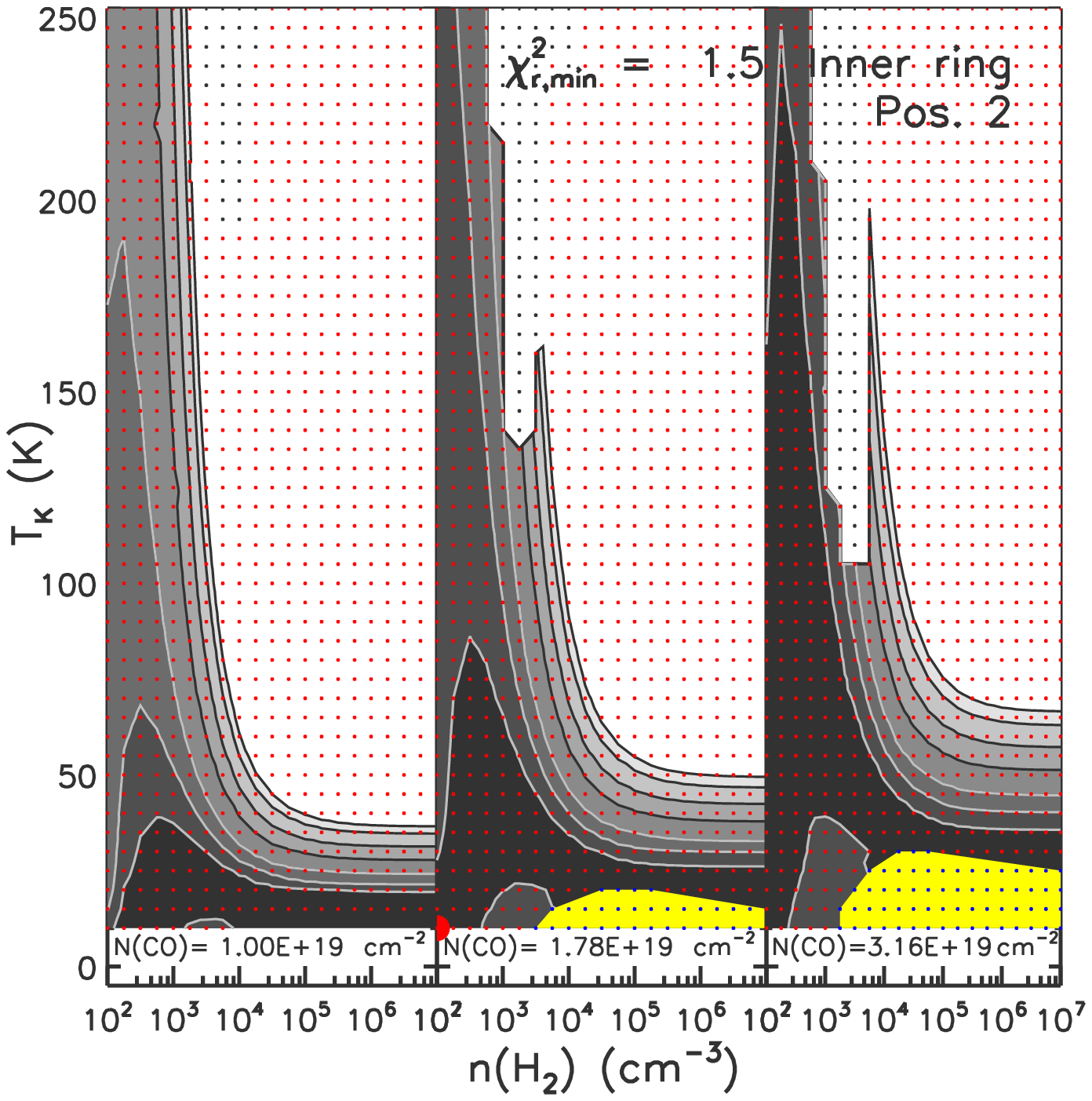}
  \hspace{-25pt}
  \includegraphics[width=7.3cm,clip=]{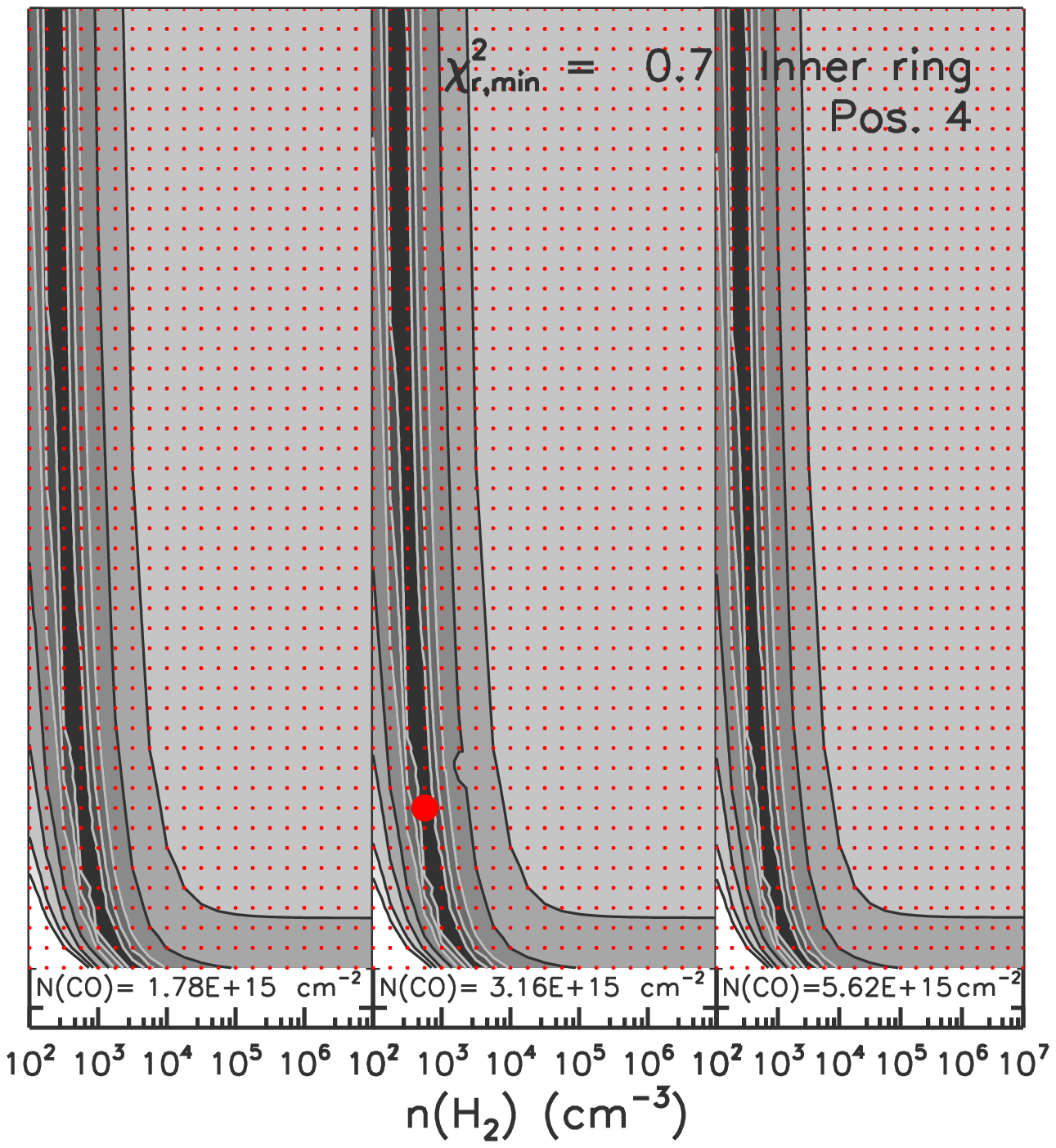}\\
  \caption{Same as Figures~\ref{fig:n4710chifc} and
    \ref{fig:n4758chifd} but for the tenuous molecular gas in the
    inner ring of NGC~4710 (positions $-4$, $-3$, $0$, $1$, $2$ and
    $4$). The area containing models with $\leq1\sigma$ confidence
    levels is much smaller than the best-fit model symbol at positions
    $-4$ and $1$ in NGC~4710.}
  \label{fig:n4710chisc}
\end{figure*}
%

%
%
\begin{table*}
  \begin{center}
    \setlength{\tabcolsep}{3pt}
    \caption{Model results for the tenuous and dense molecular gas in
      the nuclear disc and inner ring of NGC~4710.}
    \label{tab:result1}
    \begin{tabular}{cccrllrll}
      & & & & Tenuous gas & & & Dense gas &\\ 
      \hline
      Component & Position & Offset ($^{\prime\prime}$) & Parameter & $\chi^{2}$ & Likelihood &
      \hspace*{20mm} Parameter & $\chi^{2}$ & Likelihood\\
      (1) & (2) & (3) & (4) & (5) & (6) & (7) & (8) & (9)\\
      \hline
      nuclear disc & $-1$ & $-6\farcs5$ & $T_{\rm K}$ & $10^{\ast}$~K & $25^{+119}_{-14}$~K & $T_{\rm K}$ & $95$~K & $135^{+73}_{-63}$~K\\ 
      &  & & $\log$($n$(H$_2$)) & $3.5$~cm$^{-3}$ & $3.3^{+1.9}_{-0.7}$~cm$^{-3}$ & $\log$($n$(H$_2$)) & $6.5$~cm$^{-3}$ & $6.4^{+0.2}_{-0.2}$~cm$^{-3}$\\
      && & $\log$($N$(CO)) & $19.0$~cm$^{-2}$ & $19.6^{+0.6}_{-0.6}$~cm$^{-2}$ & $\log$($N$(X)) & $16.5$~cm$^{-2}$ & $14.3^{+0.8}_{-0.9}$~cm$^{-2}$\\
      \\
      & $0$ & $0^{\prime\prime}$ & $T_{\rm K}$ & $10^{\ast}$~K & $11^{+2}_{-1}$~K & $T_{\rm K}$ & $95$~K & $153^{+67}_{-58}$~K\\
      & & &  $\log$($n$(H$_2$)) & $3.7$~cm$^{-3}$ & $3.7^{+1.7}_{-1.0}$~cm$^{-3}$ & $\log$($n$(H$_2$)) & $6.5$~cm$^{-3}$ & $6.4^{+0.2}_{-0.2}$~cm$^{-3}$\\
      & & & $\log$($N$(CO)) & $19.0$~cm$^{-2}$ & $19.1^{+0.6}_{-0.2}$~cm$^{-2}$ & $\log$($N$(X)) & $16.5$~cm$^{-2}$ & $14.3^{+0.7}_{-0.9}$~cm$^{-2}$\\
      \\
      & $1$ & $+6\farcs5$ & $T_{\rm K}$ & $230$~K & $57^{+169}_{-45}$~K & $T_{\rm K}$ & $175$~K & $134^{+75}_{-70}$~K\\
      & &  & $\log$($n$(H$_2$)) & $2.7$~cm$^{-3}$ & $3.3^{+2.1}_{-0.6}$~cm$^{-3}$ & $\log$($n$(H$_2$)) & $6.3$~cm$^{-3}$ & $6.4^{+0.3}_{-0.2}$~cm$^{-3}$\\
      & & & $\log$($N$(CO)) & $20.5$~cm$^{-2}$ & $19.7^{+0.7}_{-0.7}$~cm$^{-2}$ & $\log$($N$(X)) & $15.0$~cm$^{-2}$ & $14.4^{+1.1}_{-0.9}$~cm$^{-2}$\\
      \\
      & $2$ & $+13\farcs0$ & $T_{\rm K}$ & $65$~K & $\le250$~K & & &\\
      & &  & $\log$($n$(H$_2$)) & $2.0^{\ast}$~cm$^{-3}$ & $\le7$~cm$^{-3}$ & & &\\
      & & & $\log$($N$(CO)) & $15.3$~cm$^{-2}$ & $\le21$~cm$^{-2}$ & & &\\
      \hline
      inner ring &$-4$ & $-26\farcs0$ & $T_{\rm K}$ & $10^{\ast}$~K & $\le250$~K\\
      & & & $\log$($n$(H$_2$)) & $2.0^{\ast}$~cm$^{-3}$ & $\le7$~cm$^{-3}$\\
      & & & $\log$($N$(CO)) & $19.3$~cm$^{-2}$ & $\le21$~cm$^{-2}$\\
      \\
      & $-3$ & $-19\farcs5$ & $T_{\rm K}$ & $175$~K & $\le250$~K\\
      & & & $\log$($n$(H$_2$)) & $2.0^{\ast}$~cm$^{-3}$ & $\le7$~cm$^{-3}$\\
      & & & $\log$($N$(CO)) & $17.8$~cm$^{-2}$ & $17.8$~cm$^{-2}$\\
      \\
      & $0$ & $0^{\prime\prime}$ & $T_{\rm K}$ & $65$~K & $\le250$~K\\
      & & & $\log$($n$(H$_2$)) & $2.0^{\ast}$~cm$^{-3}$ & $\le7$~cm$^{-3}$\\
      & & & $\log$($N$(CO)) & $17.5$~cm$^{-2}$ & $16.5$~cm$^{-2}$\\
      \\
      & $1$ & $+6\farcs5$ & $T_{\rm K}$ & $230$~K & $\le250$~K\\
      & & & $\log$($n$(H$_2$)) & $2.8$~cm$^{-3}$ & $\le7$~cm$^{-3}$\\
      & & & $\log$($N$(CO)) & $20.0$~cm$^{-2}$ & $\le21$~cm$^{-2}$\\
      \\
      & $2$ & $+13\farcs0$ & $T_{\rm K}$ & $10^{\ast}$~K & $\le250$~K\\
      & & & $\log$($n$(H$_2$)) & $2.0^{\ast}$~cm$^{-3}$ & $\le7$~cm$^{-3}$\\
      & & & $\log$($N$(CO)) & $19.3$~cm$^{-2}$ & $\le21$~cm$^{-2}$\\
      \\
      & $4$ & $+26\farcs0$ & $T_{\rm K}$ & $50$~K & $\le250$~K\\
      & & & $\log$($n$(H$_2$)) & $2.8$~cm$^{-3}$ & $\le7$~cm$^{-3}$\\
      & & & $\log$($N$(CO)) & $15.5$~cm$^{-2}$ & $17.5$~cm$^{-2}$\\
      \hline\\
    \end{tabular}
  \end{center}
  Notes: Likelihood results list the median values and $68\%$
  ($1\sigma$) confidence levels. A star ($^{\ast}$) indicates a
  physical parameter lying at the edge of the model grid.
\end{table*}
%

%
%
\begin{table}
  \begin{center}
    \caption{Model results for the dense gas in the nuclear disc of
      NGC~5866.}
    \label{tab:result2}
    \begin{tabular}{ccrll}
      \hline
      Position & Offset ($^{\prime\prime}$) & Parameter & $\chi^{2}$ & Likelihood\\
      (1) & (2) & (3) & (4) & (5)\\
      \hline
      $-1$ & $-6\farcs5$ & $T_{\rm K}$ & $125$~K & $\le250$~K\\
      & & $\log$($n$(H$_2$)) & $6.0$~cm$^{-3}$ & $\le7$~cm$^{-3}$\\
      & & $\log$($N$(X)) & $13.8$~cm$^{-2}$ & $\le21$~cm$^{-2}$\\
      \\
      $0$ & $0^{\prime\prime}$ & $T_{\rm K}$ & $225$~K & $102^{+103}_{-75}$~K\\
      & & $\log$($n$(H$_2$)) & $5.0$~cm$^{-3}$ & $6.6^{+0.3}_{-1.7}$~cm$^{-3}$\\
      & & $\log$($N$(X)) & $16.3$~cm$^{-2}$ & $16.3^{+1.7}_{-1.9}$~cm$^{-2}$\\
      \\
      $1$ & $+6\farcs5$ & $T_{\rm K}$ & $125$~K & $\le250$~K\\
      & & $\log$($n$(H$_2$)) & $6.0$~cm$^{-3}$ & $\le7$~cm$^{-3}$\\
      & & $\log$($N$(X)) & $13.8$~cm$^{-2}$ & $\le21$~cm$^{-2}$\\
      \hline
    \end{tabular}
  \end{center}
  Notes: Likelihood results list the median values and $68\%$
  ($1\sigma$) confidence levels.
\end{table}

\begin{figure*}
  \includegraphics[width=7.3cm,clip=]{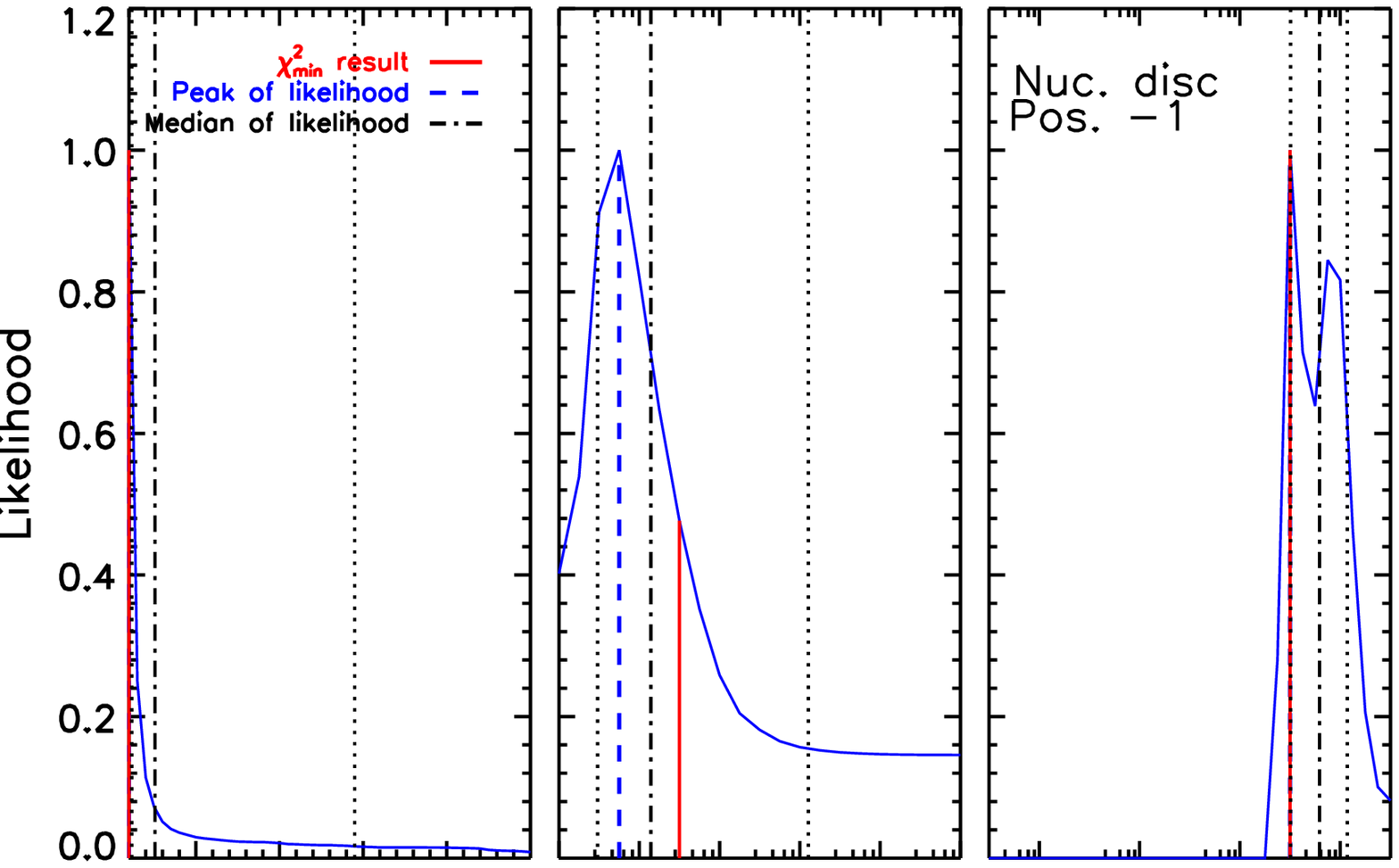}
  \hspace{-10pt}
  \includegraphics[width=7.3cm,clip=]{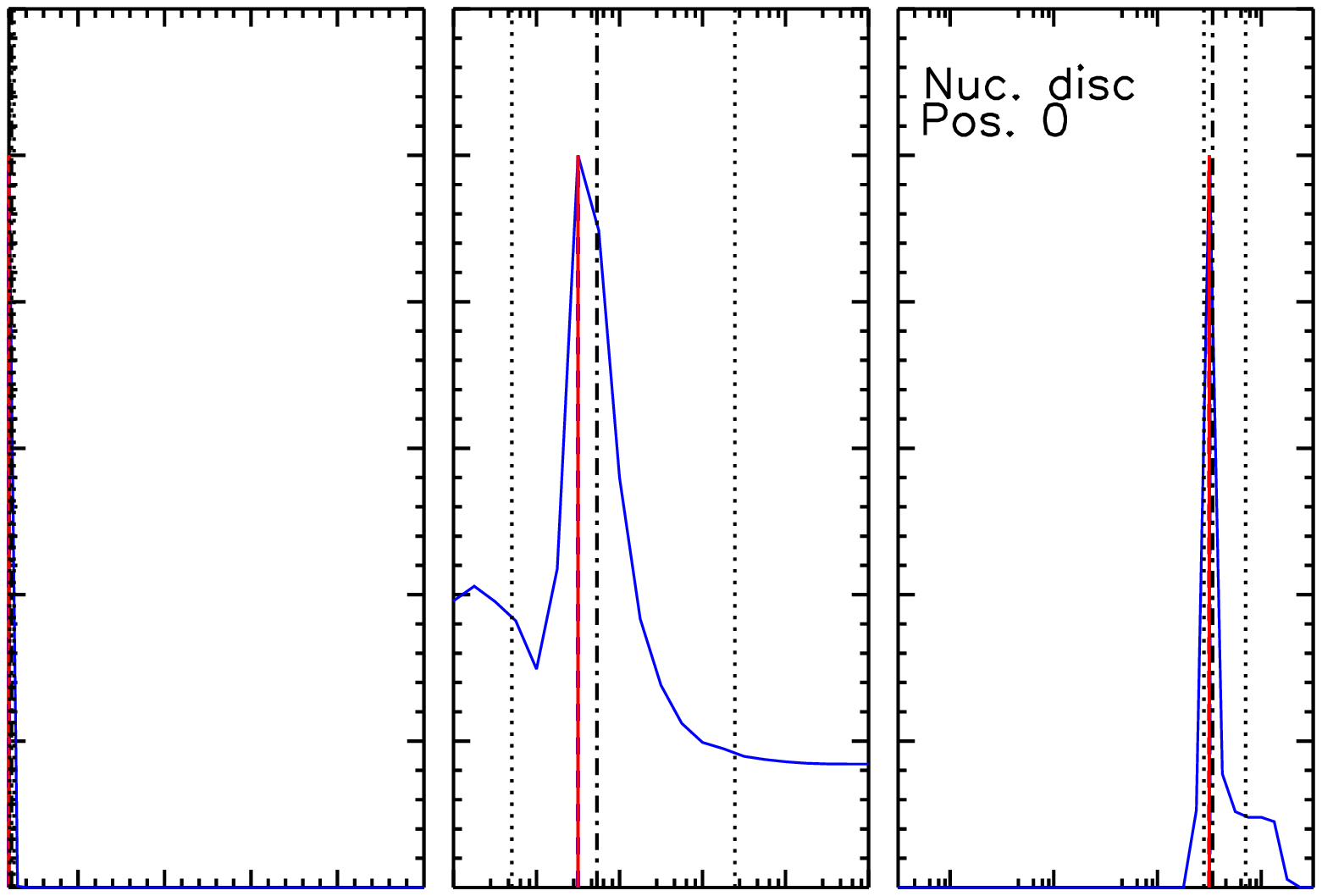}\\
  \vspace{-8pt}
  \includegraphics[width=7.3cm,clip=]{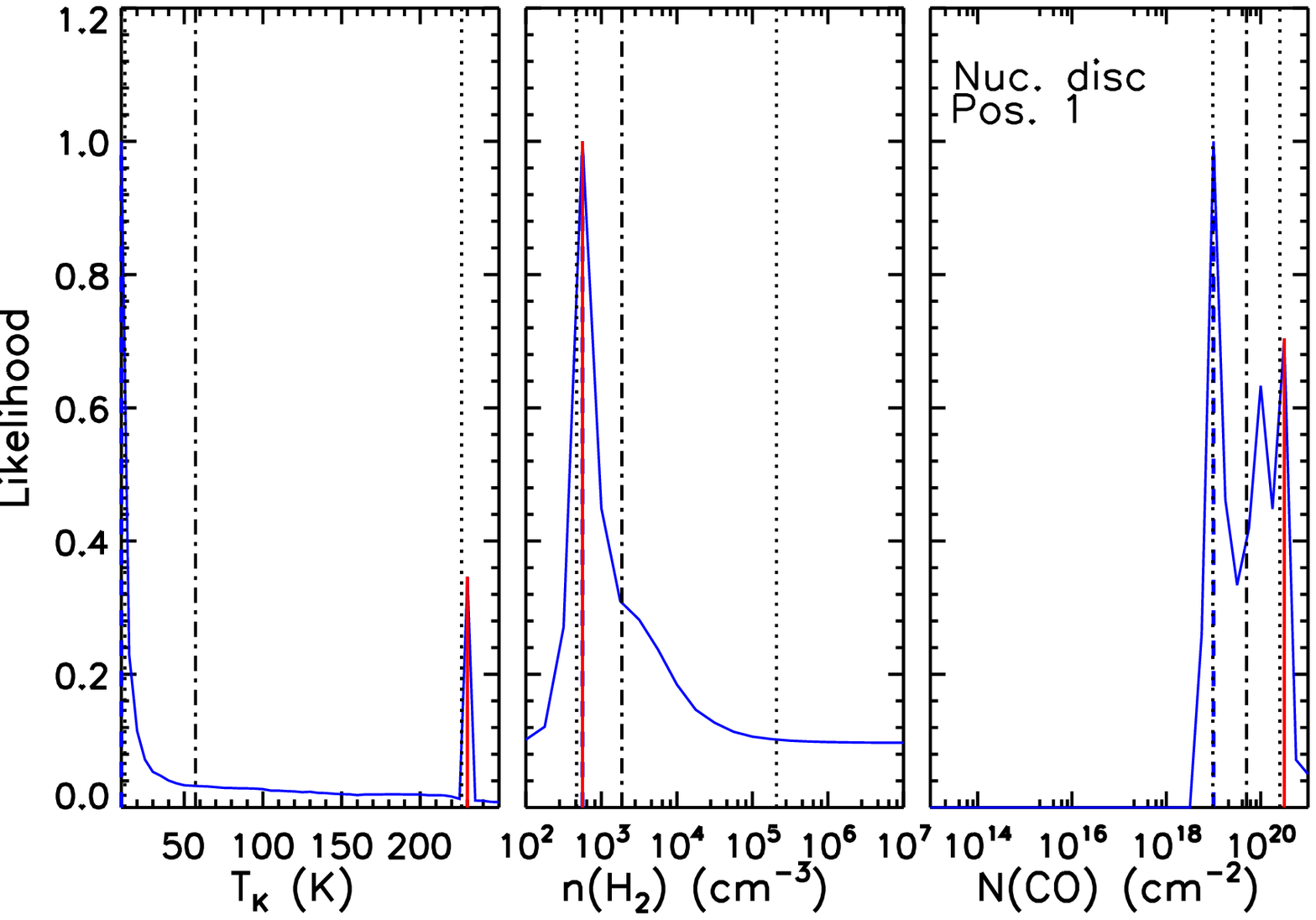}
  \hspace{-10pt}
  \includegraphics[width=7.3cm,clip=]{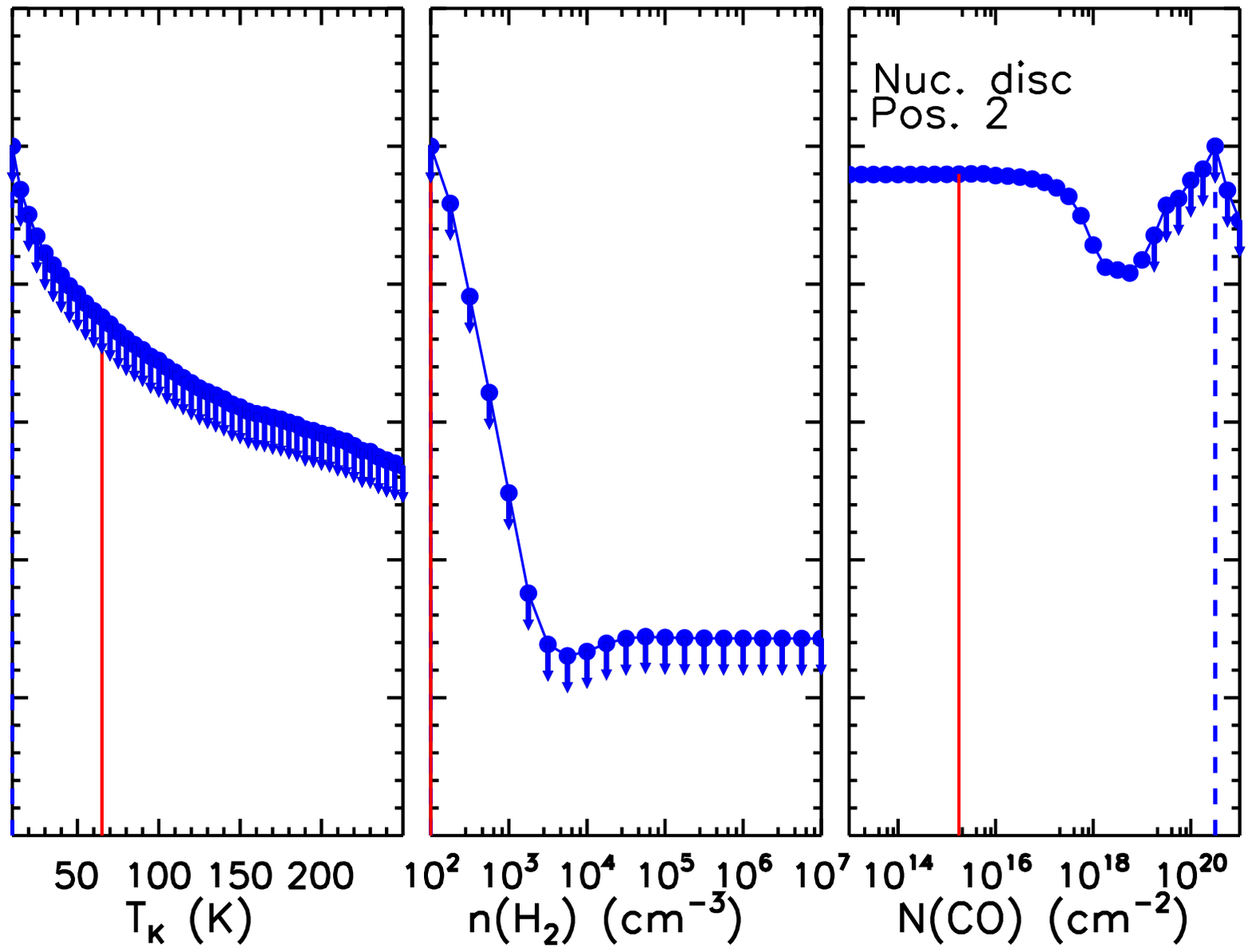}\\
  \caption{PDF of each model parameter marginalised over the other
    two, for the tenuous molecular gas in the nuclear disc of NGC~4710
    (positions $-1$, $0$, $1$ and $2$). In each PDF, the peak (most
    likely) and median value within the model grid are identified with
    a dashed blue and dashed-dotted black line, respectively. The
    $68\%$ ($1\sigma$) confidence level around the median is indicated
    by dotted black lines. The best-fit model in a $\chi^{2}$ sense is
    indicated by a solid red line. The PDFs for position~$2$ do not
    include the median value nor the $1\sigma$ confidence level, as at
    least one observed line ratio is a lower limit, resulting in PDF
    upper limits (blue dots with arrows).}
  \label{fig:n4710likefc}
\end{figure*}

We also estimate the likelihood of our models, calculating at each
position the probability distribution function (PDF) of each model
parameter marginalised over the other two. That is, for each possible
value of a model parameter within our grid, we calculated the sum of
the ${\rm e}^{-\Delta\chi^2/2}$ terms for all possible values of the
other two parameters. These PDFs are shown in
Figures~\ref{fig:n4710likefc}\,--\,\ref{fig:n4710likesc}, along with
their peaks (most likely values), medians and $68\%$ ($1\sigma$)
confidence levels around the median (the latter also listed in
Tables~\ref{tab:result1} and \ref{tab:result2}). Again, for positions
with a line ratio lower limit where a model meets the criterion for
case (2) above, the likelihoods represent upper limits. If the peak of
the PDFs is an upper limit, the most likely model is then ill defined.

%
%
%

%
%
\begin{figure*}
  \centering
  \includegraphics[width=7.3cm,clip=]{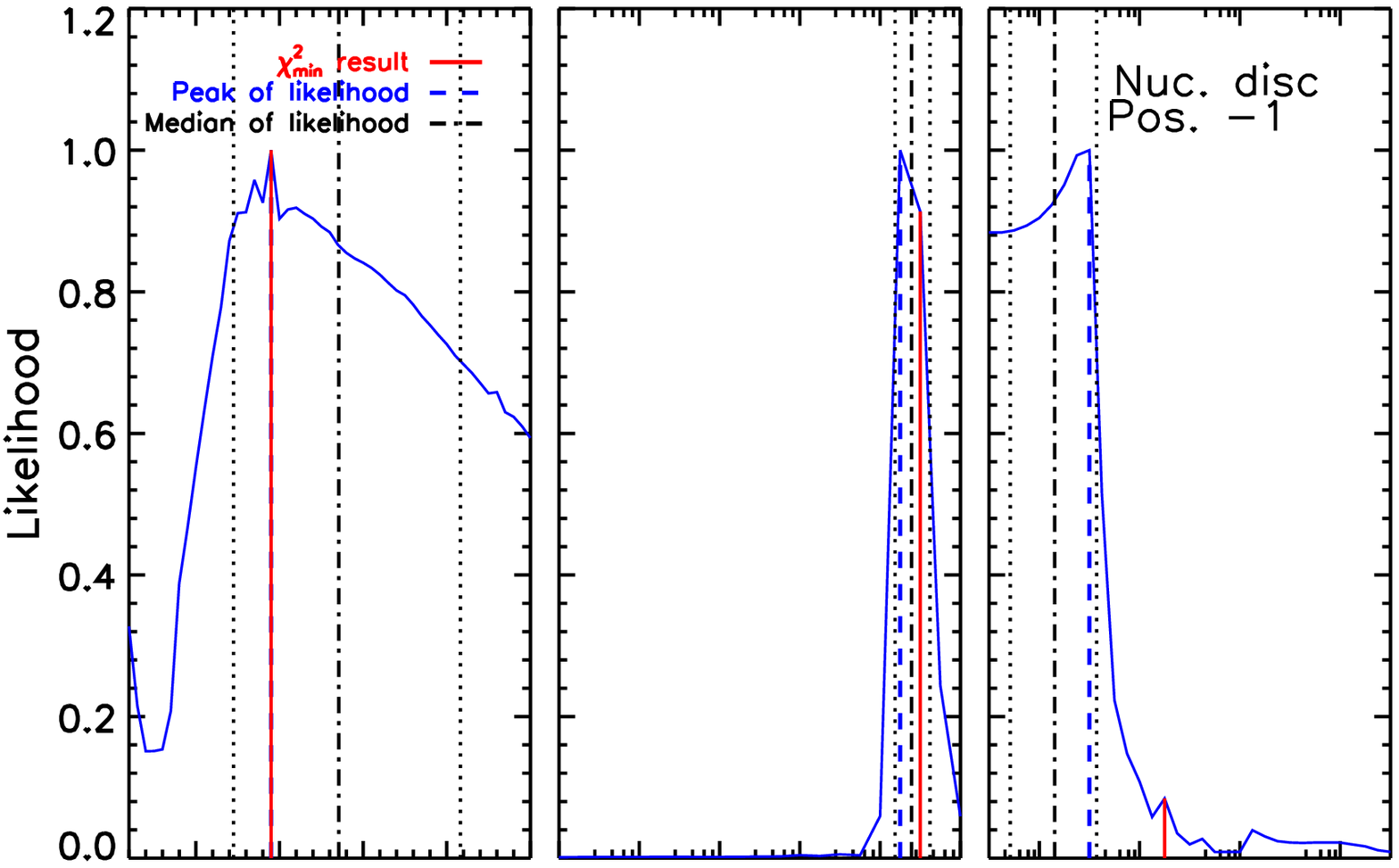}
  \hspace{-10pt}
  \includegraphics[width=7.3cm,clip=]{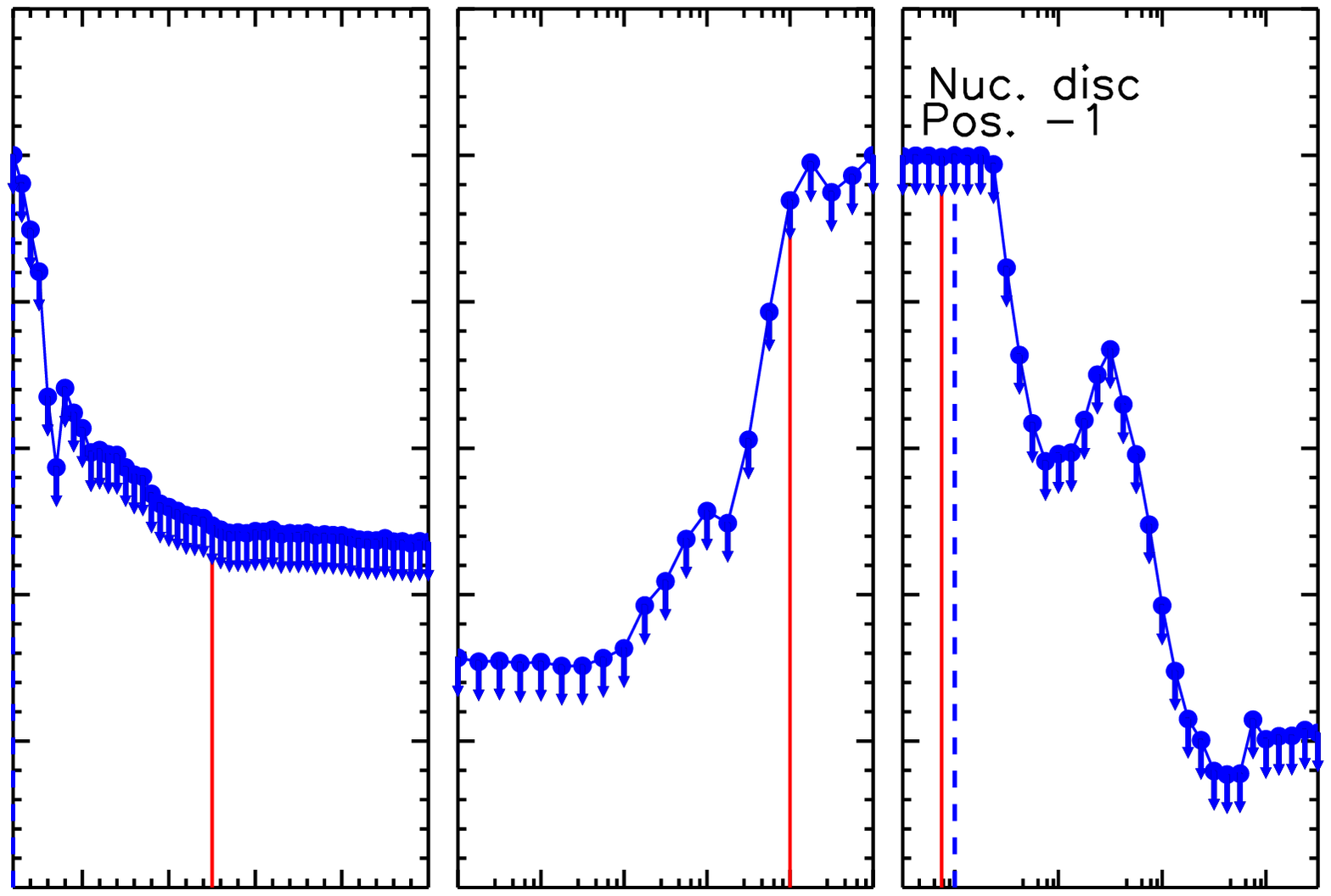}\\
  \vspace{-8pt}
  \includegraphics[width=7.3cm,clip=]{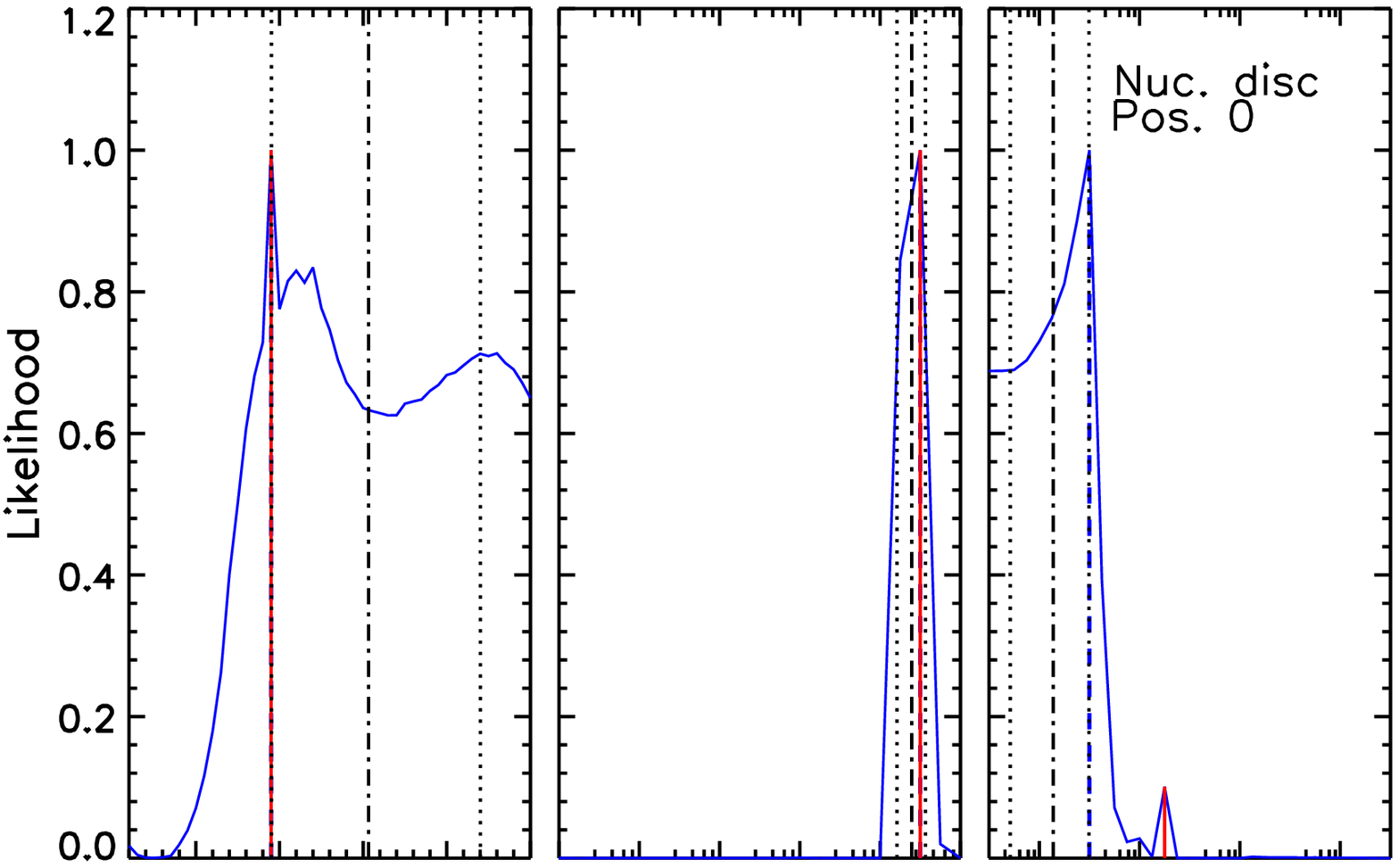}
  \hspace{-10pt}
  \includegraphics[width=7.3cm,clip=]{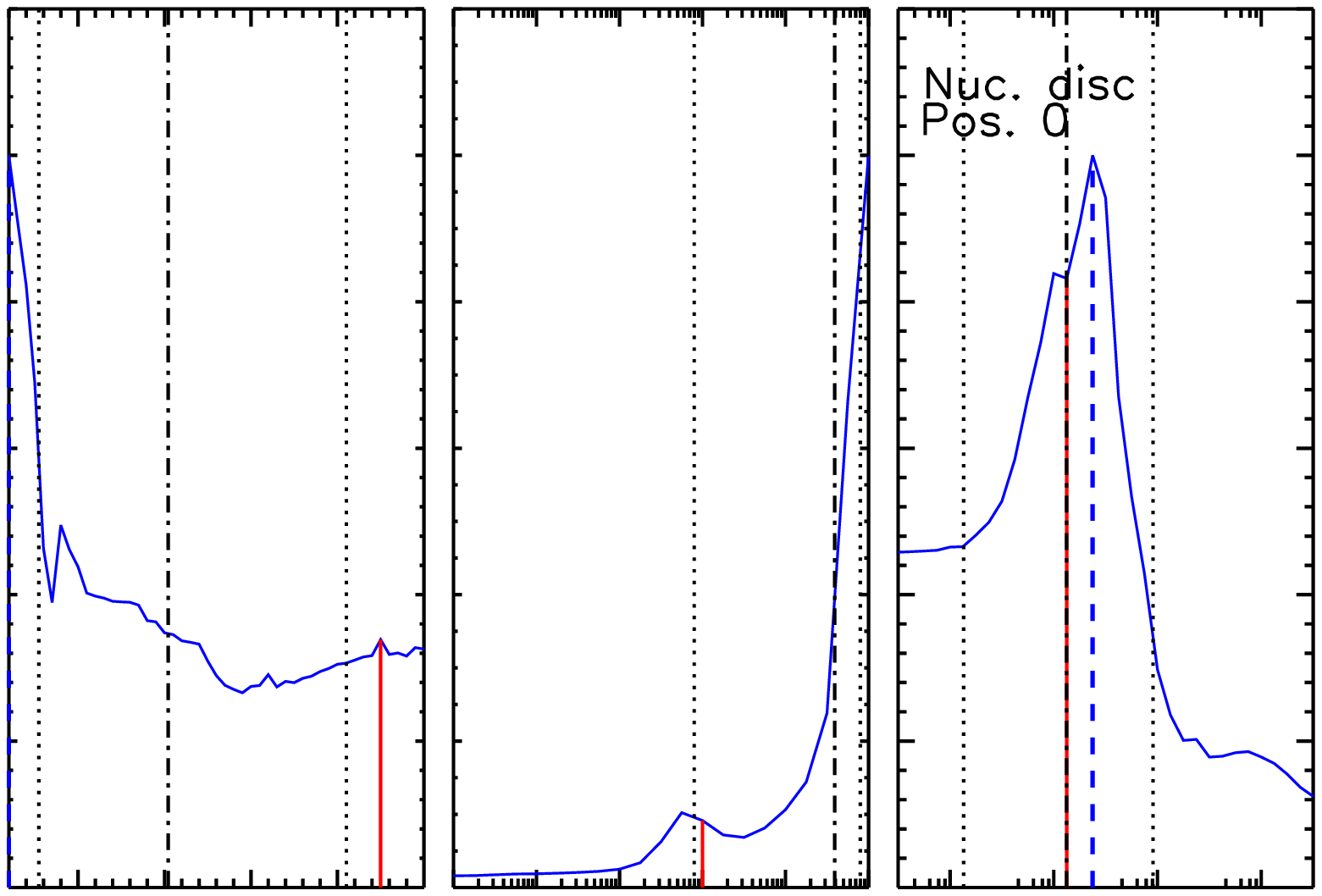}\\
  \vspace{-8pt}
  \includegraphics[width=7.3cm,clip=]{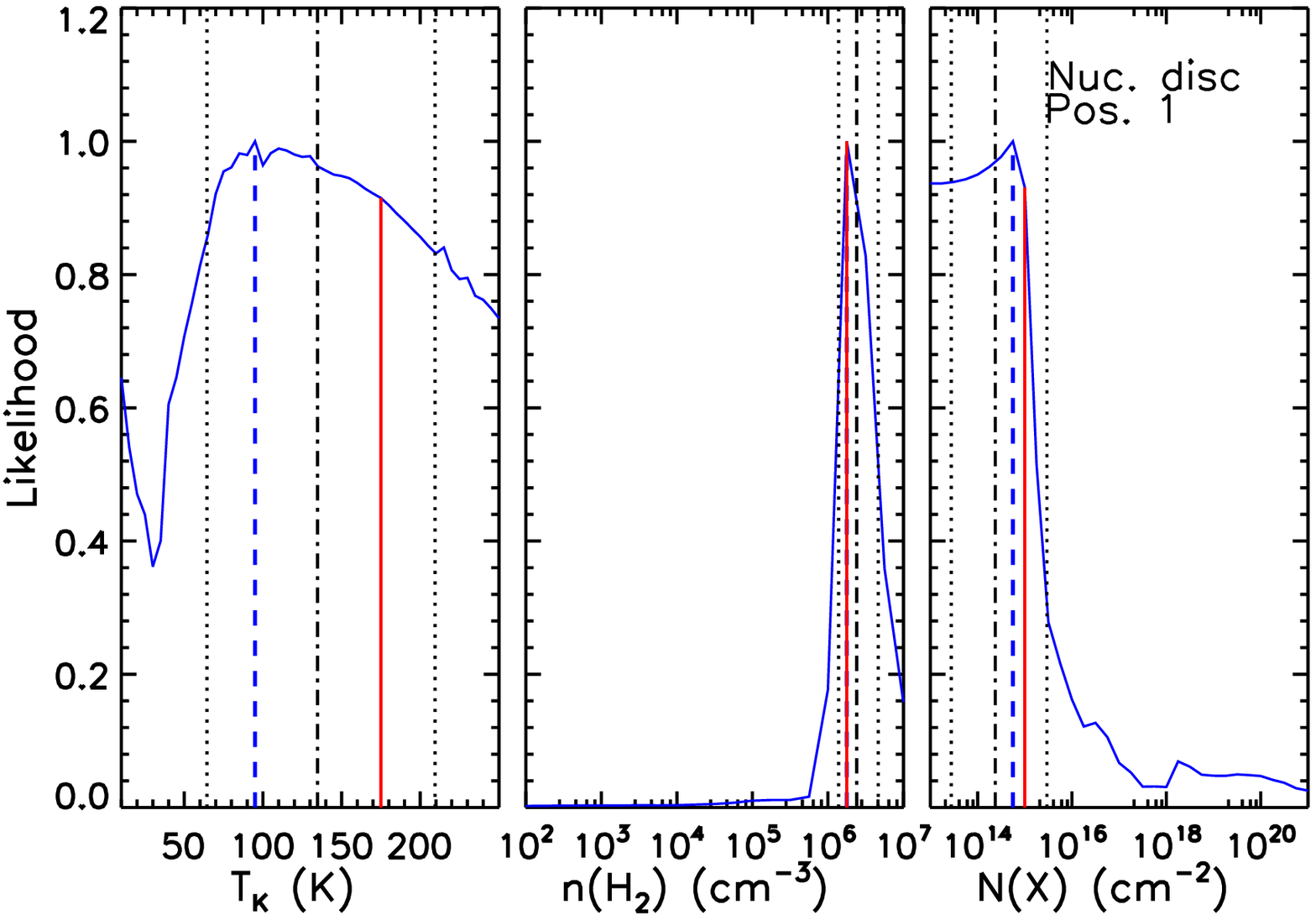}
  \hspace{-10pt}
  \includegraphics[width=7.3cm,clip=]{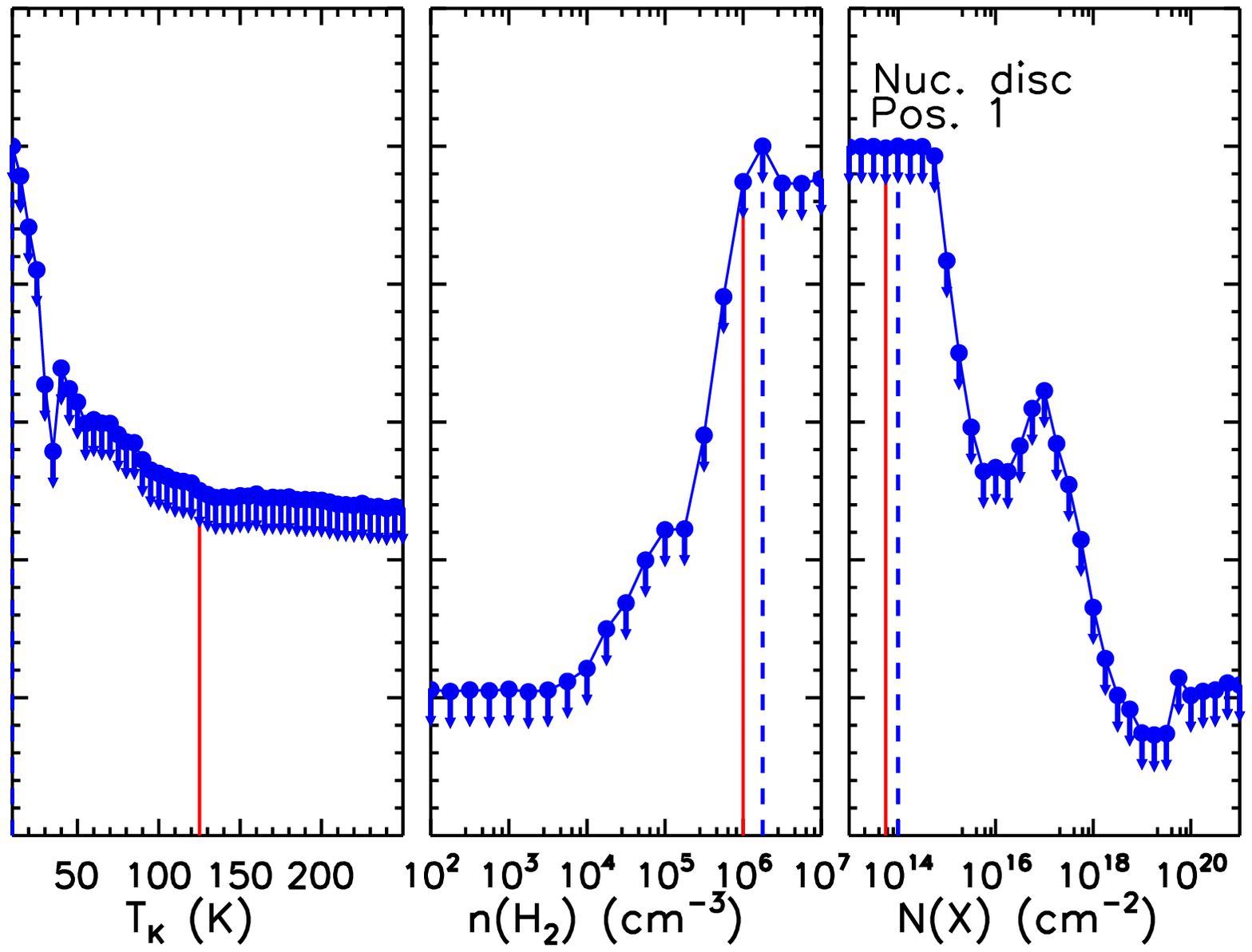}     
  \caption{Same as Figure~\ref{fig:n4710likefc} but for the dense
    molecular gas in the nuclear disc of NGC~4710 (left) and NGC~5866
    (right) (positions $-1$, $0$ and $1$). $N$(X) stands for the column
    number density of all four high density tracers. The PDFs for
    positions~$-1$ and $1$ in NGC~5866 do not include the median value
    nor the $1\sigma$ confidence level, as at least one observed line
    ratio is a lower limit, resulting in PDF upper limits (blue dots
    with arrows).}
  \label{fig:n4758likefd}
\end{figure*}
%

%
%
\begin{figure*}
  \includegraphics[width=7.3cm,clip=]{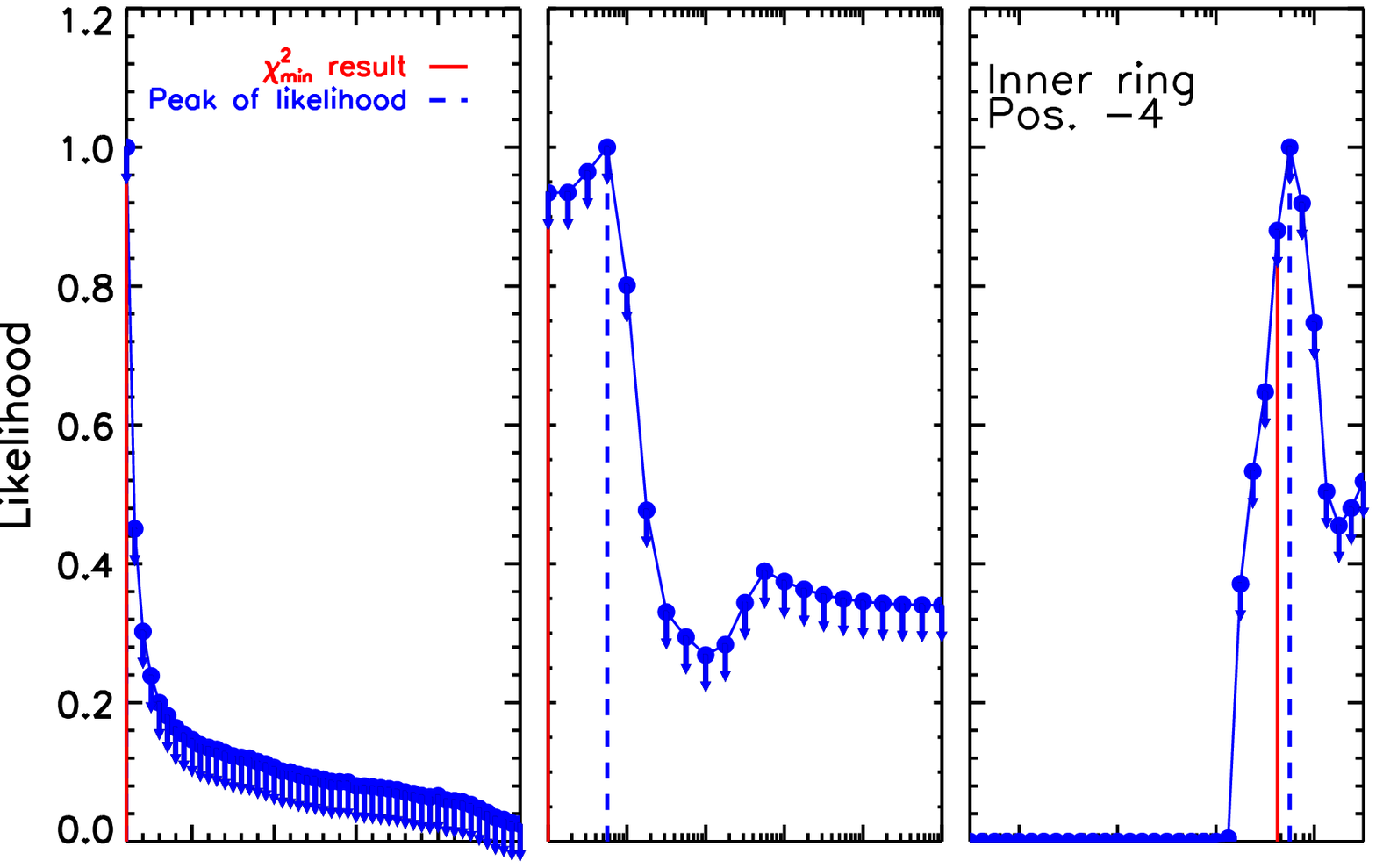}
  \hspace{-10pt}
  \includegraphics[width=7.3cm,clip=]{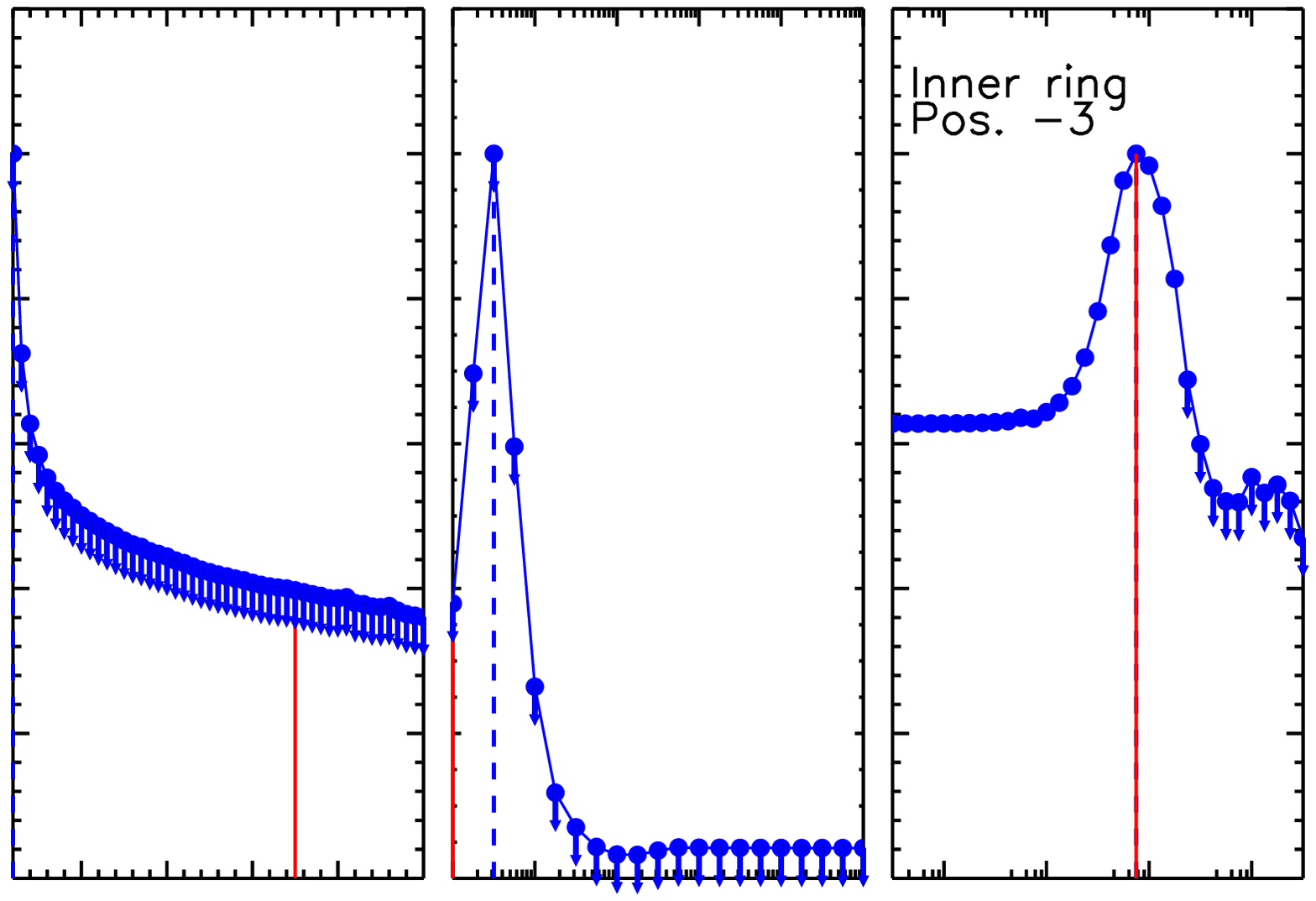}\\
  \vspace{-8pt}
  \includegraphics[width=7.3cm,clip=]{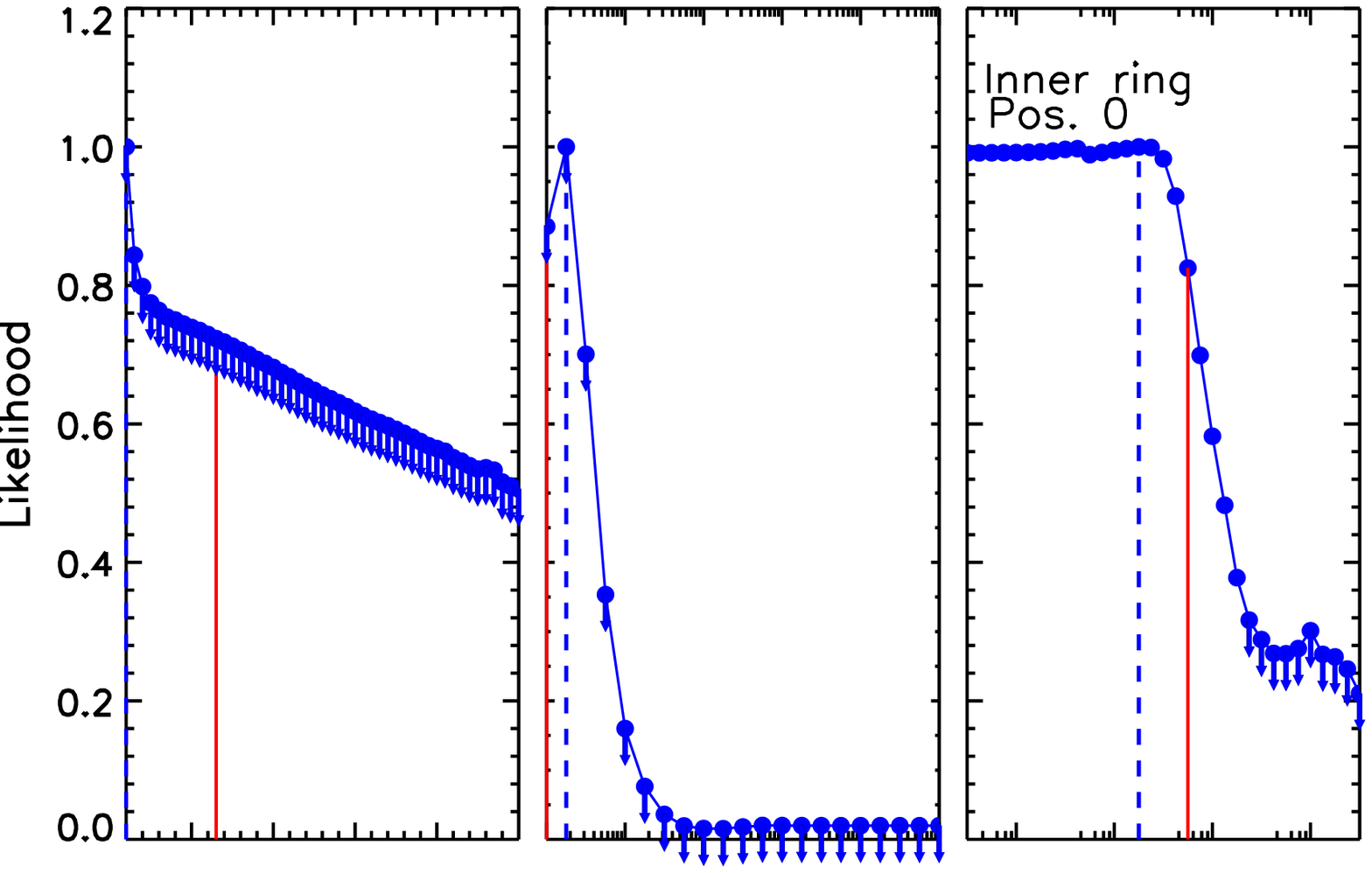}
  \hspace{-10pt}
  \includegraphics[width=7.3cm,clip=]{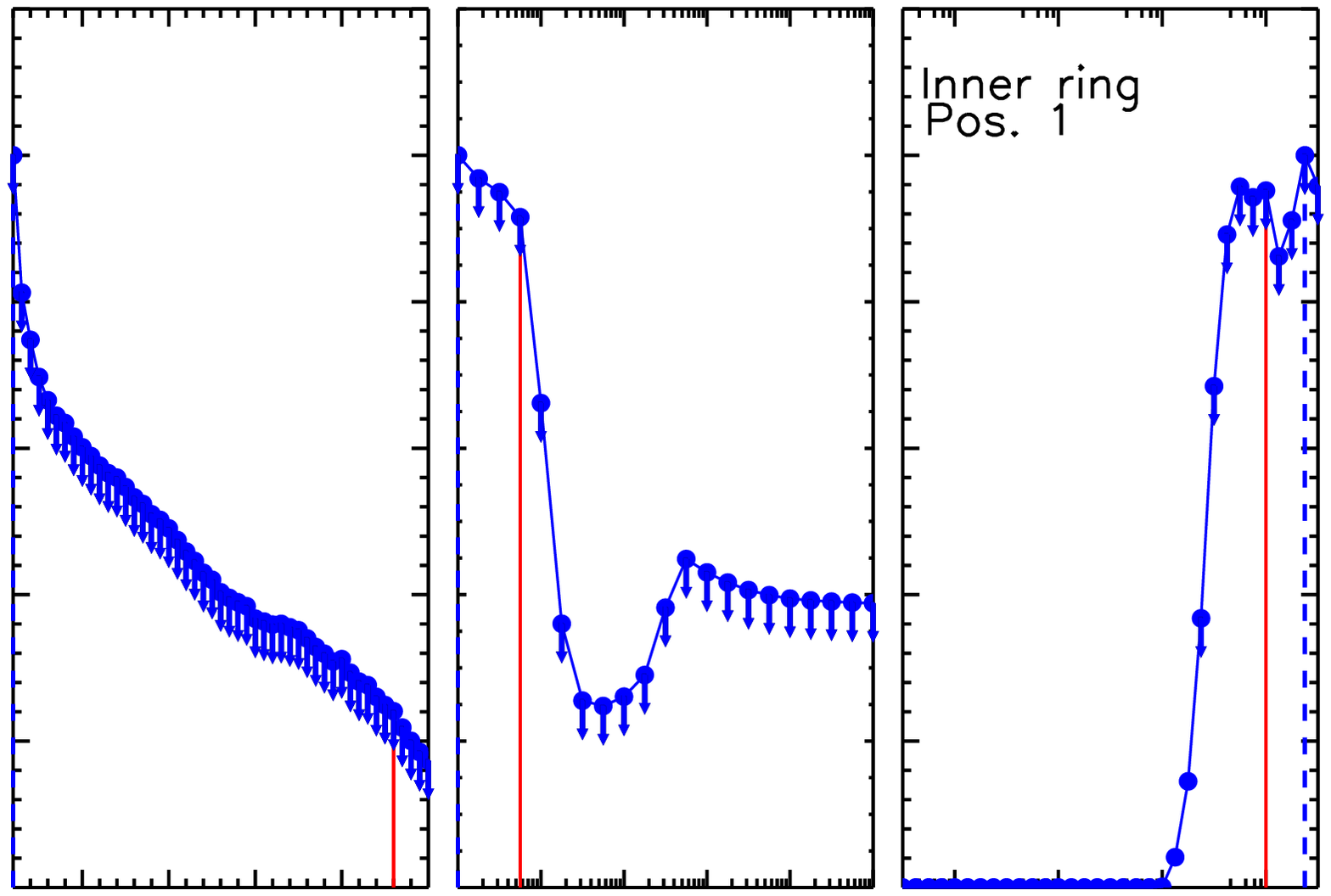}\\
  \vspace{-8pt}
  \includegraphics[width=7.3cm,clip=]{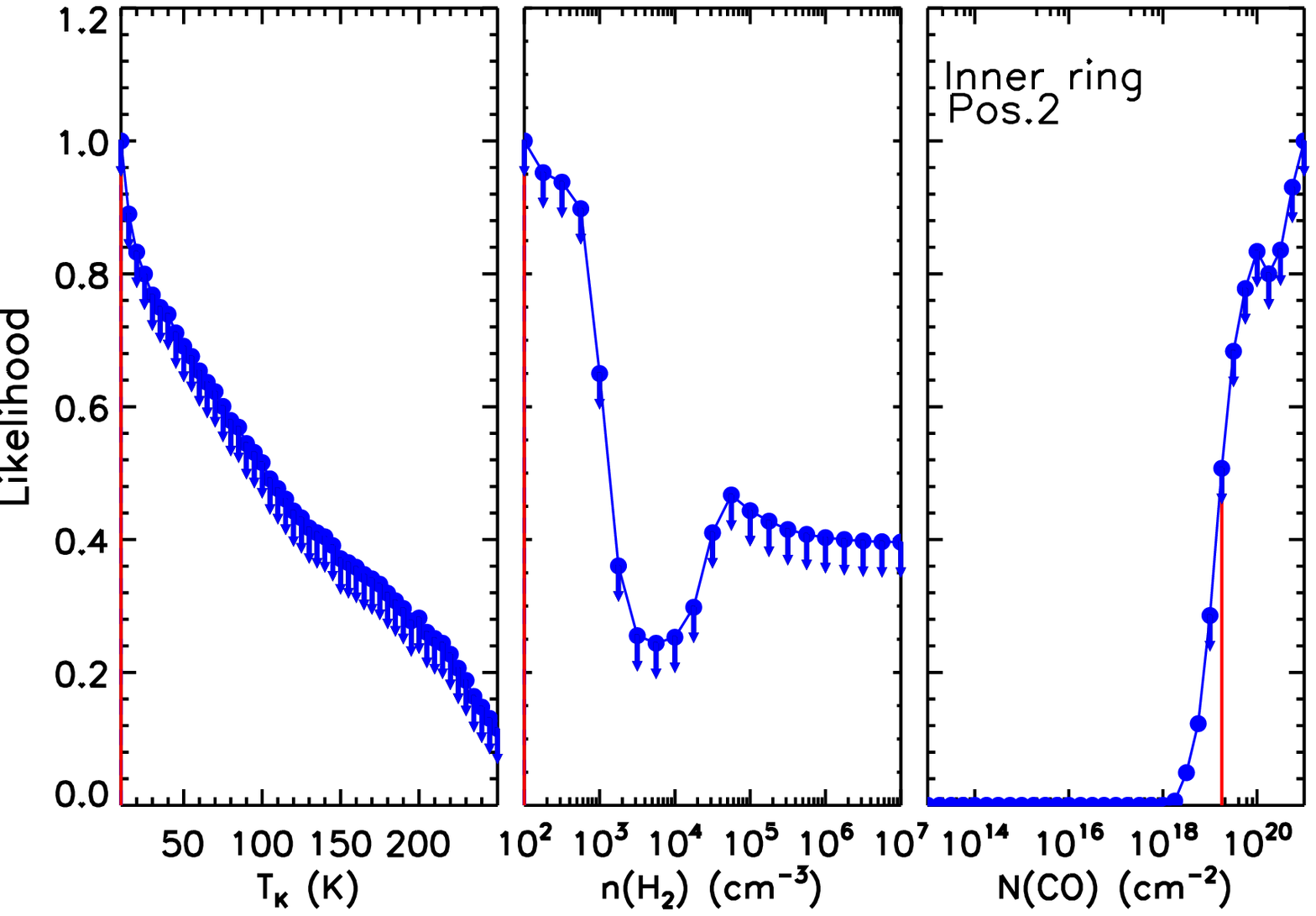}
  \hspace{-10pt}
  \includegraphics[width=7.3cm,clip=]{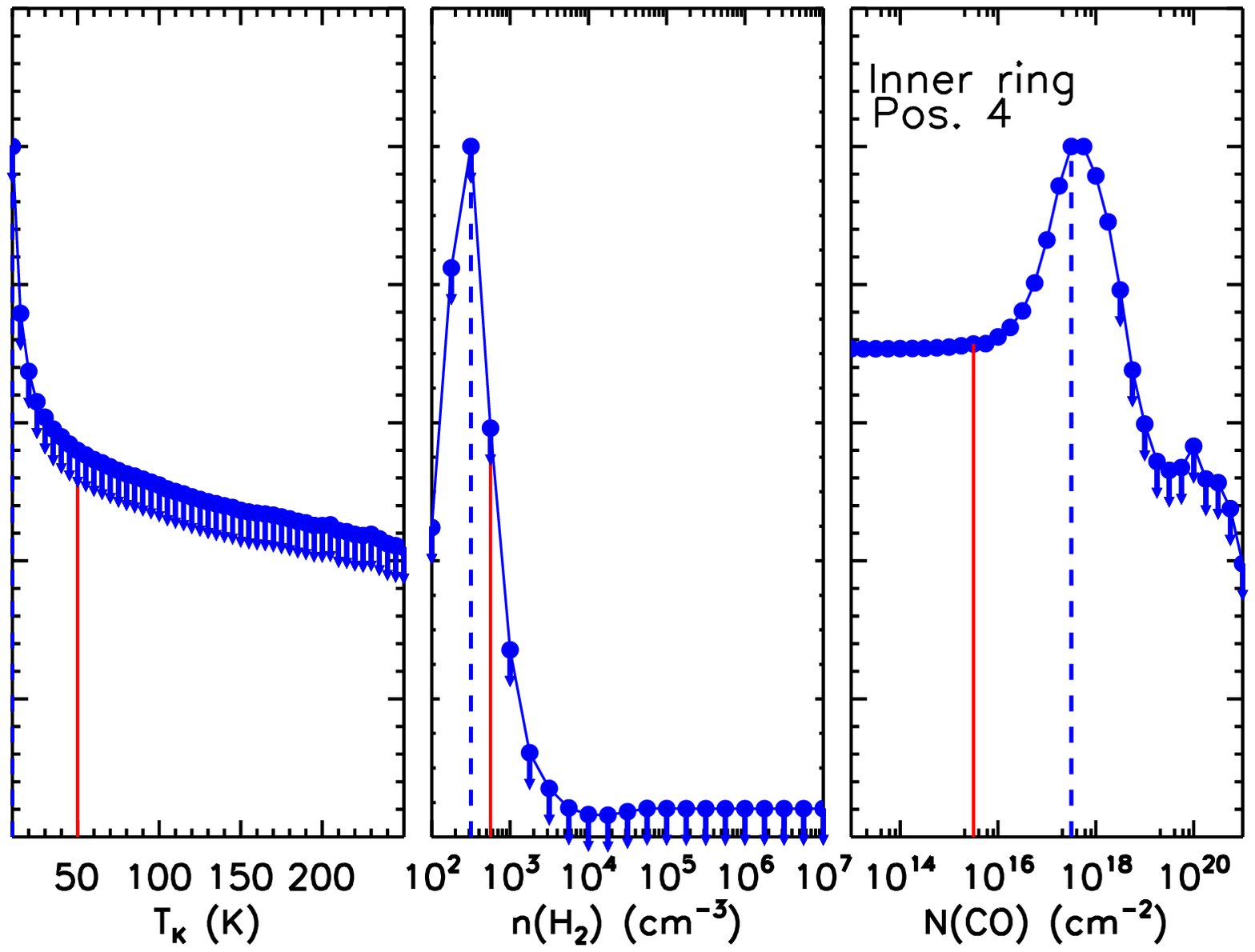}\\
  \caption{Same as Figures~\ref{fig:n4710likefc} and
    \ref{fig:n4758likefd} but for the tenuous molecular gas in the
    inner ring of NGC4710 (positions $-4$, $-3$, $0$, $1$, $2$ and
    $4$). No PDF includes the median value or the $1\sigma$
    confidence level, as at least one observed line ratio is a lower
    limit at each position, resulting in PDF upper limits (blue dots
    with arrows).}
  \label{fig:n4710likesc}
\end{figure*}
The best-fit model results generally agree well with those obtained
from the likelihood analysis, in the sense that the three best-fit
model parameters ($T_{\rm K}$, $n$(H$_2$) and $N$(CO)) are generally
contained within the $68\%$ ($1\sigma$) confidence level around the
median of the PDF (or are just outside of it; see
Figs.~\ref{fig:n4710likefc}\,--\,\ref{fig:n4710likesc} and
Tables~\ref{tab:result1}\,--\,\ref{tab:result2}).  However, because of
the flatness of the PDFs, the uncertainties in the most likely model
results are rather large (particularly for $T_{\rm K}$). Significant
differences are found between the tenuous and dense gas components for
certain parameters ($n$(H$_2$) and $N$(CO)), but unsurprisingly the
different positions within a single kinematic component are generally
statistically indistinguishable.

The line widths used for the modeling are similar to the values used
for RADEX modeling of other external galaxies in the literature
\citep[e.g.][]{rang11,ka12}. However, the observed line widths in the
nuclear discs are combinations of the intrinsic widths of the lines
and the range of rotational velocities contained within the
synthesized beams. The latter effect could be dominant in the nuclear
discs (while it is negligible in the inner rings), as the molecular
gas rotation is increasing rapidly in these components (unrelated to
the physical conditions in the molecular clouds). The true intrinsic
widths of the lines in the nuclear discs should thus lie between the
observed line widths in the inner rings (i.e.\
$\approx50$~km~s$^{-1}$) and those observed in the nuclear discs. To
test whether our modeling results are sensitive to the line widths
adopted, we ran the nuclear disc models again using both the line
widths measured in the inner rings and the average line widths of the
nuclear discs and inner rings. The best-fit model results are formally
different in a few positions only (and generally only the CO column
densities), and the results from the three different line widths
tested always agree within the uncertainties. As expected
\citep{van07}, the RADEX results are therefore only minimally affected
by the line widths assumed, and possible beam smearing effects on the
line widths adopted in the nuclear discs does not have a significant
impact on our modeling results.
\label{lastpage}
\end{document}